\documentclass[10pt]{article}

\usepackage{grffile}
\usepackage{ctable}

\usepackage{graphicx}

\usepackage{amsmath,amsfonts,amssymb}
\usepackage{amsthm}
\usepackage{color}

\usepackage[letterpaper,textheight=9in,textwidth=7in]{geometry}

\usepackage[skip=10pt,font=footnotesize]{caption}
\graphicspath{{Figures/}}

\newtheorem{Theorem}{Theorem}

{\begin{trivlist}\item[]\textbf{Proof#1 }}%
{\qed\end{trivlist}}

\newenvironment{Acknowledgment}%
 {\begin{trivlist}\item[]\textbf{Acknowledgments.}}{\end{trivlist}}

 {\begin{center}\textbf{Abstract}}{\end{center}}

\makeatletter\@addtoreset{figure}{section}\makeatother

\makeatletter \@addtoreset{equation}{section} \makeatother

\makeatletter\@addtoreset{figure}{section}\makeatother

\makeatletter \@addtoreset{equation}{section} \makeatother

\newcommand{\R}{\mathbb{R}}
\newcommand{\C}{\mathbb{C}}

\newcommand{\Z}{\mathbb{Z}}

\newcommand{\mc}[1]{\mathcal{#1}}
\newcommand{\mb}[1]{\mathbb{#1}}

\newcommand{\tl}[1]{\tilde{#1}}
\newcommand{\lp}{\left}
\newcommand{\rp}{\right}
\newcommand{\la}{\lp\langle}
\newcommand{\ra}{\rp\rangle}
\newcommand{\beq}{\begin{equation}}
\newcommand{\eeq}{\end{equation}}
\newcommand{\ba}{\begin{align}}
\newcommand{\ea}{\end{align}}
\newcommand{\fr}[2]{\frac{#1}{#2}}
\newcommand{\p}{\partial}
\newcommand{\ri}{\mathrm{i}}

\newcommand{\rlin}{\mathrm{lin}}

\newcommand{\re}{\mathrm{e}}

\renewcommand{\Re}{\mathrm{Re}}
\renewcommand{\Im}{\mathrm{Im}}

\title{Growing patterns} 
\author{Ryan Goh\thanks{Department of Mathematics and Statistics, Boston University, 665 Commonwealth Ave., Boston,  MA 02215, USA; \texttt{rgoh@bu.edu}.}, 
  Arnd Scheel\thanks{School of Mathematics, University of Minnesota, 206 Church St. SE, Minneapolis,  MN 55455, USA.}}

\begin{document}

\maketitle
\begin{abstract}
    Pattern forming systems allow for a wealth of states, where wavelengths and orientation of patterns varies and defects disrupt patches of monocrystalline regions. Growth of patterns has long been recognized as a strong selection mechanism. We present here recent and new results on the selection of patterns in situations where the pattern-forming region expands in time. The wealth of phenomena is roughly organized in bifurcation diagrams that depict wavenumbers of selected crystalline states as functions of growth rates. We show how a broad set of mathematical and numerical tools can help shed light into the complexity of this selection process.


\end{abstract}

\section{Introduction}\label{s:intro}


The formation of regular spatial structure in nature has intrigued scientists for many centuries, across nearly every physical discipline.  Researchers have sought to determine both specific mechanisms which lead to ``patterns" in a given physical setting, and also universal phenomenological and mathematical mechanisms which help describe patterns across seemingly different physical domains. 

One mathematical mechanism that is often proposed is the growth of spatially periodic modes caused by small random fluctuations of an unstable, spatially uniform, equilibrium state. For example, Turing's seminal work \cite{turing1952chemical}, showed how a stable chemical reaction and spatial diffusion can combine to induce pattern-forming instability. Here, typically as some physical parameter is varied, a finite band of spatially periodic modes become linearly unstable.  Random fluctuations in the underlying medium (such as thermal fluctuations or small impurities in the homogeneous state) can then excite the unstable modes which grow until being saturated by nonlinearities inherent in the system. 

In isotropic spatial environments, perturbation by small uniform noise will excite spatial modes of any spatial orientation, leading to the formation of patches of regular structure which are oriented randomly to each other and have spatial wavenumber close to one of the unstable modes; see for example Figure \ref{f:sh-pat} below.  Zooming out from these local patches, one observes various types of imperfections, often referred to as defects, in the regular structure, such as wavenumber and phase mismatches, disclinations, dislocations, and grain boundaries.  

From this viewpoint, the defect free nature of patterns observed in various systems across different domains seems surprising. Spatial growth and heterogeneity have been recognized for their crucial role  in mediating and selecting patterns, leading in particular to the emergence of surprisingly regular, defect-free crystalline states. More precisely, through the temporal evolution of a system boundary, through the evolution or variation of the medium itself, or through a spatio-temporal  external forcing on the system,  orientation, wavenumber, and type of pattern can be \emph{selected} and defect formation suppressed. Such pattern selection mechanisms have been observed in the patterning of various biological, chemical, and physical systems, including the regulation of digit and skeletal patterning in growing organisms \cite{digit,HISCOCK2015408}, the formation of spiral primodium arrangements on apically growing plant meristems \cite{phyllo}, the formation of crystallographic lattices on fish retinae \cite{Nunley806679}, bacterial colony growth \cite{eshel,keller1971traveling}, the formation of periodic bands of precipitate in traveling chemical reactions \cite{Droz00,thomas,hantz00,Lagzi13,keller81}, 
animal coat patterning \cite{gierer1972theory}, and even the formation of von K\'arm\'an vorticies via the perturbation of a laminar flow by a moving object \cite{achenbach_1974,huerre90,chomaz05}.  
In man-made experiments, researchers hope to use such growth processes to exert precise control over the structure formed in a given material while suppressing imperfections and defects.  Examples include the formation of nanoscale patterns via high energy ion bombardment of metal alloys \cite{bradley}, ripple formation through progressive surface erosion \cite{friedrich}, deposition of patterns via dewetting or evaporation on a surface \cite{Thiele14}, eutectic lamellar crystal growth \cite{zigzageutectic,double}, elastic surface crystals \cite{stoop2018defect}, or the directional quenching of metallic alloy melts \cite{krekhov,foard}, and other general phase separation behaviors \cite{kurita17, tsukada2019topological}.  The last example provides in fact easy intuition for this broad area of study. One begins with a stable and homogeneous liquid alloy melt, which when rapidly cooled becomes unstable to a phase separative instability. This process, known as a quench, leads to the formation of randomly oriented lamellae and "cow patch" shapes. Alternatively, a directional quench induces the self-organized formation of regular patterns in its wake by moving across the domain in a spatially progressive manner, locally cooling the alloy, and leaving behind an unstable state from which patterns can form. 

An example that particularly motivates the present work, is the spatial patterning in a light-sensing CDIMA reaction-diffusion system \cite{lengyel1992chemical}. Patterning in this diffusion limited chemical reaction can be suppressed by illumination with high-intensity light. Suddenly turning off the light throughout the system excites patterning modes of all orientations and leads to patches of randomly oriented periodic stripes  with defects spread throughout the domain. If instead, a mask that progressively blocks the light is moved across the domain, the pattern-forming instability leads to regular patterns and controlling the mask shape and motion allows for control of  patterns formed in the wake \cite{ miguez2006effect, konow2019turing}.


We focus throughout on this type of controlled growth, although examples of different forms of heterogeneity or growth mechanism abound.  In plant and developmental biology, diffusion induced pattern-formation, or ``Turing patterns" in various growth scenarios were studied with  different types of domain growth and evolution, including in particular \emph{apical} growth where material is added progressively to the boundary of the domain, while the bulk is left untouched. Specifically, this situation arises in models for plant growth, where only a collection of cells  on the boundary of the plant, the apical meristem, are able to replicate \cite{phyllo}. 
Different examples include self-similar, uniform, or isotropic growth, where all parts of the material grow uniformly, and arise for instance when modeling pattern-formation in cell colonies that are constantly dividing and  causing domain growth. In both cases, an evolving growth rate can have novel and dramatic effects on patterns formed in the domain \cite{Maini487,crampin1999reaction,crampin2002pattern}.  Without being comprehensive, we also mention boundary curvature, manifold evolution, growth anisotropy, and piecewise-constant kinetics as related examples beyond the scope of this work \cite{sanchez2019turing,krause2019one,krause2019influence, plaza2004effect,kozak19}; see also \cite{krause_review,vanGorder21} for recent reviews.
%
Beyond externally controlled growth,  the pattern-forming process may impact or even drive the growth process, with examples ranging from cell biology over combustion fronts to the evolution of growing bacterial colonies with chemotactic movement  \cite{ruppert2020nonlinear,Texier08,painter02}. 

Stepping away from these more general scenarios, we now turn back to our basic mathematical setup.

\subsection{Prototype model: the Swift-Hohenberg equation}\label{s:sh}
We consider a prototypical model of pattern formation, the Swift-Hohenberg equation
\beq\label{e:sh0}
u_t = -(1+\Delta)^2 u + f(u;\mu), \qquad u\in \R,\  \Delta = \sum_{i = 1}^n \partial_{x_j}^2, \,\, \mathbf{x} = (x_1,...,x_n)\in \R^n,\  n\in \mathbb{N},
\eeq
designed as a phenomenological model for the formation of spatially periodic convection rolls in Rayleigh-B\'enard convection \cite{sh}, where a fluid is heated from below and cooled from above, driving a turning over of the fluid. Here $u$ represents thermal perturbations from a pure conductive state and $\mu$ is a parameter related to the temperature difference between  top and bottom boundaries of the fluid that controls the onset of instability. The equation, or variants of it, has also been studied in the context of localized patterns in various physical systems \cite{knobloch15}, of plant phyllotaxis \cite{phyllo}, and of patterning of elastic surface crystals \cite{stoop2018defect}. Interestingly,  a non-local variant was considered by Turing just before his passing \cite{dawes}. We start our investigation with this equation since it both exhibits universally observed pattern-forming behavior and is well studied. Relevant phenomena include the existence of stable "Turing patterns", invasion fronts, grain boundaries and defects, zigzag and wrinkling instabilities, and localized patterns. We mostly work with a simple cubic, supercritical nonlinearity 
$
f(u,\mu) = \mu u - u^3
;$ 
see Section \ref{ss:cqsh} for some results with weakly sub-critical, cubic-quintic nonlinearity $f(u,\mu) = \mu u + \gamma u^3 - u^5$.

\begin{figure}
\centering
\includegraphics[trim = 0.5cm 0.5cm 0.5cm 0.5cm,clip,width=0.3\textwidth]{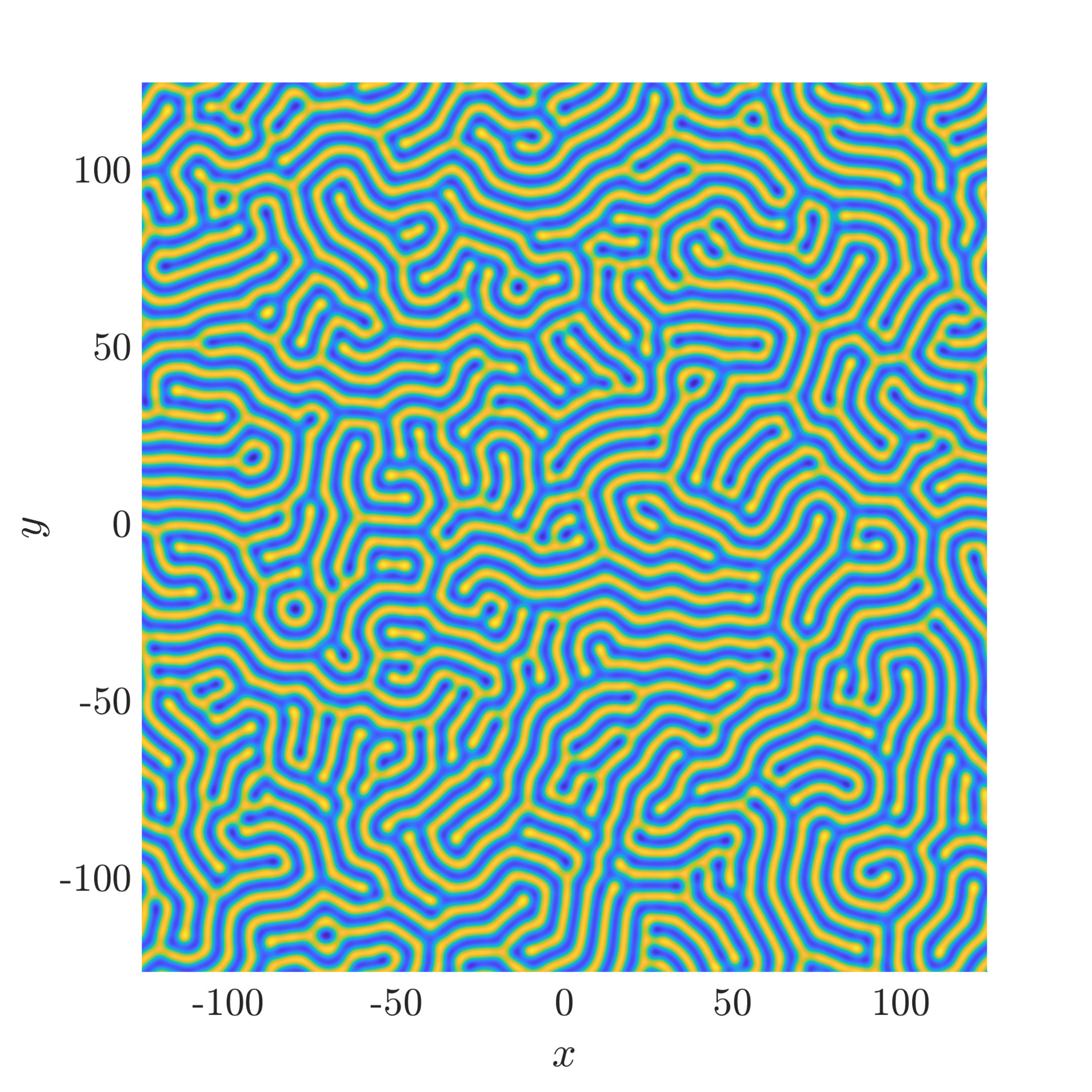}
\includegraphics[width=0.45\textwidth]{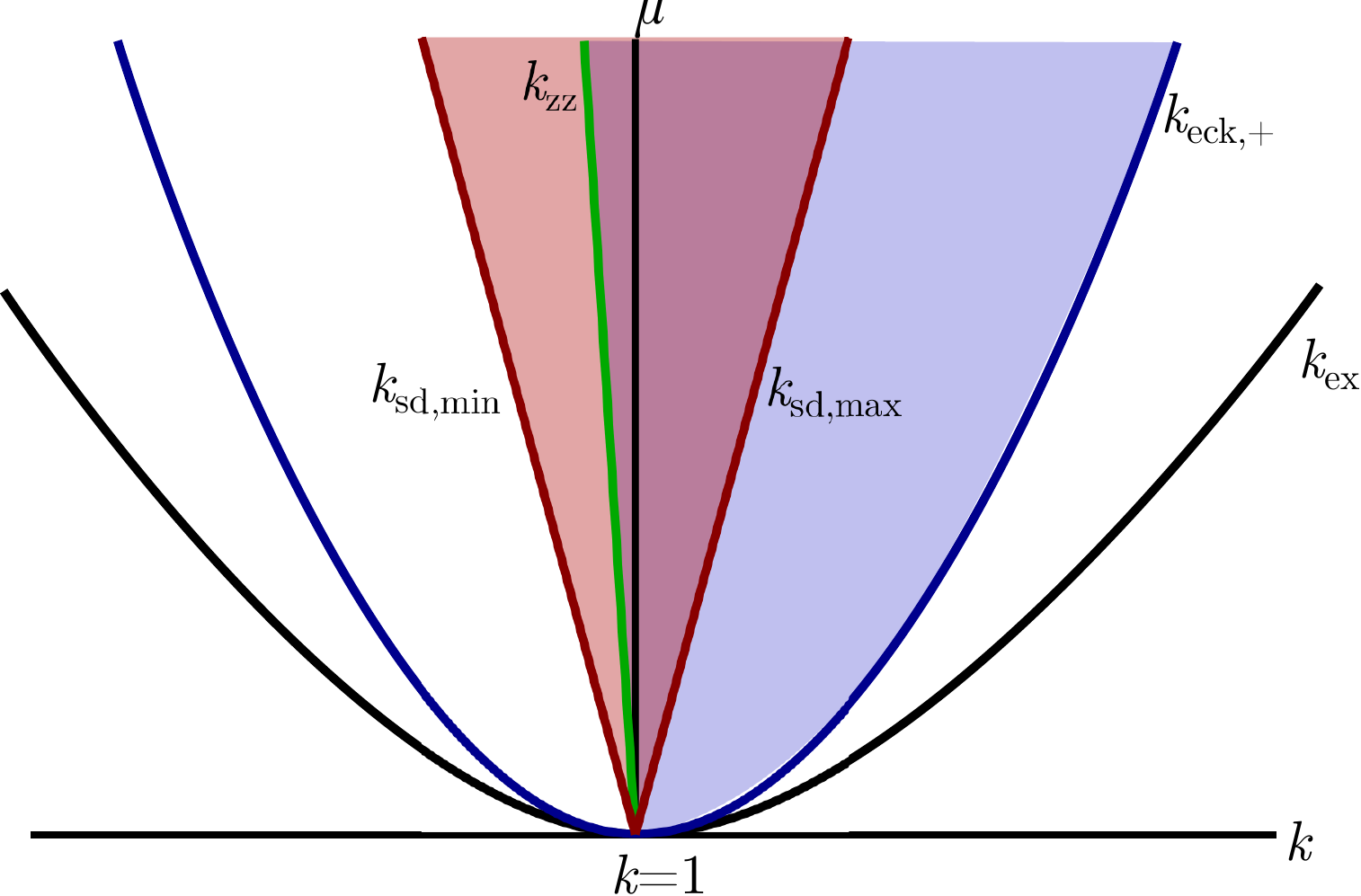}
\caption{Left: Solution of \eqref{e:sh0} with small random initial data with $\mu = 3/4$; Right: Plot of existence and stability domain in $(\mu,k)$-space near $(0,1)$. Wavenumbers exist above the black curve, are Eckhaus stable above the blue curve, zigzag stable to the right of the green curve. The red shaded region gives the wavenumbers which can be selected in 1-D by a stationary quench (see Sec. \ref{sss:st-para}) while the blue shaded region gives stable wavenumbers in two spatial dimensions. }
\label{f:sh-pat}
\end{figure}

In the linear equation $u_t = -(1+\Delta)^2 u + \mu u$, , the pattern forming instability is readily understood after Fourier-Laplace transform, setting $u = \re^{\ri \mathbf{k}\cdot \mathbf{x} + \lambda t},\quad \mathbf{k}\in \R^n$ to find the linear dispersion relation
$$
\lambda = -(1-k^2)^2 + \mu,\qquad k := |\mathbf{k}|.
$$
As $\mu$ increases through zero, a band of wavenumbers $k\sim 1$ becomes unstable, $\lambda(\mathbf{k})>0$. Note that the equation is isotropic, that is, invariant under rotations and hence exhibits no preference for any orientation of the wave vector. Indeed,
perturbations of the homogeneous equilibrium state $u\equiv 0$ for $\mu>0$ with small spatial white noise excites  various orientations of wavenumbers. Solutions grow in amplitude until   saturated by the cubic nonlinearity, leading to a labyrinth of patterns and defects; see Figure \ref{f:sh-pat}.

The simplest solutions created by linear instability and nonlinear saturation are bifurcating families of periodic equilibrium solutions $u_p(\mathbf{k}\cdot \mathbf{x};k),$ for \eqref{e:sh0}, often referred to as stripe or roll solutions. They satisfy
\beq\label{e:shroll}
0 = -(1+k^2 \p_\theta^2)^2 u_p + \mu u_p - u_p^3,\qquad u_p(\theta+2\pi;k) = u_p(\theta;k),\qquad u_p(\theta;k) =\sqrt{\frac{4}{3}(\mu - \kappa^2)} \cos(k \theta) + \mc{O}(|\mu|),
\eeq
for $0<\mu\ll1$, where $\kappa := 1-k^2$ is in the range $0\leq \kappa^2 < \mu$, thus $k\in(k_\mathrm{ex,min},k_\mathrm{ex,max})$, with $k_\mathrm{ex,max/min} = \sqrt{1\pm\sqrt{\mu}}$ at leading order in $\mu$.  Both linear and nonlinear stability, but also instability of such patterns in various regimes has been shown in one and two spatial dimensions \cite{mielke,schneider1996diffusive}. In one dimension, stable wavenumbers are determined by the Eckhaus condition 
\beq\label{e:eck}
|\kappa|< \kappa_\mathrm{eck}:= \sqrt{\mu/3} + \mathcal{O}(\mu).
\eeq 
 In higher spatial dimensions, an additional ``zigzag" condition is required for stability,
\beq\label{e:zig}
k>k_\mathrm{zz}:=1-\frac{\mu^2}{512}+\mc{O}(\mu^3).
\eeq
Instabilities induce phase-slips and dislocations for the Eckhaus instability, and wrinkling for the zigzag instability. Away from onset, existence and stability are model dependent and regions of stable patterns in $\mu-k$ space are often referred to as the Busse balloon; see for instance \cite{RevModPhys.65.851}.

We also note that \eqref{e:sh0} is an $L^2$-gradient flow with respect to the free energy
\beq\label{e:free-en}
\mc{E}[u]:= \int_{\R^n} \left((1+\Delta) u \right)^2 - \frac{\mu u^2}{2} + \frac{u^4}{2}dx.
\eeq
Clearly, the energy landscape reflects the complexity of the dynamics of defects and grain boundaries. Minimizers among periodic patterns are stripes with $k = k_\mathrm{zz}$. Growth as studied below continuously inserts energy into the system and leads to "non-equilibrium" patterns with $k\neq k_\mathrm{zz}$.

Throughout, we will focus on $\mathbf{x}\in \R^2$, that is, $N=2$, which incorporates most experimental setups mentioned and all potential instabilities of stripes.

\subsection{Quenching models of growth}\label{ss:quench}
As a simple model for a spatio-temporal quenching process, we consider a spatio-temporal jump in the bifurcation parameter $\mu=\rho(\mathbf{x},t)$,
\begin{align}\label{e:sh-q}
u_t &= -(1+\Delta)^2 u + \rho(\mathbf{x},t) u - u^3,\qquad
\rho(\mathbf{x},t)=\begin{cases}
 \mu,&\,\, \mathbf{x}\in\Omega_t\\
 -\mu,&\,\,\mathbf{x}\in\Omega_t^c
 \end{cases},\qquad
 \mu>0,\qquad \mathbf{x}=(x,y)\in\R^2,
\end{align}
for some time-dependent, evolving domain $\Omega_t\subset\R^2$ that expands in time, $\Omega_t\subset \Omega_s,\, t<s$. For $\mathbf{x}\in\Omega_t$, the base state is unstable and patterns form, while for $\mathbf{x}\in \Omega_t^c$ it is stable and patterns are suppressed. 

\begin{figure}
\centering
\includegraphics[trim = 4.4cm 4.4cm 2.5cm 2.5cm,clip,width=0.25\textwidth]{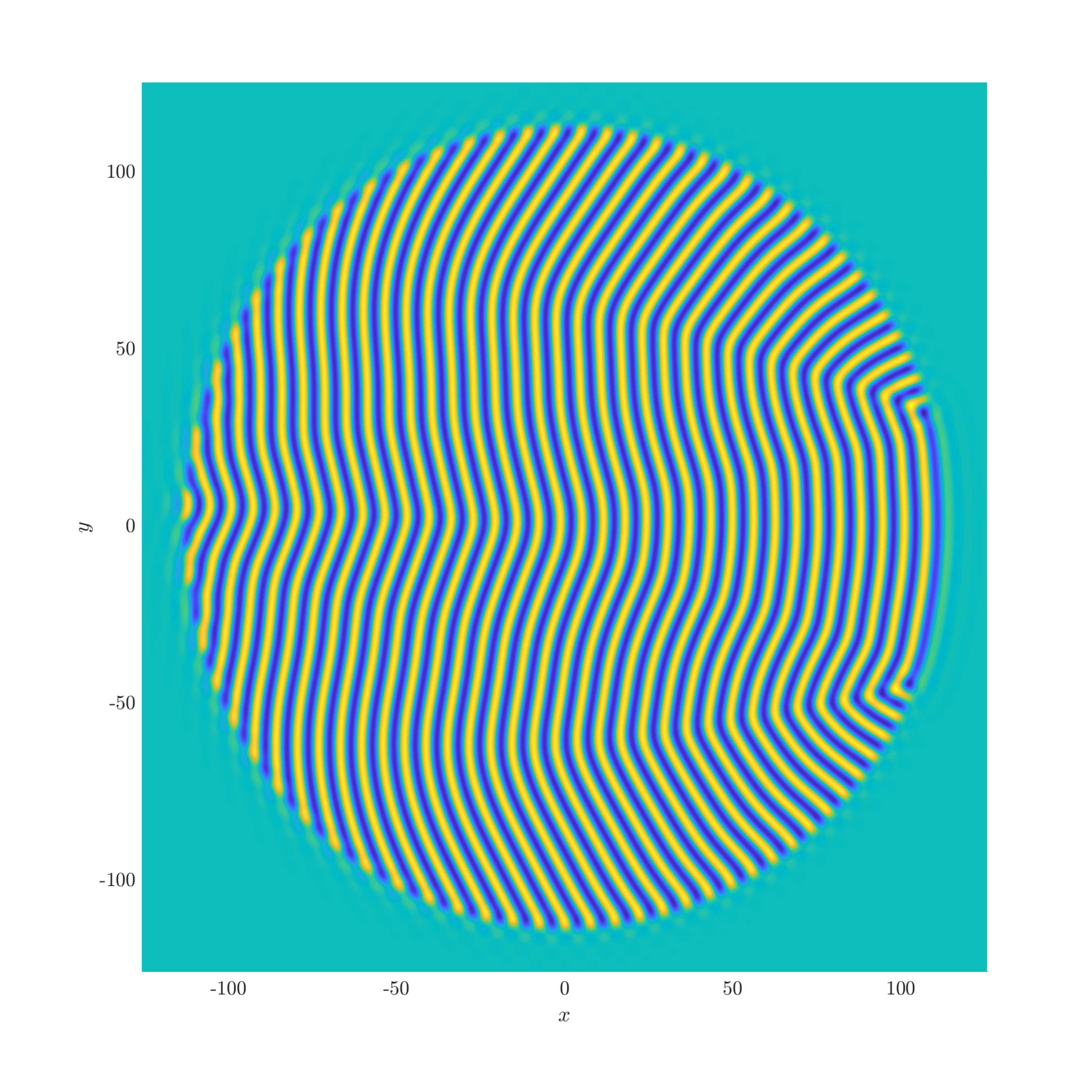}\hspace{-0.2cm}
\includegraphics[trim = 4.4cm 4.4cm 2.5cm 2.5cm,clip,width=0.25\textwidth]{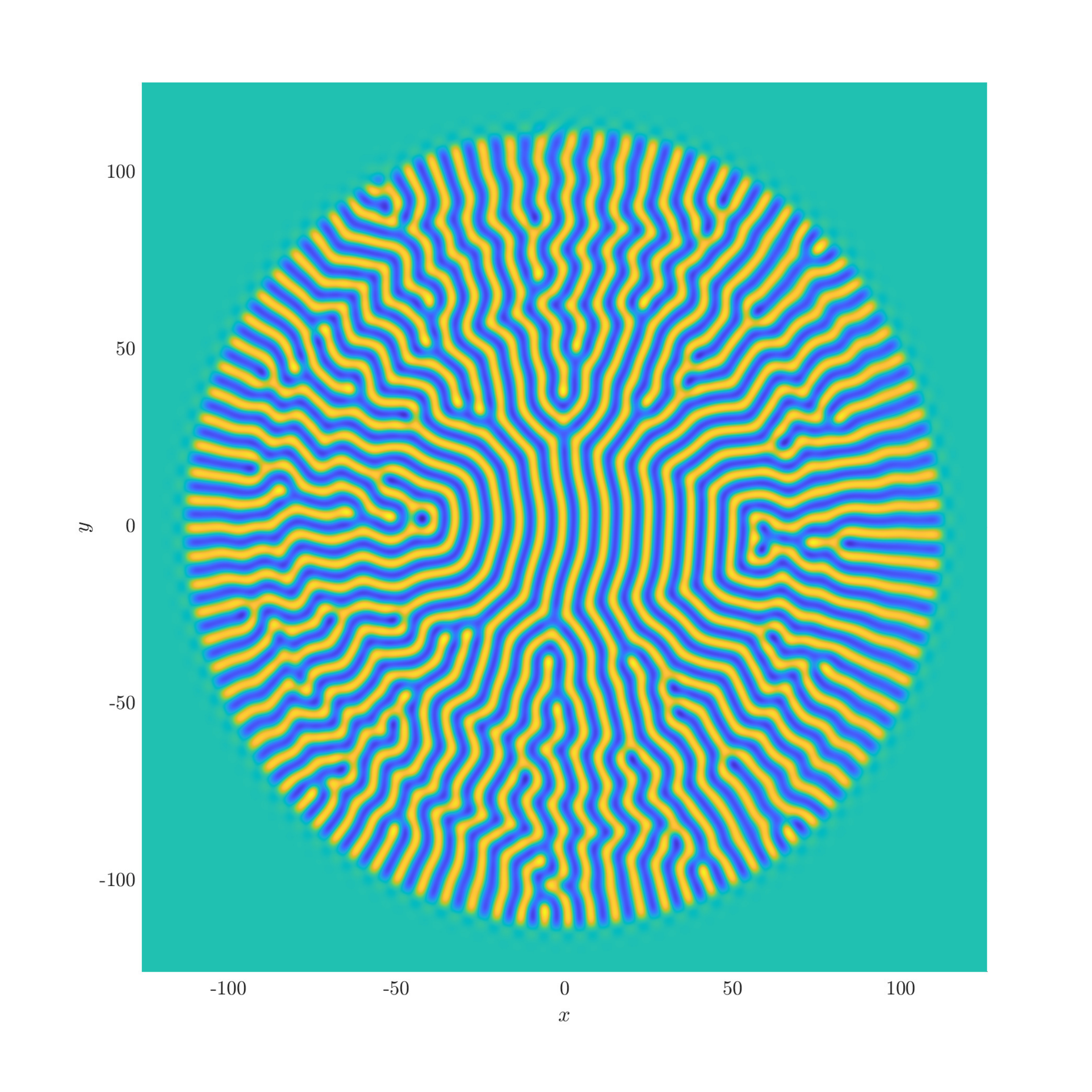}\hspace{-0.2cm}
\includegraphics[trim = 4.4cm 4.4cm 2.5cm 2.5cm,clip,width=0.25\textwidth]{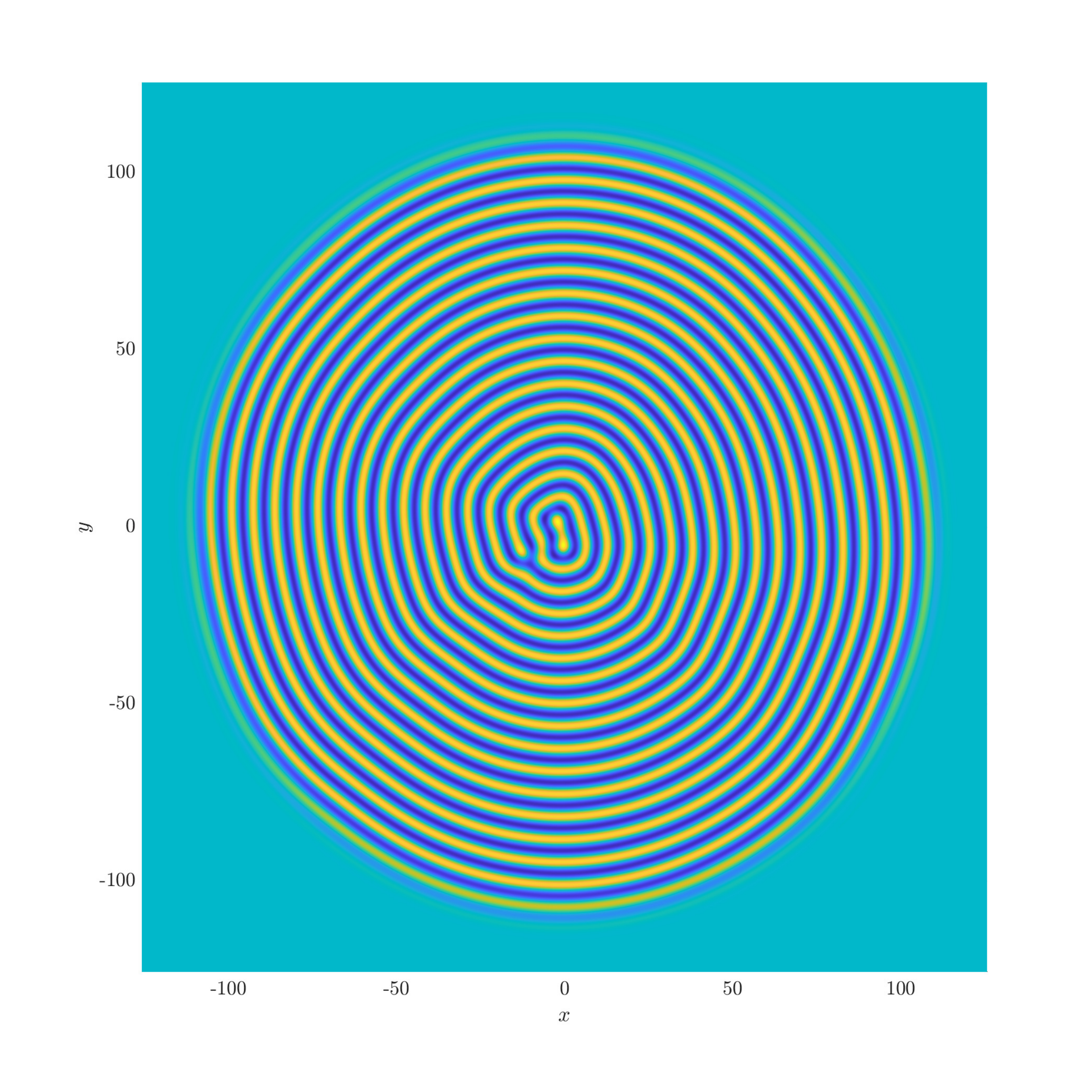}\hspace{-0.2cm}
\includegraphics[trim = 4.4cm 4.4cm 2.5cm 2.5cm,clip,width=0.25\textwidth]{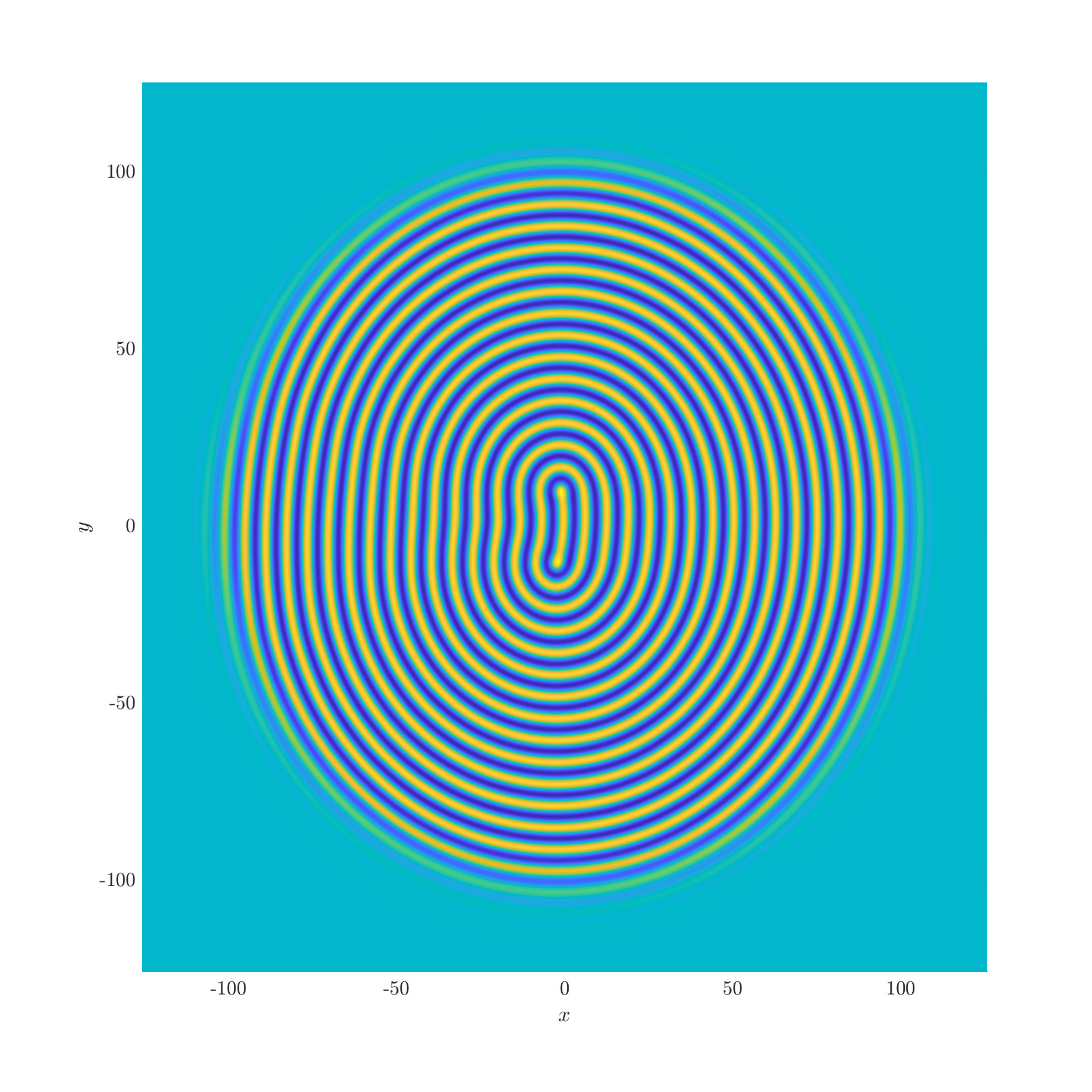}\hspace{-0.2cm}
\caption{
 Solutions of \eqref{e:sh-q} with radial quenching term \eqref{e:rad-q} at $\mu = 3/4$, with radial quenching speeds $c =0.05, 0.4, 1,3 $ from left to right. Simulations are randomly seeded at $t=0$ and run until the quenching interface is close to the boundary of the domain $[-100\pi, 100\pi]^2$. 
}\label{f:rad}
\end{figure}

\paragraph{Radial quenching.}
One interesting example is the radially expanding domain 
\beq\label{e:rad-q}
\Omega_t = \{\mathbf{x}\,|\, |\mathbf{x}|\leq ct\},\quad  t\geq0,
\eeq
with $c>0$ a parameter that denotes the \emph{growth speed} or growth rate. We envision that this parameter is controlled by the experimenter or another mechanism that is independent of $u$. The radially expanding interface $|\mathbf{x}| = ct$ organizes the pattern forming process and, after initializing the system with small uniform noise initial data, one observes a variety of solution behaviors, such as target patterns, spirals, star-like shapes, secondary wrinkling instabilities, and traveling defects for different radial speeds $c$; see Fig. \ref{f:rad}. It is interesting to note here that it is possible for several different orientations to be selected in different sectors of $\Omega_t$ for the same growth speed.  

\paragraph{Directional quenching.} 
A further simplification, which could be viewed as a large-radius or small-curvature approximation of the radial quenching process above, is a planar interface that propagates from left to right, so that
\beq\label{e:omd}
\Omega_t = \{\mathbf{x}\, |\,  x \leq c_xt\}. 
\eeq
with growth rate $c_x\geq 0$. In this case,   $\rho = \rho(x-c_xt):= -\mu\, \mathrm{sign}(x-c_xt)$, where $\mathrm{sign}(x)$ denotes the sign function.

\begin{figure}[ht]
\centering
\includegraphics[trim = 6.5cm 0.5cm 7.5cm 0.5cm,clip,width=0.25\textwidth]{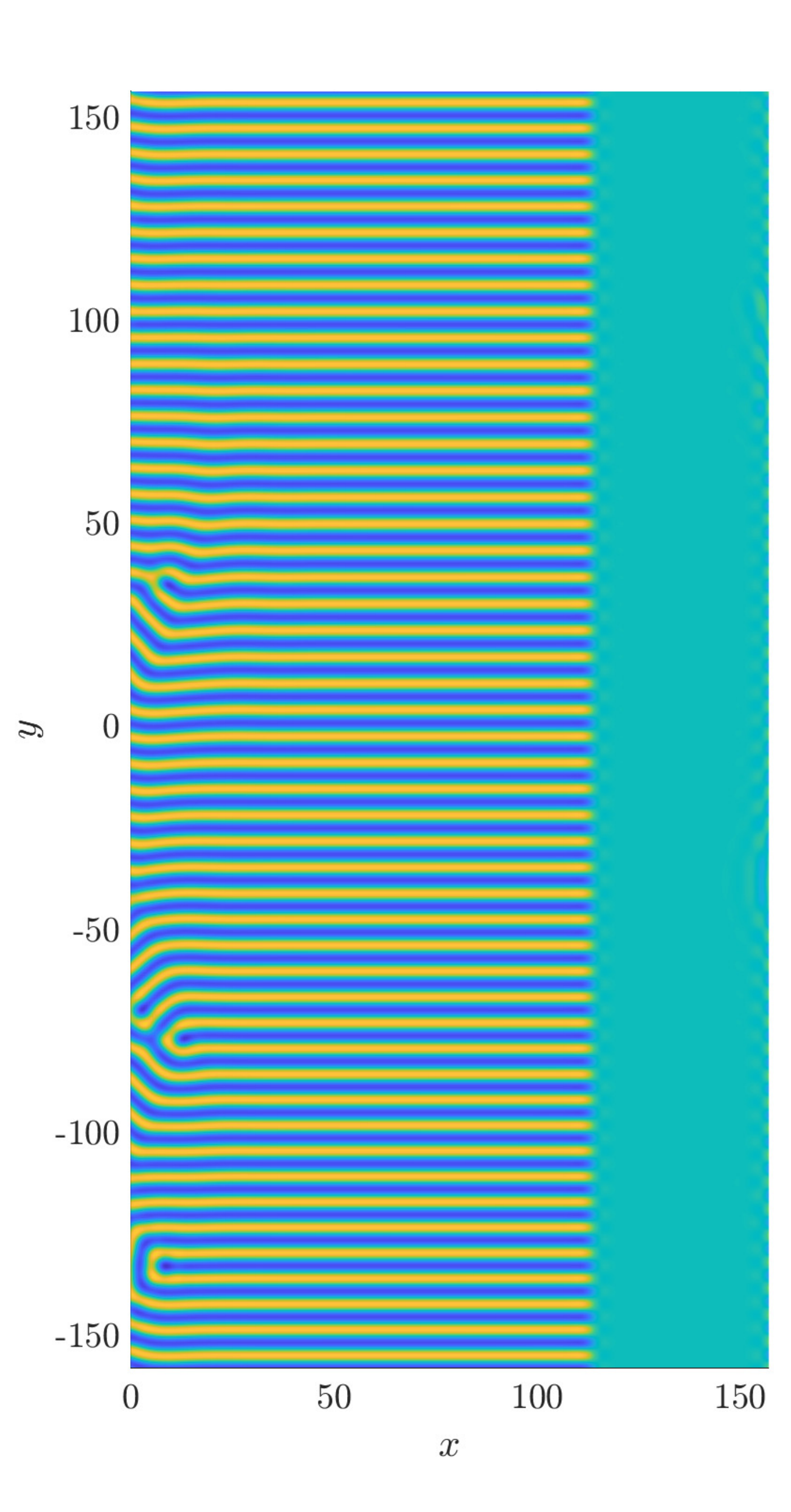}\hspace{-0.5cm}
\includegraphics[trim = 6.5cm 0.5cm 7.5cm 0.5cm,clip,width=0.25\textwidth]{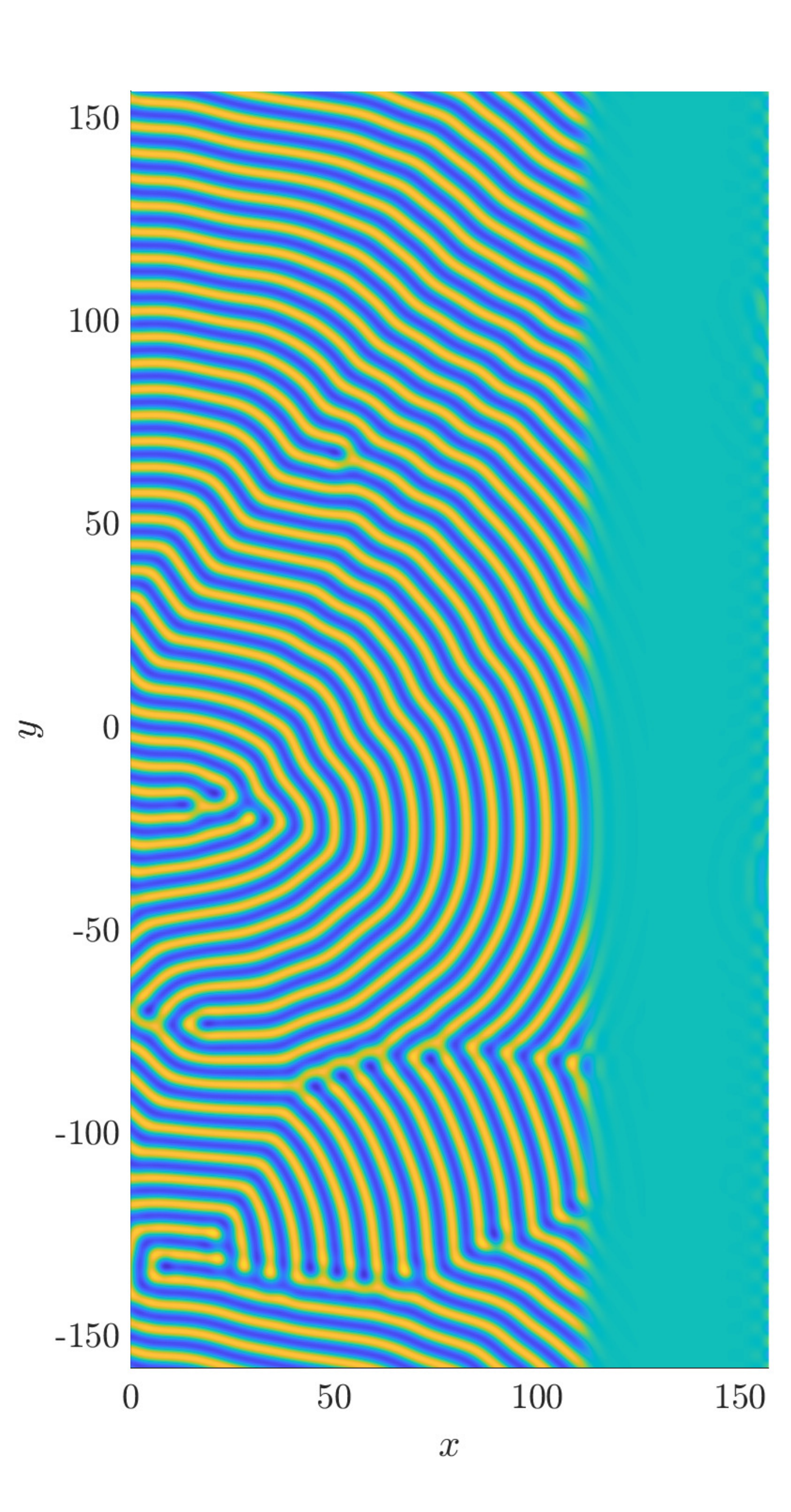}
\includegraphics[trim = 6.5cm 0.5cm 7.5cm 0.5cm,clip,width=0.25\textwidth]{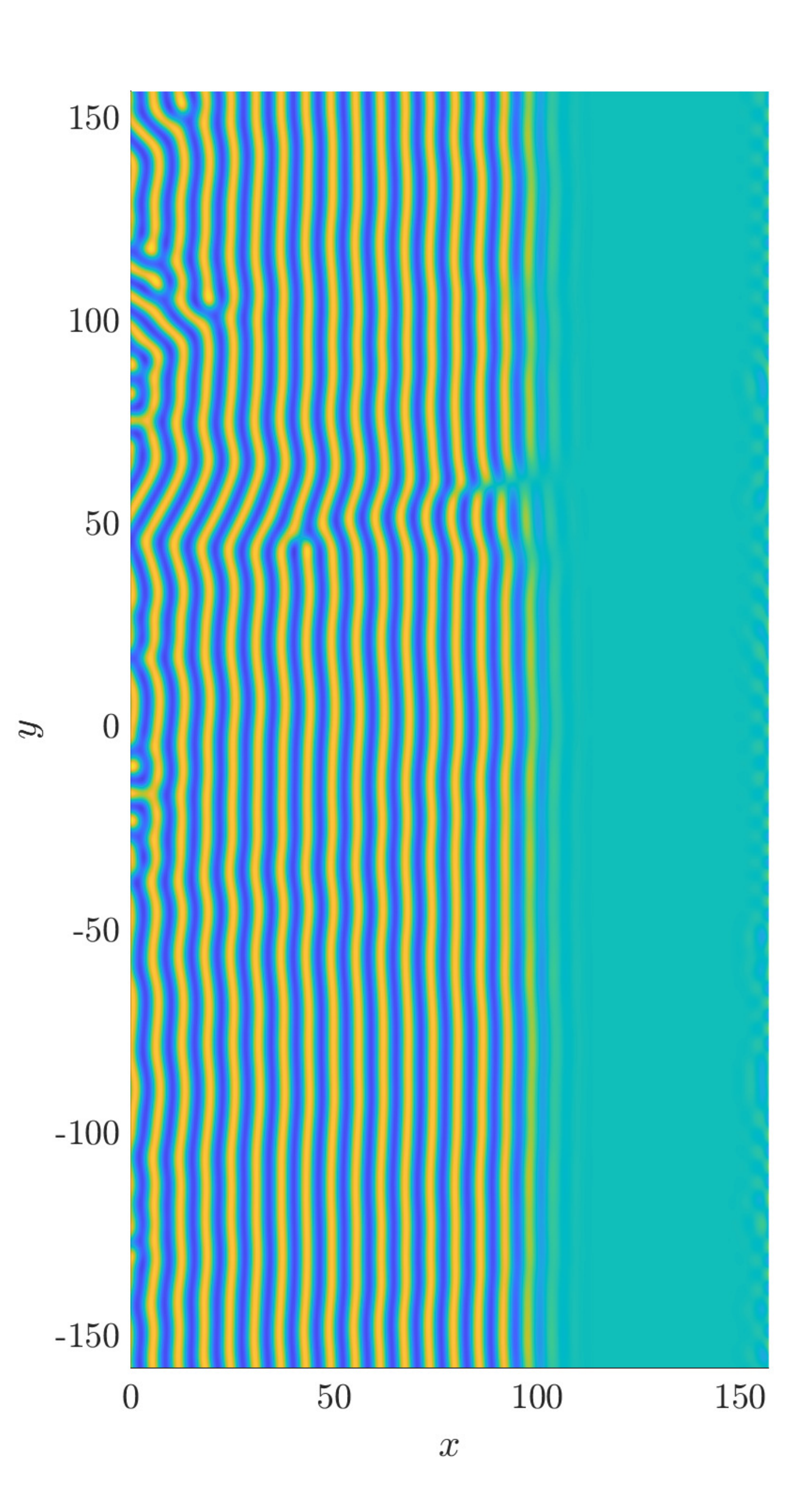}
\includegraphics[trim = 6.5cm 0.5cm 7.5cm 0.5cm,clip,width=0.25\textwidth]{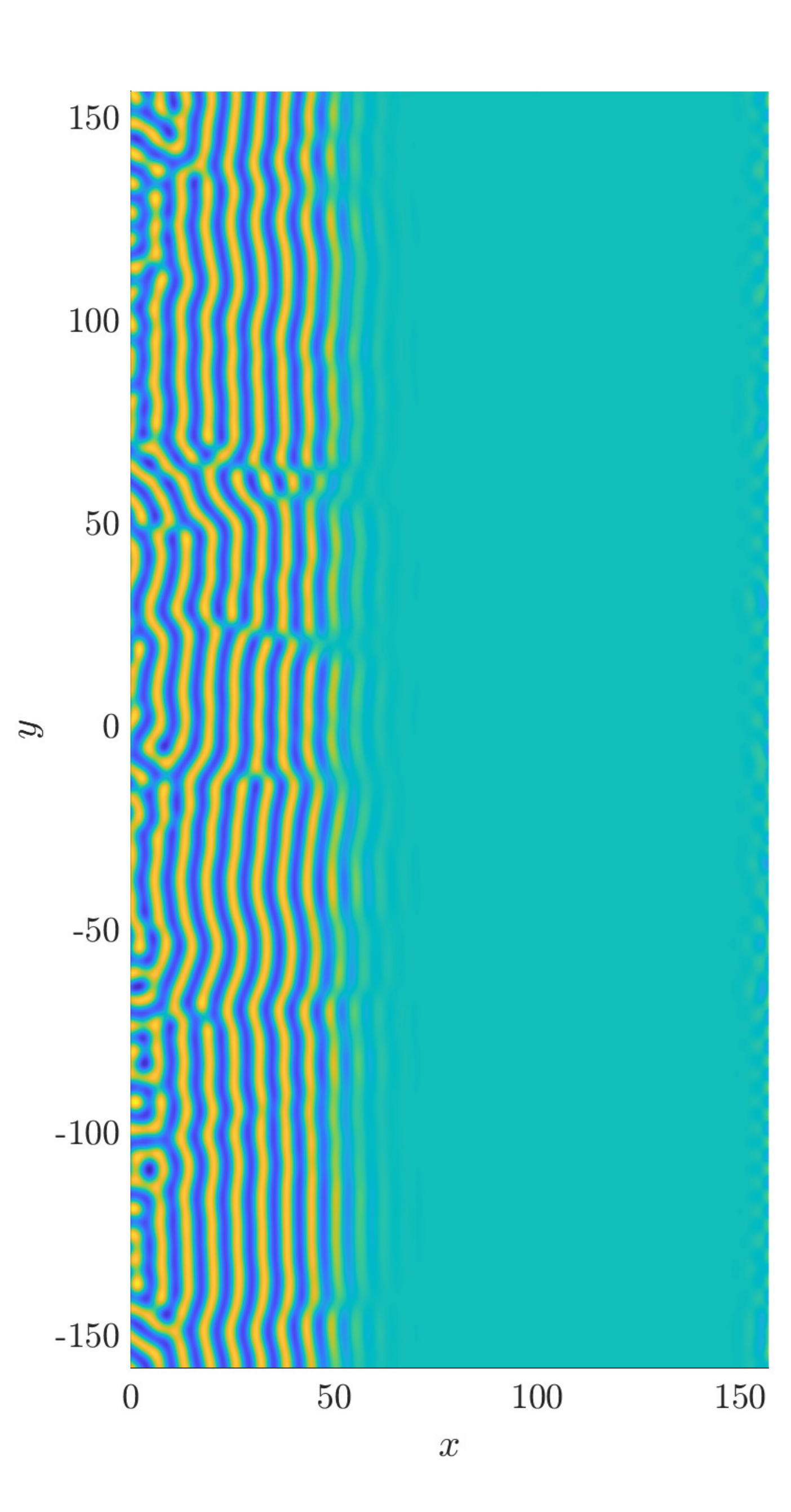}
\caption{
Solutions to \eqref{e:sh-q} with directional quench \eqref{e:omd} with $\mu = 3/4$ and a range of quenching speeds $c_x =0.4, 1, 3, 5 $ increasing from left to right, solution stopped when quenching interface $x \sim 110$ in horizontal domain.
}\label{f:dirquen}
\end{figure}

As depicted in Figure \ref{f:dirquen}, one observes for large speeds that the quenching line outpaces the patterns, setting up the unstable homogeneous state into which the patterns naturally invade at a slower speed.  As $c_x$ is decreased below this invasion speed, one first observes mostly stripes oriented \emph{parallel} to the quenching interface; for intermediate speeds, stripes which are \emph{oblique} or slanted to the interface; and for small speeds, stripes which are \emph{perpendicular} to the interface.  We remark that such a set of qualitative phenomena has been recently observed in a series of analogous experiments in the light-sensing reaction-diffusion system \cite{miguez2006effect,konow2019turing}.
There are of course many different types of quenching geometries that could be considered (see Fig. \ref{f:chev} below for a few examples) as well as other types of heterogeneity that could be introduced to model growth. We discuss  some of those in Section \ref{s:ex} but we first,  in Sections \ref{s:edge}--\ref{s:other},  focus on simple directional quenching for their motivation in experimental settings \cite{Nunley806679,zigzageutectic,double,stoop2018defect} and for their conceptual mathematical simplicity that was exploited in a series of works that build the foundation of this paper  \cite{avery2019growing,chen2021strain,goh2020spectral,gs3,gs1,gs4,gh1,weinburd,monteiro2017phase,monteiro2018horizontal, monteiro20}. 
Despite the simplicity of the setting, the ensuing wealth of phenomena is not fully understood at a rigorous, formal, or even heuristic level and we hope that this exposition will serve as motivation for further investigation and development of  novel mathematical tools. We note some of this material was presented, in an abrieviated manner, in the online article \cite{goh21a}.

\subsection{Moduli spaces of quenched patterns}\label{ss:mod}

 The mathematical understanding of the patterned solutions observed in Figure \ref{f:dirquen} has several facets, beginning with existence and local stability, instability, or metastability of front-like solutions, and expanding to continuation and bifurcation of solutions under changes in extrinsic parameters such as the quenching speed $c_x$. Of interest are then also questions of universality, that is, how much qualitative features depend on specific models, and in this context the description via amplitude or phase modulation equations. Phenomenologically, one observes at times nucleation of defects at the quenching interface and one would like to relate properties of stripe formation to the presence or absence of such defects. In specific simple cases, one may even be able to obtain global descriptions of the dynamics. 
 
 More directly, a first question one may wish to answer is if the quenching process can create a regular ``crystal'', that is: 
 \begin{center}
     \emph{For a quenching speed $c_x$, what wavenumbers and orientations of stripes can be formed in the wake of the quench? }
 \end{center}
%
Answers to this question would for instance shed light on the apparent selection of orientation in Figure \ref{f:dirquen} as well as in CDIMA experiments \cite{miguez2006effect} depending on the quenching speed $c_x$. As a further simplification, we may narrow the question to existence, only, of the simplest solutions that form stripes. That is we consider front-like solutions with whose temporal behavior can be thought of as a 1:1 resonance with the formation of perfect stripes. To make this precise, we look for solutions in the  frame moving with the quench   $\tl x = x - c_x t$. Pure stripes at $\tl x\sim -\infty$ then take the form $u_p(k_x x + k_y y;k) = u_p(k_x \tl x + k_x c_x t + k_y y;k)$,  where $\mathbf{k} = (k_x,k_y)$ is the wavevector and  $k = |\mathbf{k}| = \sqrt{k_x^2+k_y^2}$ the bulk wavenumber. 
The simplest form of solutions then is general ''heteroclinic" behavior in $\tl x$ and periodic dependence in $ k_x c_x t + k_y y$. Minimal period $2\pi$ then corresponds to a strong, 1:1 resonance of the quenching process with the crystal in the wake. We therefore introduce the scaled, $y$-comoving frame $\tl y = k_y (y - c_y t)$, in which the pure stripe solution satisfies $u_p(k_x x + k_y y; k) = u_p(k_x \tl x + \tl y; k)$.

\begin{figure}[ht!]
\centering
\includegraphics[trim = 0.1cm 0.0cm 0.0cm 0.0cm,clip,width=0.66\textwidth]{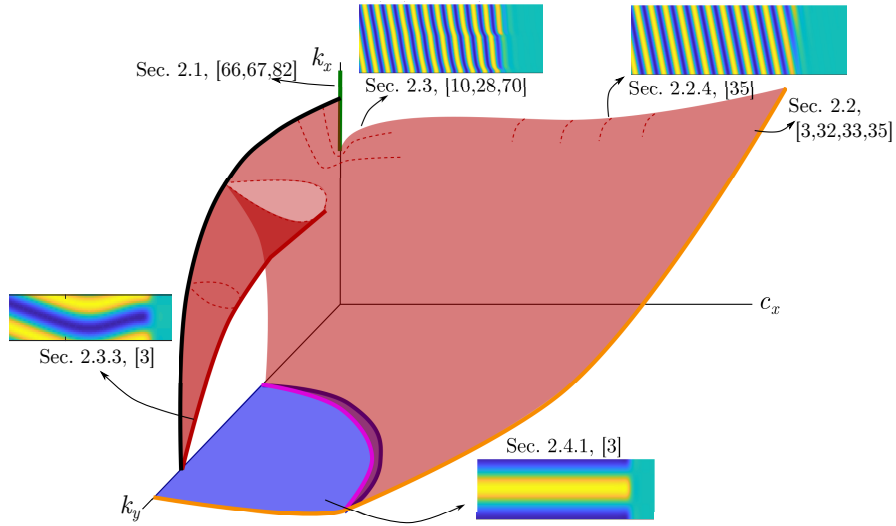}\hspace{-0.2in}
\caption{ Schematic depiction of the moduli space $\mc{M}$ with representative solution profiles in various regions, references to corresponding sections below, as well as citations to related works. See also Fig. \ref{f:mod-schem} which depicts important boundary and limit curves of $\mc{M}$. }
\label{f:mod-sch}
\end{figure}

Altogether, \eqref{e:sh-q} is then reduced to the asymptotic boundary-value problem
 \begin{align}
0&= -(1+\partial_{\tl x}^2 + k_y^2 \partial_{\tl y}^2)^2 u + \rho(\tl x) u - u^3 + c_xu_{\tl x}-k_xc_x u_{\tl y},\notag\\
0&= \lim_{\tl x\rightarrow -\infty} u(\tl x,\tl y) - u_\mathrm{p}(k_x \tl x + \tl y;k),\qquad 
0 = \lim_{\tl x\rightarrow\infty} u(\tl x,\tl y),\qquad u(\tl x,\tl y) = u(\tl x,\tl y + 2\pi);\label{e:mtw}
\end{align}
see the inserts in Figure \ref{f:mod-sch} or Figure \ref{f:kxabs} for examples of such solutions. Note that $c_x$ is an extrinsic parameter, while  $k_x$ and $k_y$ are intrinsic to the solution. Values $k_x=0$ or $k_y=0$ correspond to stripe formation perpendicular and parallel to the quenching line, respectively; nonzero values of both $k_x$ and $k_y$ correspond to oblique stripe formation.
 
\begin{figure}[ht!]
\includegraphics[trim = 0.5cm 0.25cm 0.5cm 0.5cm,clip,width=0.33\textwidth]{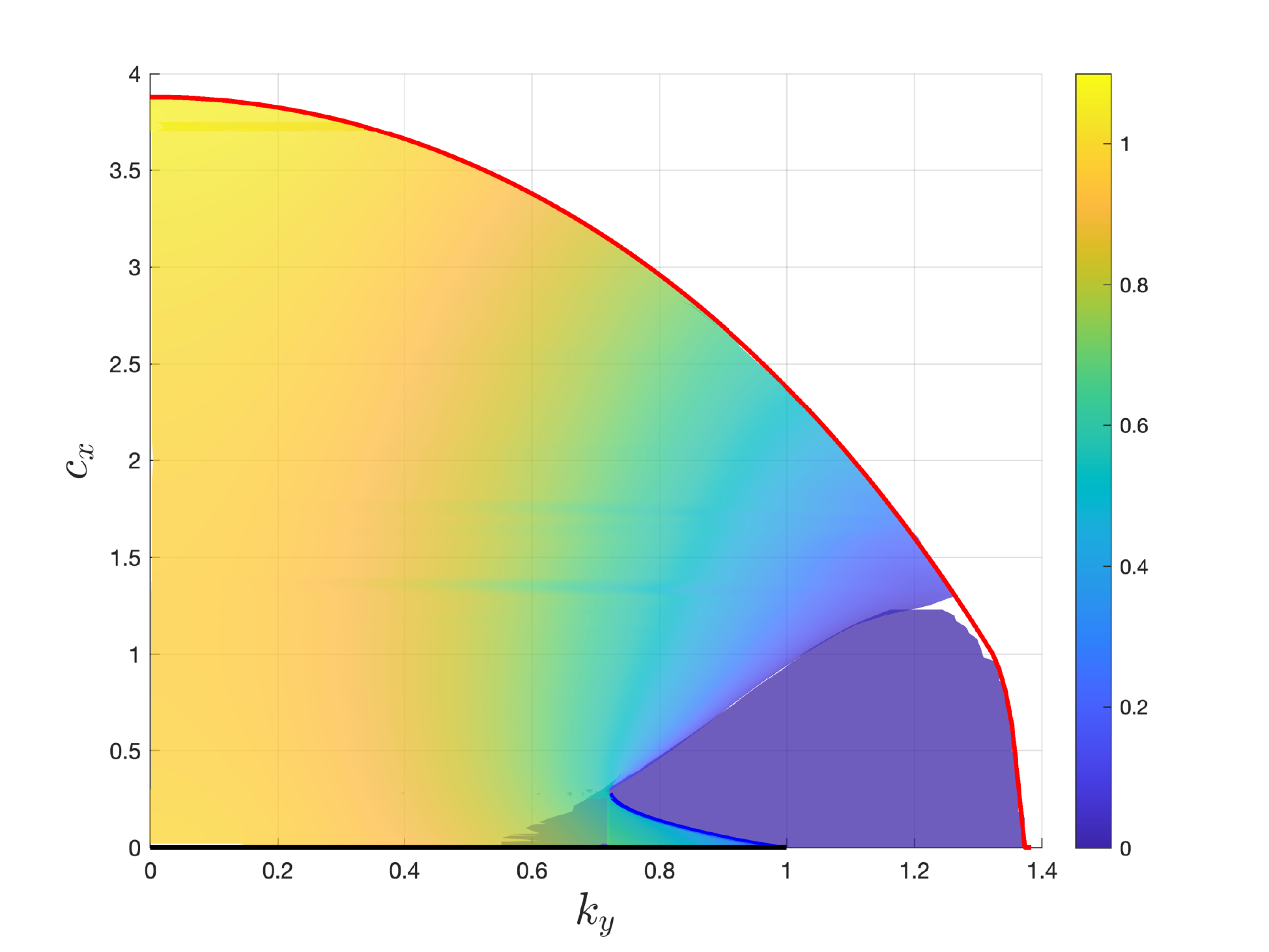}\hspace{-0.2in}
\includegraphics[trim = 0.5cm 0.25cm 0.5cm 0.5cm,clip,width=0.33\textwidth]{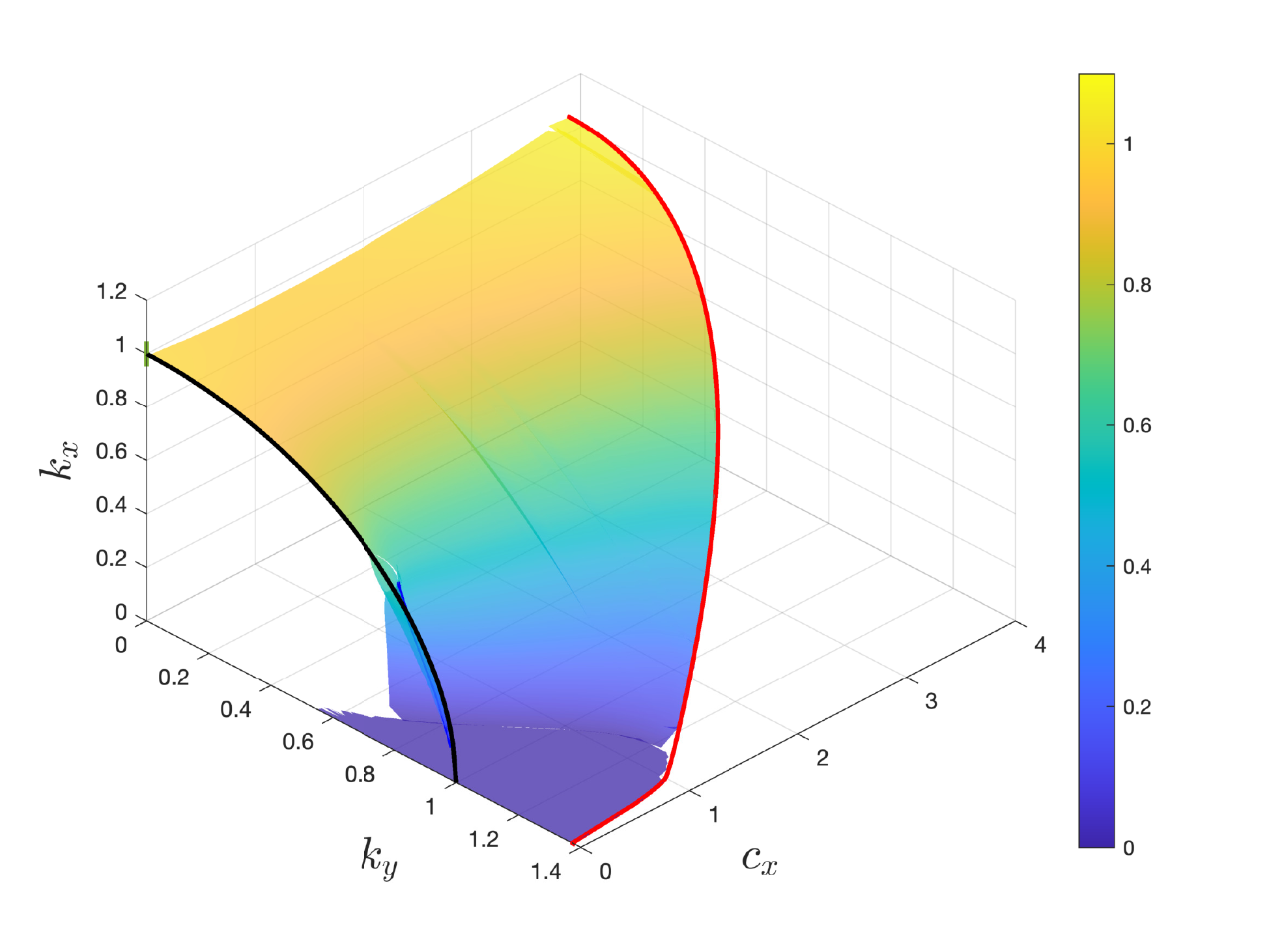}\hspace{-0.2in}
\includegraphics[trim = 0.5cm 0.25cm 0.5cm 0.5cm,clip,width=0.33\textwidth]{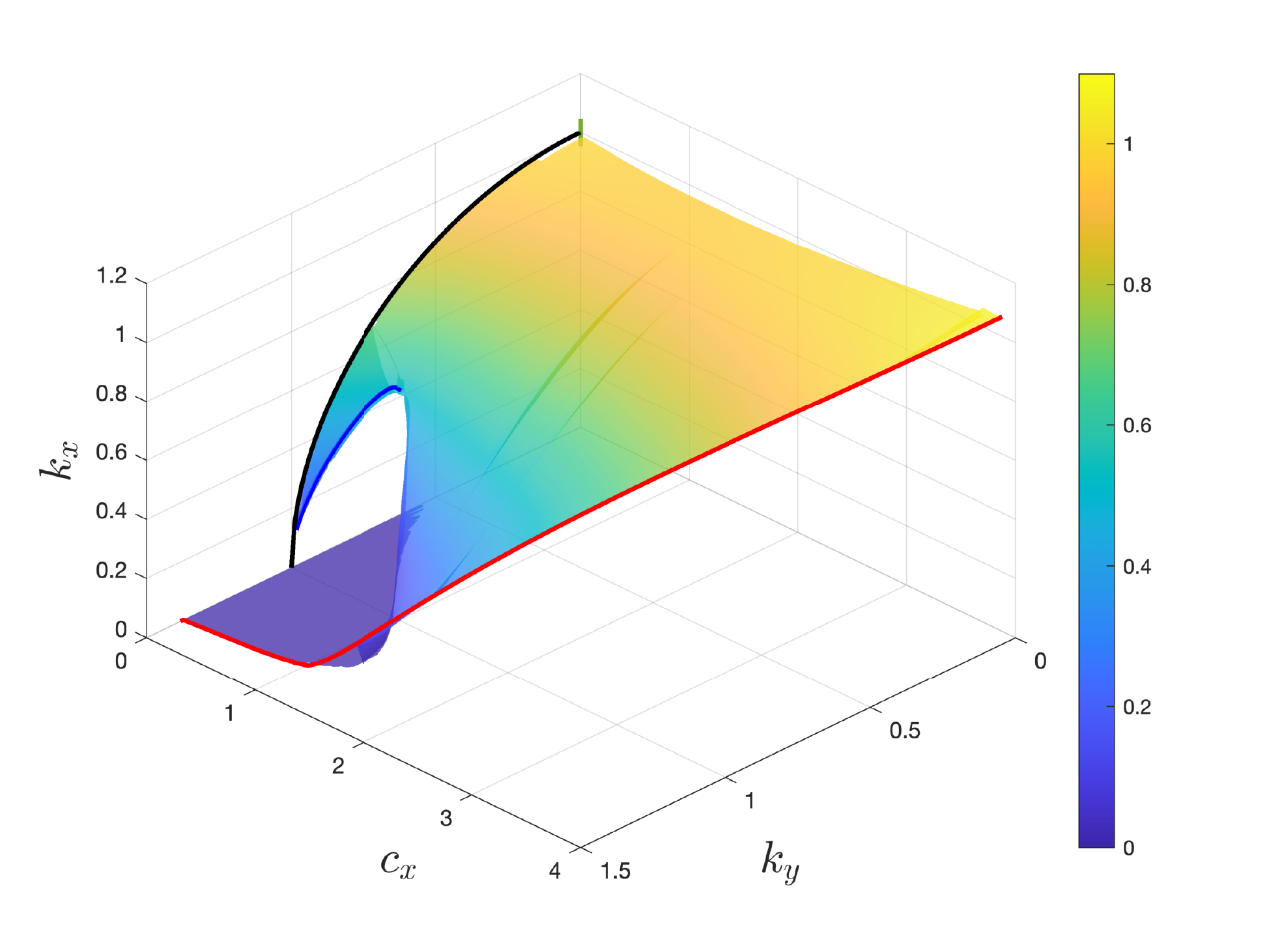}\hspace{-0.2in}
\caption{ Several views of the moduli space $\mc{M}$ for \eqref{e:mtw}, obtained from results numerical continuation (see Section \ref{s:num} below on numerical details); here $\mu = 3/4$. The red curve denotes the linear spreading curve $(k_y,c_\rlin(k_y),k_{x,\rlin}(k_y)$, given in \eqref{e:clinsh} and \eqref{e:klinsh} below for both parallel/oblique as well as perpendicult stripes. The black curve denotes the zigzag critical curve for $c_x = 0$, which satisfies $k_\mathrm{zz}^2 = k_y^2 + k_x^2$. The green curve denotes a band of wavenumbers at $k_y = c_x = 0$.}\label{f:mod}
\end{figure}

The set of parameter values $c_x,k_x,k_y$ for which solutions to \eqref{e:mtw}
exist,  
\beq
\mathcal{M} = \{(k_y,c_x,k_x)\in \R^3\,|\, \text{\eqref{e:mtw} has a solution}  \}, \label{e:mod-def}
\eeq 
naturally parameterizes the space of quenching fronts, up to possible multiplicities of solutions. It turns out that $\mathcal{M}$ is a variety with a rich structure that informs much of the understanding of the quenching process. Drawing from a classical terminology for parameterizations of solutions to (algebraic) equations  \cite{chan2021},  we refer to $\mathcal{M}$ as the \emph{moduli space} of quenched patterns. Clearly, $\mathcal{M}$ ignores multiplicities such as trivial translation symmetry in $\tl y$, but also inherent multiplicities, quotienting the structure of solutions for finite $\tl x$ and retaining only far-field information near $\tl x=-\infty$. Exploiting Fredholm properties of the linearization of \eqref{e:mtw} at solutions, one finds that $\mathcal{M}$ is generically locally a graph $k_x=k_x(c_x,k_y)$, indicating the selection of a stretching of patterns in the direction perpendicular to the quenching line; for more detail see \cite{gs3,avery2019growing} as well as Section \ref{ss:fred} below.  
We show numerical computations of $\mc{M}$ in Figure \ref{f:mod}.   Figure \ref{f:mod-sch} gives a schematic depiction with corresponding solution profiles, as as well as references to past works and sections of this work which explore a given region.  A table that summarizes various limits, singularities, and boundaries of $\mc{M}$ can be found in the beginning of Section \ref{s:edge} below. 


Broadly, one hopes to connect quantitative and qualitative  properties of $\mc{M}$ to phenomena in quenched pattern formation. Practically, the object $\mc{M}$ can be viewed as a ``cookbook" or guide for fabricating patterns, indicating which wave vectors can be selected for a given quenching speed, while also revealing locations where novel dynamic phenomena and bifurcations occur. In other words, if one can control the vertical spatial period of the experimental domain, and thus control $k_y$, the variety $\mc{M}$ indicates which quenching speeds $c_x$ can grow a pattern with horizontal wavenumber $k_x$. In a reductionist sense, $\mc{M}$ also gives effective boundary conditions in a homogenized description of the crystalline structure: averaging over the ``microstructure'', that is, the stripes, one is left with a local wave vector as an effective variable. Dynamics in such a description are usually diffusive and the relation between $k_x$ and $k_y$ gives effective boundary conditions for the vector-valued diffusion equation. 

We therefore hope that a focus on the moduli space $\mathcal{M}$, as promoted here, will help organize and guide further exploration of the interplay between growth and pattern formation, investigating in particular how $\mathcal{M}$ changes as system parameters vary, or how such moduli spaces differ among different systems, such as 
%
%
the Complex-Ginzburg-Landau equation, the CDIMA reaction-diffusion system, the Cahn-Hilliard equation \cite{foard}, phase-field equation, and other reaction-diffusion systems. More narrowly, we explore in  the following Section \ref{s:edge} various regions of $\mc{M}$ in more detail, discussing the types of solutions  observed, what physical mechanisms affect wavenumber selection, and what types of mathematical methods can be used to study solutions rigorously. We also demonstrate how  bifurcation points and singularities in $\mc{M}$ lead to qualitative changes in the full temporal dynamics of the original model \eqref{e:sh-q}.

\subsection{Overview}
We use the supercritical cubic Swift-Hohenberg equation \eqref{e:sh-q} as a testbed to explore directionally quenched patterns. By using one specific, but prototypical equation, this work seeks to review, combine, and unify the authors' previous works \cite{avery2019growing,chen2021strain,goh2020spectral,gs3,gs1,gs4,weinburd} which studied quenched patterns from a mathematical viewpoint in a variety of models. We expect many of the phenomena observed in this context to be generic, and thus observable in other pattern-forming systems. Throughout, we use the moduli space $\mc{M}$ to organize our results. We shall also indicate areas of $\mc{M}$ or phenomena which are yet to be fully understood at a rigorous or even heuristic level.

 Section \ref{s:edge} describes the moduli space for the quenched Swift-Hohenberg equation \eqref{e:sh-q} and compares stripe selecting mechanisms in different $c_x$ and $k_y$ regimes. Section \ref{s:stab} briefly discusses stability of these pattern forming fronts. Section \ref{s:other} discusses how other traveling heterogeneities, different from the steep quench, affect wavenumber selection.  Section \ref{s:ex} then presents new numerical results for the moduli space in other prototypical models of pattern formation, such as the complex Ginzburg-Landau equation, a reaction-diffusion model for the CDIMA chemical system, as well as two modified Swift-Hohenberg equations, one with spatial anisotropy, and another with a subcritical cubic-quintic nonlinearity. Appendix \ref{ss:fred} reviews the local description of $\mc{M}$ as a graph in $k_x$ over $(k_y,c_x)$ using Fredholm theory, and Appendix \ref{s:num} gives an overview of the numerical continuation approach we use to approximate pattern-forming front solutions of \eqref{e:mtw} on a finite computational domain.

\section{Qualitative properties of quenching: singularities of $\mathcal{M}$}\label{s:edge}
Singularities and boundaries of the moduli space $\mc{M}$ give important information on pattern-forming dynamics and are excellent starting points for mathematical analysis. Understanding boundaries and bifurcation points,  one can then resort to  continuation techniques to "fill in" the bulk of $\mc{M}$. On the other hand, boundaries and singularities correspond to  qualitative changes in the pattern-forming dynamics.  

\begin{figure}[ht!]
\centering
\includegraphics[trim = 0.0cm 0.0cm 0.0cm 0.0cm,clip,width=0.75\textwidth]{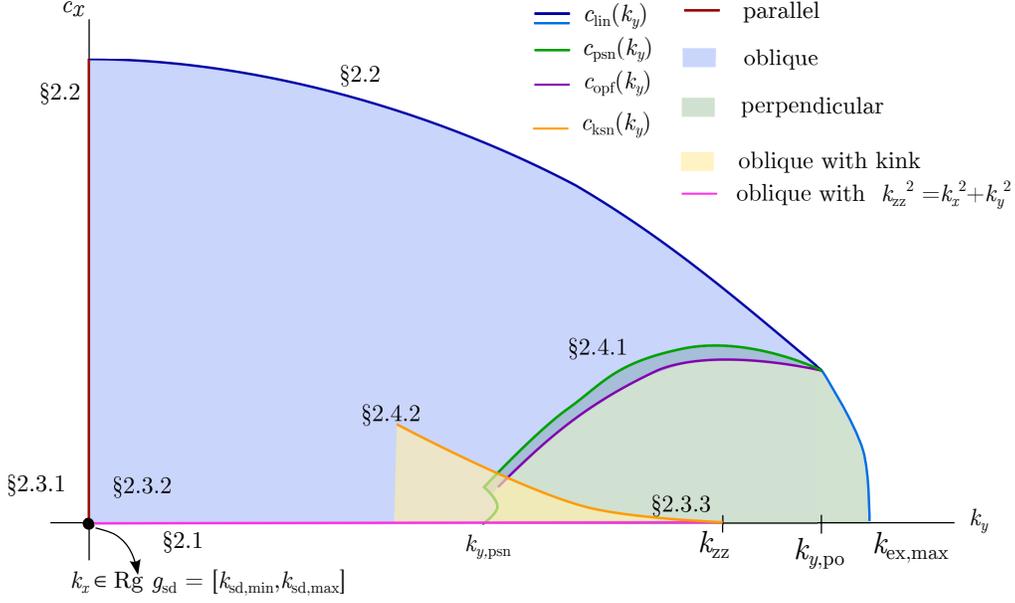}
\caption{Schematic diagram of the different regimes of stripe formation in the moduli space, along with corresponding parts of Section \ref{s:edge} where they are discussed.}
\label{f:mod-schem}
\end{figure}

The following list, together with Figure \ref{f:mod-schem}, provides a rough summary of limiting cases and singularities discussed here. All notation will be discussed throughout the following sections.
\begin{itemize}
\item \underline{\textbf{Stationary quench}, $\mathbf{c_x = 0}$}
\begin{itemize}
\item ${ k_y =  0 }$, (Sec. \ref{sss:st-para}): Range of selected wavenumbers determined by the strain displacement relation $k_x\in (k_\mathrm{sd,min},k_\mathrm{sd,max})$.
\item ${ k_y\in(0,k_\mathrm{zz}), \,\,k_x>0 }$, (Sec. \ref{sss:st-ob}): Oblique stripes have zigzag critical wavenumber $k_\mathrm{zz}^2 = k_x^2 + k_y^2$, 
\item ${ k_y\in[k_\mathrm{y,psn},k_\mathrm{ex,max} ] }, k_x = 0$,  (Sec. \ref{sss:st-perp}): perpendicular stripes. 
\end{itemize}
\item \underline{\textbf{Slow growth, $\mathbf{c_x\gtrsim0}$}}
\begin{itemize}
\item ${ k_y =  0 }$, (Sec \ref{sss:slow-para}): Monotonically increasing wavenumber curve $k_x(c_x)$ with $k_x\rightarrow k_\mathrm{sd,min}$ as $c_x\rightarrow0$.
\item ${ k_y \gtrsim  0 }$, (Sec. \ref{sss:slow-ob}): Wavenumber $k_x(c_x,k_y)$ non-monotonic in $c_x$ for fixed $k_y$, these develop a singularity at $c_x = 0$ as $k_y\rightarrow 0$.
\item ${ k_y \lesssim k_\mathrm{zz}, k_x\neq0 }$, (Sec \ref{sss:acc-ob}): Kink-dragging bubble, oblique stripes detach in a kink-forming saddle-node curve $(k_\mathrm{x,ksn},c_\mathrm{x,ksn})(k_y)$. 
\end{itemize}
\item \underline{\textbf{Fast growth}, $\mathbf{c_x\lesssim c_\mathrm{lin}(k_y)}$, (Sec \ref{ss:fg})}
\begin{itemize}
\item ${ k_y\in[0,k_{y,\mathrm{po}}}$:  For $k_y$ fixed and $k_x\neq0$, striped fronts cease to exist for $c_x\geq c_\mathrm{lin}(k_y)$. Leading order wavenumber prediction of $k_x$, for $c_x\lesssim c_\mathrm{lin}(k_y)$, given by absolute spectrum of trivial state. 
\item ${k_y\in[\sqrt{(2+\sqrt{3\mu})/2},k_\mathrm{ex,max})}$: perpendicular stripes selected, detachment for $c_x\geq c_\mathrm{lin}(k_y)$ 
\end{itemize}
\item \underline{\textbf{Intermediate growth}}:
\begin{itemize}
\item ${c_x\lesssim c_\mathrm{x,psn}(k_y), \, k_y\in(k_\mathrm{y,psn},\sqrt{(2+\sqrt{3\mu})/2)}}$: (Sec. \ref{ss:perp-ret}) Perpendicular stripe detachment through a saddle-node bifurcation, along curve $c_\mathrm{x,psn}(k_y)$.
\item ${c_x\gtrsim c_\mathrm{x,opf}(k_y)}$: (Sec. \ref{ss:perp-ret})  Oblique stripes reattach in a symmetry breaking pitchfork bifurcation at $k_x\sim0$ along the curve $c_\mathrm{x,opf}(k_y)$. 
\end{itemize}
\end{itemize}

\subsection{Stationary fronts}\label{ss:stft}
\begin{figure}[ht]
\centering
\includegraphics[trim = 0.0cm 0.0cm 0.5cm 0.5cm,clip,width=0.4\textwidth]{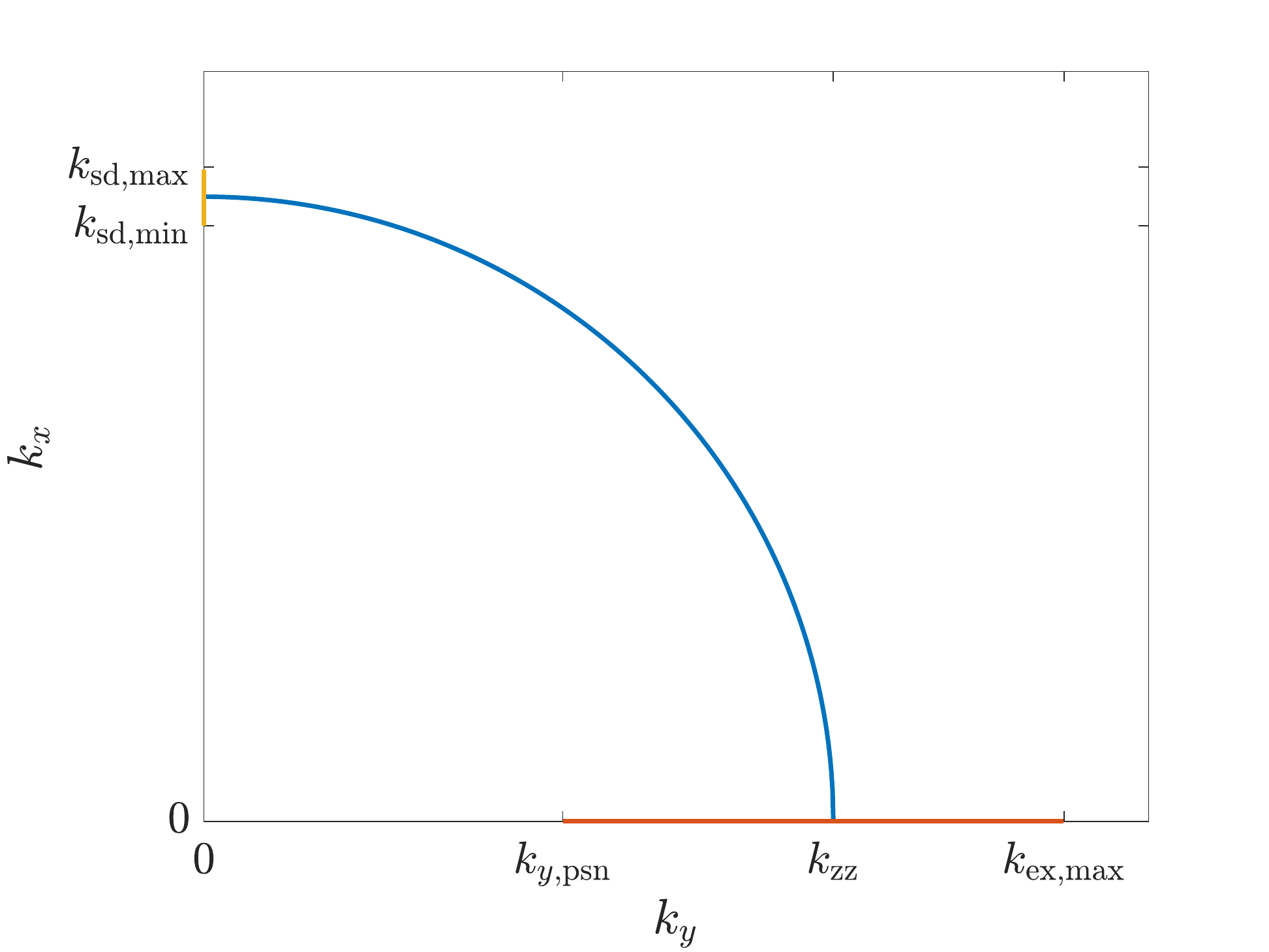}
\caption{Left: $c_x = 0$ cross section of moduli space $\mc{M}$ for $\mu = 3/4$, showing selected wavenumbers for front solutions \eqref{e:mtw-y}, showing three distinct domains: parallel stripe selection with $k_y =0, \, k_x\in(k_\mathrm{sd,min},k_\mathrm{sd,max})$ (yellow), oblique stripes with $k_\mathrm{zz}^2 = k_x^2 + k_y^2$ (blue), and perpendicular stripes with $k_x = 0$ for $k_y\in(k_{y,\mathrm{psn}},k_\mathrm{ex,max}=\sqrt{1+\sqrt{\mu}})$; see subsections below for explanation of notation\label{f:ky0}}
\end{figure}

We start with the conceptually simple case of a stationary quench, $c_x = 0$, where \eqref{e:mtw} reduces to an elliptic equation 
\beq\label{e:mtw-y}
0 = -(1+\p_x^2 + k_y^2\p_y^2)^2 u + \rho(x) u - u^3, \quad (x,y)\in \R\times (\R/2\pi\Z),
\eeq 
on an unbounded cylinder with asymptotic boundary conditions as in \eqref{e:mtw}. 
Figure \ref{f:ky0} depicts a schematic of the $c_x = 0$ cross-section of $\mc{M}$. One finds, in particular, a band of wavenumbers $k_x$ compatible with $k_y = 0$ (yellow) and a band of wavenumbers $k_y$ compatible with $k_x = 0$ (orange) limits, but a unique curve $k_x(k_y)$ when $k_x,k_y\neq 0$. We discuss these three cases separately in the following. 

\subsubsection{Parallel stripes, $k_y = 0$}\label{sss:st-para}
\begin{figure}[ht]
\centering
\includegraphics[trim = 0cm 0cm 0cm 0cm,clip,width=0.28\textwidth]{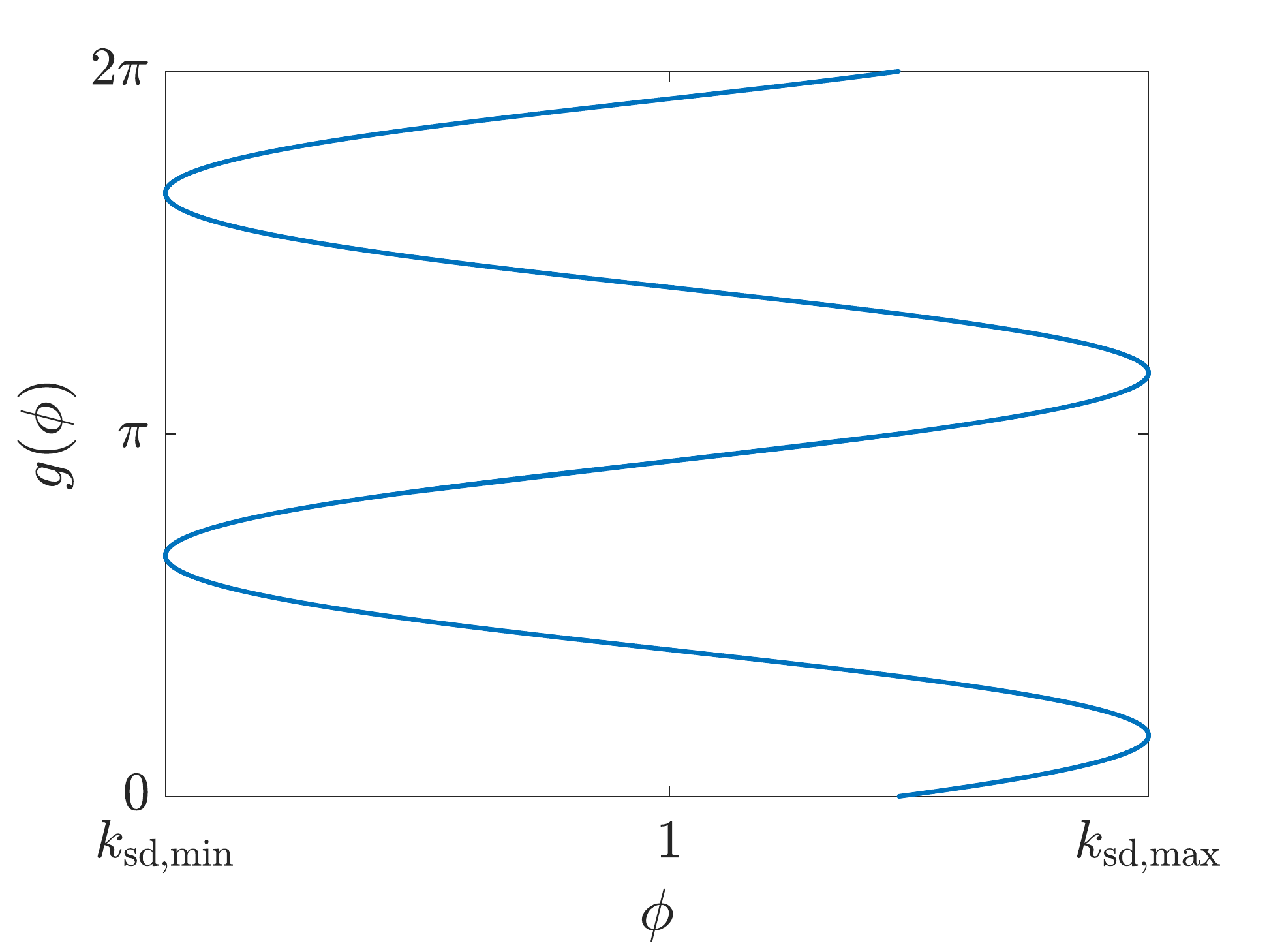}\hspace{-0.2in}
\includegraphics[trim = 0cm 0cm 0cm 0cm,clip,width=0.4\textwidth]{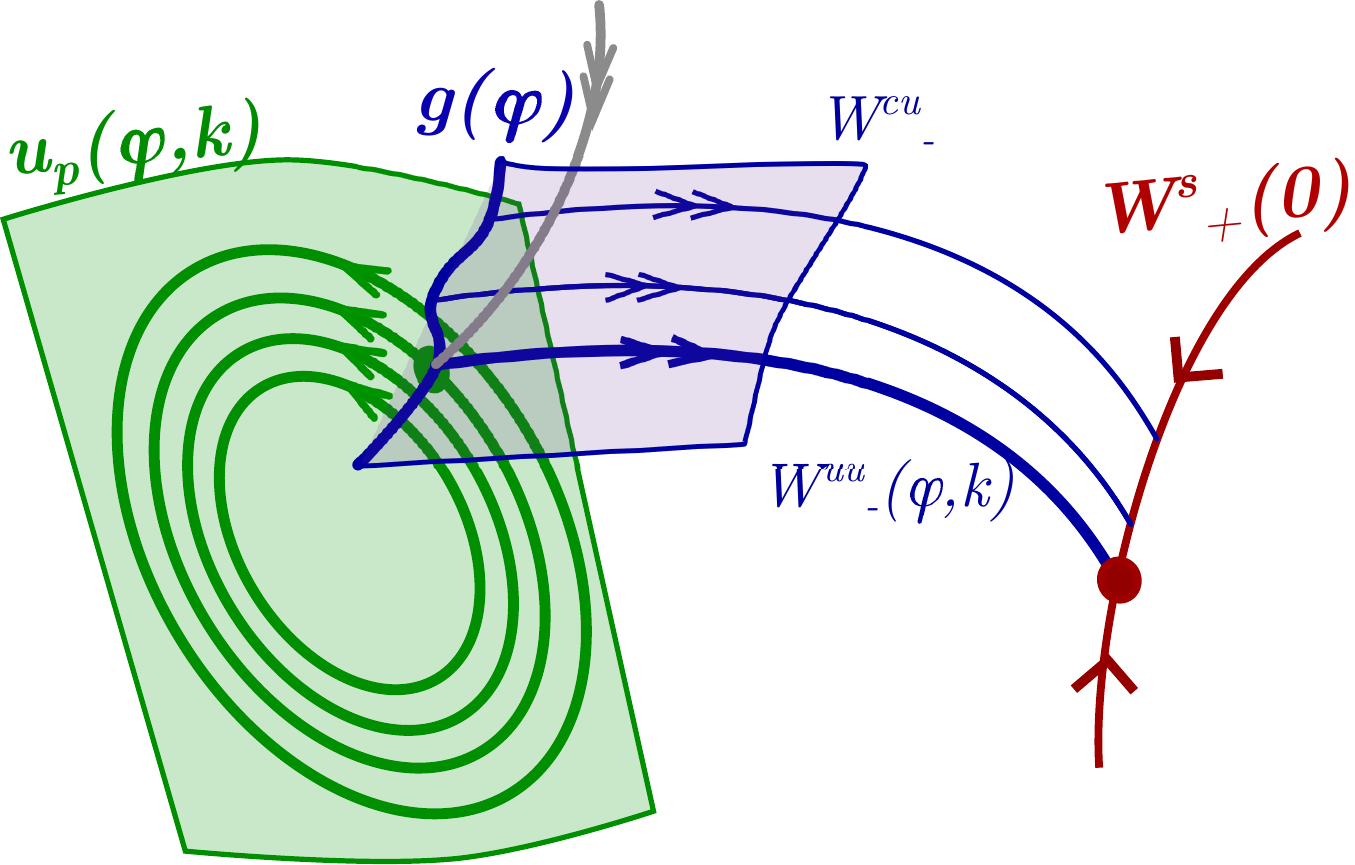}\hspace{-0.05in}
\includegraphics[trim = 0.75cm 0.3cm 1.3cm 1cm,clip,width=0.28\textwidth]{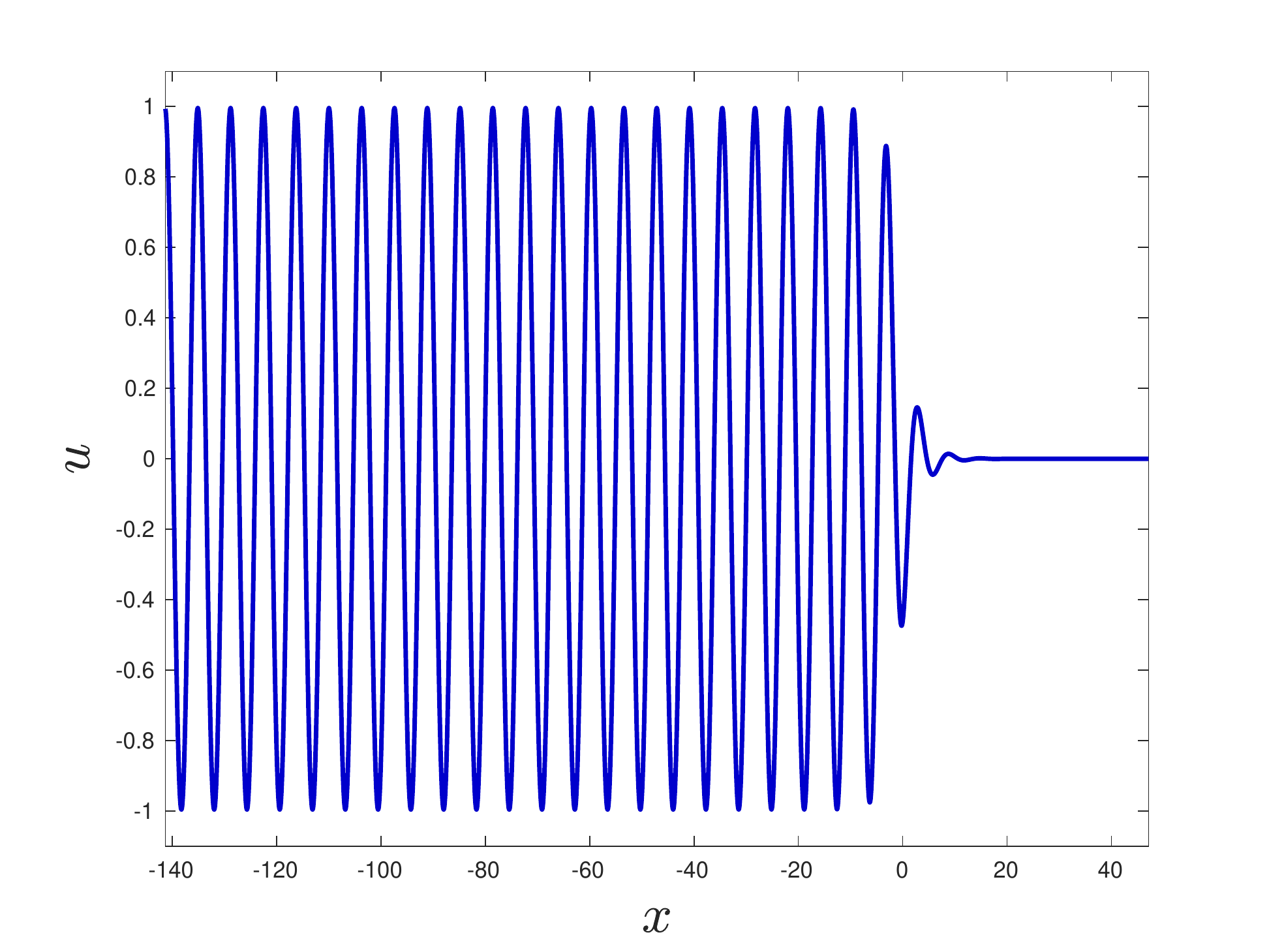}
\caption{Left: plot of strain-displacement relation \eqref{e:sh-sd} for $\mu = 3/4$; Center: schematic phase-portrait of heteroclinic intersection in \eqref{e:mtw-x}, adapted with permission from \cite[Fig. 2(a)]{beekie}. Copyrighted by the American Physical Society; Right: corresponding plot of the solution.   }\label{f:stdisp}
\end{figure}

Setting $k_y=0$,  \eqref{e:mtw-y}  reduces to a non-autonomous Hamiltonian ODE
\beq\label{e:mtw-x}
0 = -(1 + \p_x^2)^2 u + \rho(x) u - u^3.
\eeq
Quenched fronts can then be studied using \emph{spatial dynamics}, where one views this equation as a non-autonomous dynamical system with evolutionary variable $x$. Since  $\rho$ is a step-like function, the system is piecewise constant and solutions can be found from separate phase portraits with $\rho = +\mu$ for $x<0$ and with $\rho = -\mu$ for $x>0$. In both portraits, the homogeneous solution $u \equiv 0$ corresponds to an equilibrium point. In the former, stripe solutions $u_p$ within the Eckhaus stable range take the form of hyperbolic periodic orbits with 2-dimensional center-unstable manifold. The union of these manifolds over the wavenumber $k_x$ forms a 3-dimensional manifold, which we denote $W^\mathrm{cu}_-$.  In the latter phase portrait, the equilibrium $u \equiv 0$ is a hyperbolic equilibrium with 2-dimensional stable manifold, denoted as $W^\mathrm{s}_+(0)$. Patterned fronts are heteroclinic orbits that lie in the intersection $W^\mathrm{cu}_-\cap W^\mathrm{s}_+(0)$. 
Intersections in the ambient phase space $\R^4$ are then expected to occur in a one parameter family of distinct orbits, due to the broken translational invariance. From the intersection, orbits are constructed flowing the intersection point  backwards and forwards in $x$ using the $\rho = \pm \mu$ flows respectively; see Figure \ref{f:stdisp}. 
Generically, intersections can be parameterized by base points in the $W^\mathrm{cu}_-$, that is, asymptotic wavenumber $k_x$ and phase $\phi$, that is, $|u_*(x) - u_p(k_x x + \phi;k_x)|\rightarrow 0$ , as $x\rightarrow-\infty$. The one-dimensional intersection then gives a curve in $k_x-\phi$-space, which is referred to as a \emph{strain-displacement relation} relation; see \cite{morrissey} for more details and rigorous proofs.
In the specific case of the Swift-Hohenberg equation and small $\mu\gtrsim 0$, $k_x = g_\mathrm{sd}(\phi)$ for some $2\pi$-periodic function $g$ \cite{weinburd}, with wavenumber-selecting strain-displacement relation
\begin{equation}
\{(\phi,k_x)\,|\, k_x = g_\mathrm{sd}(\phi)\}\subset \mathbb{R}/2\pi\mathbb{Z} \times \R \label{e:strdis},
\end{equation}
but different $x$-dependence or boundary condition at $x = 0$ can lead to more complex dependence between $k_x$ and $\phi$ \cite{morrissey}. At small $\mu$, normal form and center manifold theory were used to rigorously establish heteroclinics in \eqref{e:mtw-x} with  leading-order expansion for the strain-displacement curve
\beq\label{e:sh-sd}
g_\mathrm{sd}(\phi) = 1 + \frac{\mu\cos(2\phi)}{16} + \mc{O}(\mu^{3/2}).
\eeq
Note that as a consequence, the  quenching interface restricts the set of possible selected wavenumbers to  $k\in(k_\mathrm{sd,min},k_\mathrm{sd,max})$ with
$$
\qquad\quad k_\mathrm{sd,min}:=\min_\phi g_\mathrm{sd},\,\, k_\mathrm{sd,max}:=\max_\phi g_\mathrm{sd}, \qquad  k_\mathrm{sd,max/min} = 1 \pm \frac{\mu}{16} + \mc{O}(|\mu|^{3/2}),
$$ 
an $\mc{O}(\mu)$-width band well within the much wider $\mc{O}(\sqrt{\mu})$-existence and Eckhaus-stability regions; see \cite{morrissey} for various boundary conditions, \cite{monteiro20} for an alternate rigorous approach, and  \cite{beekie} for other prototypical systems. To conclude, we remark that in $(c_x,k_y,k_x)$ parameter space, this family of solutions traces out a vertical line protruding out of the main surface of the moduli space at $(c_x,k_y) = (0,0)$, see Fig. \ref{f:mod}. 

\subsubsection{Oblique stripes, $k_x,k_y \neq 0$}\label{sss:st-ob}
For $c_x = 0$, we now turn to oblique stripes with $k_x,k_y\neq0$ in \eqref{e:mtw-y}. Quenched fronts now solve an elliptic PDE, so that the type of shooting arguments described in the case $k_y=0$ are not readily available. An analysis near $\mu\gtrsim 0$ could however rely on reducing to center manifolds, separately for $x>0$ and $x<0$, and normal form theory as performed in \cite{wuscheel} to mimic the analysis in \cite{weinburd}; see for instance \cite{gs3} for a related situation. One obtains coupled equations for amplitudes of modes that are compatible with the pre-imposed periodicity in $y$. In normal form and at leading order, one expects to be able to set all amplitudes to zero except for the amplitude of a single oblique mode, which can then be analyzed as in \cite{weinburd}. The normal form symmetry on this mode is however an exact symmetry, induced by $y$-translations, suggesting that only a single wavenumber is selected by the interface. Without attempting such an analysis, we present here a rationale for the selection of energy-minimizing strain, $k=k_\mathrm{zz}$, following the reasoning in \cite{lloydscheel}. 



We write \eqref{e:mtw-y} as a first-order system for $\overline{u} = (u,u_1,v,v_1)^T$ in $x$ and find
\begin{align}
u_x &= u_1,\notag\\
u_{1,x}&= v - (1 + k_y^2 \partial_y^2)u\notag\\
v_x&= v_1\notag\\
v_{1,x}&= - (1 + k_y^2 \partial_y^2)v + \rho u - u^3.
\end{align}
This defines an ill-posed Hamiltonian equation in the phase space $Y = H^3(\mb{T})\times H^2(\mb{T}) \times H^1(\mb{T}) \times L^2(\mb{T})$. Using the standard skew-symmetric matrix
$$
J = \left(\begin{array}{cc} 0 & J_2 \\J_2 & 0 \end{array}\right),\qquad J_2 =  \left(\begin{array}{cc} 0 & 1 \\-1 & 0 \end{array}\right),
$$
and the Hamiltonian and symplectic structure,
$$
H[ \overline u] = \int -\frac{v^2}{2} + u_1 v_1 + v (1+k_y^2\partial_y^2)u -\frac{\rho u^2}{2}+\frac{u^4}{4} dy,\qquad
\omega(\overline{u},\overline{v}) = \int \overline u\cdot (J \overline{v}) dy,
$$
this system can be written as
\begin{equation}
\overline{u}_x = J\nabla_{L^2} H[\overline{u}].
\end{equation}

Since \eqref{e:mtw-y} is invariant under translations $y\mapsto y+\theta$, Noether's theorem yields an associated conserved quantity, which we refer to as the momentum,
\beq\label{e:conS}
S[\overline{u}] = - \int u v_{1,y} + v u_{1,y} dy,\qquad J\nabla_{L^2} S[\overline{u}] = \partial_y \overline u.
\eeq
Thus $\frac{d}{dx} H[\overline{u}] = 0$ on $x<0$ and $x>0$, and $\frac{d}{dx} S[\overline{u}] = 0$ for all $x\in\R$ along solutions of \eqref{e:mtw-y}. Setting $\rho = \mu$, one can evaluate these quantities on a pure stripe solution $u_p(k_x x + y;k)$, obtaining
\begin{align}
H(k):= H[u_p(k_x x + y;k)] &= \int_y \left(\frac{k_y^4}{2} - \frac{3 k_x^4}{2} - k_x^2 k_y^2\right)(u_p'')^2 + (k_x^2 - k_y^2)(u'_p)^2  -\frac{\mu u_p ^2}{2} +\frac{u_p^4}{4}dy,\notag\\
S(k):= S[u_p(k_x x + y;k)] &= 2 k_x k_y \int_y k^2 (u_p'') - (u'_p)^2 dy.
\end{align}
Next, one uses that the zigzag critical mode $k = k_\mathrm{zz}$ minimizes the stripe free-energy
\beq\label{e:Ek}
\mc{E}(k):= \frac{1}{2\pi}\int_0^{2\pi} \frac{1}{2}\lp[ (1+k^2\partial_\theta^2)u_p\rp]^2 - \frac{\mu u_p^2}{2}+\frac{u_p^4}{4}d\theta,
\eeq
to conclude that that $S = \int_y |k|^2(u_p'')^2 - (u_p')^2 dy = 0$ precisely for $k = k_\mathrm{zz}$.

Along heteroclinic solutions in the heterogeneous system with $\rho = -\mu\mathrm{sign}(x)$, we find that the asymptotic condition $u\rightarrow +\infty $ at $x\rightarrow+\infty$ enforces $S[u] = 0$ along the entire heteroclinic solution. Hence, we conclude that any heteroclinic solutions of \eqref{e:mtw-y}, with $k_y\neq0$ and satisfying $u\rightarrow u_p$ as $x\rightarrow -\infty$ and $u\rightarrow 0$ as $x\rightarrow+\infty$ must either select perpendicular stripes with $k_x = 0$  or oblique stripes with zigzag critical mode $k_{x}^2 = k_\mathrm{zz}^2 - k_y^2$. Figure \ref{f:ky0} confirms this observation numerically.

\subsubsection{Perpendicular stripes, $k_x = 0,k_y\neq0$}\label{sss:st-perp} 
Following the lines of the analysis suggested in the oblique case, one can also in this case try to construct fronts within a normal form amplitude approximation also in this case, restricting for instance to solutions that are even in $y$. One then expects an existence band that is bounded above by $k_\mathrm{ex,max} = \sqrt{1+\sqrt{\mu}}$, while the lower boundary of the band, which we denote as $k_{y,\mathrm{psn}}$, is marked by a fold point, where the perpendicular stripes develop a localized anti-phase kink, related to a cross roll instability. Numerical continuation matches these predictions; see Figure \ref{f:perp-slice1} below. 
We comment in more depth on these boundaries in Sections \ref{ss:fg} and \ref{ss:perp-ret}, below, when including positive speeds $c_x>0$.

\subsection{Fast growth and stripe detachment}\label{ss:fg}
For large growth speeds $c_x\gg1$, the quenching interface renders the homogeneous state unstable but the pattern is unable to ``keep up" and invades the now unstable state with a slower speed, so that the unstable state takes up a linearly expanding region in the wake of the interface; compare the right-most plot of Figure \ref{f:dirquen}. The speed with which a pattern invades the unstable state is often referred to as the \emph{spreading} or \emph{free invasion} speed. For $c_x$ above the free invasion speed, the asymptotic wavenumber is fixed and the growth process has little affect on the asymptotic pattern.  When the growth speed is varied below the free invasion speed, patterns catch up with the growth interface and the interaction leads to a change in the asymptotic wavenumber. Hence, in the speed regime just below the spreading speed, one can seek to understand pattern-forming fronts in the quenched system \eqref{e:mtw} as perturbations of the free invasion front in the homogeneous system with $\rho \equiv \mu$. In Section \ref{sss:inv} we briefly discuss front invasion into an unstable state. Section \ref{ss:csel} then gives heuristics on how linear instability information helps predict quenched pattern-formation for $c_x$ just below the invasion speed. In Section \ref{sss:sp-ctr} and \ref{ss:obl}, we respectively discuss dynamical systems and functional analytic approaches to rigorously establishing fronts in this speed regime.

\subsubsection{Free front invasion into an unstable state} \label{sss:inv}
A wealth of results on pattern-forming invasion into an unstable state exist for various mathematical models. This includes heuristically and rigorously derived predictions for the invasion speed and asymptotic wavenumber with which a pattern invades \cite{vS,holz}. In the supercritical Swift-Hohenberg equation, the spreading speed of a pure striped pattern with a fixed vertical period can be predicted using \emph{only} the linear information near the homogeneous unstable state \cite{vS}, that is, using only the linearized equation,
\beq\label{e:shl}
v_t = Lv:=-(1+\partial_x^2+\partial_y^2)^2 v + \mu v.
\eeq 
In this case, where the linear growth ahead of the patterned state determines the invasion, the front is sometimes referred to as a \emph{pulled front}. We shall outline how to determine these linear predictions below, but refer to \cite{vS} for a general phenomenological overview, and \cite{holz} for a rigorous derivation and study of these speeds in linear systems. We also remark that if the supercritical nonlinearity $f(u) = u - u^3$ is replaced with a subcritical nonlinearity $f(u) = u + \gamma u^3 - u^5$ for $\gamma>1$, nonlinear growth accelerates the front faster than predicted by linear information \cite{pp_transition}. Such fronts are generally called \emph{pushed fronts} and their interaction with a quenching interface is discussed in Section \ref{ss:cqsh} below. 


\paragraph{Linear speeds, pinched double roots, and marginal stability criteria.}

With a focus on pulled fronts, we now derive predictions for speeds and selected wavenumbers from the linearized equation \eqref{e:shl}. Retaining the information on periodicity in $y$, we subsitute an ansatz  $v(\tl x,y,t) = e^{\ri k_y y} \tl v(\tl x,t) $, which yields
\beq\label{e:shlc}
\tl v_t =  L(k_y,c) \tl v:= -(1+\p_{\tilde x} ^2 -k_y^2)^2 \tl v + \mu \tl v + c \p_{\tilde x} \tl v.
\eeq
Following for instance the narrative in \cite{holz}, one defines a linear spreading speed  $c_\mathrm{lin}(k_y)$ as the supremum of speeds $c$ for which localized initial conditions to \eqref{e:shlc} do not decay \emph{pointwise}, or, equivalently, for which $\sup_{|x|<1} |\tl v(t,x)|\to \infty$ for $t\to\infty$. The frame moving with $c_\mathrm{lin}$ then tracks the leading edge of the spatio-temporally growing instability. The spreading speed can also be thought of as a marginal stability criterion, and the selection of fronts can be phrased more generally as a marginal stability selection; see \cite{avery2022} for a more comprehensive discussion and results towards such a general selection criterion. 

One determines pointwise growth rates from the complex dispersion relation, obtained with the ansatz $\tl v(\tl x,t) = \re^{\lambda t + \nu \tl x}$ as $0 = d(\lambda,\nu;k_y,c) = -(1+\nu^2 - k_y^2)^2 + \mu + c\nu - \lambda$. 

A stationary phase approximation gives pointwise exponential growth rates $\re^{\lambda_\mathrm{br} t}$ through the location of double roots  $(\nu_\mathrm{br},\lambda_\mathrm{br})$ of the dispersion relation, which satisfy 
\begin{align}
0 = d(\lambda_\mathrm{br},\nu_\mathrm{br};k_y,c),\qquad 0 = \p_\nu d(\lambda_\mathrm{br},\nu_\mathrm{br};k_y,c),\label{e:msc}
\end{align}
along with a ``pinching"-condition; see \cite{holz}. Spreading speeds are then obtained as $c_\mathrm{lin}(k_y) =  \sup\{ c\,:\, \mathrm{Re}\lambda_\mathrm{br}(c) \geq 0\}$. At the spreading speed, marginal stability implies that at the leading edge of the instability, one observes oscillations with frequency $\omega_\mathrm{lin}=\Im\lambda_\mathrm{br}$. Assuming a 1:1 resonance between these oscillations in the leading edge and the pattern laid down in the wake, a property sometimes referred to as node conservation, one then predicts a wavenumber $k_{x,\mathrm{lin}}=\omega_\mathrm{lin}/c_\mathrm{lin}$. 

Dependence of $c_\mathrm{lin}$ on $k_y$ is quite generally monotonically decreasing in isotropic systems \cite{holz}. In the case of the Swift-Hohenberg equation \eqref{e:shlc}, one finds explicitly \cite{avery2019growing, vS}
\begin{align}
 c_\mathrm{lin}(k_y)&=\left\{
\begin{array}{ll}
\frac{4 \left(2-2 k_y^2+\sqrt{1-2 k_y^2+k_y^4+6 \mu}\right) \sqrt{-1+k_y^2+\sqrt{1-2 k_y^2+k_y^4+6 \mu}}}{3
\sqrt{3}},& 0<k_y<  \sqrt{\frac{2+ \sqrt{3\mu}}{2}}\\
\frac{4 \sqrt{-1+k_y^2-\sqrt{4-8 k_y^2+4 k_y^4-3 \mu}} \left(-2+2 k_y^2+
\sqrt{4-8 k_y^2+4 k_y^4-3 \mu}\right)}{3\sqrt{3}},& \sqrt{\frac{ 2+\sqrt{3\mu}}{2}}<k_y<\sqrt{1+\sqrt{\mu}},
 \end{array} \right.\label{e:clinsh}\\
 k_{x,\mathrm{lin}}(k_y)&=\left\{
 \begin{array}{ll}
 \frac{3 \left(3-3 k_y^2+\sqrt{1-2 k_y^2+k_y^4+6 \mu}\right)^{3/2}}{8 \left(2-2 k_y^2+\sqrt{1-2 k_y^2+k_y^4+6
\mu}\right)},& 0<k_y< \sqrt{\frac{2+ \sqrt{3\mu}}{2}}\\
0, &\sqrt{\frac{2+ \sqrt{3\mu}}{2}}<k_y<\sqrt{1+\sqrt{\mu}}. 
 \end{array}\right.\label{e:klinsh}
\end{align}
%
%
%
%
%
%
Note in particular the change to $k_x=0$  for wavenumbers $k_y>k_{y,\mathrm{po}} = \sqrt{\frac{2+ \sqrt{3\mu}}{2}}$, indicating a selection of perpendicular stripes for those larger values of $k_y$, in contrast to the selection of oblique stripes for smaller $k_y$. 

Patterns in fact "detach" for growth speeds $c_x>c_\rlin$, so that  the piecewise-smooth curve 
$$
\{(k_y,c_{x,\rlin}(k_y),k_{x,\rlin}(k_y))\,,\, k_y\in(0,\sqrt{1+\sqrt{\mu})}\}
$$
gives the upper boundary, in $c_x$, of the moduli space $\mc{M}$; see Fig. \ref{f:kxabs} for a comparison of this algebraic prediction of the boundary with numerical results for a range of $k_y$ values.

\paragraph{Essential and absolute spectra.}
We comment briefly on a complementary aspect of the transition between pointwise growth and decay, often discussed as a distinction between absolute and convective instability, based on spectral properties of the linearization \cite{Scheel00}. 
Since the linearization $L$ has constant coefficient, its  $L^2(\R)$-spectrum consists entirely of  the essential spectrum,
$$
\sigma_{ess,L^2}(L):=\{\lambda\,:\,\mc{L} - \lambda \text{ is not Fredholm index 0 in $L^2$}\}= \{ \lambda \,:\, \mathrm{Re} \,\nu_j(\lambda) = 0,\text{ for some j = 1,...,4} \},
$$ 
where the $\nu_j(\lambda)$ are the roots of the dispersion relation $d(\lambda,\nu)$ for fixed $\lambda$.

For a fixed frame speed $c_x>0$, instabilities that are convected towards $x = -\infty$ can be identified by posing $L$ in an exponentially weighted space $L^2_{\eta,<}(\R)$ defined through the weighted norm 
\[
\|u\|_{L^2_{\eta,<}}^2:=\int_\R |\re^{-\eta\xi} u(\xi)|^2 d\xi.
\]
Since multiplication by the weight provides an isomorphism to $L^2$, we find that the essential spectrum in the weighted space is given by 
$$\sigma_{L^2_{\eta,>}}(L):=\{\lambda\,:\, \mathrm{Re}\,\,\nu_j(\lambda) = \eta,\text{ for some j = 1,...,4} \}.
$$ 
Clearly, $\Re\, \sigma_{L^2_{\eta,>}}(L)<0$ for some $\eta$ implies pointwise exponential decay for the linear equation. This leads to characterizing an in stability as "exponentially convective" if there is an exponential weight so that the spectrum is stable in this weighted norm. 

A weaker characterization of convective stability is based on the notion of \emph{absolute spectrum} \cite{Scheel00,rss}. One therefore orders roots $\nu$ to $d(\lambda,\nu)$ with respect to real part, $\Re\,\nu_1\leq \ldots\leq \Re\,\nu_{2N}$ (with $N=2$ in the case of the Swift-Hohenberg equation), and $\Re\,\nu_N<0<\Re\,\nu_{N+1}$ for $\Re\,\lambda\gg 1$. We then define the absolute spectrum as
\begin{equation}\label{e:abs}
  \sigma_\mathrm{abs}=\{\lambda\,|\,\Re\,\nu_N(\lambda)=\Re\,\nu_{N+1}(\lambda)\},
\end{equation}
where we assume that the $\nu_j$ are ordered by real part for all $\lambda$. Clearly, for any $\lambda$ not in the absolute spectrum, there is a weight $\eta$ so that $\lambda$ does not belong to the essential spectrum in this weighted space, with a consistent number of roots $\nu$ to the left and right of $\eta$. This implies for instance stability in arbitrarily large bounded domains \cite{Scheel00}, leading to using stability of the absolute spectrum as a criterion for convective stability. 

The absolute spectrum consists of algebraic curves in the complex plane that terminate in branch points, where $\nu_N=\nu_{N+1}$. Typically, these branch points form the rightmost, most unstable points of the absolute spectrum \cite{rss}, and correspond to pinched double roots introduced above \eqref{e:msc}, so that pointwise stability and stability of the absolute spectrum typically coincide.

\subsubsection{Selected patterns for growth speed $c_x\lesssim c_\mathrm{inv}(k_y)$}\label{ss:csel}
Quenched front solutions in \eqref{e:mtw}, with transverse wavenumber $k_y$ fixed, bifurcate as $c_x$ is decreased below the free invasion speed so that points in the moduli space are bounded in the $(c_x,k_y)$ plane by the curve $c_\mathrm{lin}(k_y)$. We expect this linear mechanism to determine the upper boundary of the moduli space for generic systems where the free invasion front is pulled. We now show a formal calculation which that predicts $k_x(c_x,k_y)$ close to $c_\mathrm{lin}(k_y)$ based on the absolute spectrum. Such a calculation was made rigorous through the construction of quenched fronts in the context of the supercritical complex Ginzburg-Landau equation, see \ref{ss:cgl} and \cite[\S 2]{gs1}, and we expect it also to hold more generically in pattern forming systems near pulled fronts; see also the related work \cite{gs4} for related phenomena in a 1-dimensional Cahn-Hilliard equation with a directional quenching mechanism, albeit with only a compact unstable region $\Omega_t$. 
On the other hand, if the free invasion front is pushed, wavenumber selection in the wake of a quench is more subtle. Typically, non-monotonic front locking behaviors arise and lead to patterns being ``dragged" by the quench for speeds faster than the free spreading speed. An example of this is discussed in Section \ref{ss:cqsh} where quenched fronts are studied in the Swift-Hohenberg equation with a sub-critical cubic-quintic nonlinearity.

In the quenched system, as $c_x$ decreases below $c_\mathrm{lin}(k_y)$, the rest state $u\equiv0$, becomes absolutely unstable, with the absolute spectrum $\sigma_\mathrm{abs}(L)$ crossing the imaginary axis at the complex conjugate branch points $\lambda_\mathrm{br},\overline{\lambda}_\mathrm{br}$.  This crossing indicates that perturbations will grow pointwise in $x<0$, leading to a pattern which grows and saturates the domain. In this sense, this bifurcation can be viewed as a perturbation of the free-invasion front with selected wavenumber $k_{x,\mathrm{lin}}(k_y)$, as $c_x$ decreases below $ c_\mathrm{lin}(k_y)$. It turns out that at leading order, the quenched front oscillates with frequency $\omega_\mathrm{abs}(c)$ given by the intersection of the absolute spectrum with the imaginary axis. Heuristically, this frequency corresponds to temporally neutral oscillations supported by the background state. From this frequency, predictions for the horizontal spatial wavenumber $k_x$ can be determined assuming a  1:1-resonance in the dispersion relation, $\omega = k_x c_x$. 

\begin{figure}[ht]
\centering
\includegraphics[trim = 0cm 0cm 0cm 0cm,clip,width=0.3\textwidth]{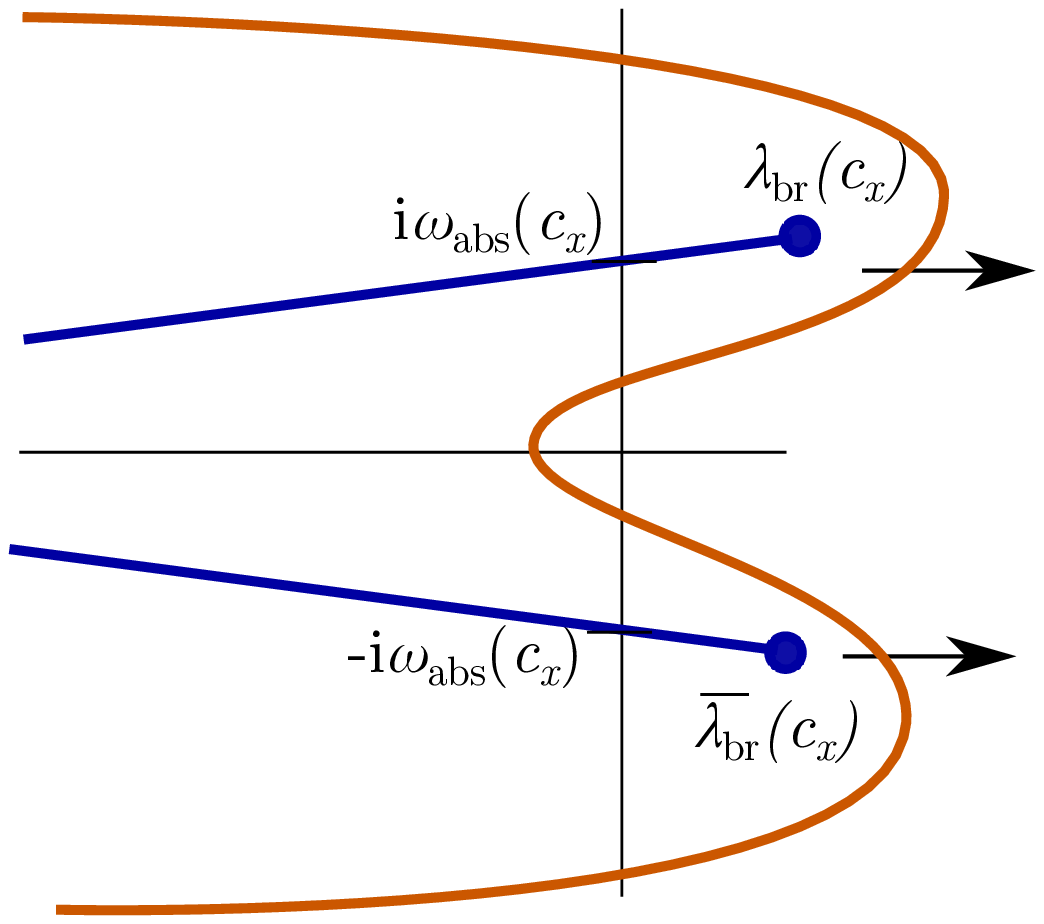}\quad
\includegraphics[trim = 0.5cm 0.0cm 0.5cm 0.5cm,clip,width=0.4\textwidth]{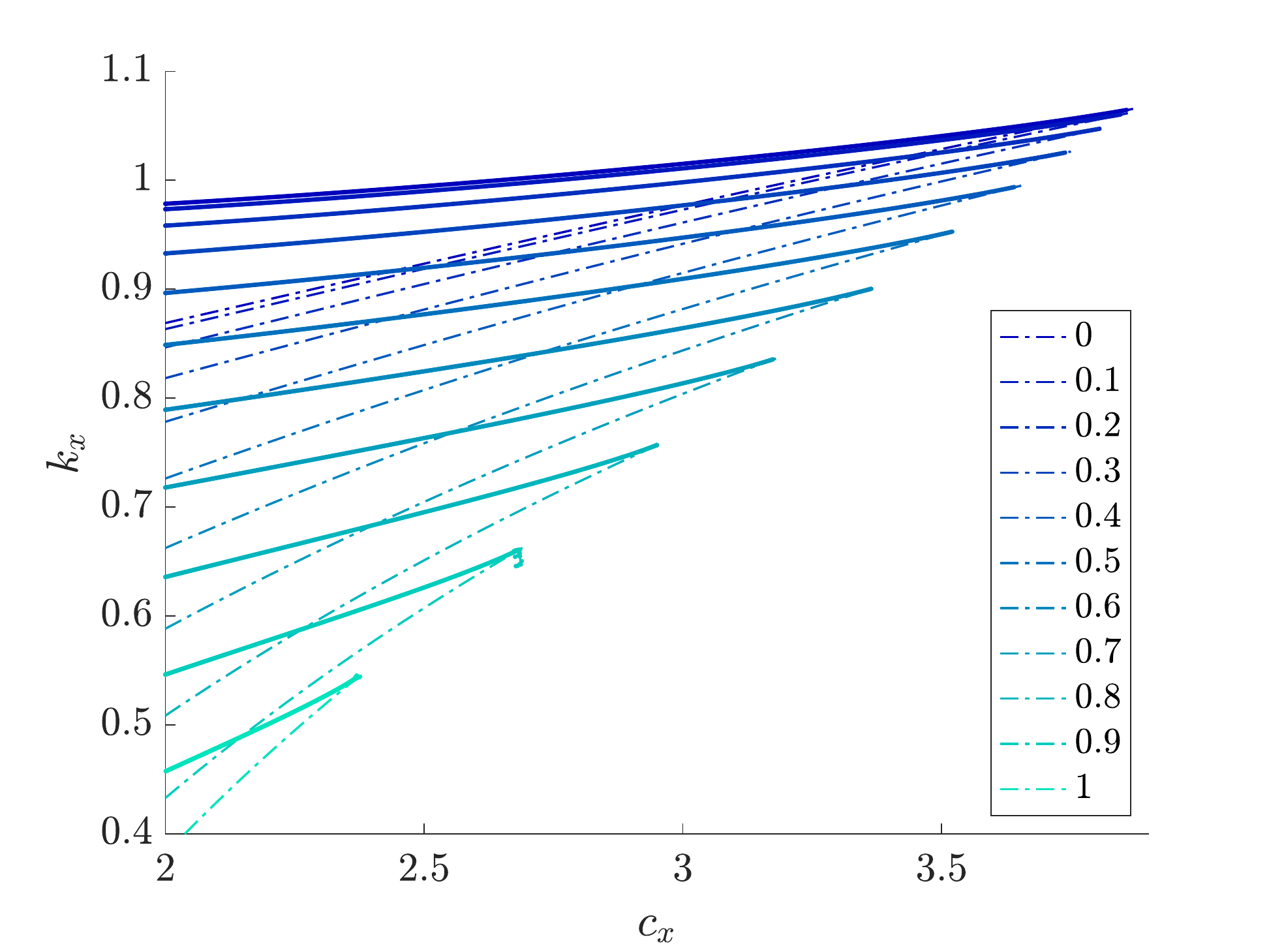}\\
\includegraphics[trim = 0.cm 0.0cm 0.cm 0.cm,clip,width=0.28\textwidth]{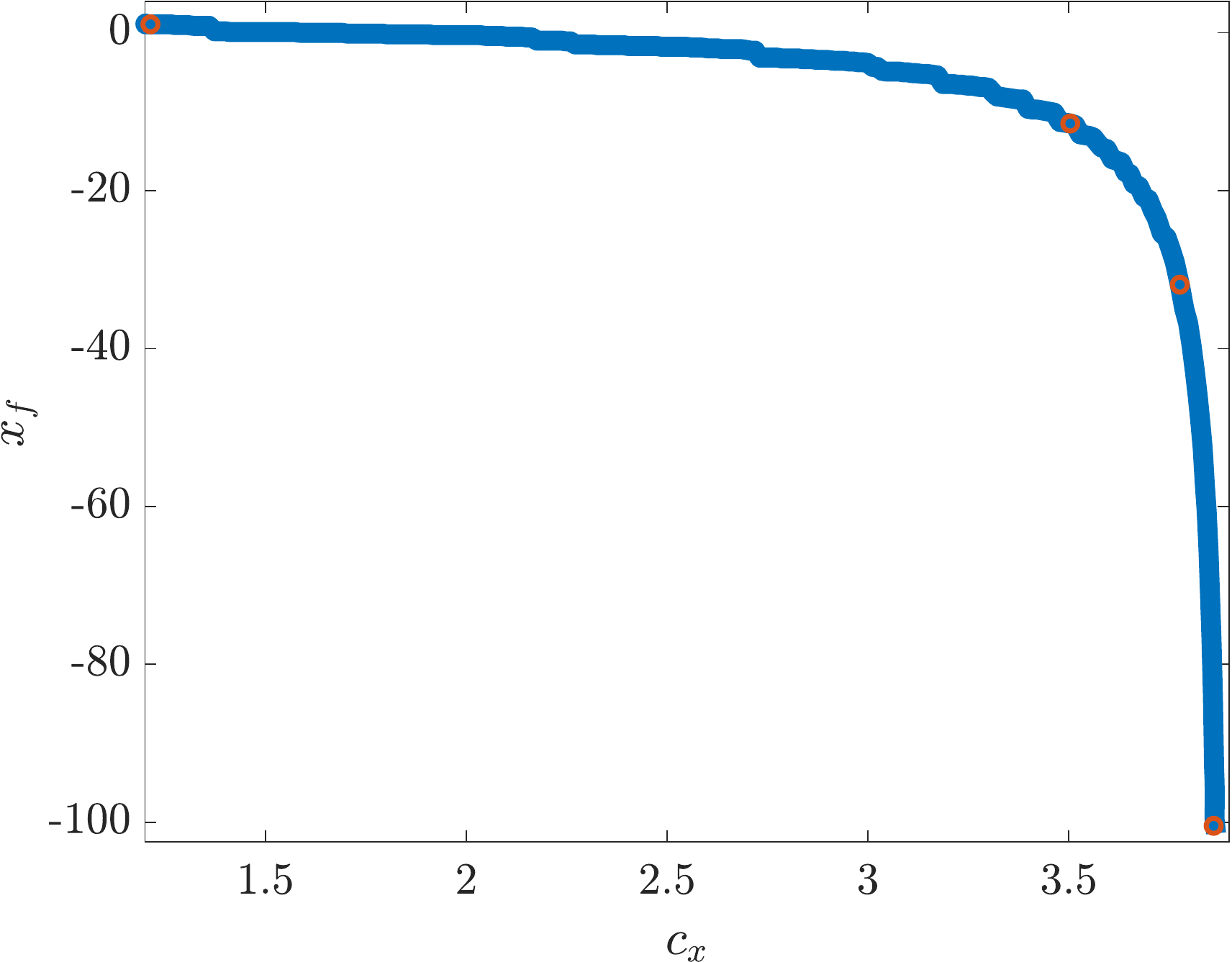}
\includegraphics[trim = 0.cm 0.0cm 0.cm 0.cm,clip,width=0.28\textwidth]{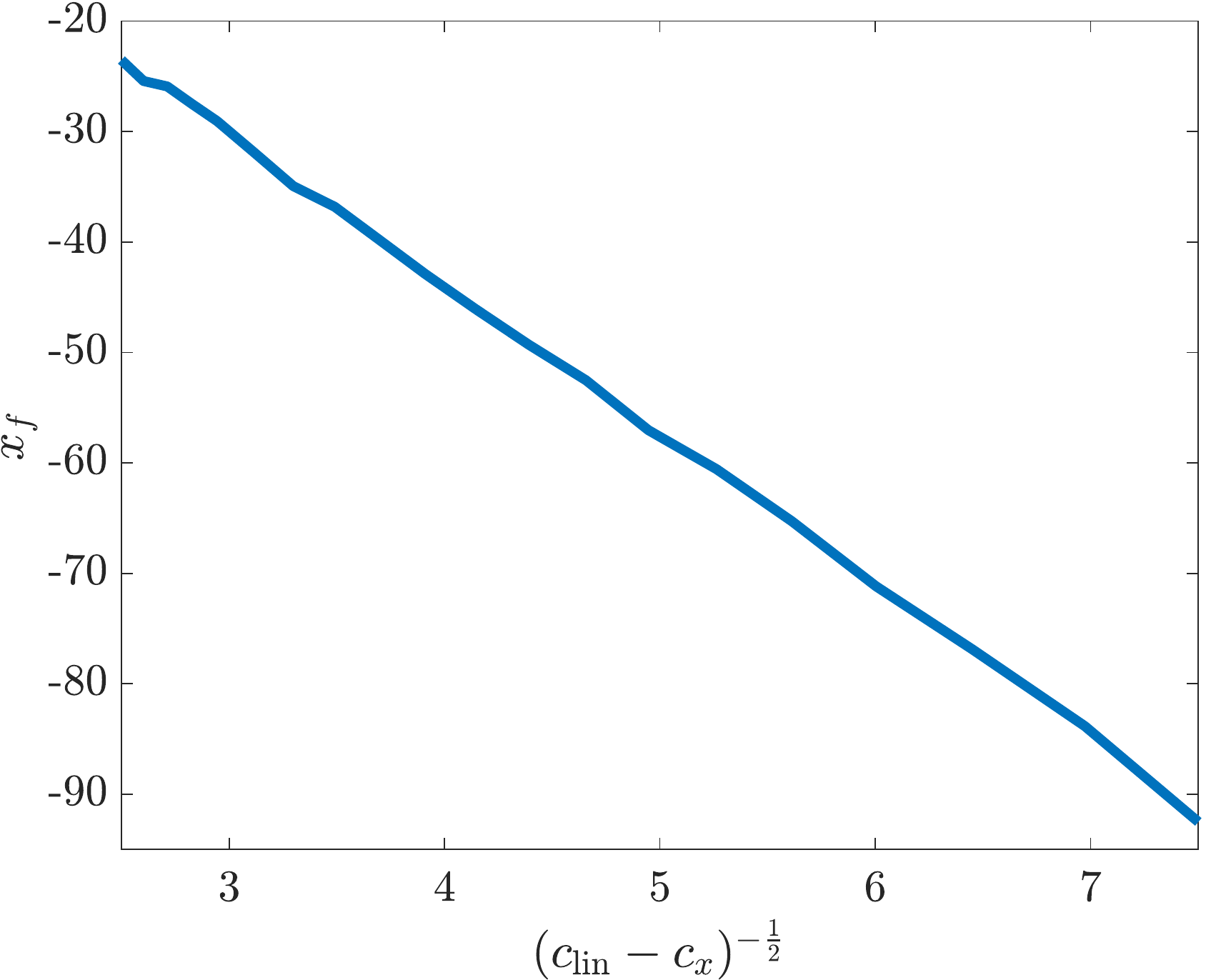}
\includegraphics[trim = 0cm 0cm 0cm 0cm,clip,width=0.32\textwidth]{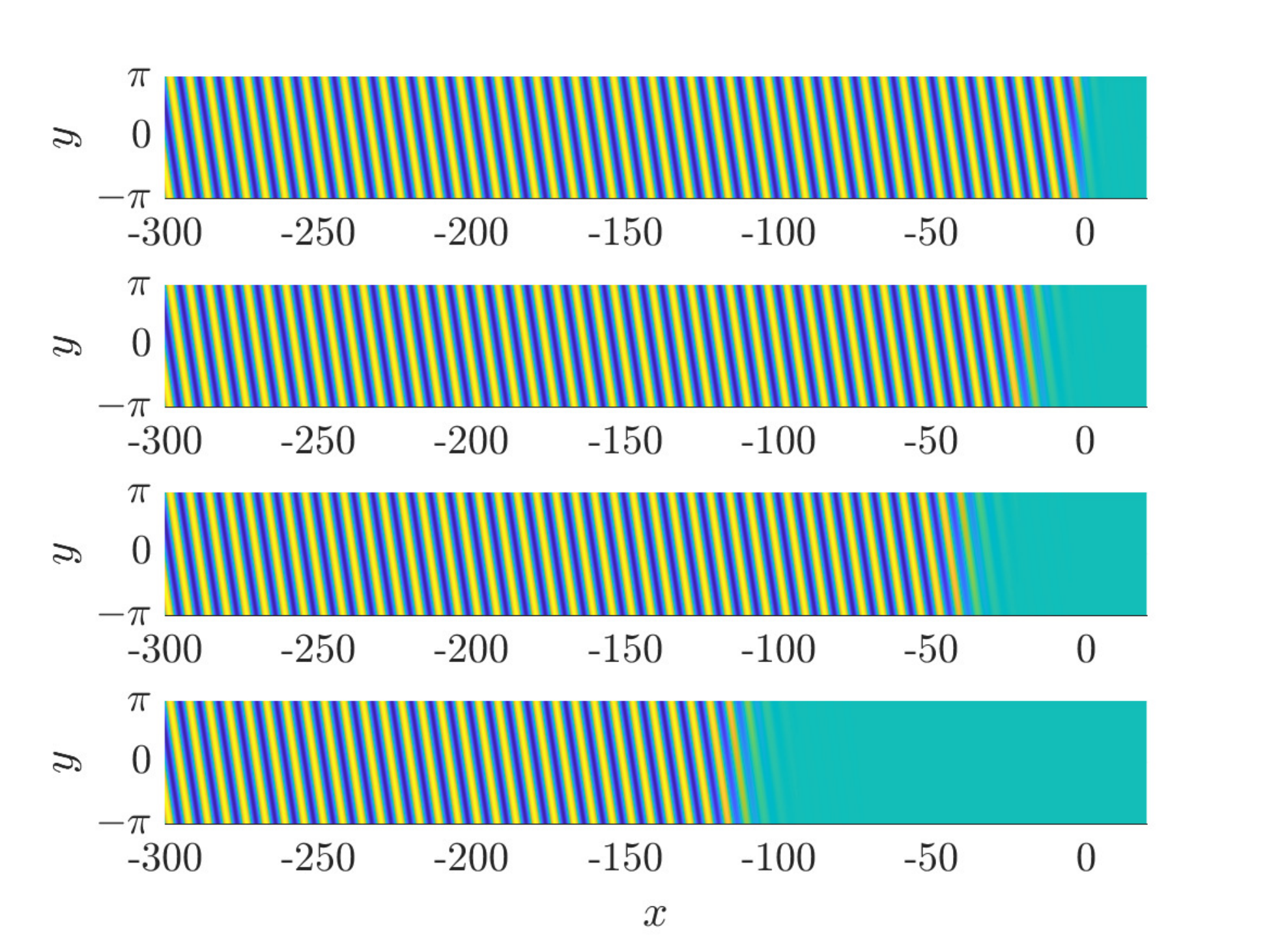}
\caption{Top Left: Schematic of the essential (orange) and absolute spectrum (blue) of $L(k_y,c)$, the linearization about the homogenous rest state with $\rho\equiv \mu$. Dots denote the branch points $\lambda_\mathrm{br}(c),\overline{\lambda}_\mathrm{br}(c)$ which cross the imaginary axis $\ri\R\subset \C$ as $c$ is decreased through $c_\mathrm{lin}$. Top Right: Comparison of selected wavenumbers $k_x(k_y,c_x)$ in \eqref{e:mtw} (solid) near the all-stripe detatchment line $\{(k_y,c_x,k_x)\,:\, (k_x,c_x) = (k_{x,\mathrm{lin}}(k_y),c_{x,\mathrm{lin}}(k_y),k_y\in[0,\fr{\sqrt{2+\sqrt{3\mu}}}{2}\}$  with leading-order predictions (dash-dot) \eqref{e:kxabs} using linear spreading speed and the absolute spectrum, for a range of fixed $k_y$ values (different colors). Bottom left and center: Plot of front interface location $x_f$ against $c_x$ and $(c_\rlin - c_x)^{-1/2}$ for $k_y = 0$; Bottom Right: corresponding solution plots for a few speeds, corresponding to orange dots in bottom left figure. }\label{f:kxabs}
\end{figure}

In order to obtain expansions for this intersection, note that in a neighborhood of $\lambda_\mathrm{br}$, the absolute spectrum generically takes the form of a curve emanating leftwards from each branch point. Hence, for $c_x$ just below  $ c_\mathrm{lin}(k_y)$, curves of absolute spectrum intersect the imaginary axis at unique locations $\pm \ri \omega_\mathrm{abs}(c_x)$ with $\ri \omega_\mathrm{abs}(c_\rlin) = \lambda_\mathrm{br}(c_\rlin)$; see Figure \ref{f:kxabs}. 

We calculate a linear approximation to the intersection $\sigma_\mathrm{abs}(L(k_y,c_x))\cap \ri \R$ by expanding near the branch point for $0<c_\mathrm{inv} - c_x \ll1$. For curves of absolute spectrum near a generic branch point, we solve
$$
\sigma_\mathrm{abs}(L) = \{ \lambda\,:\, 0 = d(\lambda,\nu;k_y,c_x) = d(\lambda,\nu+\ri\gamma;k_y,c_x), \gamma\in \R\},
$$
so that $\sigma_\mathrm{abs}$ consists of curves $\{\lambda_\mathrm{abs}(\gamma;k_y,c_x)\,:\,\gamma\in \R\}$ ending at the branch point when  $\gamma = 0$. Restricting to the specific curve with $\lambda_\mathrm{abs}(0;k_y,c_x) = \lambda_{br}(k_y,c_x)$, 
we expand near $\gamma = 0$,
\begin{align}
\lambda_\mathrm{abs}(\gamma;k_y,c_x) &= \lambda_{br}(k_y,c_x + \frac{\p_\gamma^2\lambda_\mathrm{abs}(0;k_y,c_x)}{2}\gamma^2 + \mc{O}(\gamma^3)\notag\\
&= \lambda_{br}(k_y,c_x)-\frac{\partial_\nu^2 d( \lambda_\mathrm{abs}(0;k_y,c_x),\nu;k_y,c_x)}{2}\gamma^2+ \mc{O}(\gamma^3).
\end{align}
since $\partial_\gamma\lambda_\mathrm{abs}(0;k_y,c_x) = 0$ and $\partial_\gamma^2\lambda_\mathrm{abs}(0;k_y,c_x) = \p_\gamma^2 d = -\partial_\nu^2 d$. Truncating at second-order, the intersection of $\lambda_\mathrm{abs}(\gamma;k_y,c_x)$ with $\ri\R$ is obtained by setting $\gamma_*^2/2 = \mathrm{Re} \lambda_{br}(k_y,c_x)/\mathrm{Re}\partial_\nu^2 d$, so that
$$
\omega_\mathrm{abs}(\gamma_*,k_y,c_x) \approx \mathrm{Im}\,\lambda_{br}(k_y,c_x) - \mathrm{Im}\,\partial_\nu^2 d\lp( \frac{\mathrm{Re} \lambda_{br}(k_y,c_x)}{\mathrm{Re}\partial_\nu^2 d}\rp).
$$
Each of the quantities in the above expression can be explicitly calculated, and the leading-order prediction for the selected wavenumber is thus given as
\beq\label{e:kxabs}
k_{x}(k_y,c_x) =  \omega_\mathrm{abs}(\gamma_*,k_y,c_x)/c_x + \mc{O}(|c_\mathrm{lin}(k_y) - c_x|^2);
\eeq
see Figure \ref{f:kxabs} for a numerical corroboration of this prediction for a range of $k_y$ values.  

We also observe that as $c_x\rightarrow c_\rlin$ from below, the location of the front interface, defined as $x_f = \inf\{ \xi \,:\, |u(x,y)|<\delta,\,\, \text{for}\,\, x>\xi\}$ with $\delta>0$ fixed and small, recedes from the quenching line. In other words, as the quenching speed approaches $c_\rlin$ from below the pattern locks farther and farther away from the quenching interface, leaving a plateau state near $u\equiv0$ in between.  In particular we find that 
\beq\label{e:xf}
x_f(c_x) \sim (c_\rlin - c_x)^{-1/2},\qquad\quad c_x\lesssim c_\rlin,
\eeq
see Figure \ref{f:kxabs} bottom row. This is consistent with the rigorous expansion \eqref{e:kcgl_sel2} of $x_f$ for the complex Ginzburg-Landau equation found in \cite{gs1} and discussed in Section \ref{ss:cgl} below.



\subsubsection{Spatial dynamics formulation and center manifold approach}\label{sss:sp-ctr}
Existence of quenched fronts with $k_y=0$ was rigorously established near onset, $\mu\gtrsim 0$,  for all speeds $c<c_\mathrm{lin}$. 
\begin{Theorem}\label{t:ps}\cite[Thm. 2]{gs3}
Let $k_y = 0$. Then for all $\mu>0$ sufficiently small, there exists a $\delta>0$ such that for all quenching speeds $c_x$ with $c_\mathrm{lin}(0) -\delta \leq  c_x < c_\mathrm{lin}(0)$, there exists a $k_x$ such that \eqref{e:mtw} has a solution. Furthermore, this front is non-degenerate, having linearization $\mb{L}$ which is Fredholm of index 0 with an algebraically simple eigenvalue $\lambda = 0$ when posed in the weighted space $L^2_\eta$ for all $\eta>0$ sufficiently small. 
\end{Theorem}
The theorem is proved using a multiple-scales analysis and a pseudo-center manifold reduction on the spatial dynamics formulation of the problem with ideas originating in \cite{eckmann1991propagating}. We sketch the idea of the proof, here.  

\begin{figure}
\centering
\includegraphics[trim = 0.cm 0.0cm 0.0cm 0.cm,clip,width=0.8\textwidth]{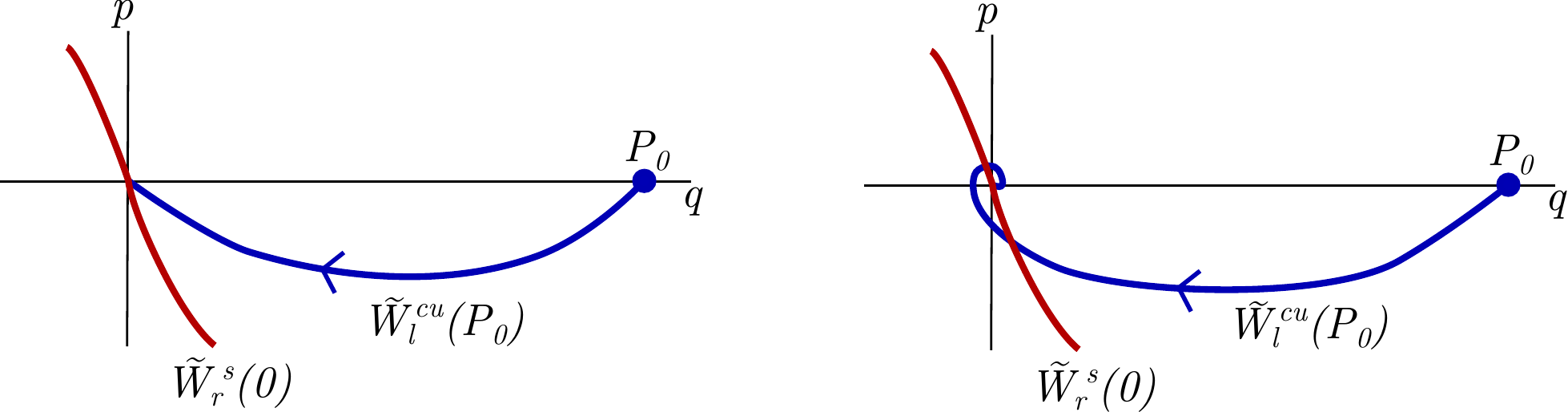}
\caption{Phase portraits in the real subspace of the leading-order equations on the center-manifold \eqref{e:sh-wc} for both $\tl\mu>0$ (blue) and $\tl \mu<0$(red) for different speeds $\tl c$ above and below the free invasion speed. Reproduced with permission from \cite[Fig. 3.2]{gs3}. Copyrighted by John Wiley and Sons.}
\label{f:shctrpp}
\end{figure}

One scales $\mu = \epsilon^2 \tl\mu, c_x = \epsilon \tl c, k_x = 1 + \epsilon \tl \gamma$ and looks for solutions of \eqref{e:sh-q} of the form $u(x,y,t) = W(x-c_x t, x)$ which are $2\pi/k_x$-periodic in the second variable. Note, this is a different, but equivalent, solution ansatz to that of \eqref{e:mtw}.  To construct fronts, one considers the phase-portraits for the $\xi:=x-c_xt > 0$ and $\xi<0$ dynamics separately. Decomposing $W$ into Fourier series in $x$, $W(\xi,x) = \sum_{n\in\Z} W_n(\xi) \re^{-\ri n kx}$, and inserting into the equation with $\rho \equiv \pm\mu$ one obtains
\beq
\lp(-(1+(\partial_\xi - \ri k n)^2)^2 +\epsilon^2 \tl\mu_{\mathrm{r/l} } + \epsilon\tl c \partial_\xi\rp) W_n(\xi) =
\sum_{p+q+r = n} W_p(\xi) W_q(\xi) W_r(\xi),\quad  
\eeq
with $\tl\mu_\mathrm{r/l} = \mp \tl\mu$ for $x\gtrless0$ and some $\tl\mu>0$ fixed.
Writing each equation as a first-order system in $\xi$ and linearizing about $W\equiv0$, the infinite dimensional system decouples into a countable set of four-dimensional complex linear systems each with spectrum determined by the characteristic polynomial
$$
0=p_n^\mathrm{r/l}(\nu) = (\nu - \ri(kn+1))^2(\nu - \ri(kn - 1))^2 - \epsilon \tl c \nu +\tl\mu_\mathrm{r/l} \epsilon^2.
$$
For $\epsilon = 0,$ each linearization has a pair of geometrically simple and algebraically double eigenvalues $\nu = \ri(kn \pm 1)$. Perturbing in $0<\epsilon \ll1$, all eigenvalues $\nu$ move off $\ri\R$ with speed $\mc{O}(\epsilon^{1/2})$ except for the $n = \pm1$ pairs which are $\mc{O}(\epsilon)$,
$$
\nu_\mathrm{r/l}^\pm = \epsilon\lp( \frac{-\tl c\pm \sqrt{\tl c^2-16(\tl\mu_\mathrm{r/l} + \ri \tl c\tl\gamma)})}{8} +\ri \tl \gamma \rp) +\mc{O}(\epsilon^2).
$$
One can then apply Theorem A.1 of \cite{eckmann1991propagating} to obtain local center manifolds $W^c_{\pm}(0)$ which are complex two-dimensional and tangent to the aforementioned $\mc{O}(\epsilon)$ eigenspace.  These manifolds contain the set of bounded solutions near the origin for both the $\mu_\mathrm{l}$ and $\mu_\mathrm{r}$ phase portraits. Strong stable and unstable local foliations of the normal hyberbolic dynamics near the origin collapse the infinite-dimensional dynamics on to $W^{c}_\pm(0)$ so that the desired heteroclinic is determined at leading-order by analyzing the following system for coordinates $(p,q)\in \mb{C}^2$ on the center manifold,
\begin{align}\label{e:sh-wc}
\frac{dq}{d\zeta}&= p + \mathcal{O}(\epsilon),\notag\\
\frac{dp}{d\zeta}&= \frac{1}{4}\lp( \tl \mu\, \mathrm{sign}(\zeta) q - \tl c p + 3q|q|^2 + \tilde \gamma (2\ri p + \tl \gamma q) + \mathcal{O}(\epsilon) \rp).
\end{align}
The origin $(0,0)$ is a hyperbolic equilibrium in the $\zeta>0$ phase portrait, while the $\zeta<0$ phase portrait with $\tl c>0$ has heteroclinic orbits between the family of fixed points $\mathcal{P} = \{ \re^{\ri\theta} (1/\sqrt{3},0)\,:\, \theta\in[0,2\pi) \}$ and the origin. Overlaying these two portraits, a phase-plane analysis shows an intersection of the unstable manifold of $\mathcal{P}$ in the $\zeta<0$-dynamics with the stable manifold of $(0,0)$ in the $\zeta>0$ dynamics for $0<4 - \tl c \ll1$; see Figure \ref{f:shctrpp}.
%
%
An intersection in the full systems is then found by using a Melnikov integral to show that these manifolds are transversely unfolded in the speed $\tl c$ and wavenumber parameter $\tl\gamma$.


\subsubsection{Perturbing parallel fronts}\label{ss:obl}
One could consider the existence problem for oblique stripes, $k_y>0$, fixed, in a fashion similar to Section \ref{sss:sp-ctr}. Difficulties arise however in the limit $k_y\to 0$ for small $\mu$. In order to investigate the regime of small $k_y$ we therefore investigated a weak bending problem, perturbing from a parallel front to find a weakly oblique front in \cite{gs3}, under suitable stability and non-degeneracy conditions that happen to be satisfied for the fronts found in Theorem \ref{t:ps}. 
The perturbation result can be stated as follows.
\begin{Theorem}\label{t:os}\cite[Thm. 2]{gs3}
Suppose there exists a solution $(u_*,k_{x,*})$ of the modulated traveling wave equation \eqref{e:mtw} with $k_y = 0$ for some fixed $c_x>0, \mu>0$. Further, suppose this solution is non-degenerate as in Theorem \ref{t:ps}. Then there exists a family of oblique striped front solutions $(u_\mathrm{tr},k_{x,\mathrm{tr}})$ to \eqref{e:mtw} depending on $k_y\sim0$, sufficiently small, which are $C^2$ smooth in $k_y$ measured in $C_\mathrm{loc}^0(\R\times \mb{T})$. At leading order, the horizontal wavenumber satisfies $k_x(k_y) = k_{x,*} - \frac{b_y}{c_x} k_y^2 + \mc{O}(k_y^4)$, with 
\begin{equation}
b_y := \la -2\partial_y^2 (1+\partial_x^2) u_*,e_*\ra_{L^2},
\end{equation} 
where $e_*$ is a function spanning the cokernel of the linearization $\mb{L}$ about $u_*$ and which satisfies $\la e_*, \partial_y u^* \ra_{L^2_\eta} = 1.$ 
\end{Theorem}

The two main technical challenges in the proof of this result are the presence of neutral essential spectrum  of the $L^2$-linearization of the parallel striped front, and the singular limit $k_y = 0$. The spatial dynamics approach, as described above, has been historically useful to address the former, but leads to difficulties when attempting to addressing the latter.  In particular, one would try to use a variational equation to study the phase space near the unperturbed front and exponential dichotomies to construct perturbed invariant manifolds for $k_y\neq0$ and locate heteroclinic intersections. This becomes difficult as $k_y\neq0$ changes the domain on which asymptotic linearizations are closed densely-defined operators.

The result in \cite{gs3} therefore relies on a functional analytic approach to address these difficulties. One separates the asymptotic behavior from the interfacial dynamics using a \emph{far-field core decomposition} of the front solution 
\beq
u(x,y) = w(x,y) + \chi(x) u_p(k_xx+ y;k),\quad k = \sqrt{k_x^2+k_y^2}. \quad 
\label{e:coreff}
\eeq
The core perturbation $w$ satisfies $w\in L^2_\eta(\R\times\mb{T})$ while $\chi$ is a smooth step function with $\chi \equiv 1$ for $x\leq -d-1$ and $\mathrm{supp}\chi \subset \{x<-d\}$ for some $d>0$ fixed.  This decomposition enforces the desired far-field behavior, controlled by the wavenumbers $k_x,k_y$, while the exponentially localized perturbation $w$ glues the far-field pattern to the asymptotically constant state ahead of the quench. Inserting this ansatz into \eqref{e:mtw} and subtracting off the expression $\chi\lp[-(1+\partial_x^2+k_y^2\partial_y^2)^2u_p + c_x(\partial_x + k_y \partial_y)u_p + \mu u_p - u_p^3\rp]$, which is identially equal to zero, one obtains a nonlinear equation $\mathcal{F}(w;k_x,k_y,c_x) = 0$ for the localized core variable. 

One then sets $w_* = u_* - \chi u_p(k_x x + y;k_x)$ to be the core-perturbation given by the $k_y=0$ front so that $\mathcal{F}(w_*;k_{x,*},0,c_{x,*}) = 0$. The key advantage of substituting an exact solution into the far field is that this equation now is well-posed on spaces of exponentially localized functions, where the linearization is Fredholm, albeit with negative index; see  Appendix \ref{ss:fred} for details on Fredholm indices in this context. One compensates for the negative Fredholm index by viewing the selected wavenumber $k_x$, inserted through the farfield ansatz, as an additional variable. 

To address the singular-limit, an approach similar to \cite{rademacher2007saddle} was used to precondition the nonlinear problem
\begin{equation}
0 = \tl F(w;k_x,k_y,c_x):= \mc{P}(k_x,k_y) \circ F(w;k_x,k_y,c_x)
\end{equation}
where $\mc{P}$ is a Fourier multiplier with symbol $\widehat{\mc{P}}(k_x,k_y):= (-1 - (1-\ell^2 - k_y^2 m^2)+c_x\ri(\ell - k_y m))^{-1}, \quad m\in \Z, \ell\in\R.$ This allows one to obtain sufficient smoothness of $\tl F$ in $(w,k_x,k_y)$ near $(w^*,k_x^*,0)$. Then, the genericity of the front implies that $\p_{k_x}\mathcal{F}\not\in \mathrm{Rg}\, \mb{L}$ so that the joint linearization $\p_{w,k_x}\mc{F}$ is Fredholm index 0 with trivial kernel and thus invertible, allowing one to solve for $(w,k_x)$ in terms of $k_y$. Expanding in $k_y$ then gives the leading order behavior of $k_x$ in $k_y$. 

As a simple consequence, we find that for fixed $c_x>0$ the horizontal wavenumber $k_x$ depends quadratically on $k_y$ in a neighborhood of 0; see Figure \ref{f:kysm} for a numerical depiction of this via fixed $c_x$ cross-sections of $\mc{M}$.




\subsection{Slow speeds and modulational approximations}\label{ss:slow} 
We next consider the slow growth regime $c_x\gtrsim0$ of the moduli space $\mc{M}$. Figures \ref{f:snap}, \ref{f:cxsm}, and  \ref{f:kysm} reveal several qualitatively different regimes for the horizontal wavenumber $k_x(k_y,c_x)$ as $k_y$ and $c_x$ vary. For $k_y = 0$, that is for parallel striped fronts, the wavenumber selection curve $k_x(0,c_x)$ is monotonically increasing in $c_x>0$, with a minimum at $c_x = 0$, equal to  $k_{sd,min}:= \min_\varphi g(\varphi)$, the minimum of the stationary strain-displacement relation.  Next, for $k_y>0$ fixed small, we find $k_x(k_y,c_x)$ is non-monotonic in $c_x$, first decreasing from the zigzag critical wavenumber $k_{x,\mathrm{zz}} = \sqrt{k_\mathrm{zz}^2-k_y^2}$ at $c_x = 0$, reaching a local minimum, and then increasing again.  Alternatively, we also find $k_x(c_x,k_y)$ is non-monotonic in $k_y$ for $c_x$ fixed and small, passing through a series of local maxima and minima as $k_y $ is decreased to 0. For strongly oblique stripes with $k_y\lesssim k_\mathrm{zz}$, we find that the front undergoes a fold bifurcation as $c_x$ is increased, where the solution developes a localized kink (or wrinkle) near the quench interface. The curve of folds in $k_y$ touches down on the $k_x=0$-plane, connecting with purely perpendicular stripes with zigzag critical wavenumber $(k_x,k_y) = (0,k_\mathrm{zz})$; see Figure \ref{f:kinkbub}. Continuing the other direction in $k_y$, this curve of folds collides with the main body of the moduli space leading to a hyperbolic catastrophe. We discuss these various regions in more detail below.

\subsubsection{Parallel stripes, $k_y = 0$, $c_x\gtrsim 0$}\label{sss:slow-para}
\begin{figure}
\centering
\includegraphics[trim = 0.1cm 0.0cm 0.05cm 0.0cm,clip,width=0.36\textwidth]{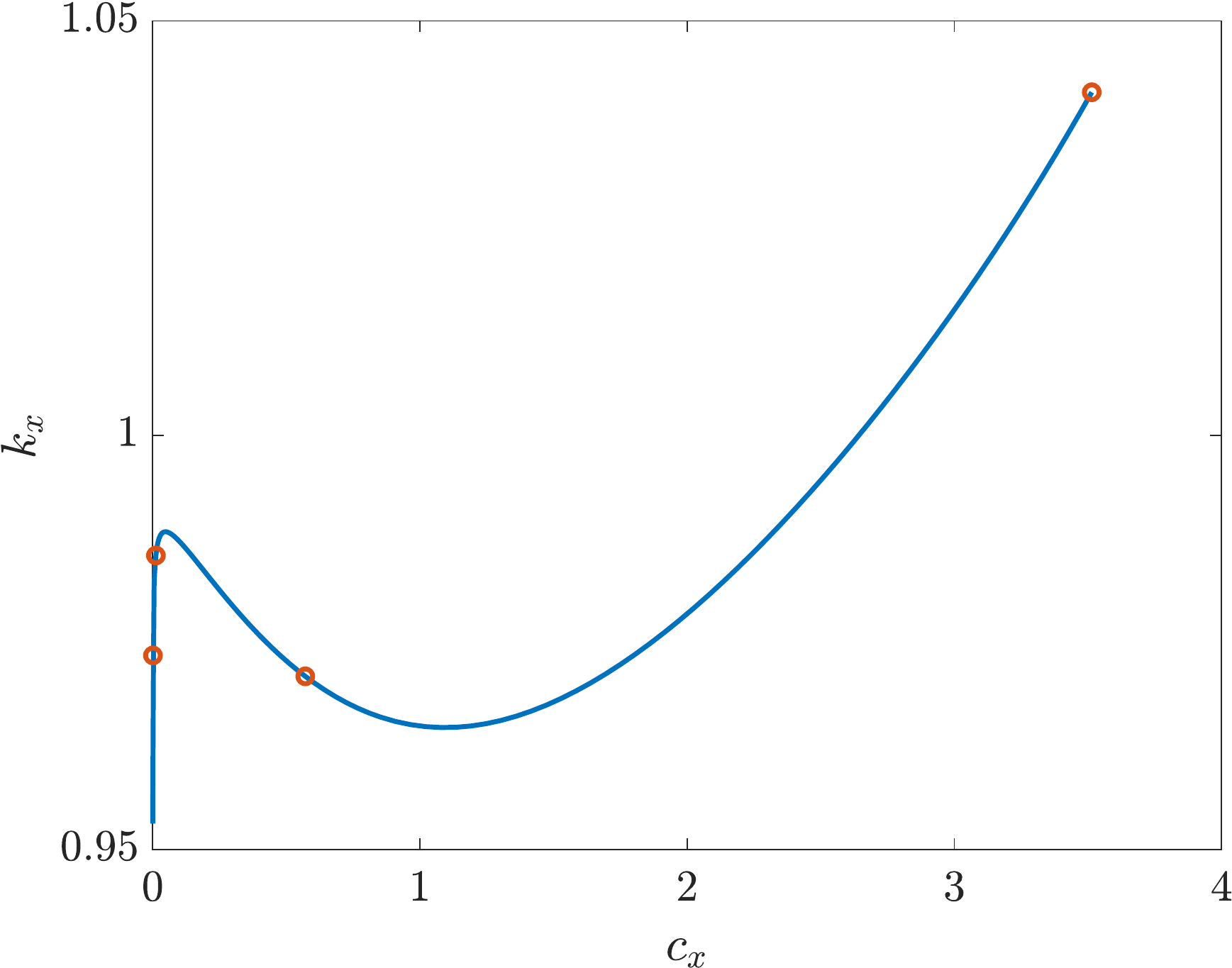}
\includegraphics[trim = 0.1cm 0.0cm 0.5cm 0.5cm,clip,width=0.4\textwidth]{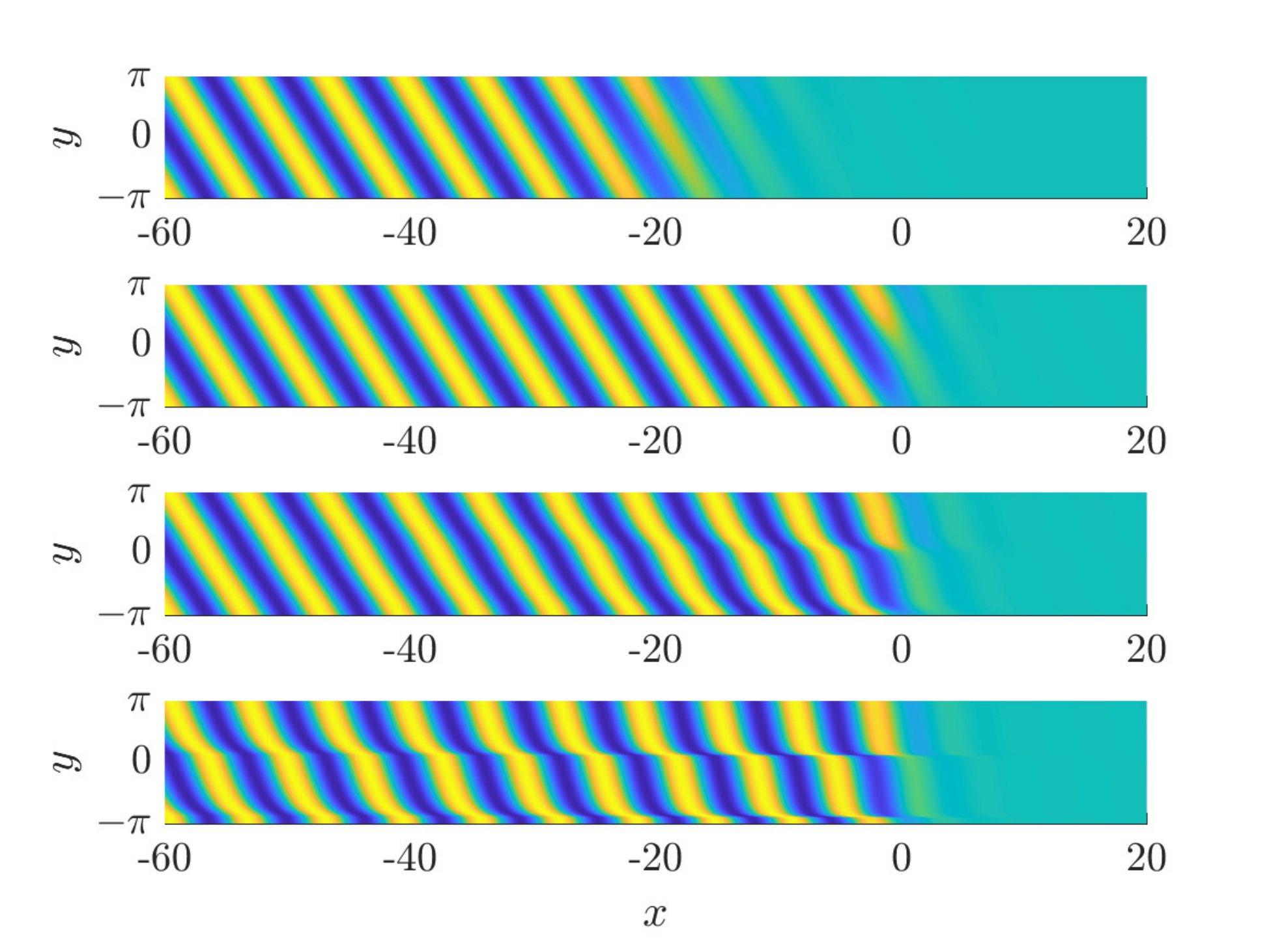}
\caption{ Left: $k_y = 0,c_x>0$ cross-section of $\mc{M}$, for $\mu = 3/4$, depicting singularity at $c_x = 0$ where $k_x(c_x)\searrow k_\mathrm{sd,min}$ as $c_x\searrow0$, end point of the curve gives the upper boundary $c_\mathrm{lin}(0)$; Right: Solution profiles $u(x,y)$ for select points on this curve, decreasing in speed from top to bottom. Recall that $y=\omega t$, $\omega=k_xc_x$ is a scaled variable, so that for small speeds $c_x$,  represents a larger time  interval.} \label{f:snap}
\end{figure}

At zero speed, parallel stripes are compatible with the boundary condition for an interval of wavenumbers determined by the strain-displacement relation, $k\in [k_{sd,min},k_{sd,max}]$. Slowly moving the boundary, one passes through this strain-displacement relation, changing the phase and wavenumber of the pattern and effectively stretching the pattern, until a minimum of the strain-displacement relation is reached. At this point, further stretching is impossible and one sees a snapping event, where a half-period of the pattern is added at the boundary in a process similar to the depinning transition of interfaces between patterned and unpatterned regions \cite{maknobloch}; see Figure \ref{f:snap} for a depiction of these dynamics (depicted in the $y$ direction of the plots).  

This periodic stretch-snap behavior leads to a perturbation in the asymptotic wavenumber of the pattern. When $c_x$ is increased from zero, more energy is inserted into the local phase allowing it to overcome the local pinning effect, leading to a weaker deformation of the asymptotic pattern and hence an increase in the wavenumber from $k_{sd,min}$.   

One can begin to understand these dynamics analytically using a simplifying modulational approximation. In the one dimensional case, since the transverse zigzag instability is suppressed, one finds for $\mu\gtrsim0$ that wavenumbers $[k_\mathrm{sd,min},k_\mathrm{sd,max}]$ lie inside the Eckhaus stability region \eqref{e:eck} so that they are spectrally, linearly, and nonlinearly diffusively  stable \cite{schneider1996diffusive}. Stripe dynamics are well approximated by a phase diffusion modulation equation \cite{dsss}. Most easily, one reduces \eqref{e:sh0} in $n = 1$ with  the parabolic scaling $\mu = \epsilon^2,X = \epsilon x, T = \epsilon^2 t$, and an ansatz $u(x,t) = \epsilon A(X,T) \re^{\ri x} + \mathrm{c.c.}$ at leading order to the  Ginzburg-Landau amlitude equation 
\begin{align}\label{e:RGL0}
A_T = 4 A_{XX} + A - 3A|A|^2.
\end{align}
In polar coordinates $A = R\re^{\ri\tl\phi}$, expanding near $R = 1/\sqrt{3}, \tl\phi = 0$, one obtains a linear phase diffusion equation
\begin{align}
\phi_T = 4\phi_{XX}. 
\end{align}
The quenching term can be modeled by  posing the equation on a half-line $\{X\leq0\}$ in a comoving frame with speed $\tl c_x$, with a mixed nonlinear boundary condition that relates the phase $\phi$ to the local wavenumber $\phi_x$ through the strain-displacement relation \eqref{e:strdis},
\begin{align}\label{e:pd0}
\phi_T &= 4\phi_{XX} + \tl c_x\phi_X, \quad X< 0,\qquad \phi_X = g(\phi),\quad X = 0. 
\end{align}
Here, $2\pi$-periodicity of $g$ implies a discrete gauge-symmetry $\phi\mapsto \phi+2\pi$. Pattern-forming fronts are represented by asymptotically linear profiles which are time-periodic with period $T_p = 2\pi/\omega$ up to this symmetry,
$$
\phi(X,T+T_p) = \phi(X,T) + 2\pi,\qquad |\phi(X,T) - (k_x X - \omega T)| \rightarrow 0, \quad X\rightarrow-\infty, \qquad\omega = \tl c_x k_x.
$$
Such solutions were studied in \cite{beekie}, using a asymptotic inner and outer  expansions in terms of Fourier-Laplace modes.  As a result, one finds a leading-order expansion for the wavenumber selection curve $k_x(\tl c_x)$ for $\tl c_x\gtrsim0$ of the form
\beq
k_x(\tl c_x) = k_{sd,\mathrm{min}} + k_1 \tl c_x^{1/2} + \mathcal{O}(c^{3/4}), \quad k_1 =\zeta(1/2)\sqrt{2 k_{sd,\mathrm{min}}/d_\mathrm{eff}},
\eeq
where $d_\mathrm{eff} = 4$ is the effective diffusivity of phase perturbations of patterns with wavenumber $k = 1$, and $\zeta(s)$ is the Riemann-Zeta function analytically continued onto the critical strip $\mathrm{Re}\, s = 1/2$. Hence, the phase-diffusion approximation shows that the selected wavenumber is smoothly dependent on the square root of the speed $c_x$, with leading-order coefficient dependent on the strain displacement relation and the stability properties of a pure stripe. We also mention that comparison principle type arguments were used to rigorously establish existence and stability of these solutions in \cite{pauthier} but existence of such slowly quenched fronts in the full Swift-Hohenberg equation has not been established.

%
%
%

\subsubsection{Weakly oblique stripes, $c_x\gtrsim0$ and $k_y\in (0, k_\mathrm{zz})$ fixed}\label{sss:slow-ob}
In this regime, we find that wavenumbers $k_x(c_x,k_y)$ depend smoothly on $c_x$. Fixing $k_y$, curves $k_x(c_x)$ limit on the energy minimizing wavenumber $k_x = k_{x,\mathrm{zz}}(k_y)$ as $c_x \to 0$. For non-zero $c_x$, the slow movement of the quench imposes a strain on the striped phase, stretching the pattern, and decreasing the wavenumber. It would be interesting to quantify and interpret this strain through a perturbation analysis. 

\begin{figure}[ht!]
\centering
\includegraphics[trim = 0.0cm 0.0cm 0.0cm 0.0cm,clip,width=0.33\textwidth]{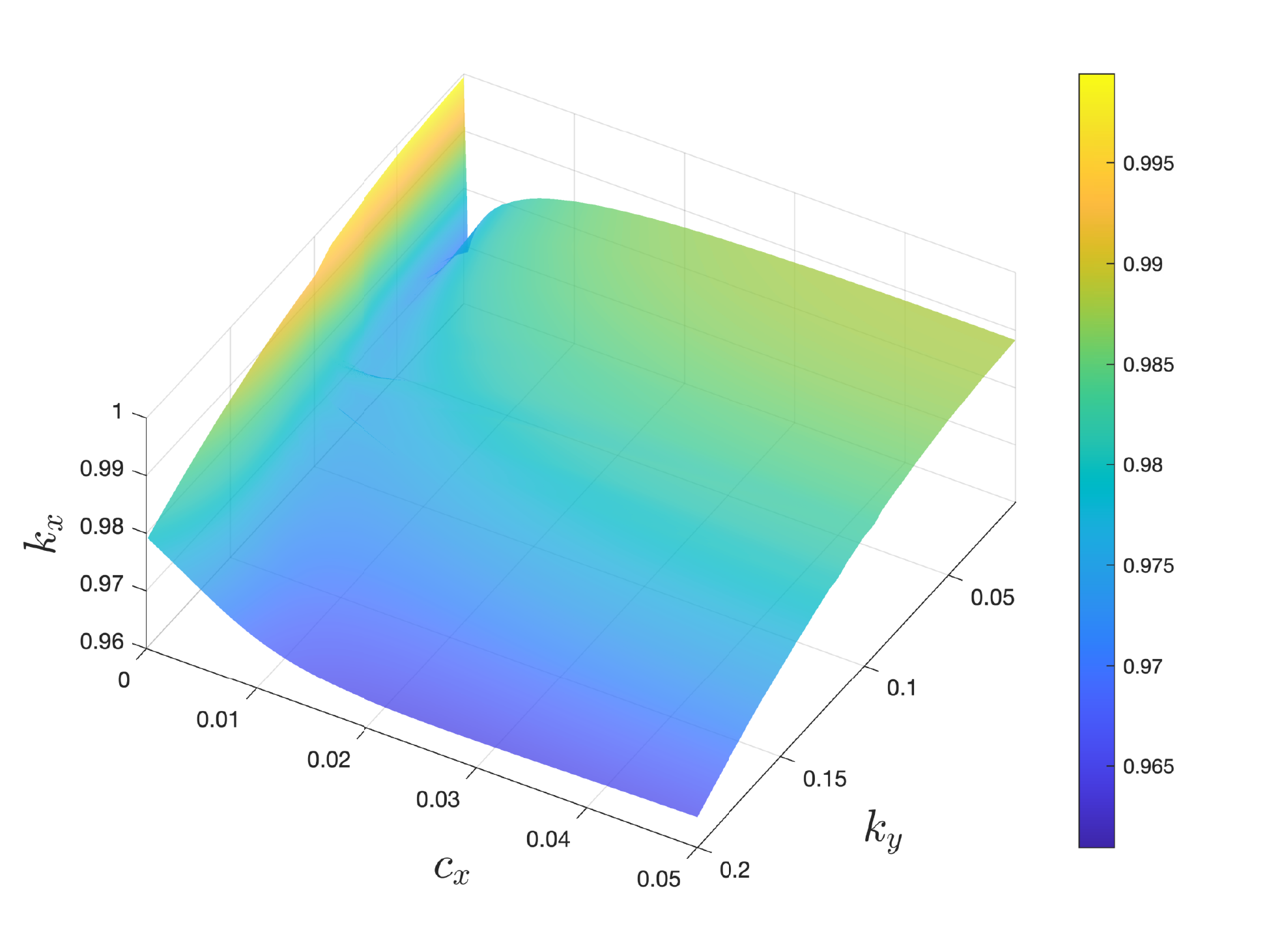}
\includegraphics[trim = 0.0cm 0.0cm 0.0cm 0.0cm,clip,width=0.33\textwidth]{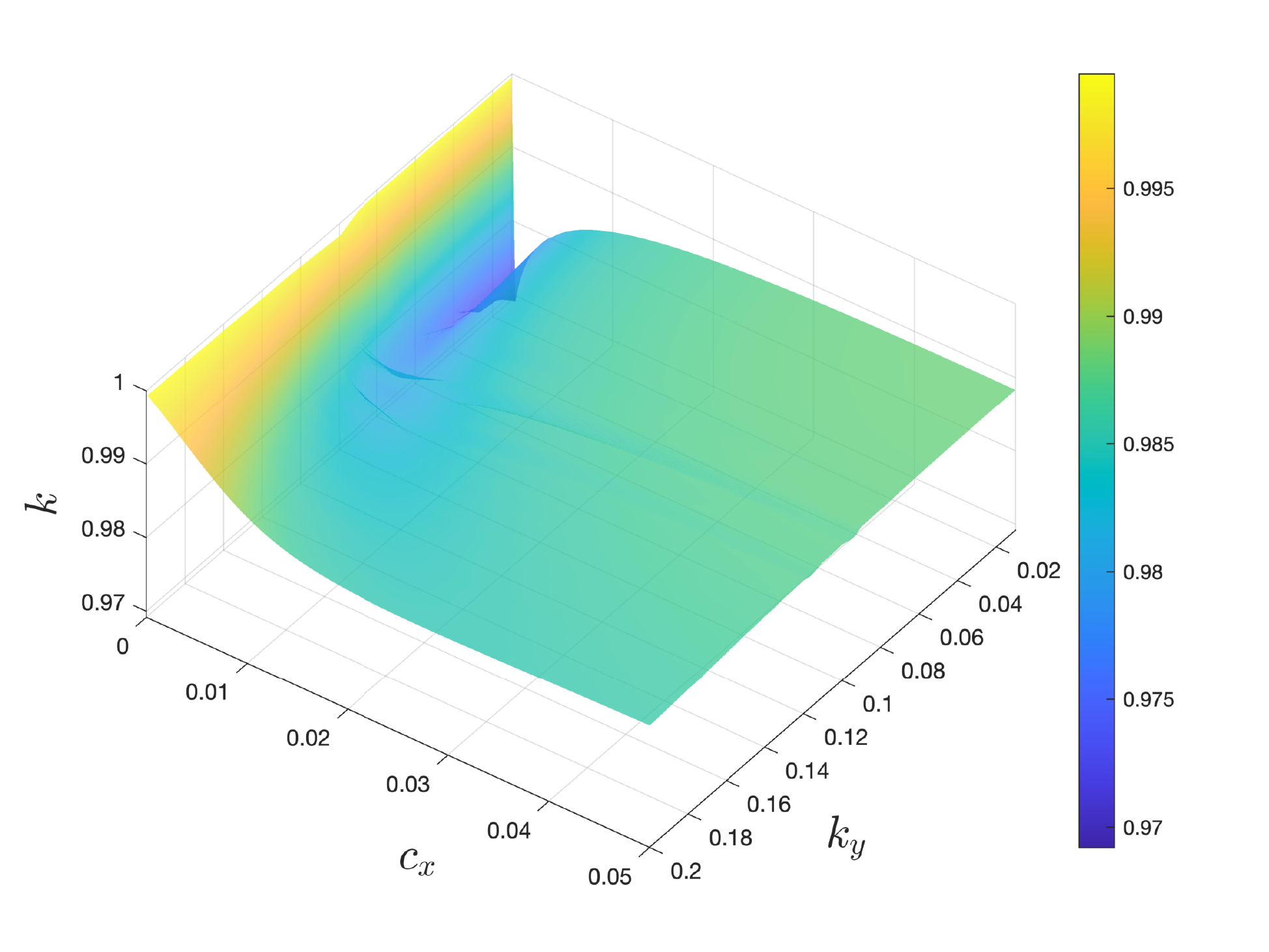}
\includegraphics[trim = 0.cm 0.0cm 0.0cm 0.0cm,clip,width=0.3\textwidth]{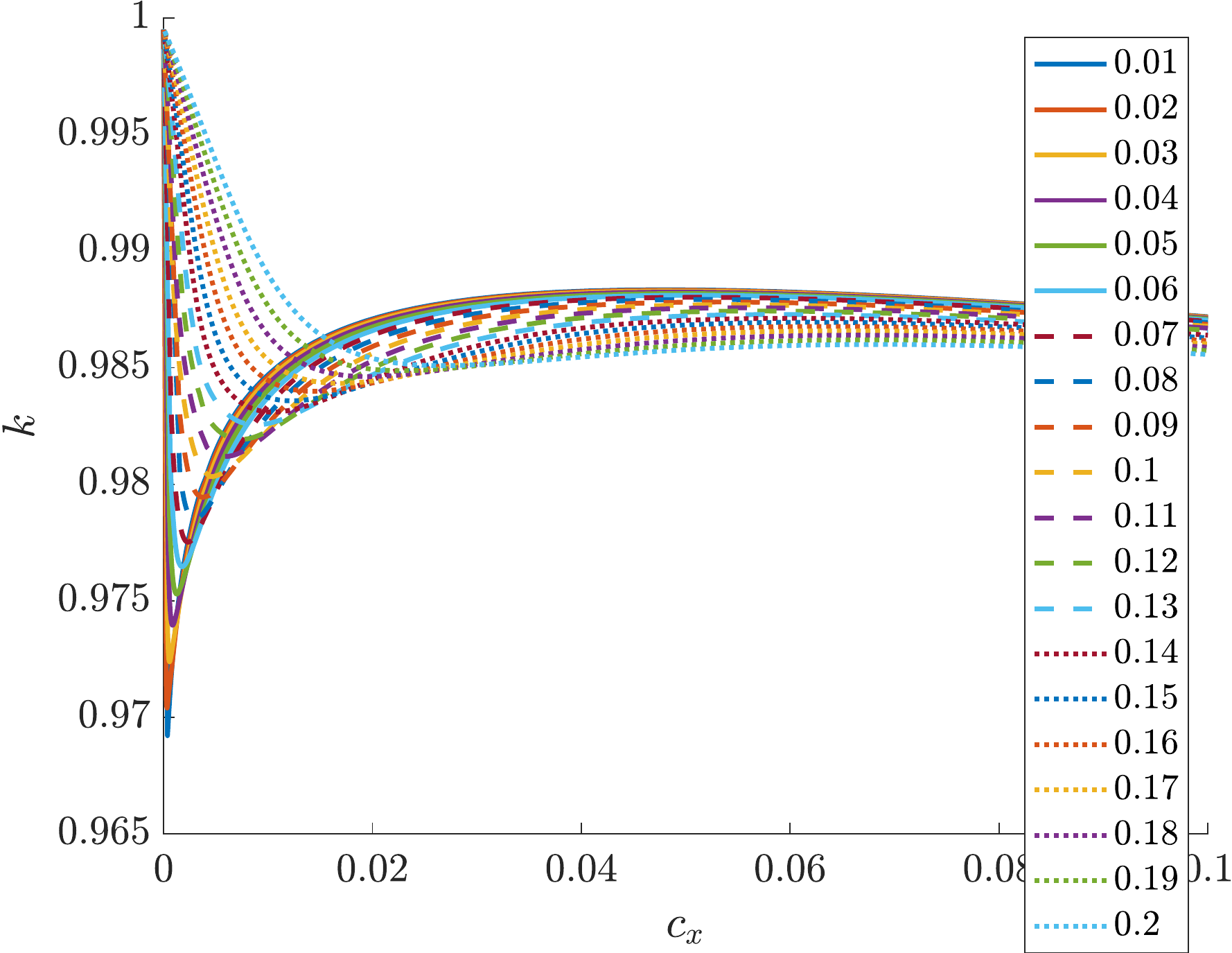}
\caption{Moduli surface for $c_x,k_y\sim0$, continuing in $c_x\rightarrow0$ for a range of fixed $k_y$ values, plots of $k_x$ (left) and $k$ (center), as well as cross sections in $k_x$ for fixed $k_y$. }\label{f:cxsm}
\end{figure}

Taking in addition the limit $k_y\rightarrow0$, the curves $k_x(c_x;k_y)$ limit set-wise on the $k_y = 0$-cross section of $\mc{M}$ which consists of the locally monotonically decreasing curve $k_x(c_x,0)$ for $c_x>0$, and the vertical line segment $\{k_y = c_x = 0, k_x\in[k_\mathrm{sd,min},k_\mathrm{zz}]\}$; see Figure \ref{f:cxsm}. In particular, curves develop a singularity at $c_x =k_y =  0$, with local slope in $c_x$ proportional to $-1/k_y$ as $k_y\rightarrow 0$. 

This steepening indicates that slow growth imposes a stronger strain on weakly oblique stripes, that is, on stripes that are almost parallel to the interface. A quantitative analysis in this regime would need to take the development of a point defect at the quenching interface into account; see Section \ref{sss:pdax} and \cite{avery2019growing}.   



\begin{figure}[ht!]
\centering
\includegraphics[trim = 0.5cm 0.25cm 0.5cm 0.5cm,clip,width=0.45\textwidth]{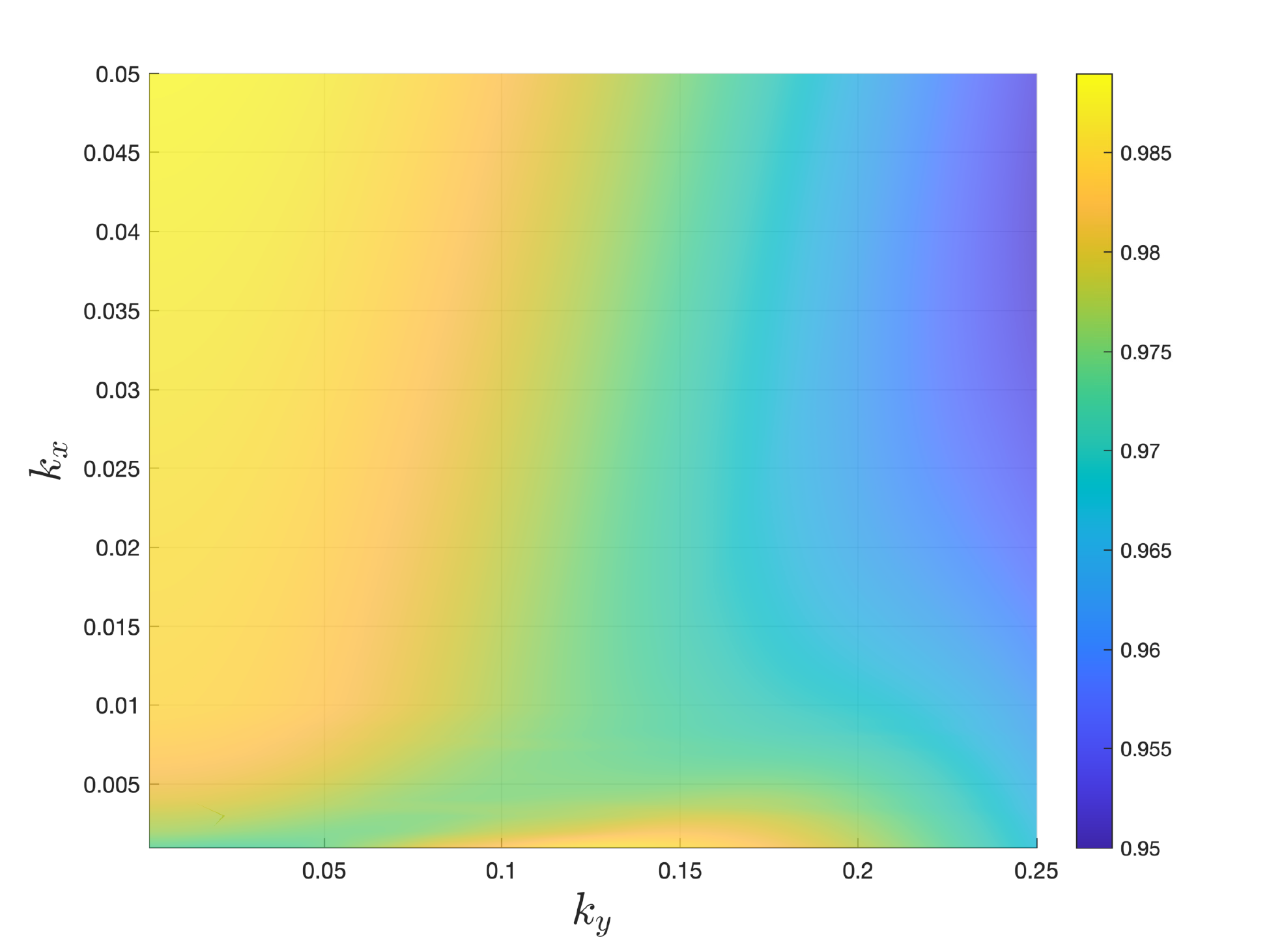}
\includegraphics[trim = 0.0cm 0.0cm 0.0cm 0.0cm,clip,width=0.4\textwidth]{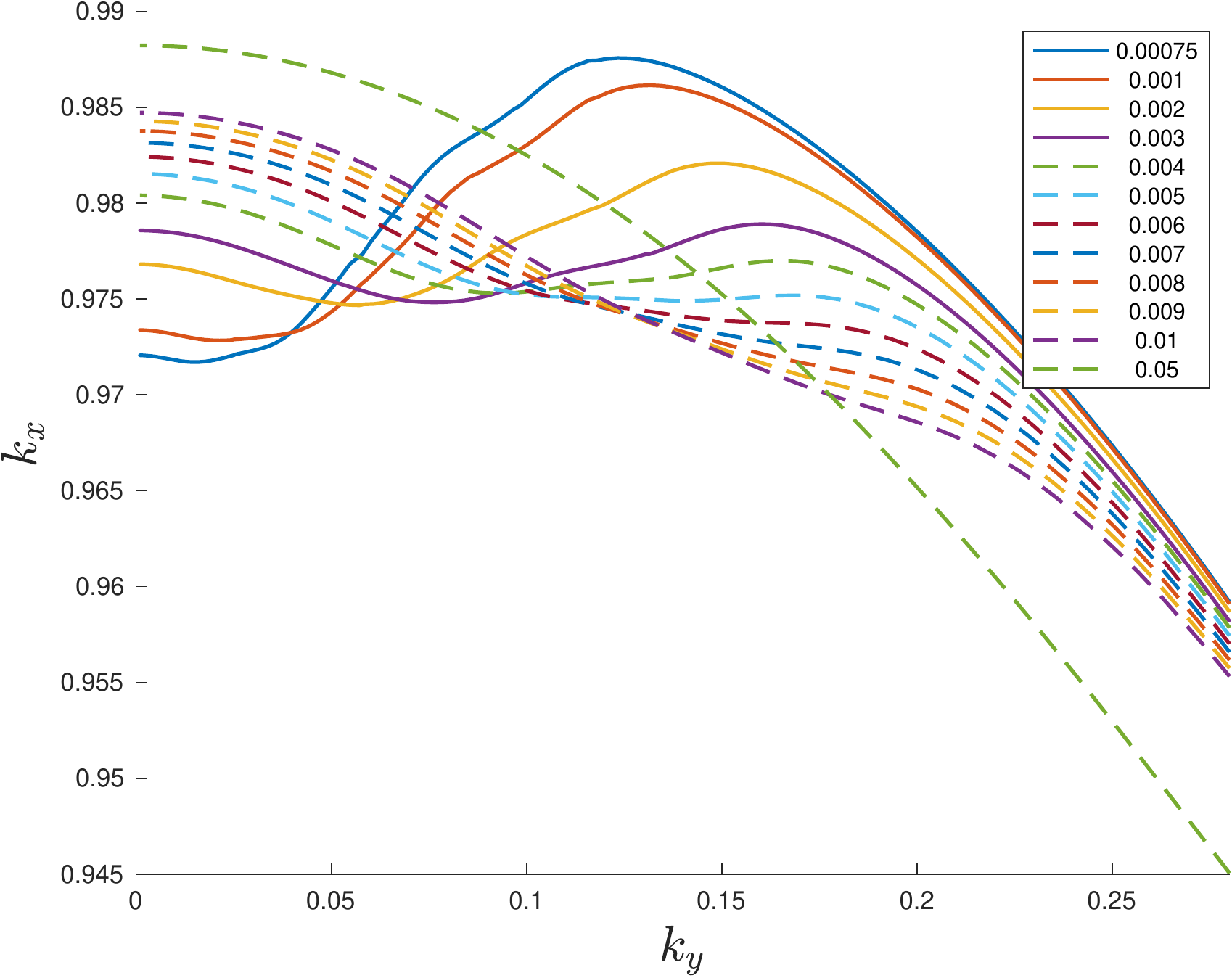}\\
\includegraphics[trim = 0.5cm 0.25cm 0.5cm 0.5cm,clip,width=0.45\textwidth]{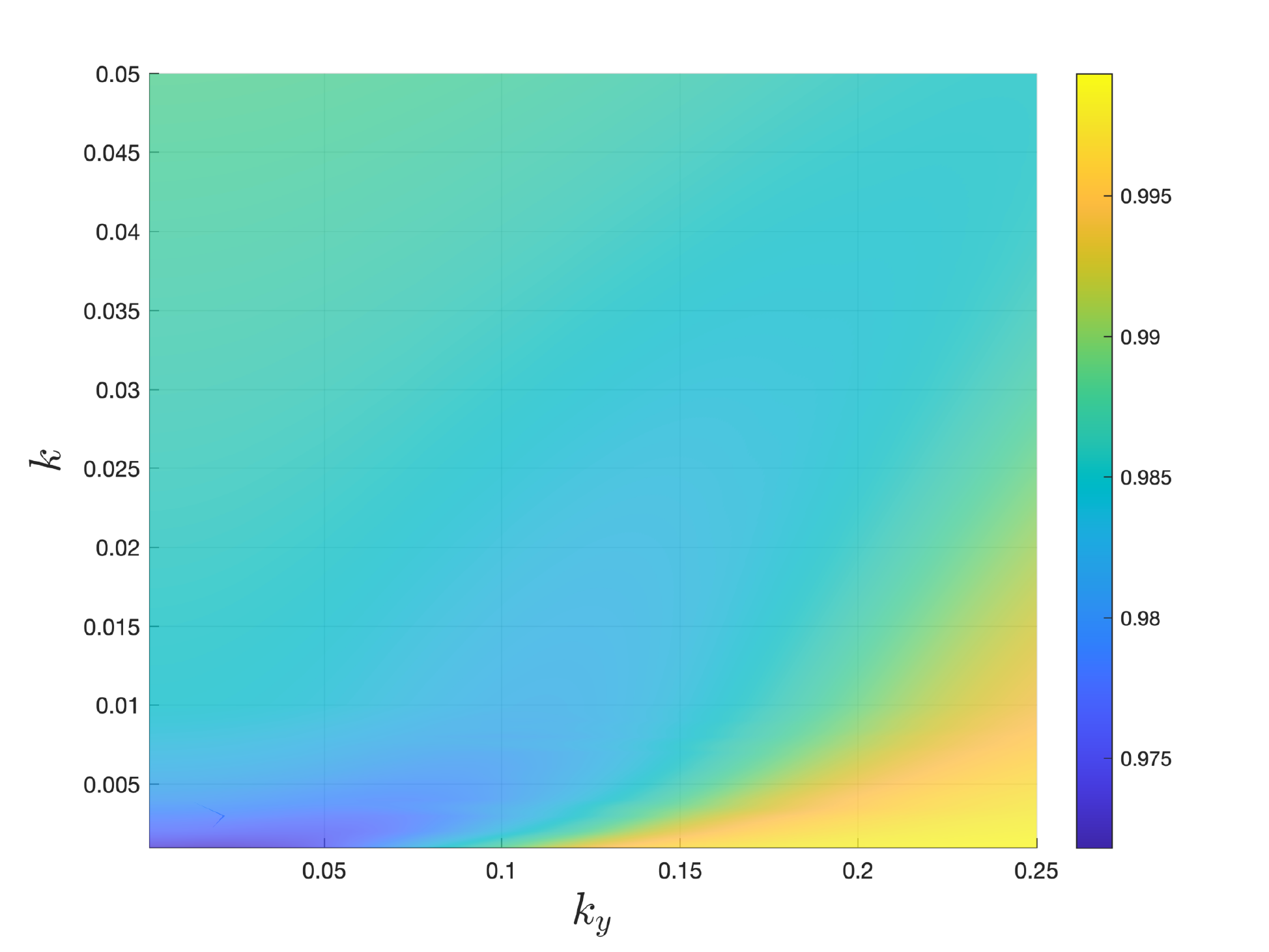}
\includegraphics[trim = 0.cm 0.cm 0.cm 0.0cm,clip,width=0.4\textwidth]{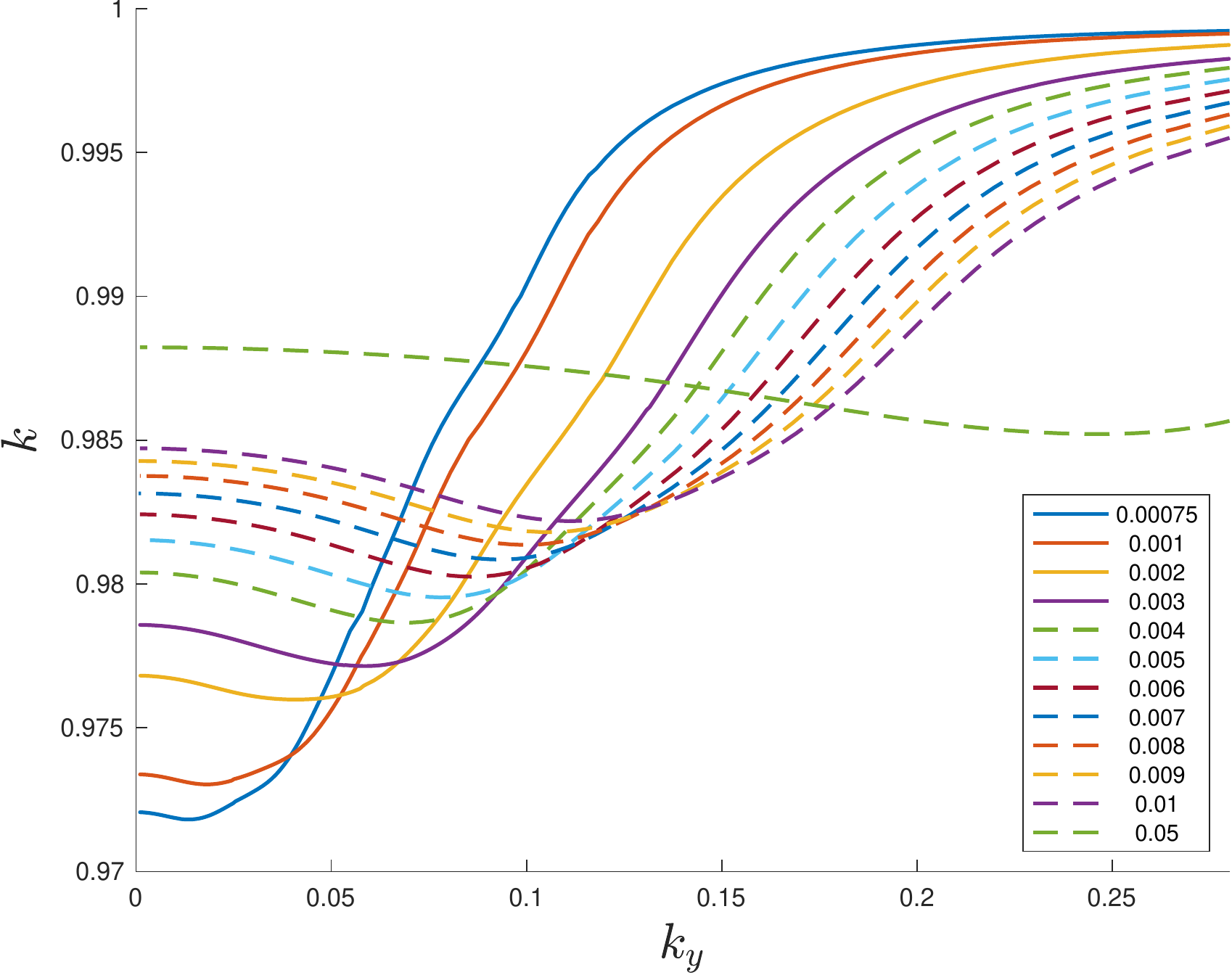}
\caption{Moduli surface for $c_x,k_y\sim0$, continuing in $k_y\rightarrow0$ for $c_x$ fixed, ranging from $10^{-3}$ to $5*10^{-2}$.  Top left: plot of surface $k_x(k_y,c_x)$; Top right: $k_y$ slices of $k_x$-surface  Bottom row: similar plots but for the bulk wavenumber $k(k_y,c_x)$. }\label{f:kysm}
\end{figure}

Alternatively, one can fix $c_x\sim0$ and continue in $k_y\sim0$. From Theorem \ref{t:os} above, one expects $k_x$ to be quadratically dependent on $k_y$ near $k_y = 0$. Moving further out from $k_y = 0$, one observes non-monotonic curves where $k_x(c_x,k_y)$ has a series of minima and maxima, the number of which depends on the magnitude of $c_x$; see Figure \ref{f:kysm}. For larger fixed $c_x$ values the first local minimum disappears, leading to a monotonically decreasing curve in $k_y$. 
A modulational analysis for $c_x,k_y\sim0$, where $k_x$ is near the zigzag critical wavenumber $k_{x,\mathrm{zz}}$, would yield a negative effective diffusivity in the $y$-direction so that higher-order terms must be included. Thus one expects to obtain a Cross-Newell equation \cite{hoyle} for modulations of the striped pattern, paired with an appropriate boundary condition to represent the quench.

\subsubsection{Acute oblique stripes and the kink-dragging bubble}\label{sss:acc-ob}
\begin{figure}[!ht]
\includegraphics[trim = 0.5cm 0.5cm 0.5cm 0.5cm,clip,width=0.29\textwidth]{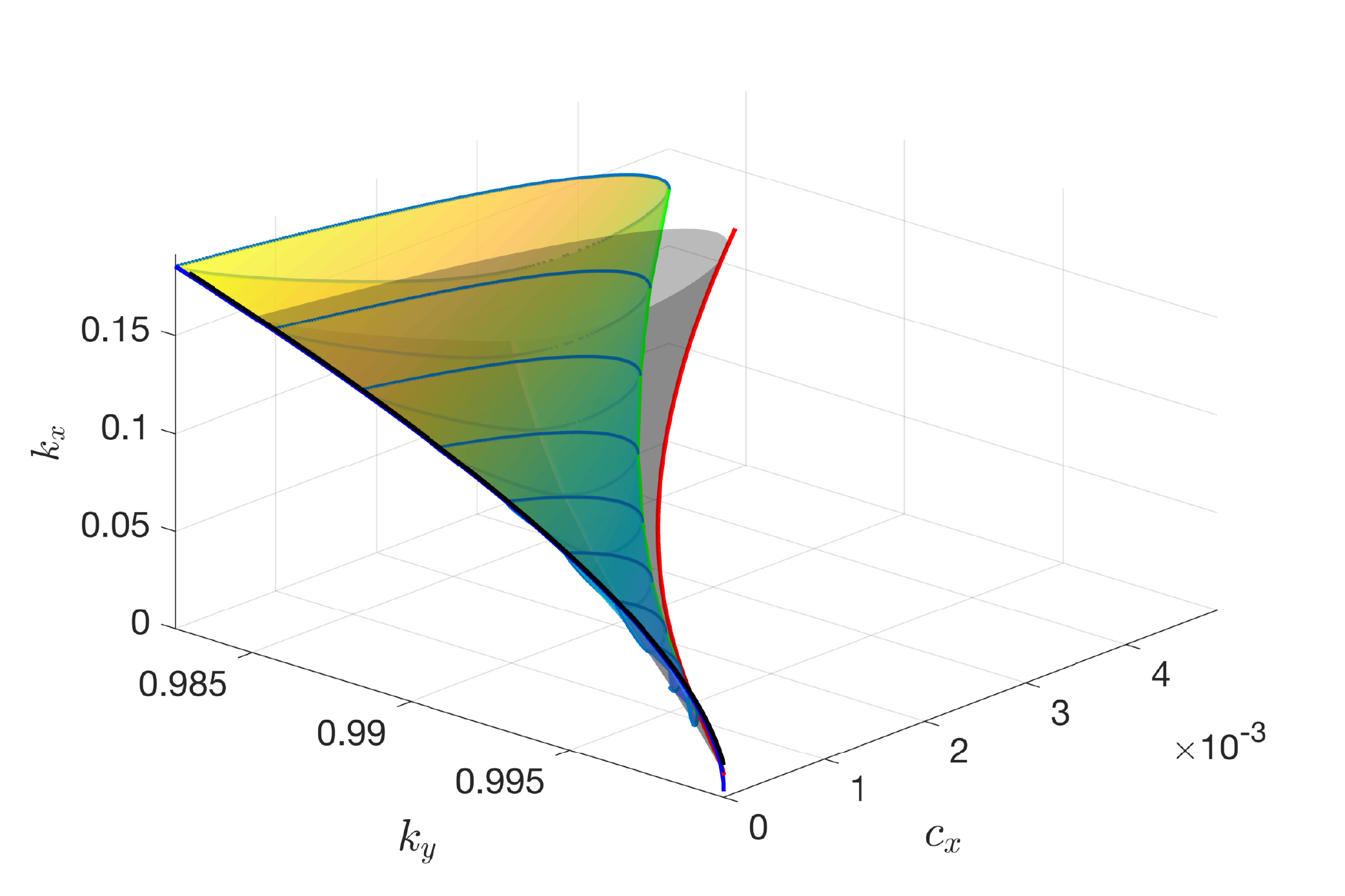}\hspace{-0.2in}
\includegraphics[trim = 0.cm 0.cm 0.cm 0.0cm,clip,width=0.23\textwidth]{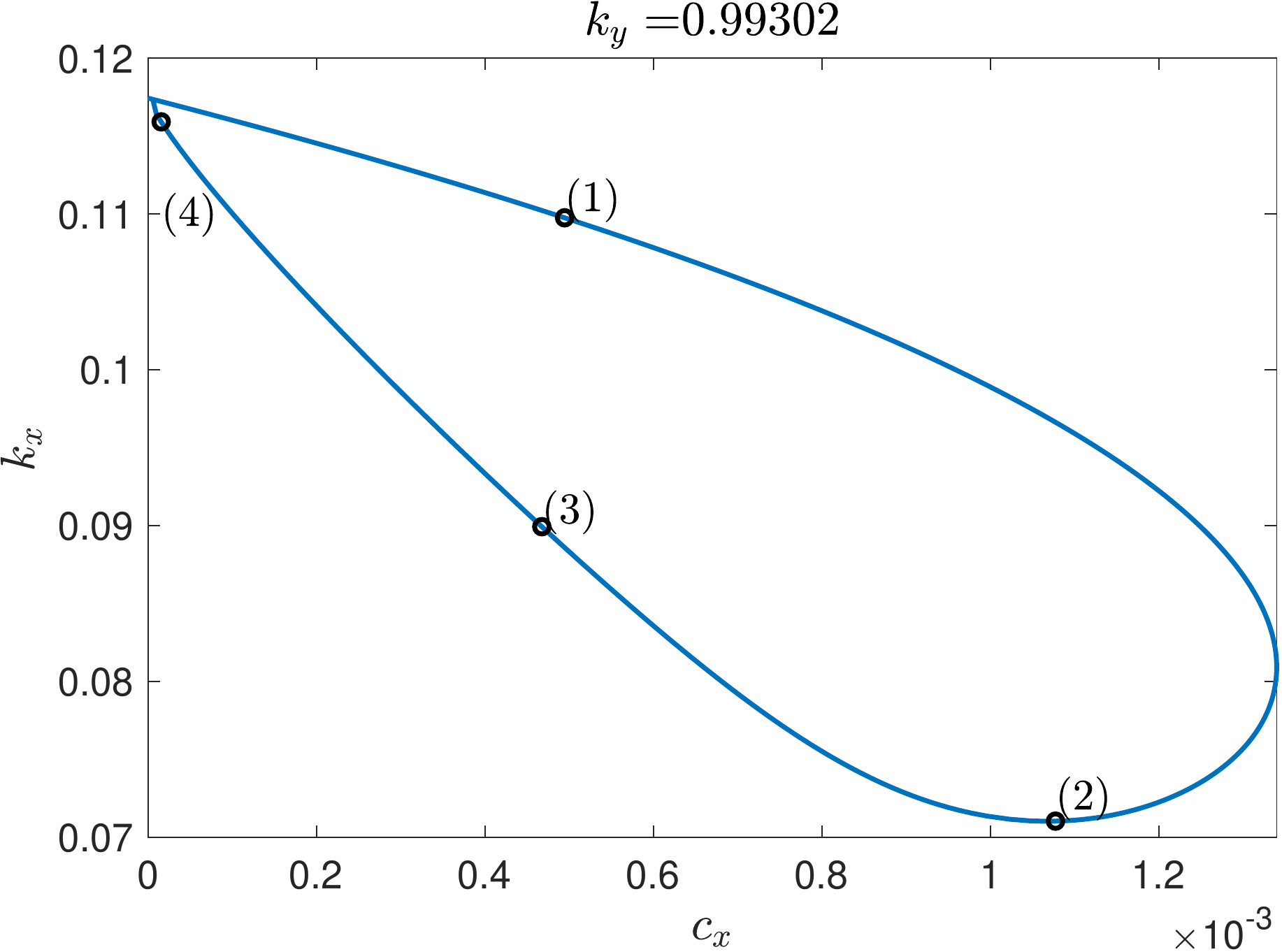}
\includegraphics[trim = 0.cm 0.0cm 0.0cm 0.0cm,clip,width=0.23\textwidth]{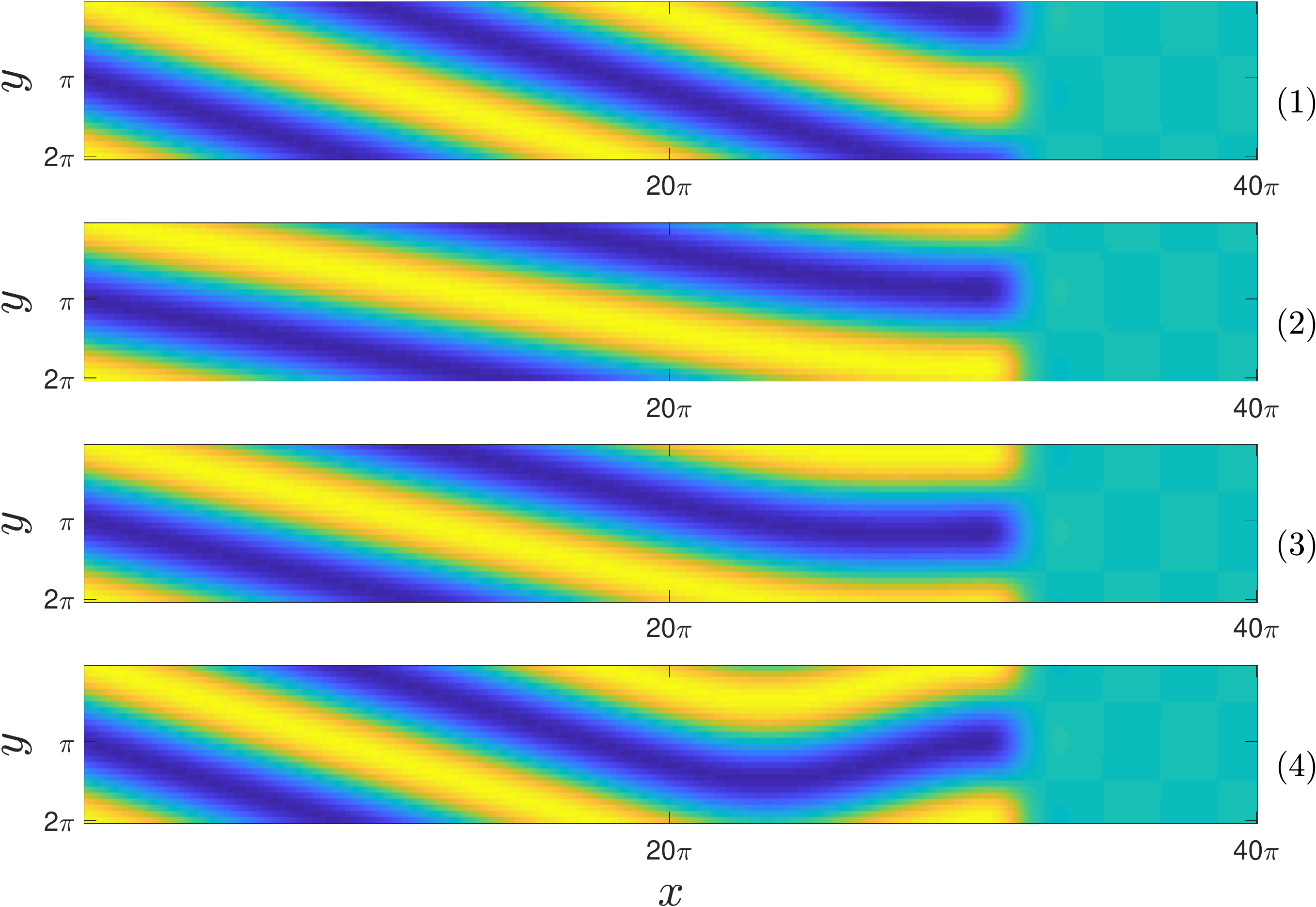}\hspace{-0.2in}
\includegraphics[trim = 0.cm 0.cm 0.cm 0.0cm,clip,width=0.25\textwidth]{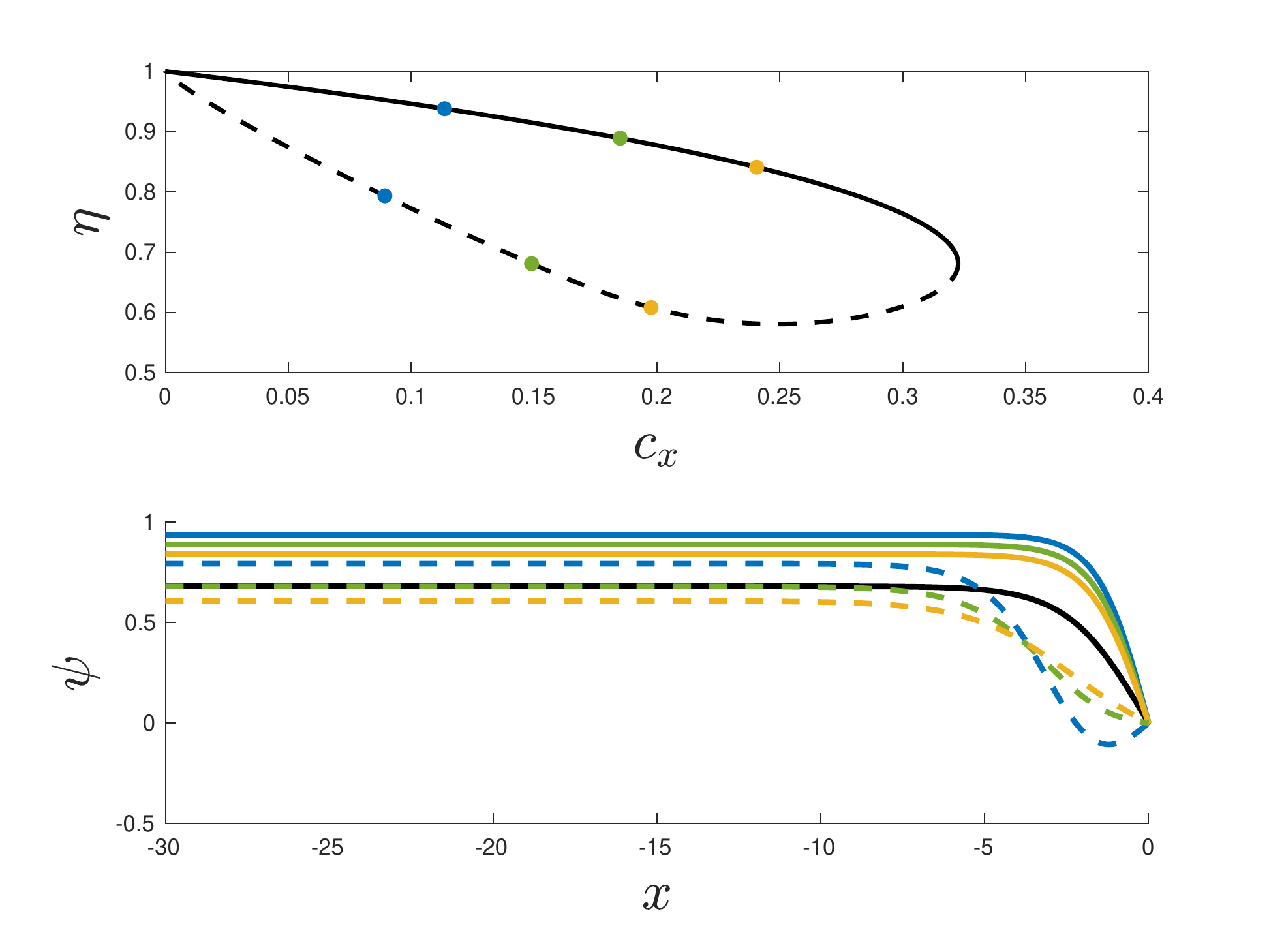}
\caption{Left: Zoom in on the kink-dragging bubble in $\mc{M}$ with $\mu = 1/4$ along with the associated bubble from continuing fronts in \eqref{e:ch1} (grey); Center-left: cross section of bubble in $\mc{M}$ for a fixed $k_y$; Center-right: Solution profiles of \eqref{e:mtw} along this cross section; Right: Kink-dragging bubble bifurcation diagram in \eqref{e:ch1}, comparing front asymptotic value $\eta$ with front speed $c_x$, solid line denotes stable solutions while dashed line gives unstable solutions, bottom inset gives example profiles $\psi_*(x)$ for points along the stable (solid) and unstable (dashed) branches. First three plots reproduced with permission from \cite[Fig. 16, 20]{avery2019growing}. Copyrighted by SIAM. }\label{f:kinkbub}
\end{figure}

We observe a qualitatively different regime when slowly grown and accutely oblique stripes are grown near the point $(k_x,k_y,c_x) = (0,k_\mathrm{zz},0)$ which corresponds to a zigzag critical perpendicular stripe. As mentioned earlier,  for fixed $k_y\lesssim k_{zz}$, and $k_x\neq 0$, curves emanate smoothly from $k_{x,\mathrm{zz}}(k_y)$ for $c_x>0$ until undergoing a fold bifurcation at the point $(c_{x,\mathrm{ksn}}(k_y),k_y,k_{x,\mathrm{ksn}}(k_y))$, where it folds back underneath itself; see Figure \ref{f:kinkbub}. Through this transition, the corresponding front solution develops an anti-phase ``kink" or ``wrinkle" near the growth interface which has the same local wavenumber as in the far-field but with opposite orientation in $y$. The curve $\{(c_{x,\mathrm{ksn}}(k_y),k_y,k_{x,\mathrm{ksn}}(k_y)),k_y\leq k_\mathrm{zz}\}$ of fold points, depicted in green in the left plot of Figure \ref{f:kinkbub}, emanates from $(0,k_\mathrm{zz},0)$ into the positive octant and reconnects with the main body of $\mc{M}$ in a hyperbolic catastrophe; see Figure \ref{f:hypcat} below. 

\begin{figure}[!ht]
\centering
\includegraphics[trim = 3cm 0.2cm 3.5cm 0.2cm,clip,width=0.48\textwidth]{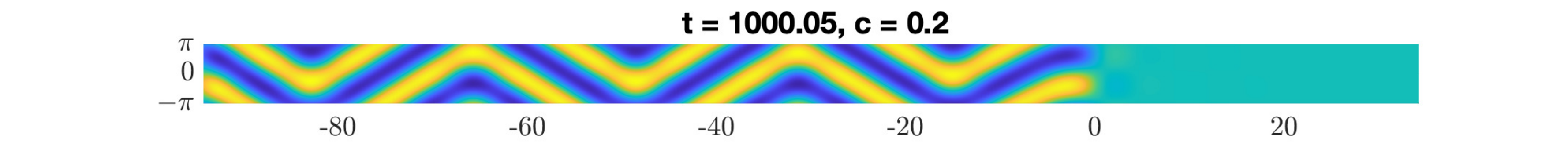}\
\includegraphics[trim = 3cm 0.2cm 3.5cm 0.2cm,clip,width=0.48\textwidth]{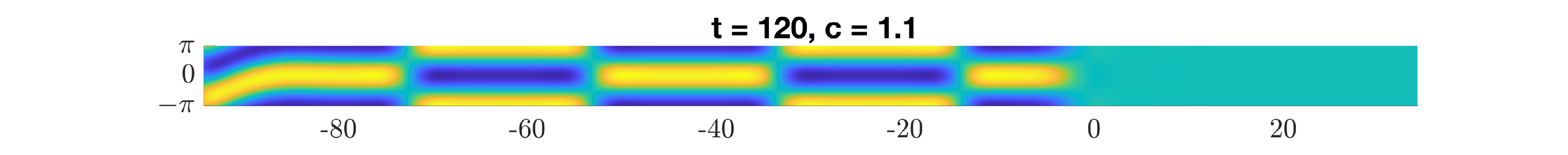}\\
\includegraphics[trim = 3cm 0.2cm 3.5cm 0.2cm,clip,width=0.48\textwidth]{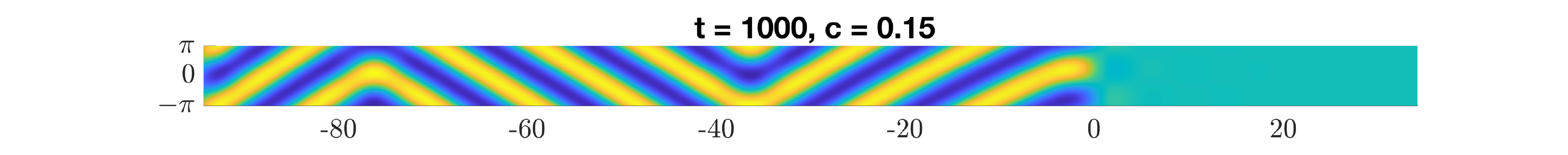}
\includegraphics[trim = 3cm 0.2cm 3.5cm 0.2cm,clip,width=0.48\textwidth]{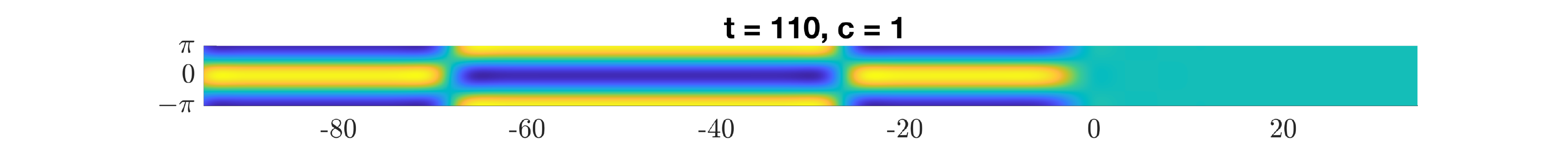}\\
\includegraphics[trim = 3cm 0.2cm 3.5cm 0.2cm,clip,width=0.48\textwidth]{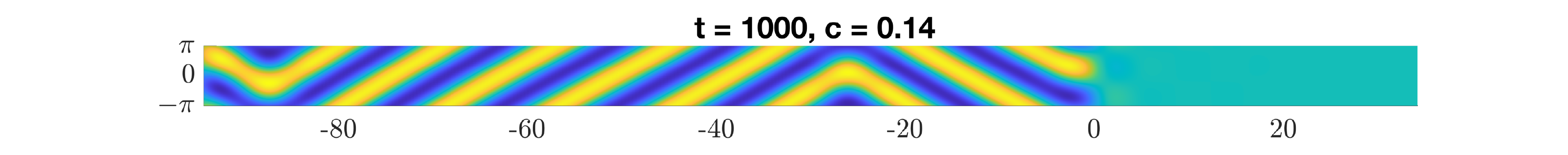}
\includegraphics[trim = 3cm 0.2cm 3.5cm 0.2cm,clip,width=0.48\textwidth]{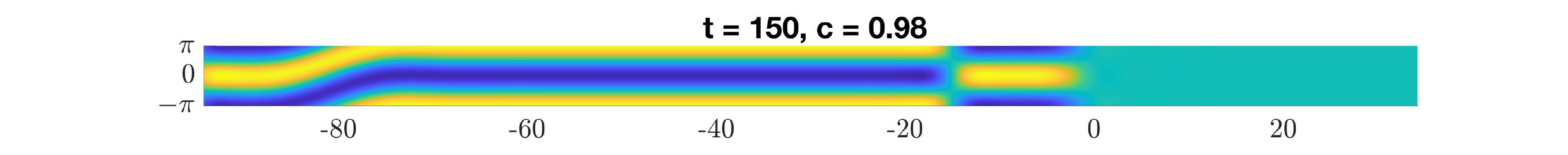}\\
\caption{Time-periodic solution profiles for the directionally quenched time-dependent equation \eqref{e:sh-q} posed in a comoving frame $x\mapsto x- c_xt$ and scaled by the vertical wavenumber $y\mapsto k_y y$, in a long cylinder $(x,y)\in[-50\pi,50\pi]\times[-\pi,\pi]$ with periodic boundary conditions and $\mu = 3/4$. Period increases as $c_x$ is decreased to the fold value. Left: Oblique stripes with $k_y = 0.8$ with speeds $c_x - c_\mathrm{x,osn}(k_y)\sim 0.07,0.015,0.005$ from top to bottom with $c_\mathrm{x,osn}(k_y)\approx 0.1355$. Right:
Perpendicular stripes with $k_y =1$, and  $c_x- c_{x,\mathrm{psn}}(k_y)\sim 0.12, 0.04,0.02$ and $ c_{x,\mathrm{psn}}(k_y)\sim 0.96$.}\label{f:snlc}
\end{figure}

The fold curve delineates a qualitative transition in the dynamics of slowly grown oblique stripes in the original equation \eqref{e:sh-q}. For growth speeds $c_x$ past the fold value $c_{x,\mathrm{ksn}}$, direct numerical simulations show saddle-node on a limit cycle dynamics, with time-periodic kink-shedding at the interface with period $T_\mathrm{kink}\sim  (c_x- c_{x,\mathrm{ksn}})^{-1/2}\nearrow\infty$ as $c_x\searrow c_{x,\mathrm{ksn}}$; see Figure \ref{f:snlc}. 

One can use a modulational approximation near a perpendicular stripe with critical wavenumber $k_y = k_\mathrm{zz}$ to understand these dynamics in a reduced model.  In particular, through the ansatz  $u(x,y,t) = \epsilon \re^{\ri k_y y} A(\epsilon x, \epsilon t) + \mathrm{c.c.}$ in the bulk domain with $k_y = 1 - \epsilon$, $\rho \equiv \mu$, one obtains at leading order in $\epsilon>0$ small, the Newell-Whitehead-Segel equation \cite{hoyle}
\beq
A_T = -(\partial_{XX} + 2\epsilon - \epsilon^2)^2 A + \mu A - 3A|A|^2.
\eeq
Expanding again in the phase $\phi$  near $R = \sqrt{\mu/3}$ in polar coordinates $A = R\re^{\ri\phi}$, one obtains
\beq
\phi_T = -c_4 \phi_{XXXX} - c_1 \epsilon \phi_{XX} + c_3 \phi_X^2 \phi_{XX},\qquad c_4 = 1, c_1 = 4, c_3 = 6.
\eeq
Through subsequent scaling and transforming to a comoving frame $X\mapsto X - c_x T$, one finds the Cahn-Hilliard equation \cite{cahn1958free} for the local vertical wavenumber $\psi = \phi_x$,
\beq
\psi_T = -(\psi_{XX} + \psi - \psi^3)_{XX} + c_x \psi_X.
\eeq
Note that $\psi\equiv \eta \neq0$ represents an oblique stripe.  Exploiting a Hamiltonian structure at $c_x=0$, one finds an effective boundary condition induced by the parameter step in Swift-Hohenberg, $\psi = \psi_{XX} = 0$ at $X = 0$; see \cite[\S 2.3]{avery2019growing}. One therefore  wishes to study
\beq\label{e:ch1}
\psi_T = -(\psi_{XX} + \psi - \psi^3)_{XX} + c_x \psi_X,\quad X<0,\qquad \psi = \psi_{XX} = 0,\qquad X = 0.
\eeq 
The striped traveling wave solutions are represented by equilibrium solutions $\psi_*(X)$ which satisfy $\psi_*(X)\rightarrow \eta, X\rightarrow-\infty$. For $c_x = 0$, they take the explicit form $\psi_d(X) = \pm \tanh(X/\sqrt{2})$. A functional analytic approach was then used in \cite[Thm. 3.1]{avery2019growing} to continue these fronts to $c_x\neq0$, determining the selected wavenumber as a function of $c_x$. For larger $c_x$, numerical continuation was used to continue the fronts $\psi_*$ in $c_x$ through the saddle-node bifurcation. If we let $(c_{x,CH},\eta_{CH}(c_{x,CH}))$ denote the fold curve obtained from the Cahn-Hilliard equation, we can obtain a prediction for the Swift-Hohenberg equation through the curve
\begin{equation}
\zeta = k_\mathrm{zz} - k_y,\qquad k_x = \sqrt{2}k_\mathrm{zz}\eta_{CH}(c_{x,CH}) \zeta^{1/2},\qquad c_x = 8 c_{x,CH} \zeta^{3/2}.
\end{equation}
The left plot of Figure \ref{f:kinkbub} gives the comparison of this prediction (red) to the numerically observed fold curve (green).
The work \cite{avery2019growing} also computed local saddle-node coefficients and predicted  limit-cycle frequencies for time-periodic solutions of \eqref{e:ch1} depicted in Figure \ref{f:snlc} with speed just above the fold speed.

\subsection{Intermediate growth regions}\label{ss:perp-ret}
\subsubsection{Perpendicular stripes and oblique stripe reattachment}\label{sss:perp-ret-1}

\begin{figure}[!ht]
\centering
\includegraphics[trim = 0.cm 0.cm 0.cm 0.cm,clip,width=0.225\textwidth]{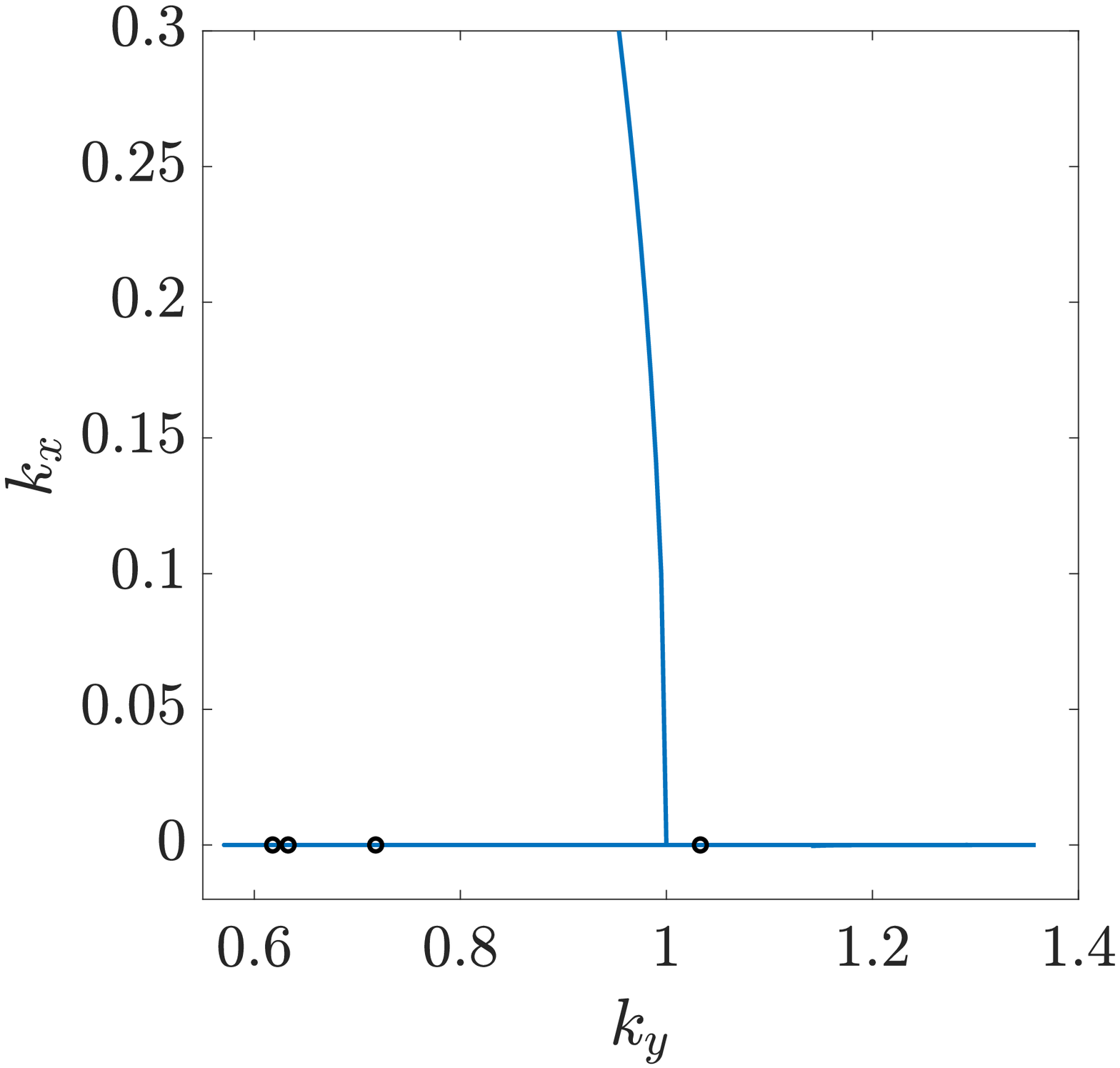}\hspace{-0.0in}
\includegraphics[trim = 0.5cm 0.5cm 0.5cm 0.5cm,clip,width=0.225\textwidth]{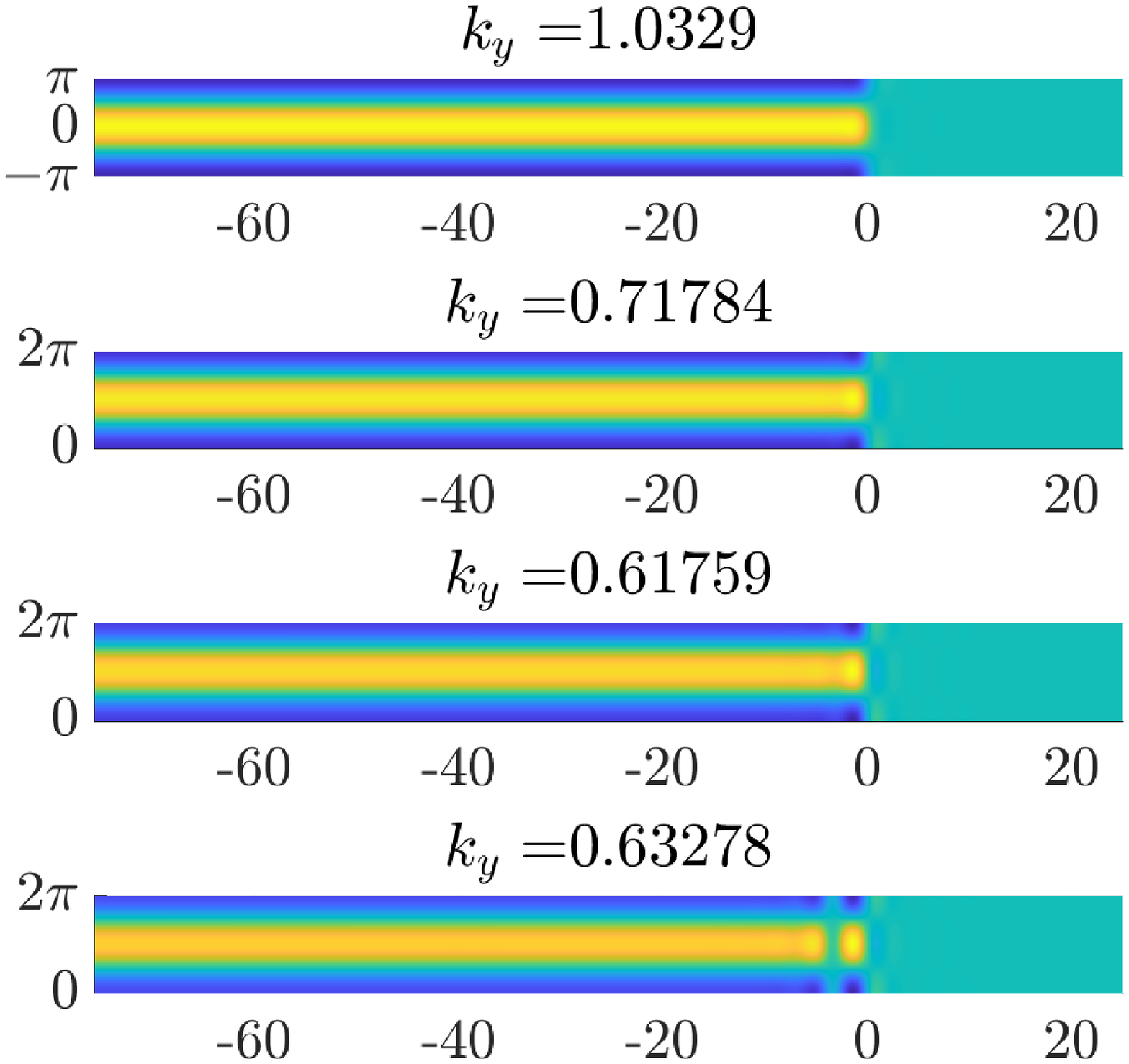}\hspace{-0.12in}
\includegraphics[trim = 0.cm 0.cm 0.cm 0.0cm,clip,width=0.275\textwidth]{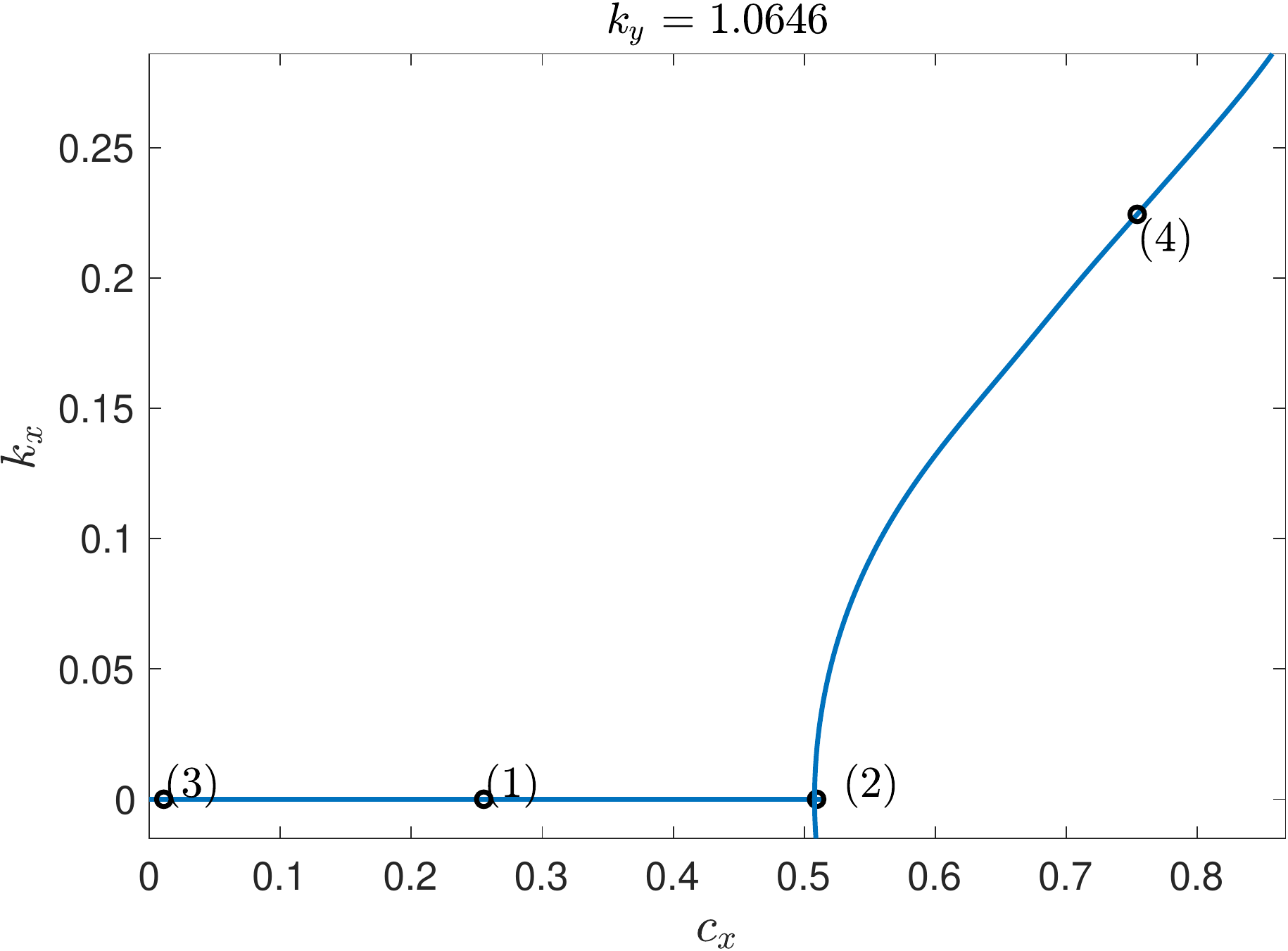}\hspace{-0.025in}
\includegraphics[trim = 0.cm 0.0cm 0.0cm 0.0cm,clip,width=0.275\textwidth]{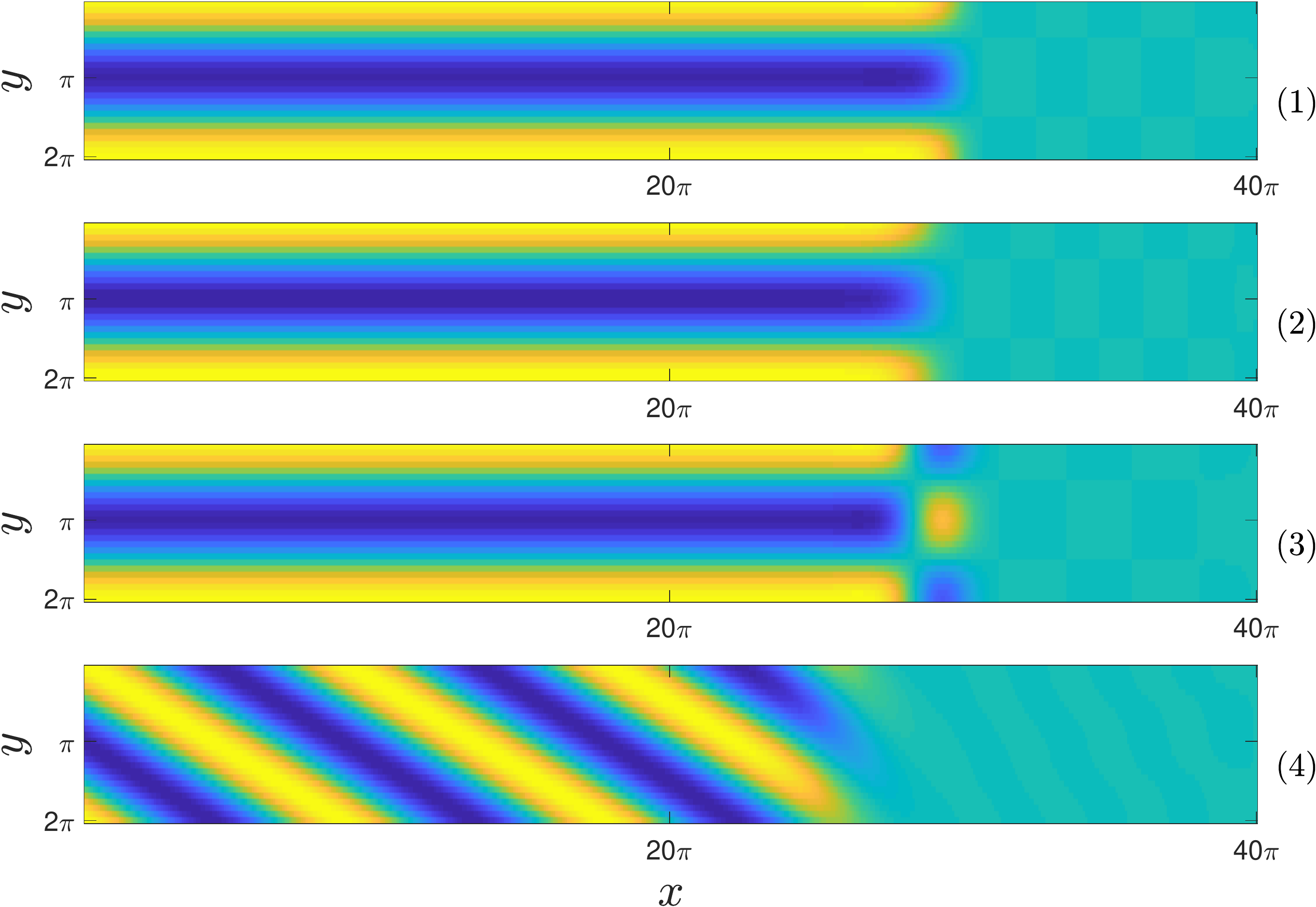}\hspace{-0.04in}
\caption{Left two plots: Zoom in of $c_x = 0$ slice of $\mc{M}$ near $(k_y,k_y) = (0,1)$, black dots correspond to adjacent perpendicular stripe solution profiles for a range of $k_y$ values on either side of the fold point $k_{y,\mathrm{sn}}$; Right two plots: cross-section of $\mc{M}$ with fixed $k_y = 1.0646$ illustrating the perpendicular to oblique bifurcation with adjacent solution profiles along this cross-section; Reproduced with permission from \cite[Fig. 19]{avery2019growing}. Copyrighted by SIAM.}\label{f:perp-slice1}
\end{figure}

\begin{figure}[!ht]
\centering
\includegraphics[trim = 0.5cm 0.0cm 0.5cm 0cm,clip,width=0.425\textwidth]{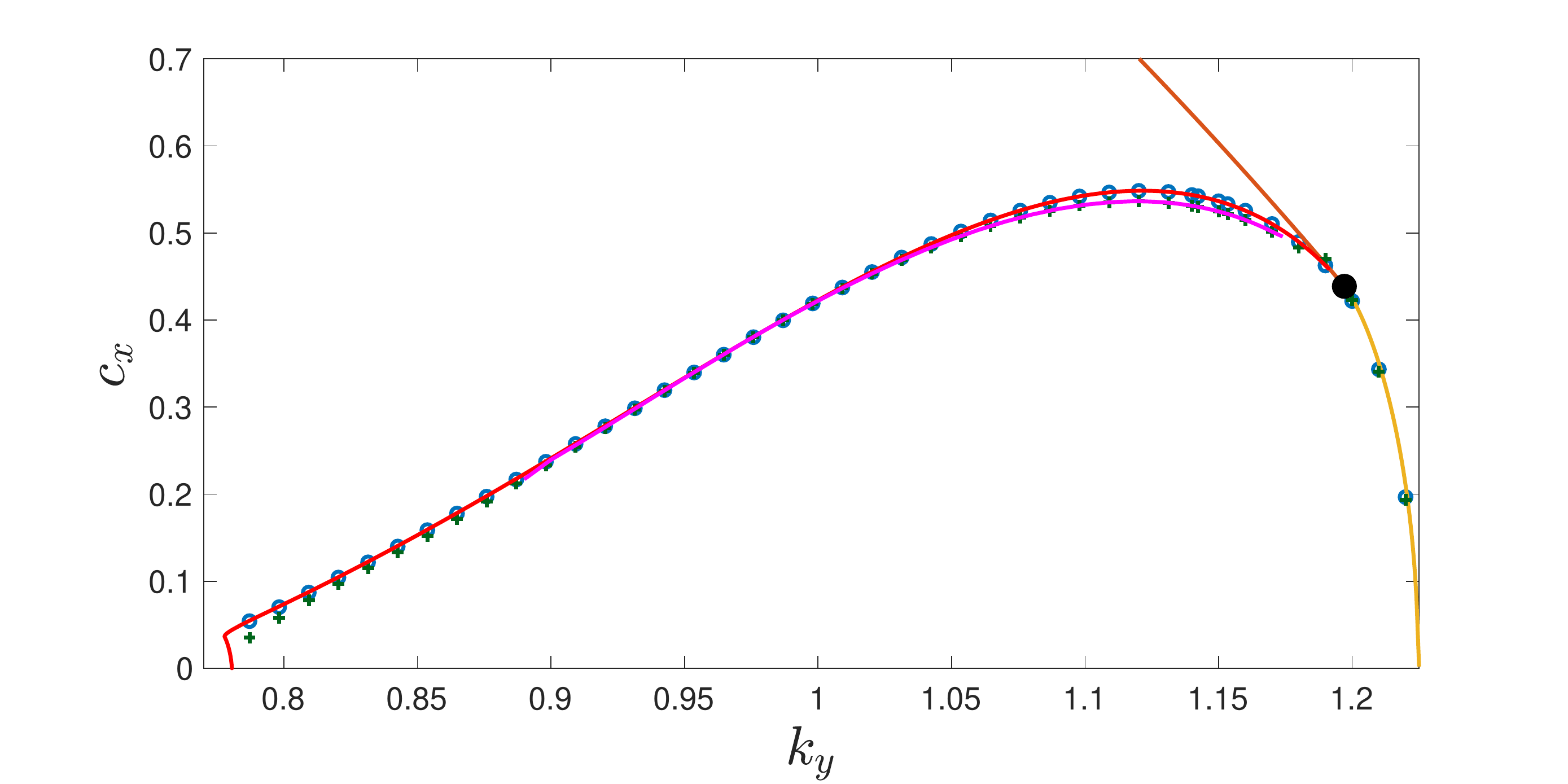}\hspace{-0.2in}
\includegraphics[trim = 20.cm 0.cm 0.cm 0.0cm,clip,width=0.35\textwidth]{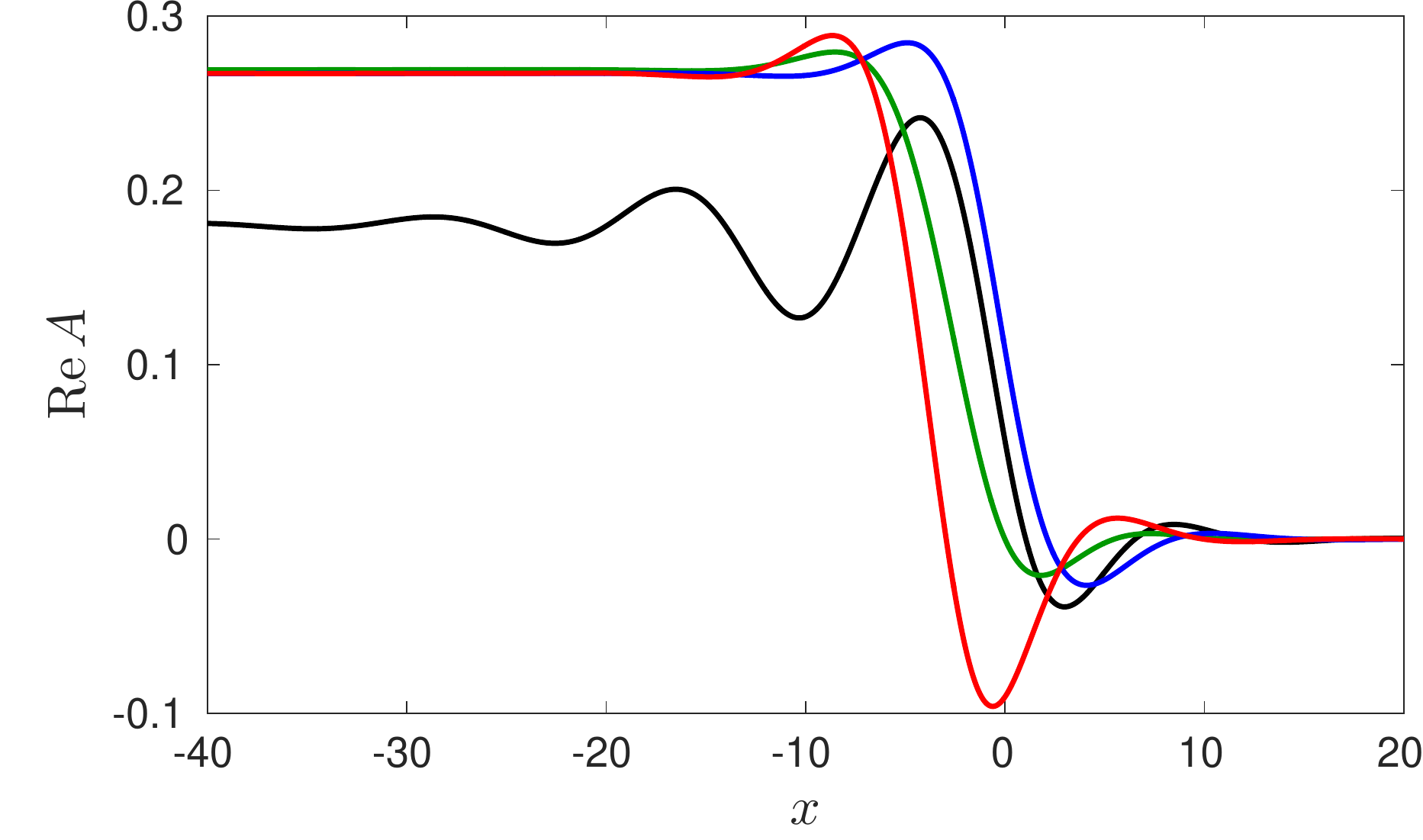}\hspace{-0.3in}
\caption{$\mu = 1/4$, Left: Domain of existence of perpendicular stripes predicted by \eqref{e:nws}, bounded above by bifurcation curves $c_\mathrm{x,psn}(k_y), c_\mathrm{x,opf}(k_y)$ (red and magenta), blue and green points indicate numerically measured bifurcation values in \eqref{e:mtw}; 
Right: front solutions of \eqref{e:nws} representing perpendicular stripes for various $c_x,k_y$ values label in center-left plot; $\mu = 1/4$ throughout, reproduced from \cite[Fig. 14, 17]{avery2019growing}. Copyrighted by SIAM.  }\label{f:perp-slice2}
\end{figure}

Similar to oblique stripes, perpendicular stripes perturb regularly as $c_x$ increases from zero. We observe that the domain of supported wavenumbers $k_y$ shrinks as $c_x$ increases. For $k_y>k_{y,\mathrm{po}}$, stripe detachment limits the range of $k_y$-wavenumbers from above, or, equivalently, the range of $c_x$-values; see Section \ref{sss:inv}. For $k_y<k_{y,\mathrm{po}}$ the free invasion calculation predicts oblique stripes near detachment, so that one expects a transition from perpendicular to oblique stripes for finite $c_x$ before detachment. 
Indeed, we find fronts undergo a kink forming saddle-node bifurcation at some finite speed $c_x = c_{x,\mathrm{psn}}(k_y)$; see Figure \ref{f:perp-slice1}. Direct numerical simulations show that solutions exhibit time-periodic kink-shedding just as in the oblique stripe case for $c_x\gtrsim c_{x,\mathrm{psn}}$, with period blow-up as $c_x$ approaches $c_{x,\mathrm{psn}}$ from above; see Figure \ref{f:snlc}.  Prior to the saddle-node, for $c=c_{x,\mathrm{ppf}}<c_\mathrm{x,psn}(k_y)$, perpendicular stripes destabilize in a pitchfork bifurcation that breaks the $y$-reflection symmetry, leading to oblique stripes; see Figure \ref{f:perp-slice1}. 
The location of these curves of saddle-node and pitchfork bifurcations can be approximately located using amplitude equations. One inserts $u(x,y,t) = A(x,t) e^{\ri y} + \mathrm{c.c.}$ into \eqref{e:mtw}, detuning by the $y$-frequency $A\mapsto A \re^{\ri k_x c_x t}$, and truncating at lowest order in $e^{\ri y}$, to obtain the Newell-Whitehead-Segal equation,
\begin{equation}\label{e:nws}
A_t = -(1+\partial_x^2 - k_y^2)^2 A + \rho(x)A - 3A|A|^2  + c_x A, \quad A\in \C.
\end{equation}
Setting $\rho\equiv  \mu$ for the wake of the quench, plane waves $A = r\re^{\ri (k_x x + \omega t)}$ with $r^2 = \frac{\mu - (1-k_y^2)^2}{3}, \omega = k_x c_x$ represent oblique stripes for $k_x \neq0$ and perpendicular stripes for $k_x = 0$. Figure \ref{f:perp-slice2} gives the results of numerical continuation of traveling wave solutions, which are equilibrium solutions to \eqref{e:nws}, connecting a plane wave solution with the trivial state as $x$ increases, representing perpendicular and oblique striped fronts. Continuing in $c_x$ with $k_y$ fixed we find perpendicular fronts destabilize in a saddle-node bifurcation and oblique stripes bifurcate in a nearby pitchfork bifurcation. Numerically continuing the fold and pitchfork points in $k_y$ we find good agreement with numerical results in the full equation \eqref{e:mtw}.

%
%

\subsubsection{Hyperbolic Catastrophe}\label{sss:hyp-cat}

\begin{figure}
\centering
\includegraphics[trim = 0.5cm 0.25cm 0.5cm 0.5cm,clip,width=0.45\textwidth]{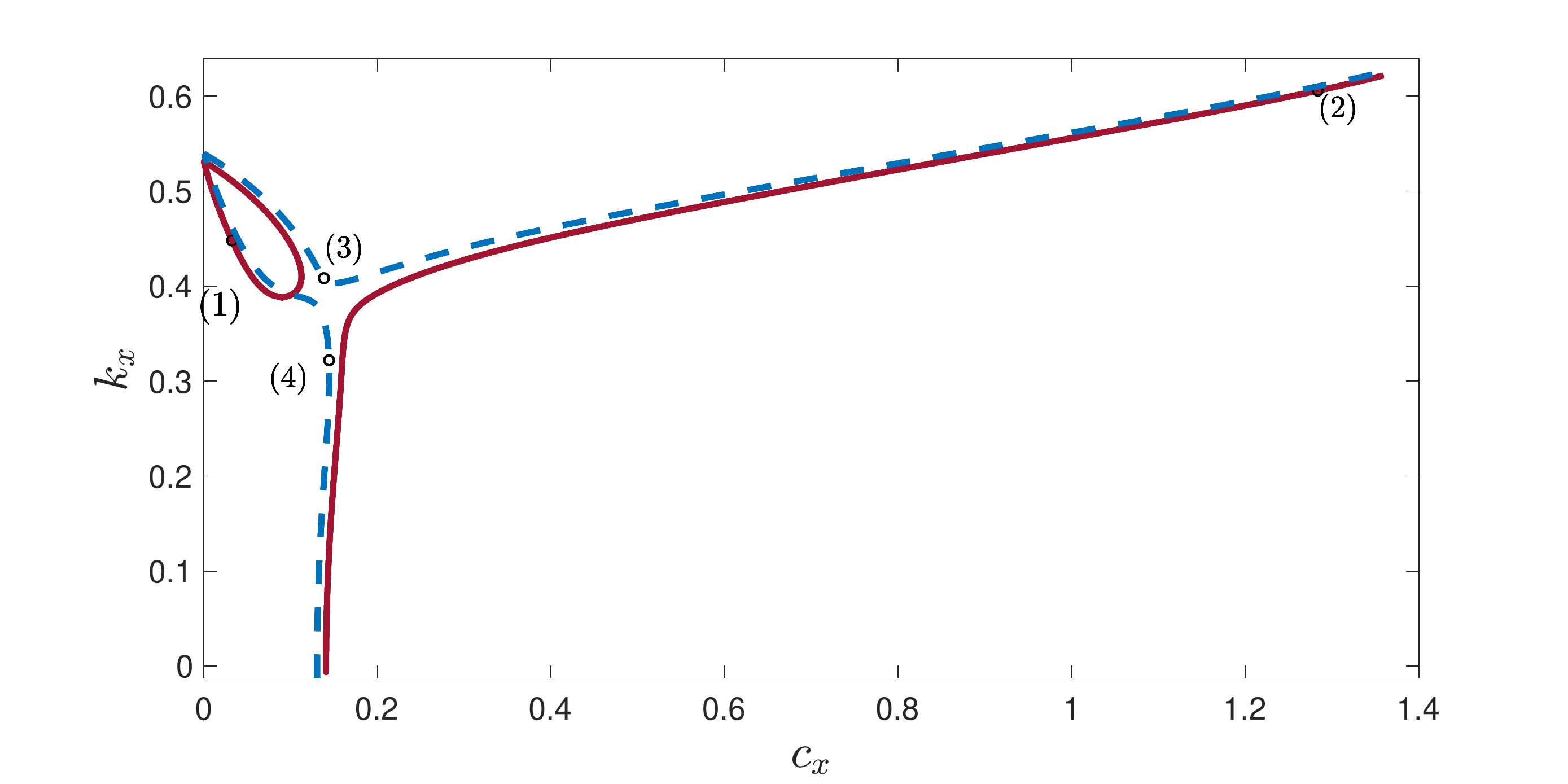}
\includegraphics[trim = 0.0cm 0.0cm 0.0cm 0.0cm,clip,width=0.4\textwidth]{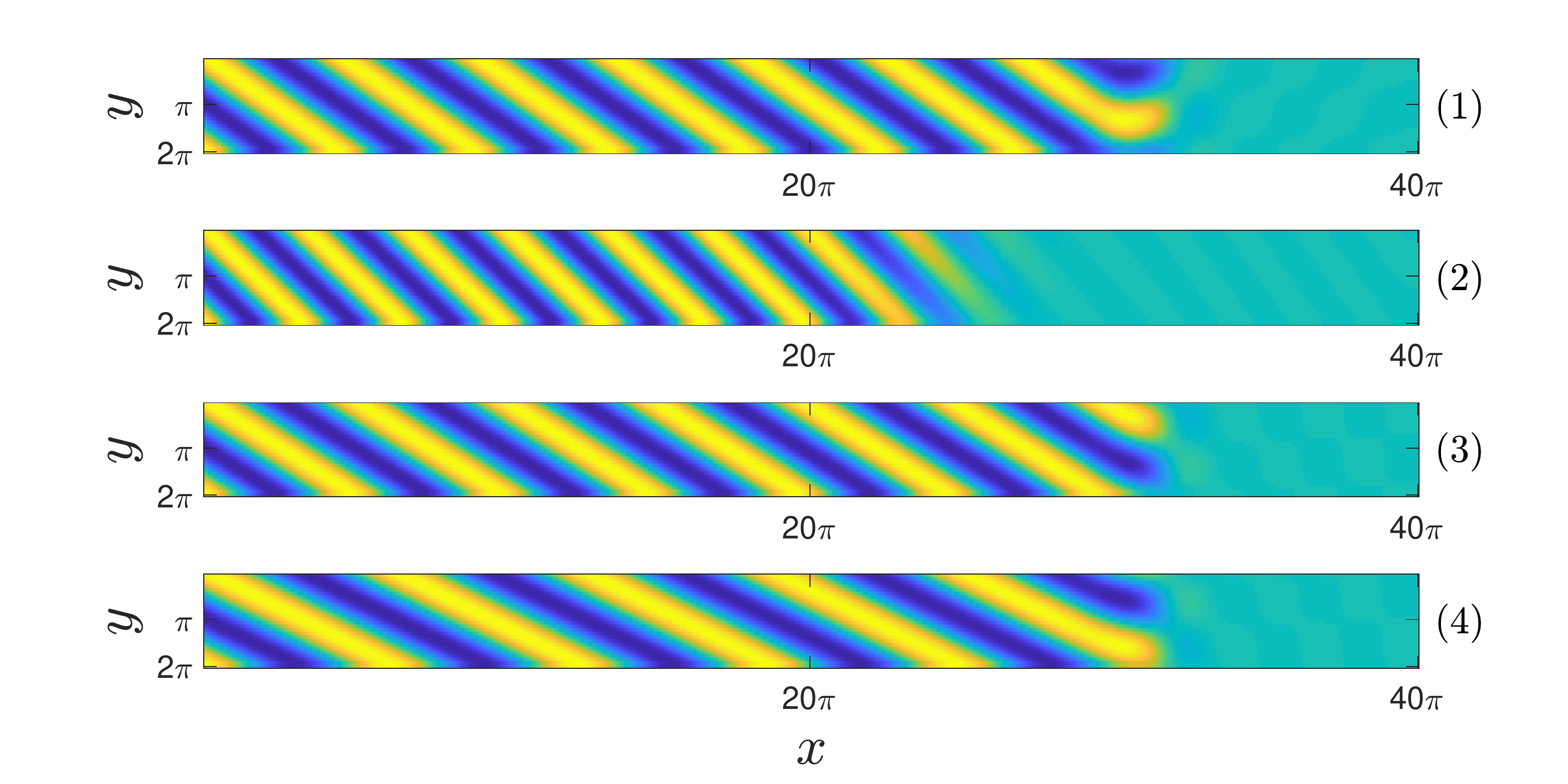}\\
\caption{Left: Slices of the $\mc{M}$ for $\mu = 1/4$, for $k_y = 0.842$ (dashed blue) and $k_y = 0.847$ (solid red). Right: Solution profiles at various points (1) - (4) along the two slices. Reproduced with permission from \cite[Fig. 21]{avery2019growing}. Copyrighted by SIAM. }\label{f:hypcat}
\end{figure}

As the kink-dragging bubble discussed in Section \ref{sss:acc-ob} expands for decreasing $k_y$, it eventually collides with the main surface of oblique stripes in a hyperbolic catastrophe. That is the two branches of oblique stripes on either side of the kink-forming saddle node separate, with one continuing upwards in $c_x$ towards the all-stripe detachment boundary $c_\rlin(k_y)$ while the other branch has a rapid drop in $k_x$ and connects with the perpendicular stripe surface; see Figure \ref{f:hypcat}. Very little is known about this singularity.

\section{Stability and dynamics of patterns}\label{s:stab} 

Stability of quenched fronts is poorly understood. We discuss here briefly a general approach and some limiting scenario where partial results are available. Quenched fronts are either equilibria in an appropriately comoving frame or time-periodic. One therefore needs to investigate properties of the linearization in a comoving frame, a constant-in-time parabolic equation, or possibly the period-map to a periodically forced parabolic equation. In either situation, spectral properties of the linearization largely determine stability. Spectra decompose into essential and point spectrum, where the former is entirely determined by the states at spatial infinity. Since the state in the leading edge is typically assumed stable, the stability of the crystalline state in the wake determines stability of essential spectra. The stability of these simple periodic solutions are amenable to a Floquet-Bloch wave analysis, and it is in principle possible to determine spectral properties in many of the situations discussed thus far. A slight complication is the possibility of a convective instability of the pattern created in the wake. In fact, the transition from convective to absolute instability is likely at the origin of much of the complexity in the transitions between perpendicular and oblique fronts at intermediate speeds discussed in Section \ref{ss:perp-ret}. The discrete part of the spectrum of the linearization, the point spectrum, is often more difficult to access. Controlling both real and imaginary part of the spectrum often allows for nonlinear stability results; see for instance \cite{gsu,Avery2021}. 

In the remainder of this section, we describe examples where some understanding of point spectrum is available. The arguably simplest example are quenched fronts at $c_x=k_y=0$, which are simply solutions to a four-dimensional ODE. As pointed out in \cite{morrissey}, the monotonicity of the strain-displacement relation gives a parity index on the number of unstable eigenvalues. Numerics suggest that this number is minimal, 0 or 1, in the present case of the Swift-Hohenberg equation with a small quench. On the other hand, \cite{morrissey} outlines many other examples of pattern-forming systems where at times stability information may be more immediately accessible, in particular the Ginzburg-Landau equation and the phase diffusion equation, discussed at several instances above as an approximation. In the phase-diffusion equation, stability is readily accessible through the sign of the strain-dispersion relation. Stability in the Ginzburg-Landau setting appears to be related to monotonicity of the amplitude profile; see \cite{morrissey}. In particular, real solutions of the Ginzburg-Landau equations are stable against real perturbations precisely when the real part is monotone; see \cite{monteiro2017phase}. 

For finite speed, the phase-diffusion approximation still allows for a quite complete existence and stability analysis, with strain-displacement relation incorporated into the boundary condition; see \cite{pauthier}. In addition to spectral stability, the results there include more global convergence to the time-periodic pattern-forming solutions. Local stability in this approximation has also been established for oblique stripes in \cite{chen2021strain}. It would be interesting to incorporate the possibility of zigzag instabilities with a Cahn-Hilliard approximation for local wavenumbers as in \cite{avery2019growing}. 

Stability also appears to be accessible near the detachment limit, where one can study stability of the quenched fronts as a perturbation of the free invasion front. Spectral and at times nonlinear stability of free invasion fronts is known in many examples, including the Ginzburg-Landau amplitude approximations and to some extent the Swift-Hohenberg equation \cite{Avery2021,eckmann_schneider}. Recently, spectral stability has been established near this detachment limit in the context of the complex Ginzburg-Landau equation; see the discussion in Section \ref{ss:cgl} and \cite{goh2020spectral}.

\section{Other types of growth and quenching}\label{s:other}

In addition to the simple parameter step which allows or precludes patterns depending on the side of the interface discussed above, there are of course many other types of quenching mechanisms and heterogeneities. We briefly mention a few specific cases of interest.

\subsection{Slow parameter ramps}\label{ss:slow-ramp}
\begin{figure}[!ht]
\centering
\includegraphics[trim = 0.5cm 0.1cm 0.4cm 0.4cm,clip,width=0.4\textwidth]{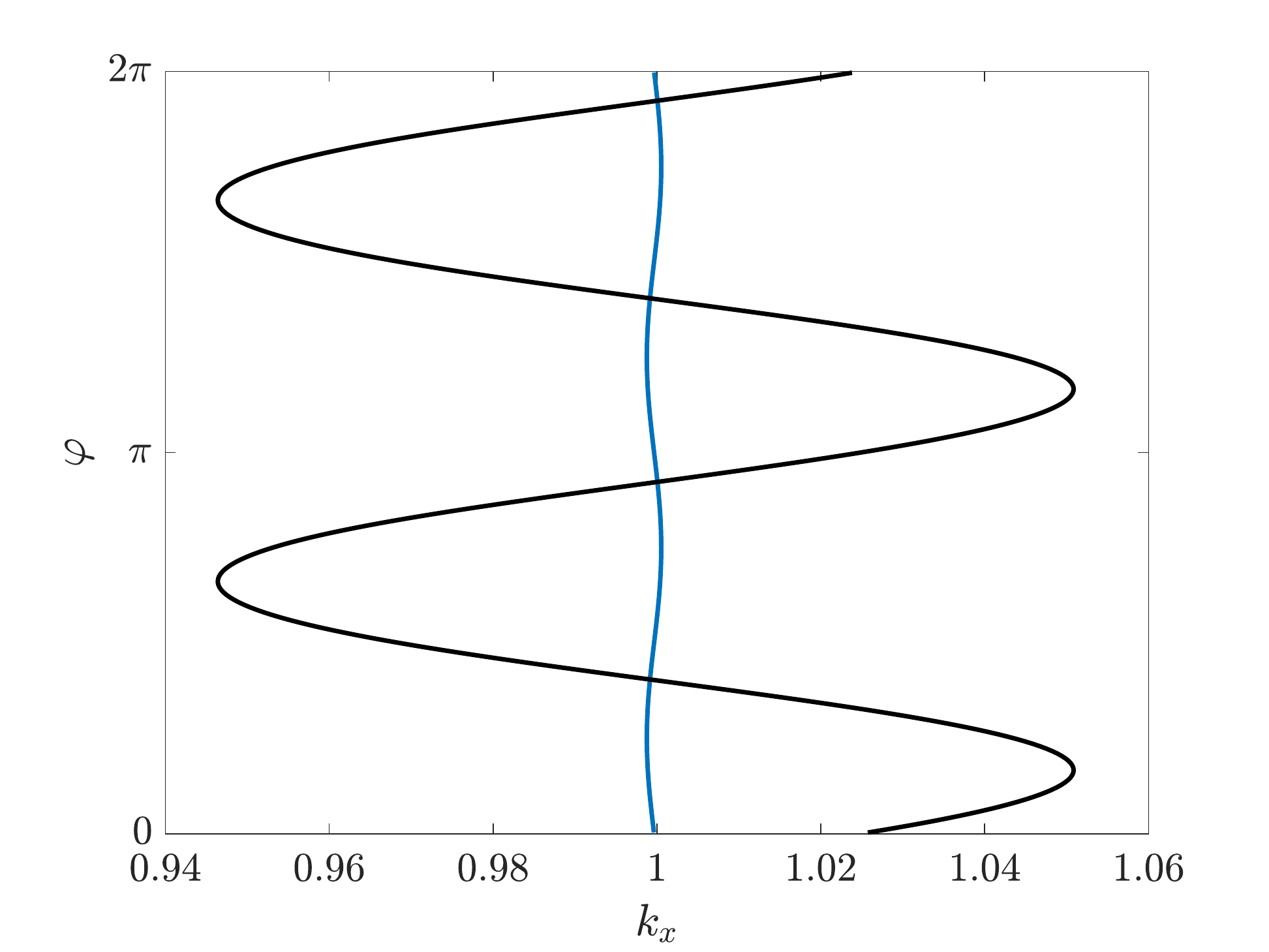}\hspace{-0.2in}
\includegraphics[trim = 0.2cm 0.1cm 0.4cm 0.4cm,clip,width=0.4\textwidth]{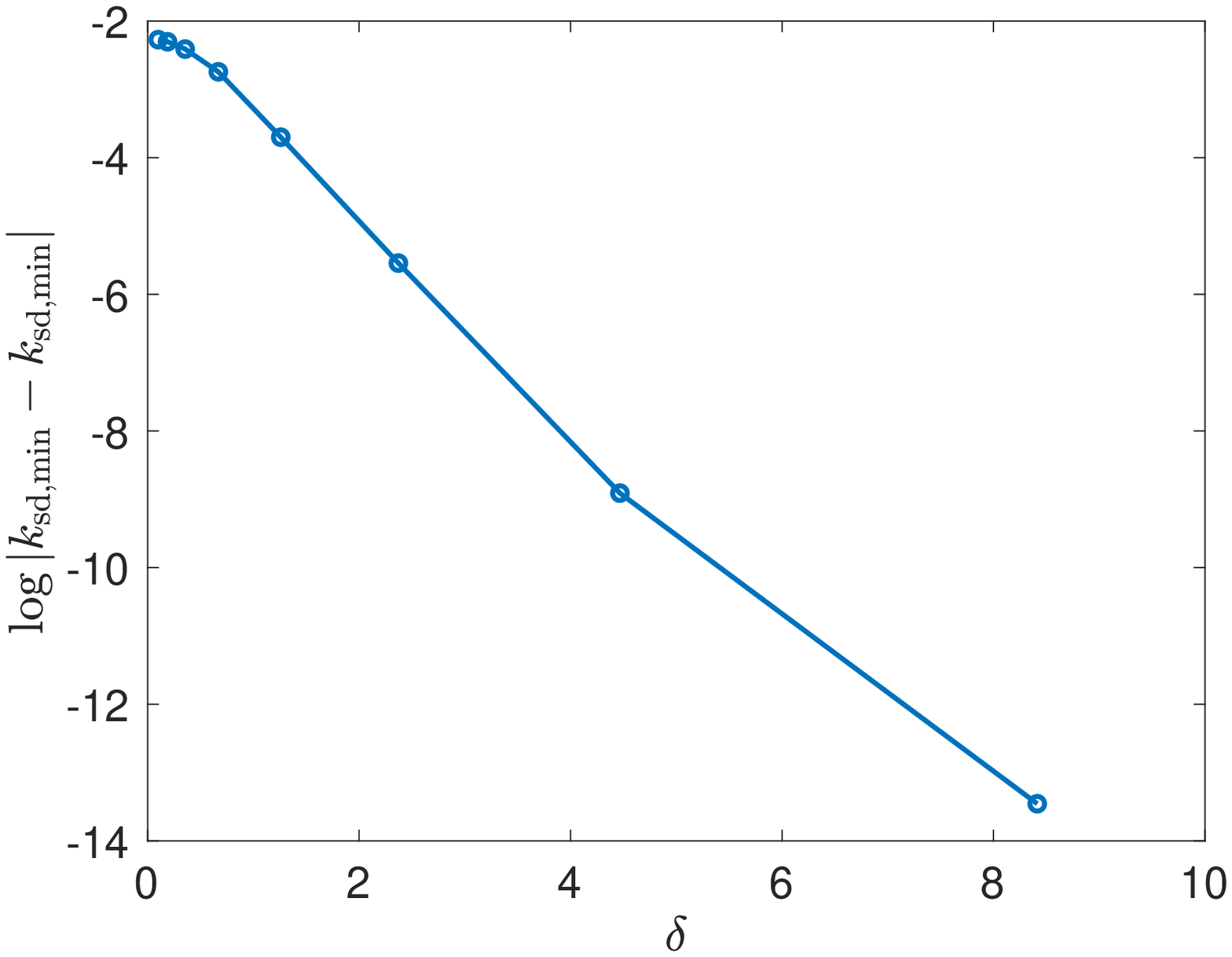}\hspace{-0.2in}\\
\includegraphics[trim = 0.5cm 0.0cm 0.4cm 0.4cm,clip,width=0.4\textwidth]{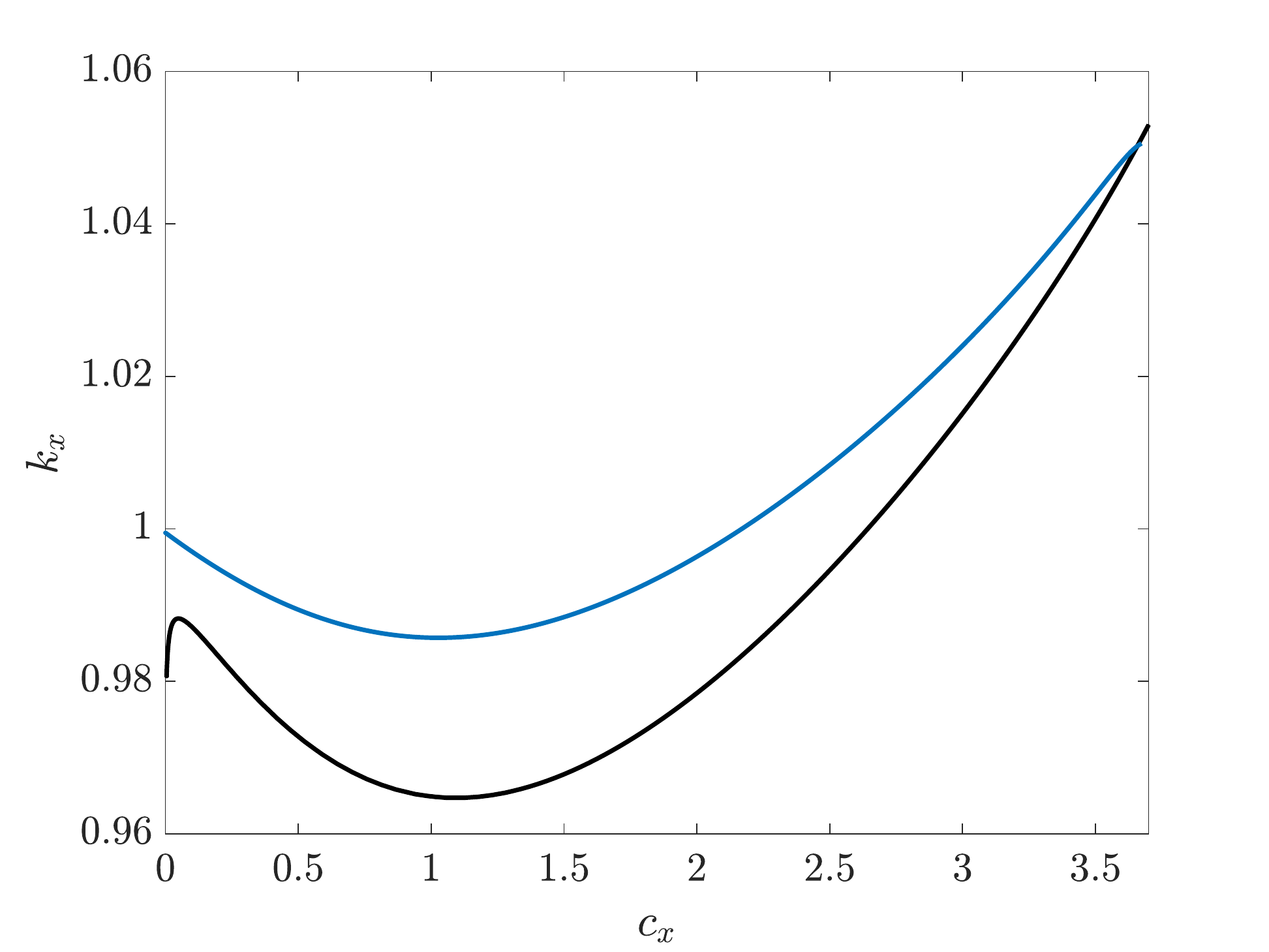}
\includegraphics[trim = 0.5cm 0.0cm 0.4cm 0.0cm,clip,width=0.4\textwidth]{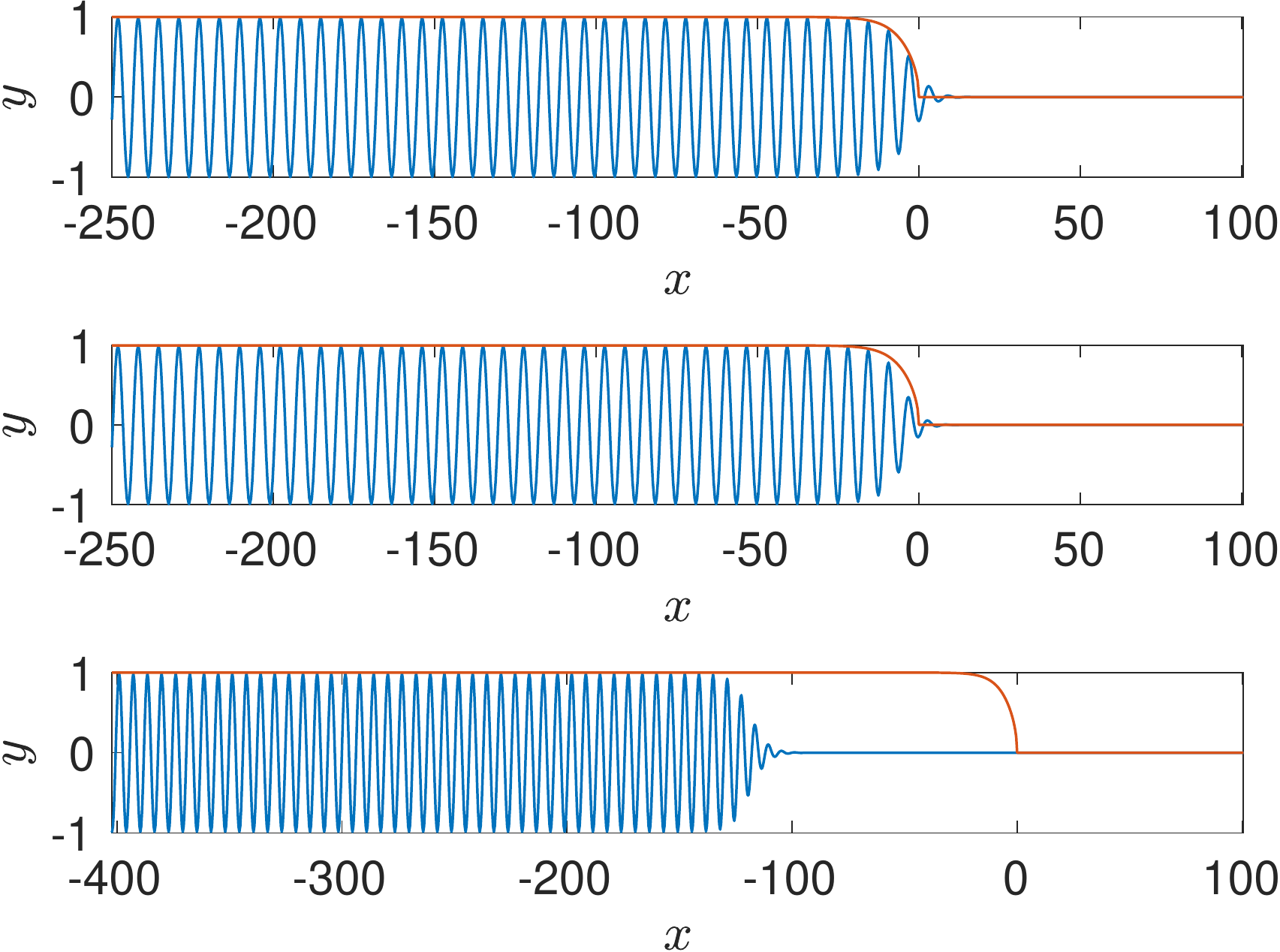}
\caption{Upper left: Strain-displacement relations for the stationary problem $c_x= 0$ in both the step (black) and slow ramp with $\delta = 10$ (blue) parameter heterogeneities, $\mu = 3/4$; upper right:  semi-log plot of the numerical range of the strain-displacement relation for several ramp slopes $\delta\in(0.1,8)$ indicating an exponential decay for large $\delta$,  Bottom left: $k_y=0$ slice of the moduli space for the parameter step (black) and slowly varying ramp with $\delta = 10$ (blue), Bottom right: Slices of solution profiles (blue) for a fixed $y$-value with $\delta = 10$ for speeds $c = 0.002, 0.851, 3.82$, plotted against the curve $\sqrt{4\rho(x)/3}$ (orange) for $x<0.$  }\label{f:sh-sd}
\end{figure}

Contrasting the rapid change in parameters modeled by the step-function $\rho(x)$, one could ask for the effect of slowly varying $\rho(x)$. In fact, smooth but rapidly varying quenches yield  qualitatively similar results to the case of step-function like parameter quenches discussed thus far. On the other hand, slow quenches  have been studied in the past with both stationary \cite{pomeauzaleski} and moving \cite{PIER200149} interfaces. In the former, it was found that the band of selected wavenumbers inside $(k_\mathrm{ex,-},k_\mathrm{ex,+})$ is narrowed significantly compared with the range of the strain-displacement relationship for the steep parameter step discussed in Sec. \ref{ss:stft} above.  Figure \ref{f:sh-sd} gives numerical results plotting the strain displacement relationship for \eqref{e:mtw-x} with $c_x = 0$ and $\rho(x) = -\mu\tanh(x/\delta)$ for $\delta\gg1$. The amplitude $k_\mathrm{sd,max} - k_\mathrm{sd,min}$ is exponentially small in $\delta$. 
The bottom left plot also shows the $k_y = 0$ cross section of the moduli space for varying $c_x$ for two values of $\delta$. We find the local maximum for $c_x$ small practically disappears as $\delta$ increases.  Heuristically, the slow ramping of the parameter suppresses the slow-fast stick/slip phase-pinning effect discussed in Section \ref{sss:slow-para} above. The bottom right plot also depicts solution profiles for several quench speeds. We find that for $x$ large negative, the pattern amplitude goes like $\sqrt{4\rho(x)/3}$. As $c_x$ increases, the front location decreases away from $x = 0$ where $\rho$ switches from negative to positive. This can be understood as follows, for a given fixed quenching speed, perturbations near $x = 0$, where $\rho$ is small, grow but are convected leftwards until they reach the $x$ location where the value $\rho(x)$ renders the trivial state absolutely unstable. In other words, we expect the front interface to be located, at leading order, near the maximum $x$-value where the value $\rho(x)$ makes the trivial state $u = 0$ absolutely unstable for the given quenching speed $c_x$.  Since $\rho$ is slowly varying we expect the next-order correction for the front location to be determined by a dynamic slow fold bifurcation coming from the Jordan block mediating the absolute/convective instability transition. More rigorously, one would transform the system into normal form, with slowly varying coefficients, thus obtaining a Ginzburg-Landau approximation with a slow quench, mimicking the rapid quench construction in \cite[\S 4(c)]{weinburd}. The real part of this equation is analyzed rigorously in \cite{gksv} using geometric singular perturbation theory, locating in particular leading-order asymptotics for the location of the front interface relative to the quench position. This work also shows that the stationary case $c_x = 0$ is governed by a slow-passage through a pitchfork bifurcation with inner solution determined by a unique connecting solution of Painlev\'{e}'s second equation.

While we did not attempt to compute the full moduli space in this case we expect that slow quenches $\delta\gg 1$ drastically restrict the range of selected wavenumbers compared with the sharp parameter jump along the quench.  To our knowledge no rigorous proof of pattern existence and wavenumber selection in any regime has been obtained so far. 

\subsection{Temporal and diffusive quenches}
\begin{figure}[!ht]
\centering
\includegraphics[trim = 4.4cm 4.4cm 0.5cm 2.5cm,clip,width=0.3\textwidth]{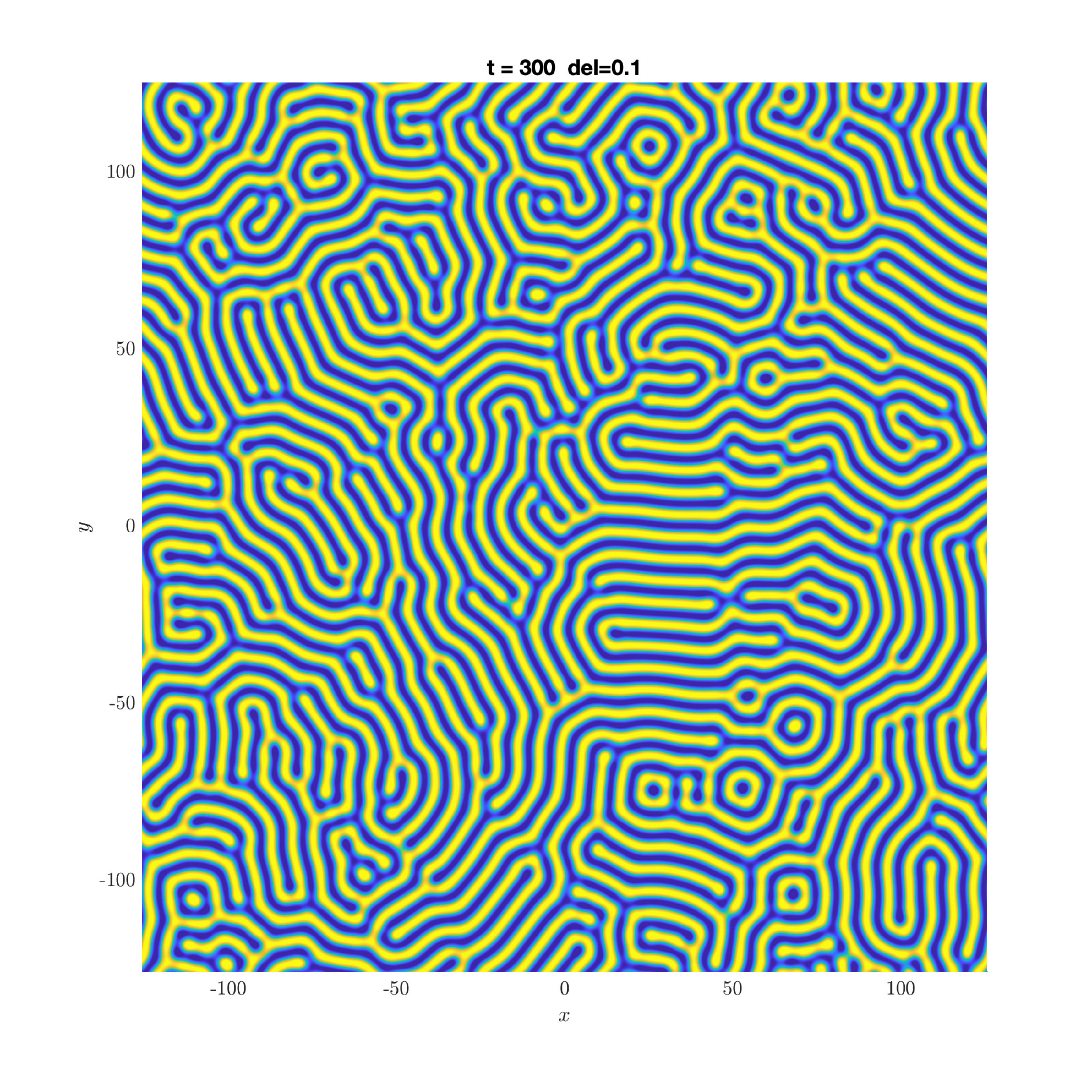}\hspace{-0.2in}
\includegraphics[trim = 4.4cm 4.4cm 0.5cm 2.5cm,clip,width=0.3\textwidth]{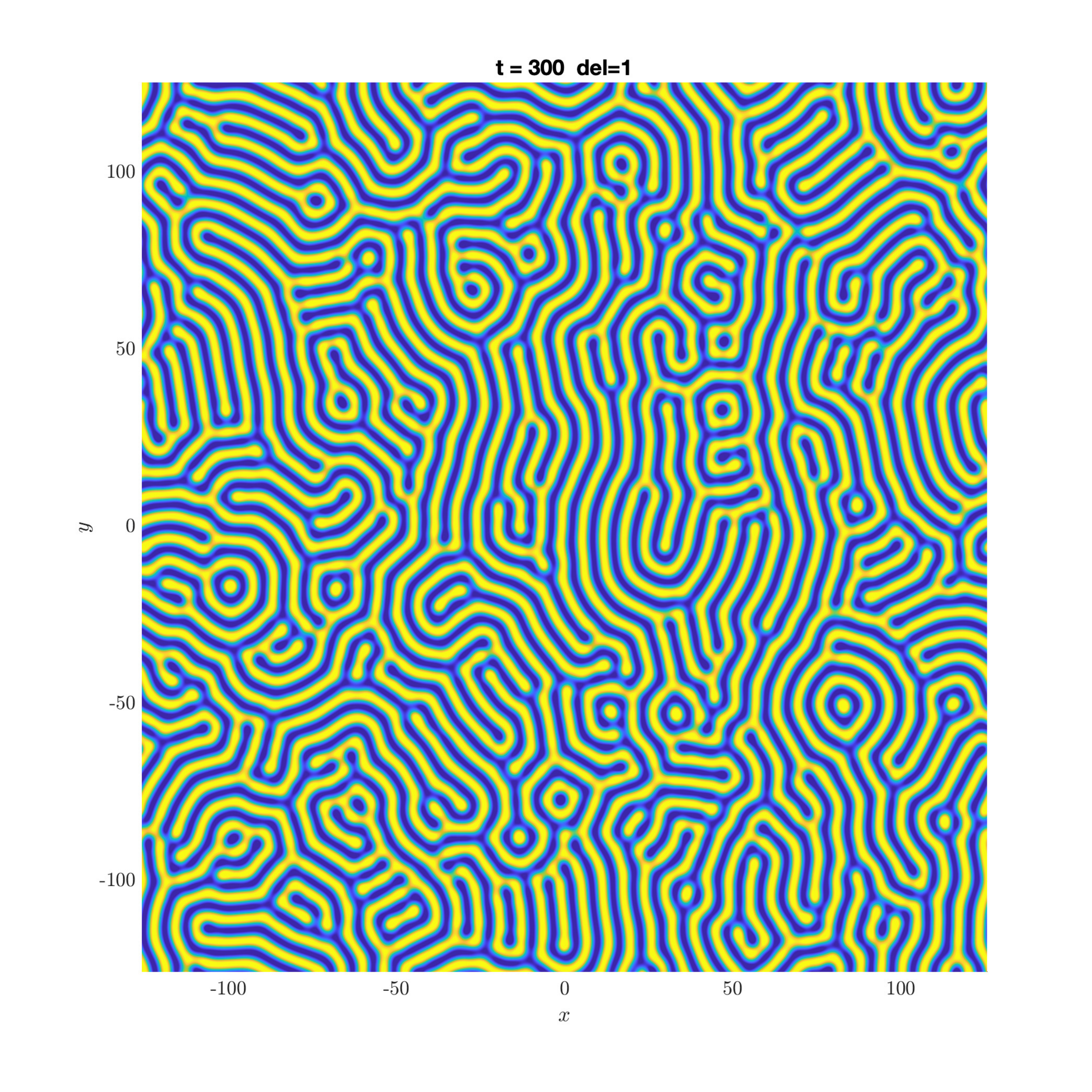}\hspace{-0.2in}
\includegraphics[trim = 4.4cm 4.4cm 0.5cm 2.5cm,clip,width=0.3\textwidth]{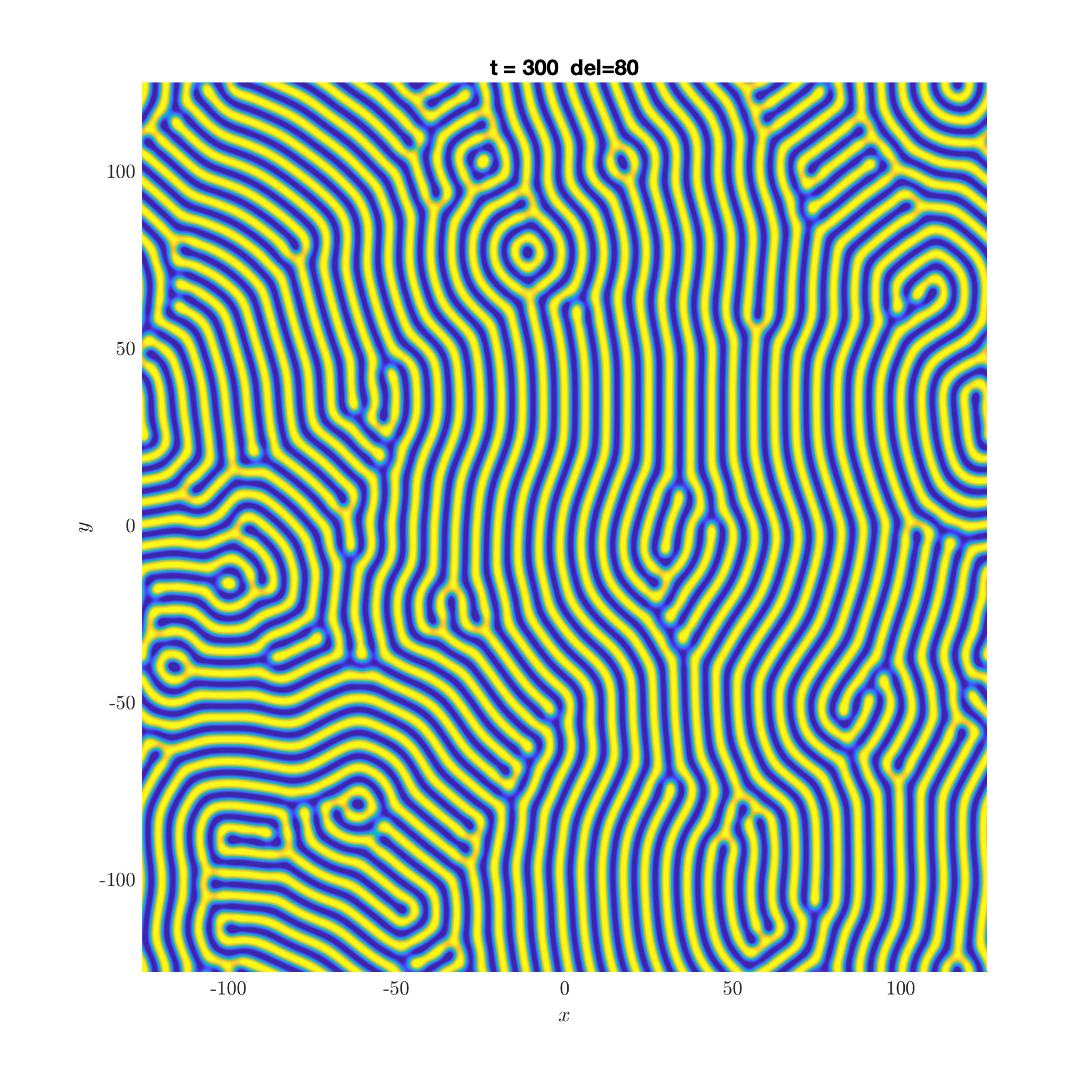}
\caption{Solutions of Swift-Hohenberg with spatially homogeneous slow quench at time $t = 0$ with $t_0 = -300$, and $\delta = 0.1, 1, 80$ left to right; domain is $[-40\pi,40\pi]^2.$ }\label{f:slow-homog}
\end{figure}

We mention two variations of the simple spatio-temporal quench $\rho(x-c_x t)$. First, consider the limit of infinite speeds $c_x$, which can be recast as a purely temporal quench,
\beq
\rho(t) = -\mu\tanh( t/\delta).
\eeq
As apparent in our earlier discussion, we do not expect pattern formation to be governed by coherent front solutions, such as solutions periodic in $x$ and heteroclinic in $t$, since the "infinite" speed $c_x$ here is clearly above the linear spreading speed. Of interest is then in how far this quench still leads to reduced presence of defects as did the directional quench that we have studied thus far. There are in fact a number of heuristics, known as the Kibble-Zurek mechanism \cite{zurich85,kibble1976topology}, that predict  defect densities that vary inversely with the parameter ramp speed when initializing with white noise initial data. 
Amplitude analysis near onset shows that this density varies as $\delta^{-\frac{1}{2}}$ for $\delta$ large \cite{stoop2018defect}; see Figure \ref{f:slow-homog} for the direct simulation results with this heterogeneity for a few values of $\delta$ and random white noise initial data. Note the qualitative difference in the resulting pattern as the quench rate, which roughly varies inversely proportionally to $\delta$, is decreased. A brief qualitative view of these results reveals the formation of fewer point defects, and larger domains of pure stripes, for larger $\delta$.

\begin{figure}[!ht]
\centering
\includegraphics[trim = 0.5cm 0.25cm 0.5cm 0.5cm,clip,width=0.45\textwidth]{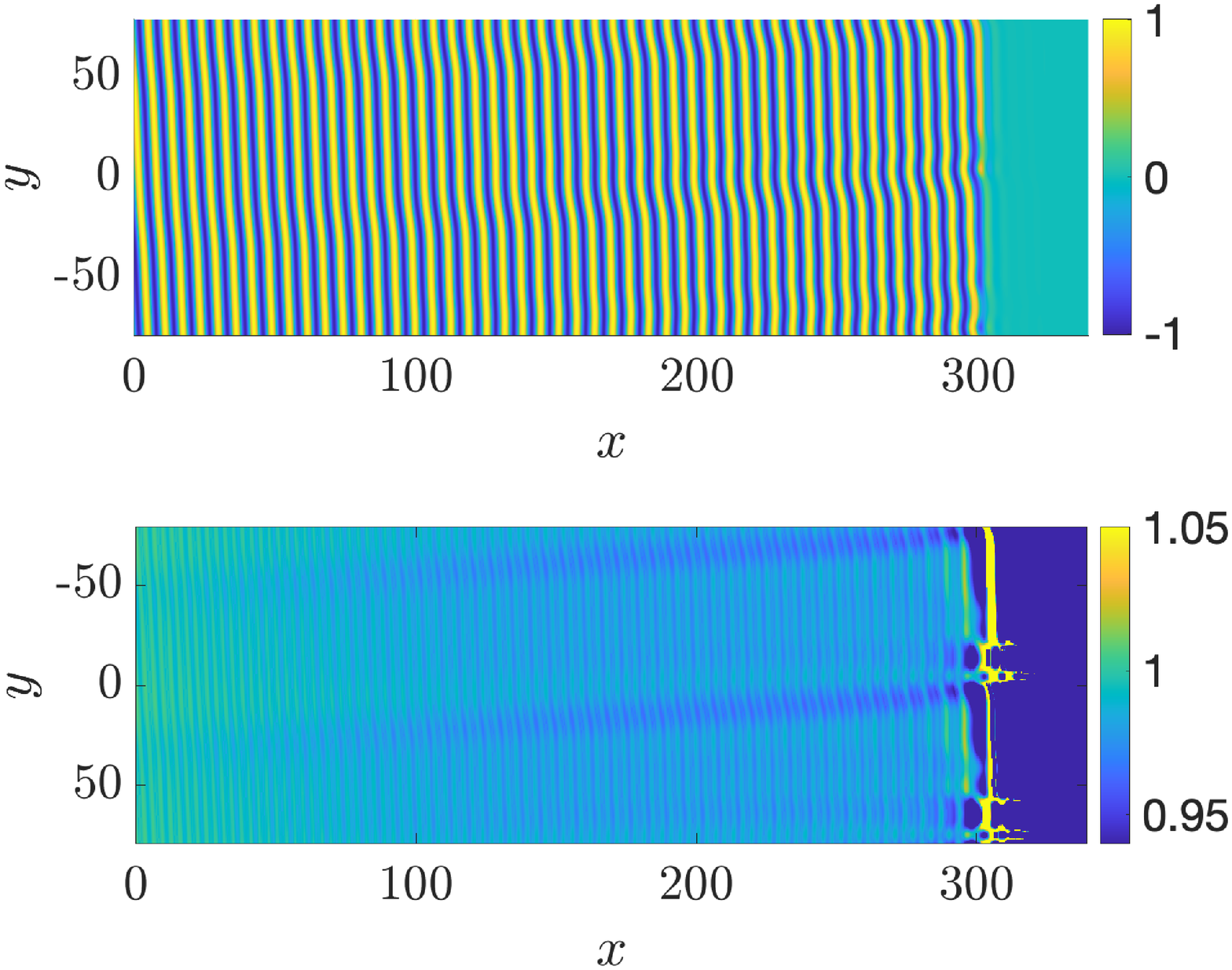}
\includegraphics[trim = 0.2cm 0.25cm 0.5cm 0.5cm,clip,width=0.45\textwidth]{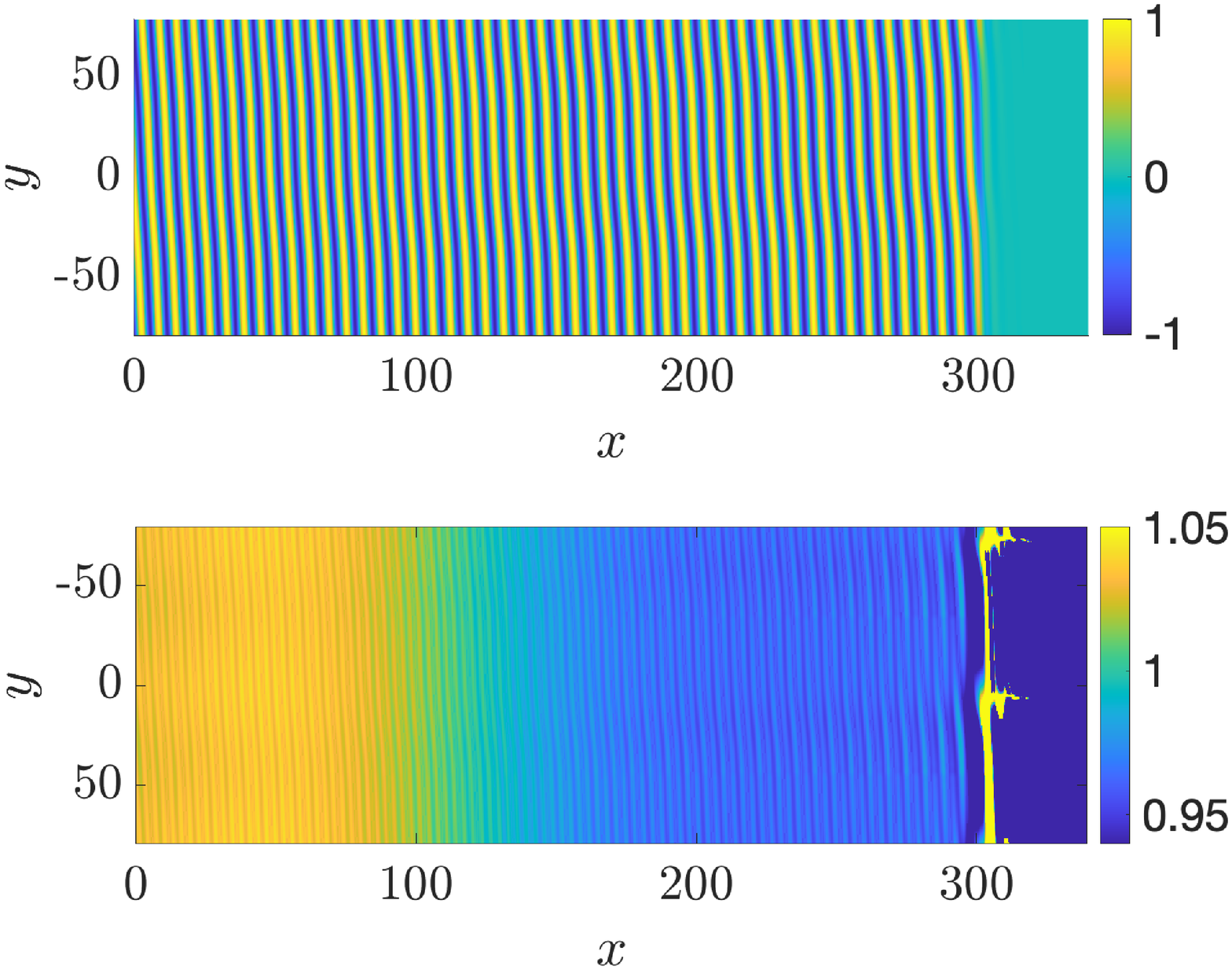}
\caption{Result of pattern selected in the wake of the diffusively propagating quench, for $d = 40$ (left) and $d = 8$ (right), top figures give resulting pattersn at the final time of the simulation, determined when the quench reaches four-fifths of the domain; bottom figures gives the local wavenumber, measured using a Hilbert transform approach (see \cite[pg. 5]{lloydscheel}). }\label{f:sh-diffq}
\end{figure}

In a different direction, the parameter quench is at times given by the propagation of a diffusive signal, leading to  quenches of the form \cite{goh2011spatial}, 
 $$
 \rho(x,t) = -\mu\mathrm{sign}(x - d\sqrt{ t}).
 $$
One expects a patterned state with non-uniform wavenumber in the wake of the quench since instantaneous speeds vary as $\frac{d}{2\sqrt{t}}$ so that stripes are grown quickly for small times, and then progressively slowly as the quench moves forward.  We observe that the quench dynamically explores different regions of the moduli space as time evolves. Figure \ref{f:sh-diffq} shows the result of a diffusively traveling quench, seeded with a weakly oblique stripe with $k_y\sim0$, where for $t\gtrsim0$ the quench travels faster than the linear spreading speed $c_\rlin(k_y)$, selecting a large wavenumber in the horizontal direction. Later on the pattern catches up with the quench and a much smaller wavenumber is selected, which continues to decrease as the growth speed decreases. Also note that the glide-dislocation defect discussed in Sections \ref{sss:slow-ob} and \ref{ss:an-sh} begins to develop for progressively slow speeds.

\begin{figure}[!ht]
\centering
\includegraphics[trim = 4.4cm 4.4cm 4.4cm 2.05cm,clip,width=0.33\textwidth]{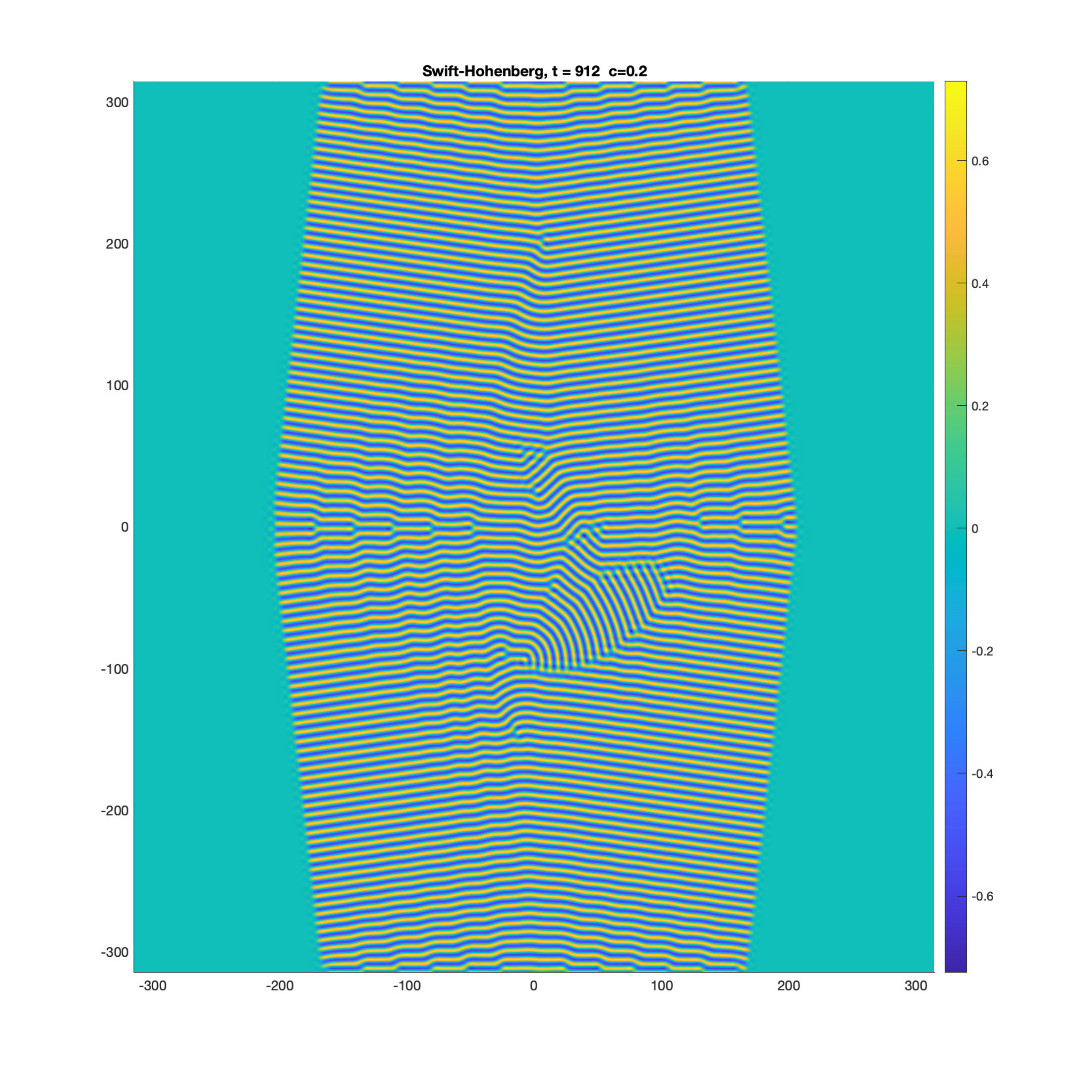}\hspace{-0.01in}
\includegraphics[trim = 4.4cm 4.4cm 4.4cm 2.05cm,clip,width=0.33\textwidth]{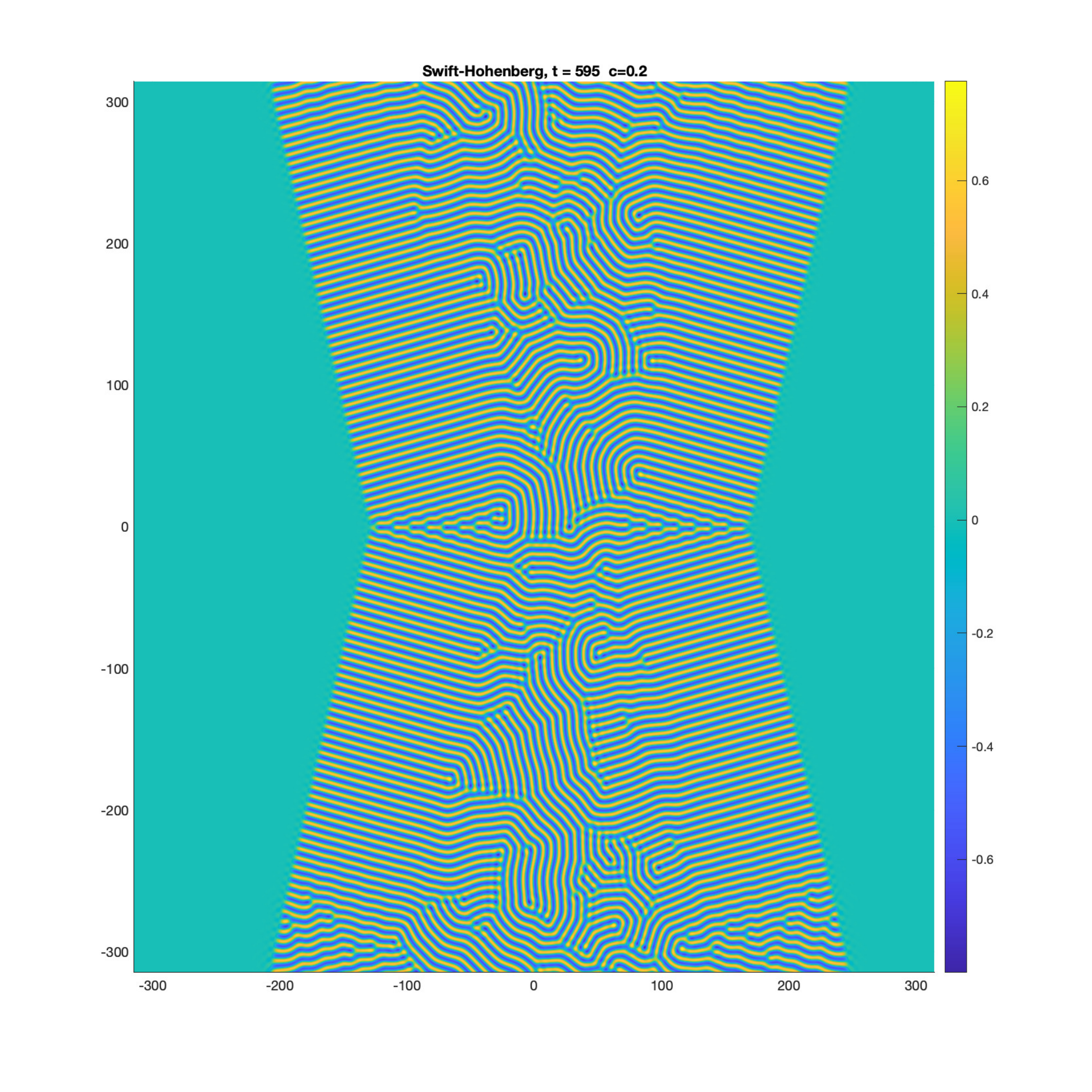}\hspace{-0.01in}
\includegraphics[trim = 4.4cm 4.4cm 4.4cm 2.05cm,clip,width=0.33\textwidth]{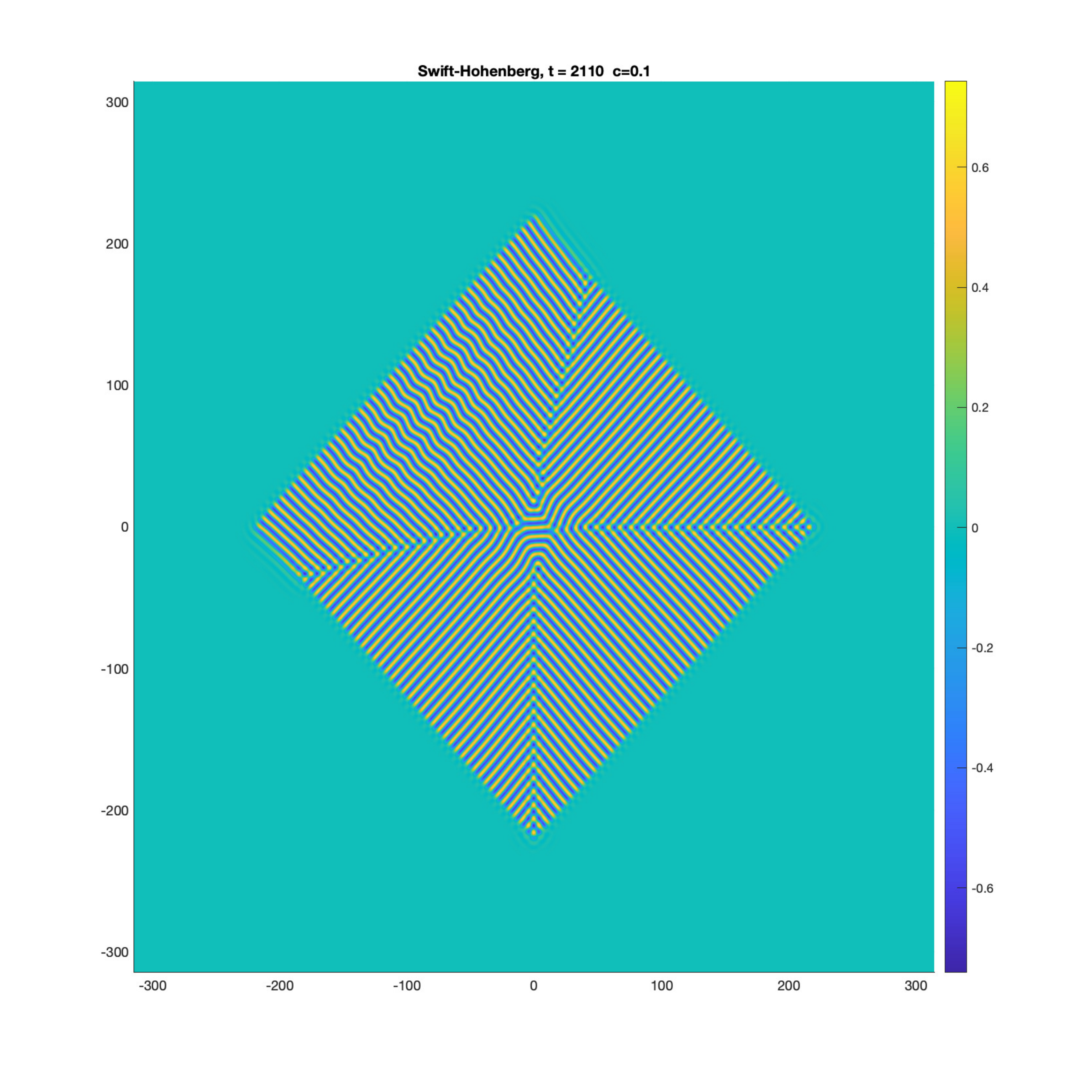}\hspace{-0.01in}\\
\includegraphics[trim = 4.4cm 4.4cm 4.4cm 2.05cm,clip,width=0.33\textwidth]{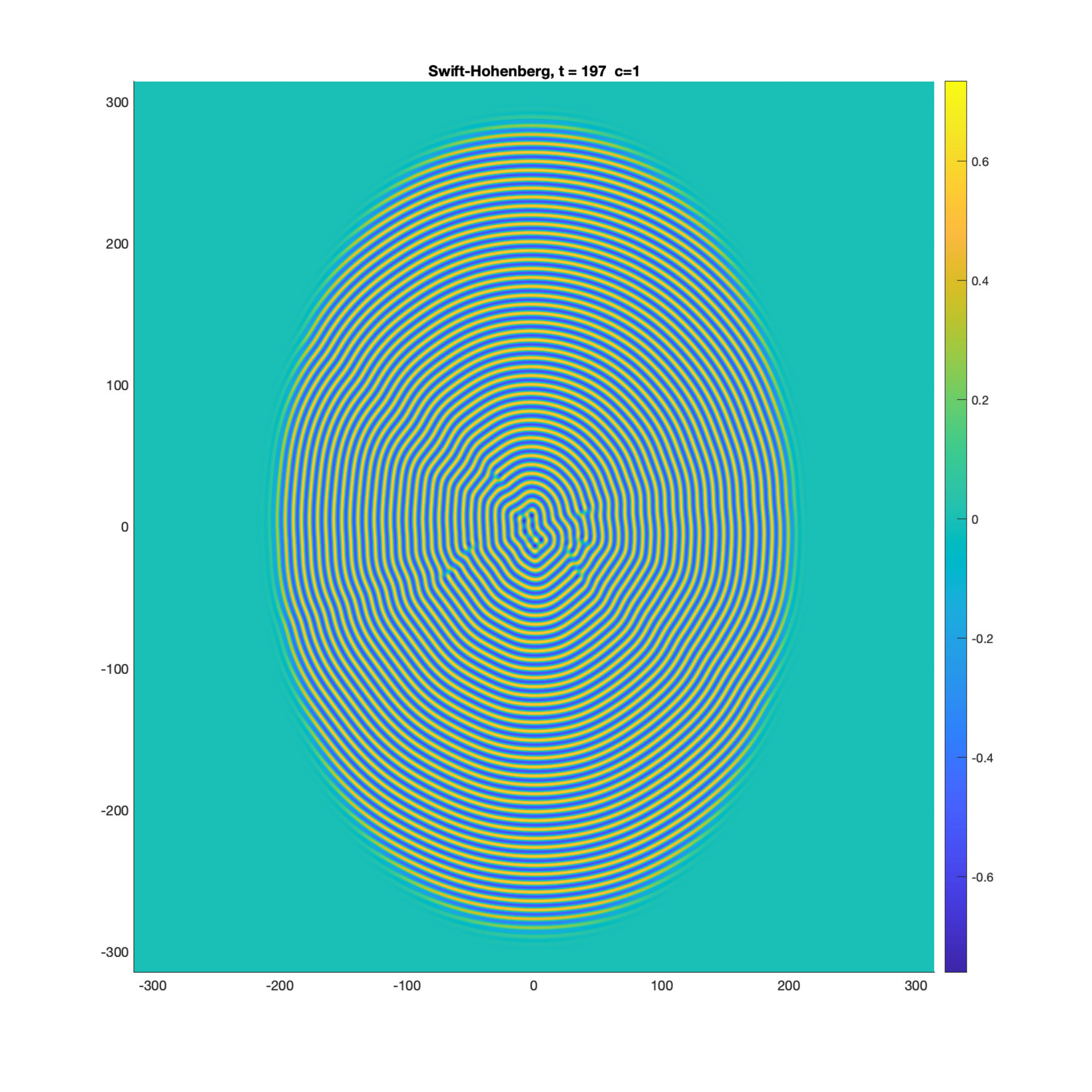}\hspace{-0.01in}
\includegraphics[trim = 4.4cm 4.4cm 4.4cm 2.05cm,clip,width=0.33\textwidth]{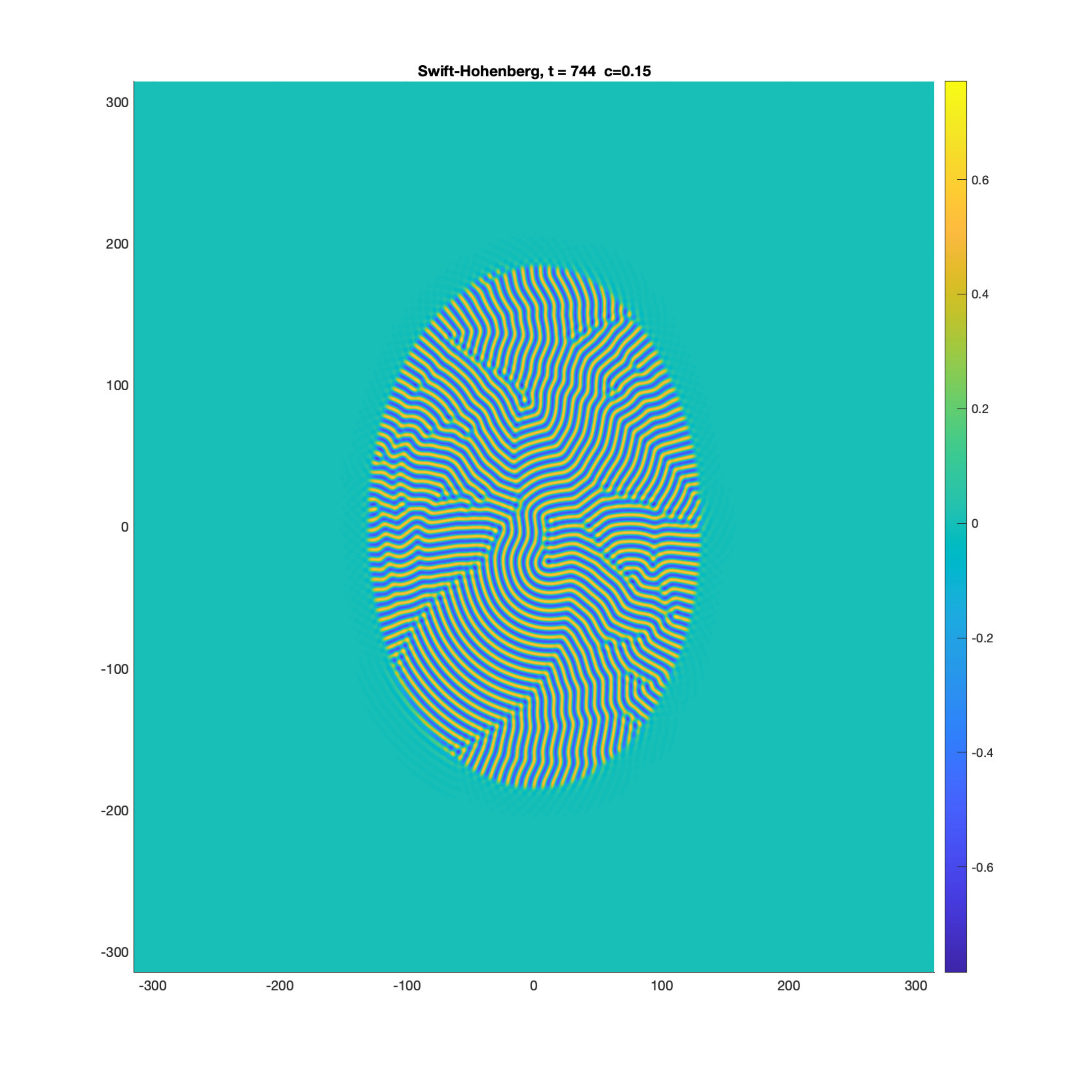}\hspace{-0.01in}
\includegraphics[trim = 4.4cm 4.4cm 4.4cm 2.05cm,clip,width=0.33\textwidth]{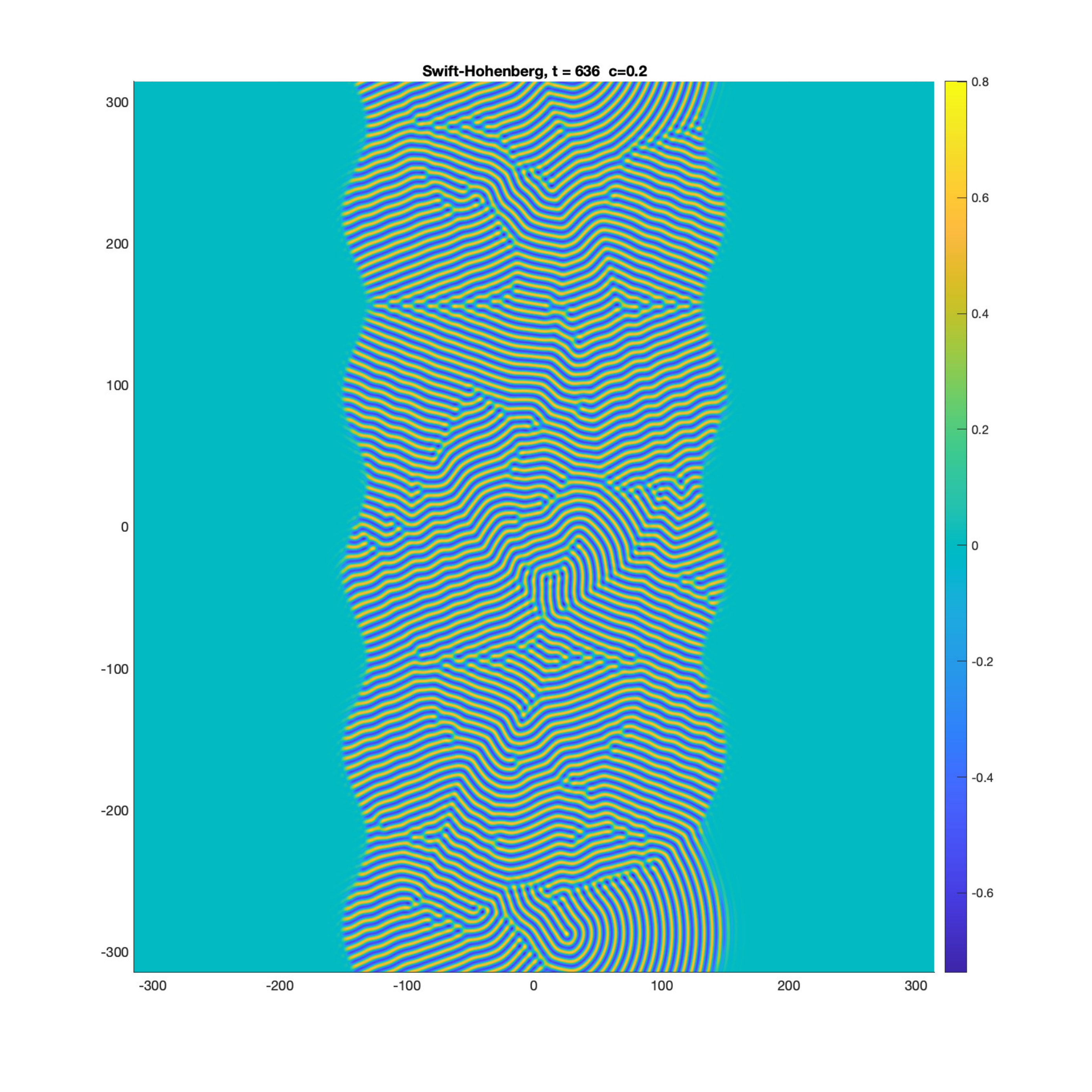}\hspace{-0.01in}
\caption{A range of different geometries for $\partial\Omega_t$, all of which move outward preserving the same geometry. From left to right top to bottom: chevron, reverse chevron, square, ellipse for both fast and slow expansion speed, and sinusoidal interfaces; domain is $[-100\pi,100\pi]^2$. }\label{f:chev}
\end{figure}

For quenching heterogeneities, one could also consider the effect of curvature using a non-directional quench as discussed in \eqref{e:sh-q}, where $\partial \Omega_t$ expands throughout the spatial domain.  Figure \ref{f:rad} depicts patterns in a radially quenched domain, but one could imagine many other interesting domains, such as elliptical, polygonal, or chevron type boundaries. See Figure \ref{f:chev} for a few examples.

Different from the "heteroclinic" quench $\rho\sim -\tanh(x-c_x t)$, one could also consider "homoclinic" quenches, 
$
\rho(x-c_x t) = \mu + h(x-c_x t),
$
with $h$ exponentially or algebraically localized, or even step-like quenches $\rho(x - c_x t) = \mu_l + (\mu_r - \mu_l)\chi(x - c_xt)$, with $\chi(\xi)$ a step-function, equal to 1 for $\xi>0$ and $0$ otherwise, for two positive values $\mu_l,\mu_r$. In this last example, the heterogeneity would mediate an interface between two patterns to the left and right of the quenching interface. Additionally, instead of a parameter heterogeneity, one could add a $u$-independent term of the form $g(x)$ to the equation; see \cite{jaramillo2019effect,jaramillo2015deformation} for related works studying the effect of localized imperfections on asymptotic patterns, a heterogeneous linear differential operator $-(1+\rho_1(x,y)\p_x^2+\rho_2(x,y)\p_y^2)^2+\mu$, or posing the equation on a bounded, or semi-bounded domain with boundary conditions \cite{morrissey,beekie}.

%
%

\section{Moduli surfaces in other prototypical models}\label{s:ex}
We highlight wavenumber selection under directional quenching in several other prototypical models of pattern formation including several alterations of the supercritical Swift-Hohenberg equation discussed above, as well as the complex Ginzburg-Landau, reaction-diffusion, and Cahn-Hilliard equations. 

\subsection{Subcritical cubic-quintic Swift-Hohenberg equation}\label{ss:cqsh}

Different nonlinearities in the Swift-Hohenberg equation can also induce novel wavenumber selection behaviors. For example, a subcritical cubic-quintic nonlinearity
\begin{equation}\label{e:sh-cq}
u_t = -(1+\Delta)^2 u + \rho u + \gamma u^3 - u^5,\quad \rho(x,t) = -\mu\,\mathrm{sign}(x - c_x t),\quad \gamma >1,
\end{equation}
induces novel, non-monotonic wavenumber selection behavior.
\begin{figure}[!ht]
\centering
\includegraphics[trim = 0.5cm 0.25cm 0.5cm 0.5cm,clip,width=0.7\textwidth]{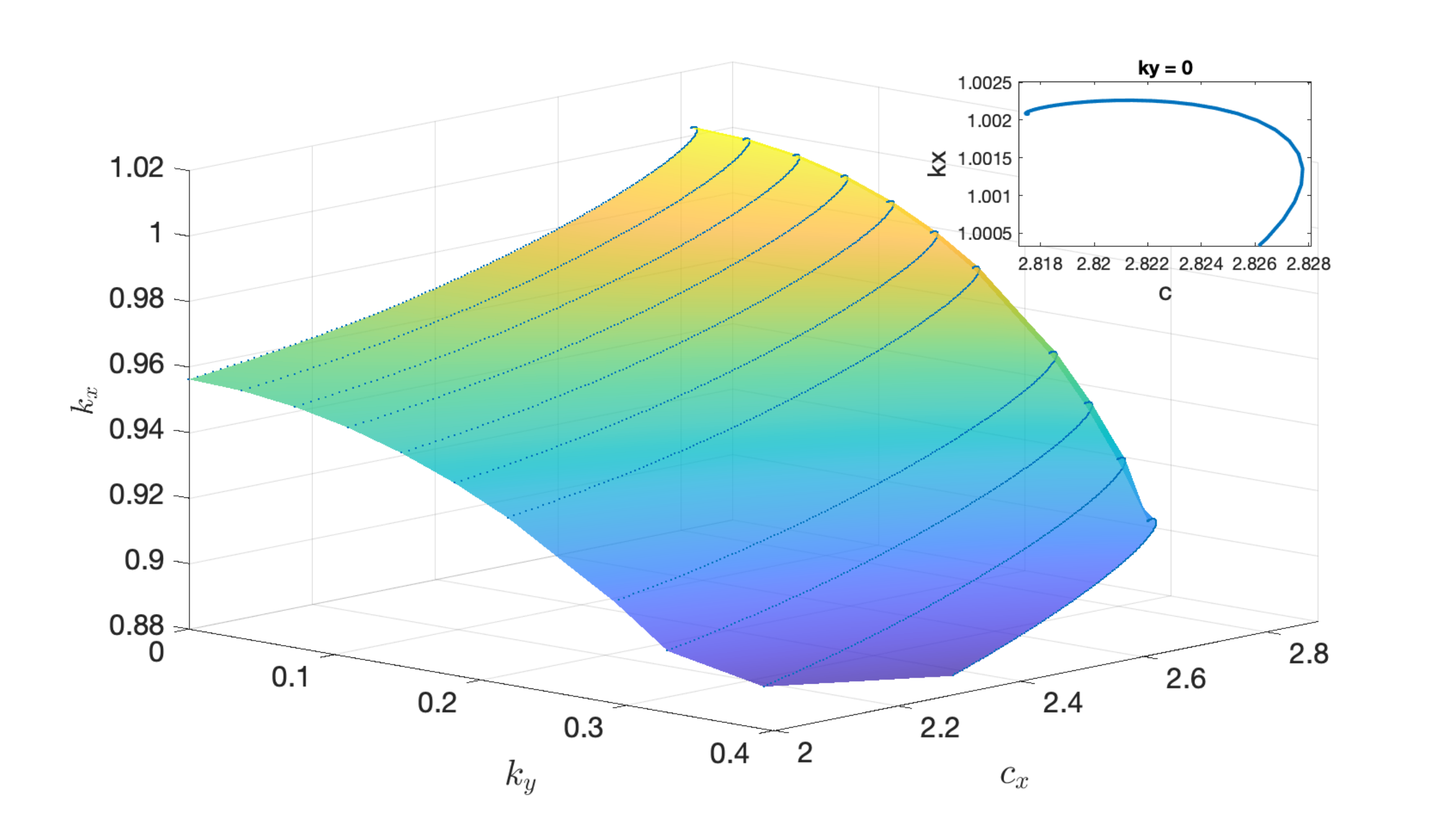}\hspace{-0.2in}
\caption{Part of the moduli space for stripe-forming fronts in \eqref{e:sh-cq} for $\gamma = 2$ and $\mu = 1/4$. Inset gives zoom in of the $k_y = 0$ slice near the nonlinear pushed spreading parameters $(c_\mathrm{p},k_\mathrm{p})$.  }\label{f:sh-push}
\end{figure}
In the corresponding homogeneous equation with $\rho\equiv \mu$, one observes \cite{vS} that the free-invasion, or spreading speed, $c_\mathrm{p}$, of the front formed by the spread of compactly supported perturbations of the unstable base state $u\equiv0$, is faster than the linear spreading speed $c_\mathrm{lin}$ and the patterned selected in the wake has wavenumber, $k_\mathrm{p}$, different than the linear prediction $k_\mathrm{lin}$. Here, the strong nonlinear growth causes perturbations of the unstable state $u\equiv0$ to grow and invade faster than the linear dynamics ahead of the front predict.  Thus, these are often called \emph{pushed fronts}.  

In the one-dimensional case, quenching mechanisms interact with the steep oscillatory tail of the free-invasion front to form a wavenumber selection curve which is not a function of $c_x$ but a logarithmic spiral in $(c_x,k_x)$ space, with center at the free-invasion parameters $(c_x,k_x) = (c_\mathrm{p},k_\mathrm{p})$; see \cite[Thm. 1]{gs2} and Fig. \ref{f:sh-push}. As a consequence, for quenching speeds $c_x\sim c_\mathrm{p}$, a discrete set of wavenumbers are selected. The multi-stability is induced by locking of oscillatory front tails into the position of the quenching line. We highlight in particular that this mechanism induces the existence of fronts for speeds above the free-invasion speed $c_x\gtrsim c_\mathrm{p}$. The result in \cite{gs2} also gives asymptotics for the ``tightness" of the spiral, at leading  order through the complex difference between the strong-stable eigenvalues of the linearization which control the decay of the front and weakly stable spatial eigenvalues of the linearization. The results of numerical continuation in Fig. \ref{f:sh-push} show such non-monotonic wavenumber curves persist for oblique stripes with $k_y>0$ leading to a spiral scroll moduli surface for large quench rates. We remark that similar behaviors were observed in one-dimensional Cahn-Hilliard and complex Ginzburg-Landau equations with similar subcritical nonlinearities \cite{gs2,goh2020spectral}.

\subsection{Quenching in anisotropic pattern-forming systems }\label{ss:an-sh}

\subsubsection{Anisotropic Swift-Hohenberg Equation}\label{sss:an-sh}

Introducing spatial anisotropy into pattern formation can actually simplify spatio-temporal dynamics by restricting the range of available orientations for patterns. As a simple example we consider the Swift-Hohenberg equation 
with strong linear damping in say the vertical direction and quenching in the horizontal direction,
\begin{equation}\label{e:sh-an}
u_t = -(1+\Delta)^2 u + \beta \partial_y^2 u + \rho u -u^3,\quad \rho(x,t) = -\mu\mathrm{sign}(x-c_xt),\quad \beta>0.
\end{equation} 
Here, for $\rho\equiv\mu$, and $\beta >0$, such damping selects stripes roughly parallel to the quenching interface and suppresses the zigzag instability of stripes. It also reduces the presence of defects, apparently eliminating point defects such as disclinations, and line defects such as grain boundaries, leaving dislocations as the main source of disorder. 

\begin{figure}
\centering
\includegraphics[trim = 0.5cm 0.25cm 0.4cm 0.5cm,clip,width=0.33\textwidth]{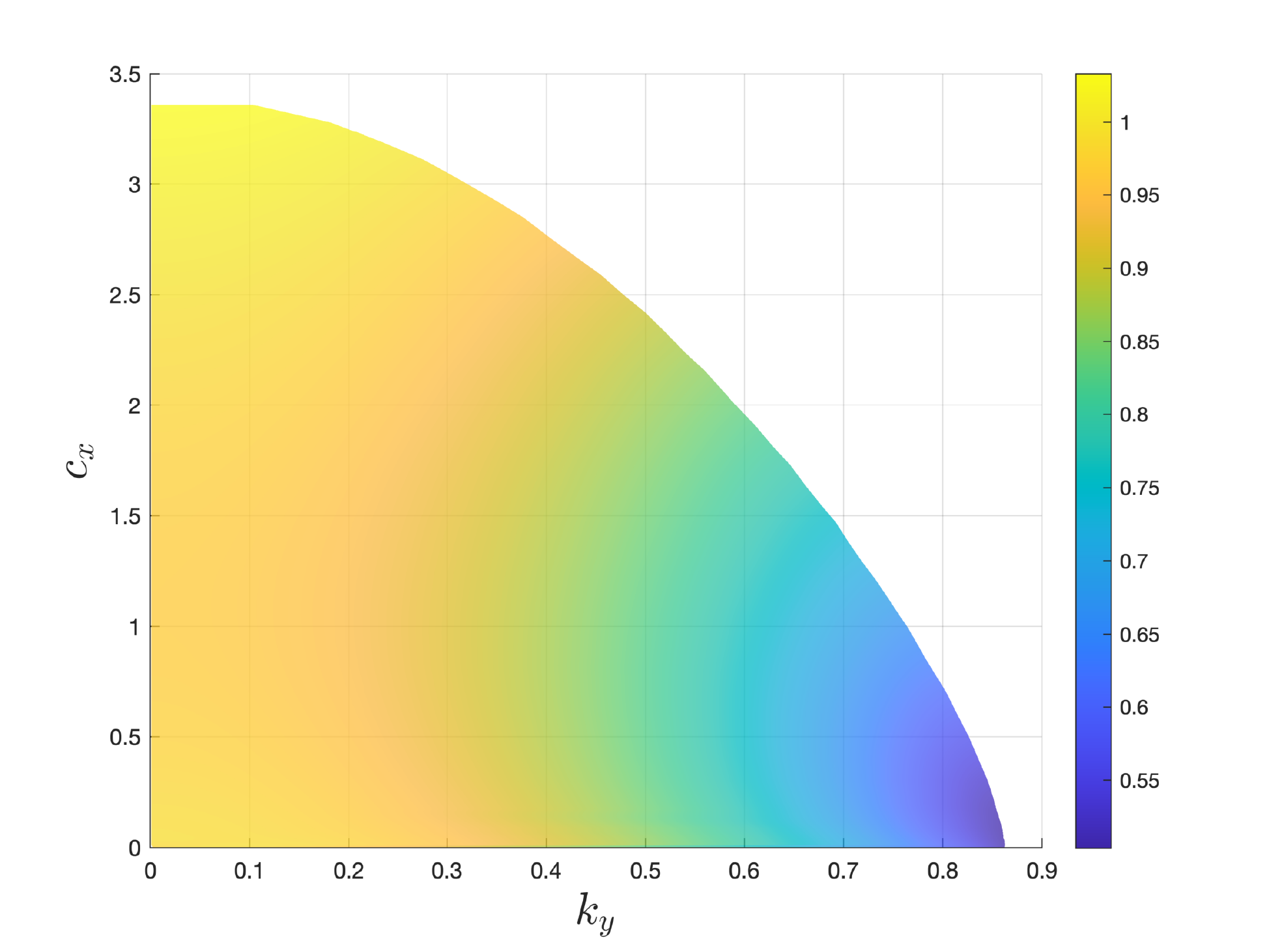}\hspace{-0.2in}
\includegraphics[trim = 0.5cm 0.25cm 0.4cm 0.5cm,clip,width=0.33\textwidth]{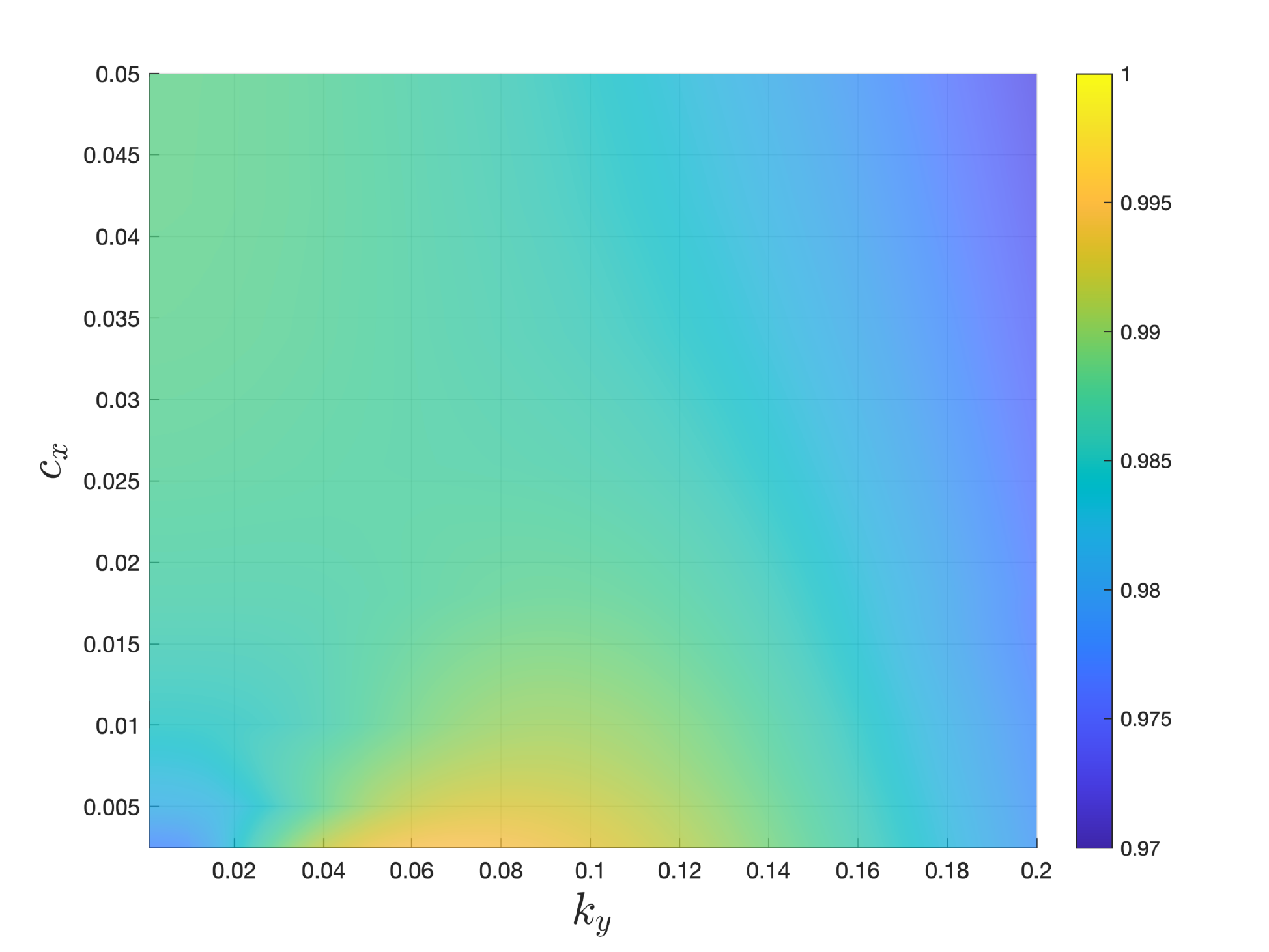}\hspace{-0.2in}
\includegraphics[trim = 0.cm 0.cm 0.cm 0.0cm,clip,width=0.33\textwidth]{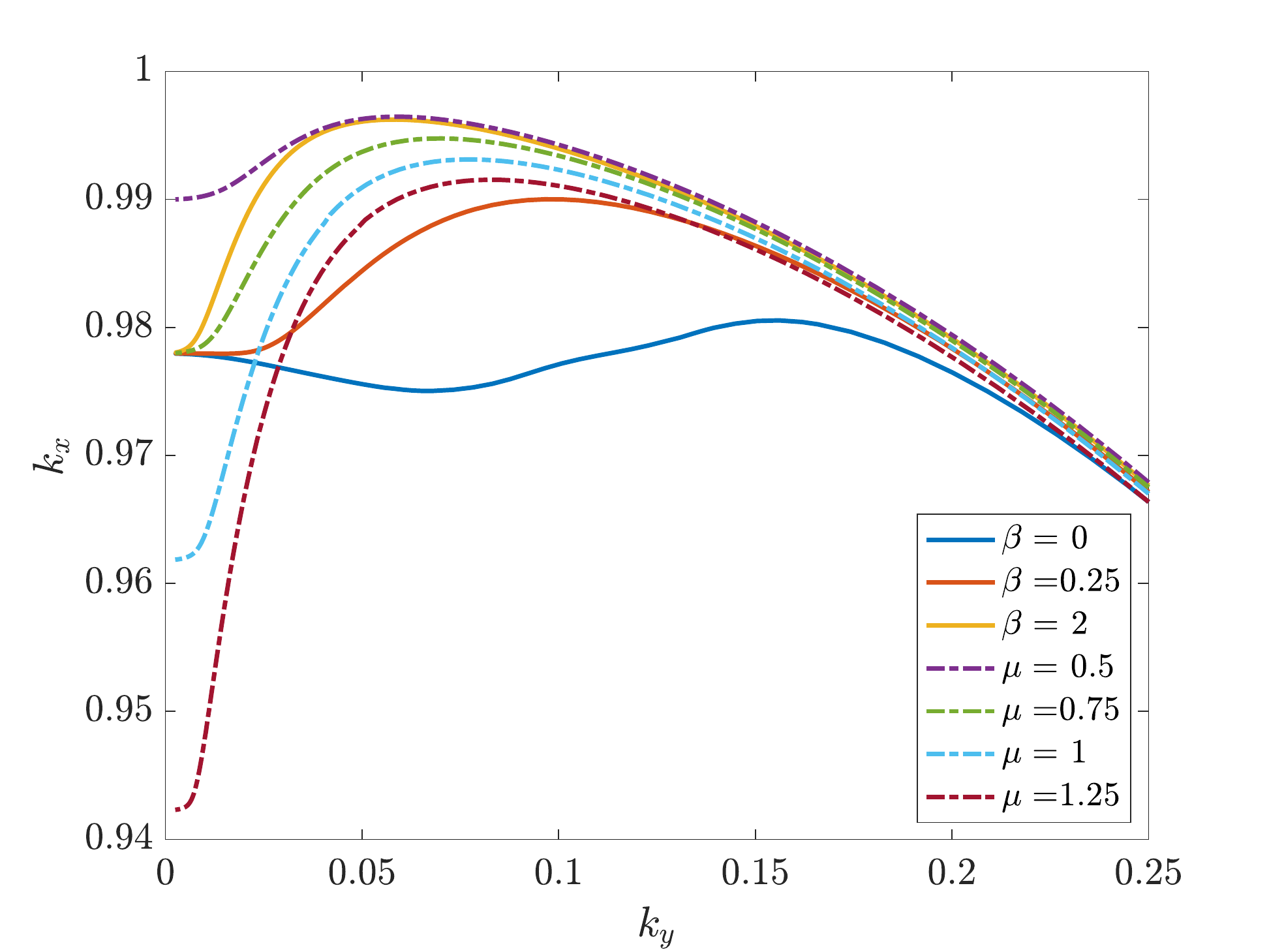}
\caption{Left: Top-down view of moduli space associated with \eqref{e:sh-an} with $\mu = 3/4, \beta = 1$, color denotes $k_x$ value; Center: zoom-in of moduli surface for $k_y,c_x\sim0$ with same $\mu,\beta$ values. Right: Moduli curves for $c_x = 0.0025$ fixed and $\mu$ and $\beta$ varied.  }\label{f:an-mod}
\end{figure}

As a consequence, the structure of the moduli space is significantly simpler in the anisotropic scenario, lacking all transitions to perpendicular stripes and the related zigzag and cross roll instabilities. Figure \ref{f:an-mod} where $\beta = 1$, $\mu = 3/4$,  shows that  the ``kink-dragging" bubble for nearly perpendicular stripes, which was induced by perturbing zigzag critical oblique wavenumbers, is not present and  only horizontal wavenumbers with $k_x\sim 1$ are supported.  Analogous predictions, not included in the figures,  from the linear spreading speed \eqref{e:klinsh} and absolute spectrum \eqref{e:kxabs} using the altered dispersion relation $d(\nu,\lambda;c,k_y) = -(1+\nu^2-k_y^2)^2 -\beta k_y^2 + \mu + c \nu - \lambda$, accurately predicted the upper boundary in $c_x$ of the moduli curve, and leading order wavenumber dependence for $c_x\lesssim c_\rlin(k_y)$ for each $k_y$. 
Zooming into the $c_x,k_y\sim0$ region for slowly growing, weakly oblique stripes,  the moduli surface possesses different monotonicity properties in $k_y$ and $c_x$ compared with the isotropic case; see Figure \ref{f:an-mod}, center and Figure \ref{f:an-mod} right. For $c_x$ fixed and small, we find $k_x(c_x,k_y)$ attains a local minimum at $k_y = 0$, a subsequent local maximum for increasing $k_y$ before decreasing monotonically for $k_y \sim \mc{O}(1)$.  It is also instructive to consider the behavior of the bulk wavenumber $k(c_x,k_y)$.  Here, with $k_y$ fixed, and $c_x$ varying small, we find the wavenumber curves interpolate between the equilibrium strain $\frac{1}{2\pi}\int_0^{2\pi} g_\mathrm{SH}(\phi)d\phi$ at $c_x = 0, k_y\neq0$ and the monotonically increasing wavenumber curve $k_x(c_x,0)$, which is the same as in the isotropic case; see Fig. \ref{f:snap} above.  That is, for $k_y\sim0$ fixed and $c_x$ increasing from $0$, curves $k(c_x,k_y)$ decrease from the equilibrium strain, attain a global minimum, and then monotonically increase.  For moderately larger $k_y$, we find this minima disappears leaving a monotonically decreasing $k(c_x,k_y)$. 
\begin{figure}
\centering
\includegraphics[trim = 0.5cm 0.25cm 0.5cm 0.5cm,clip,width=0.33\textwidth]{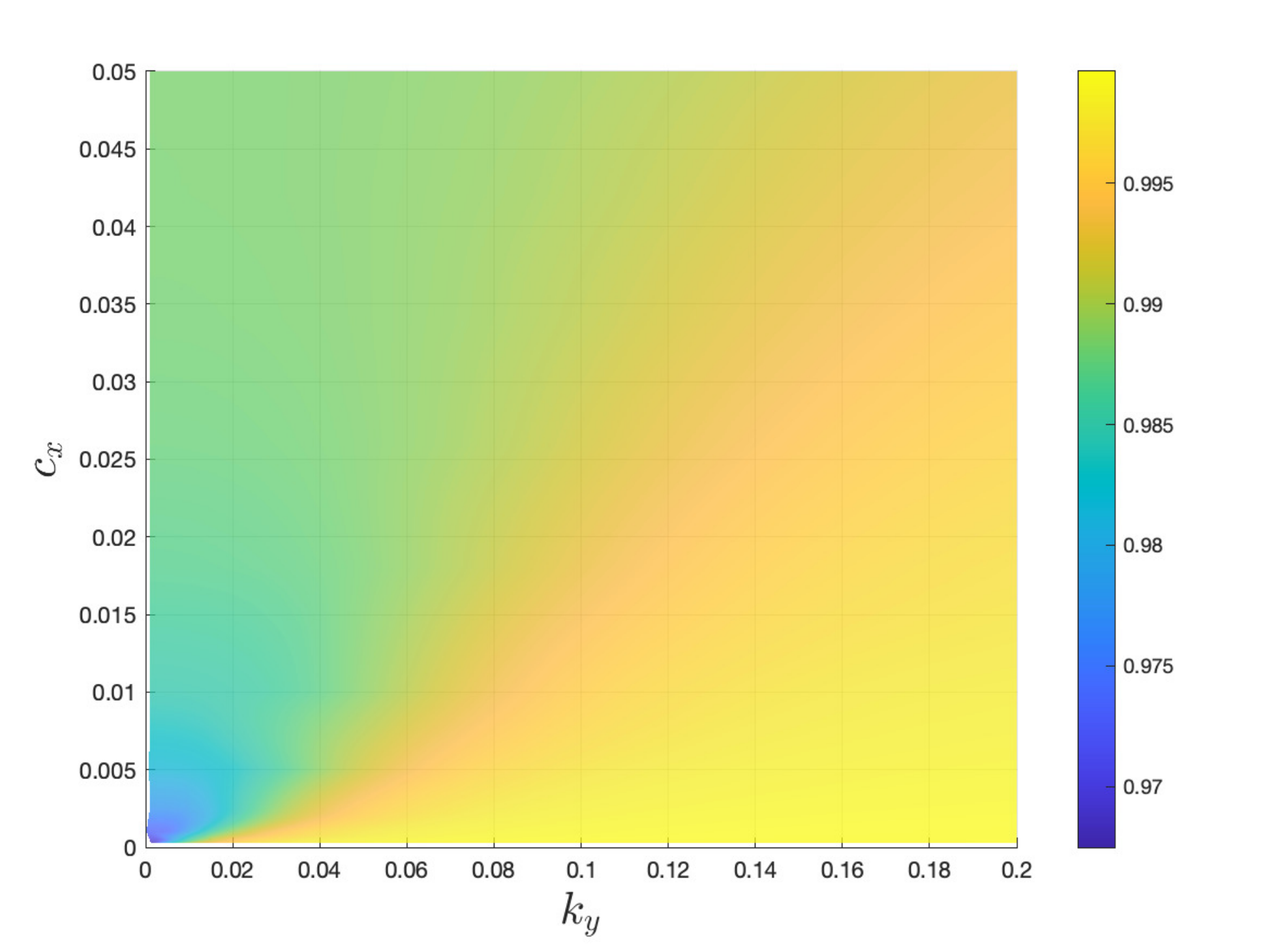}\hspace{-0.0in}
\includegraphics[trim = 0.25cm 0.0cm 0.5cm 0.5cm,clip,width=0.33\textwidth]{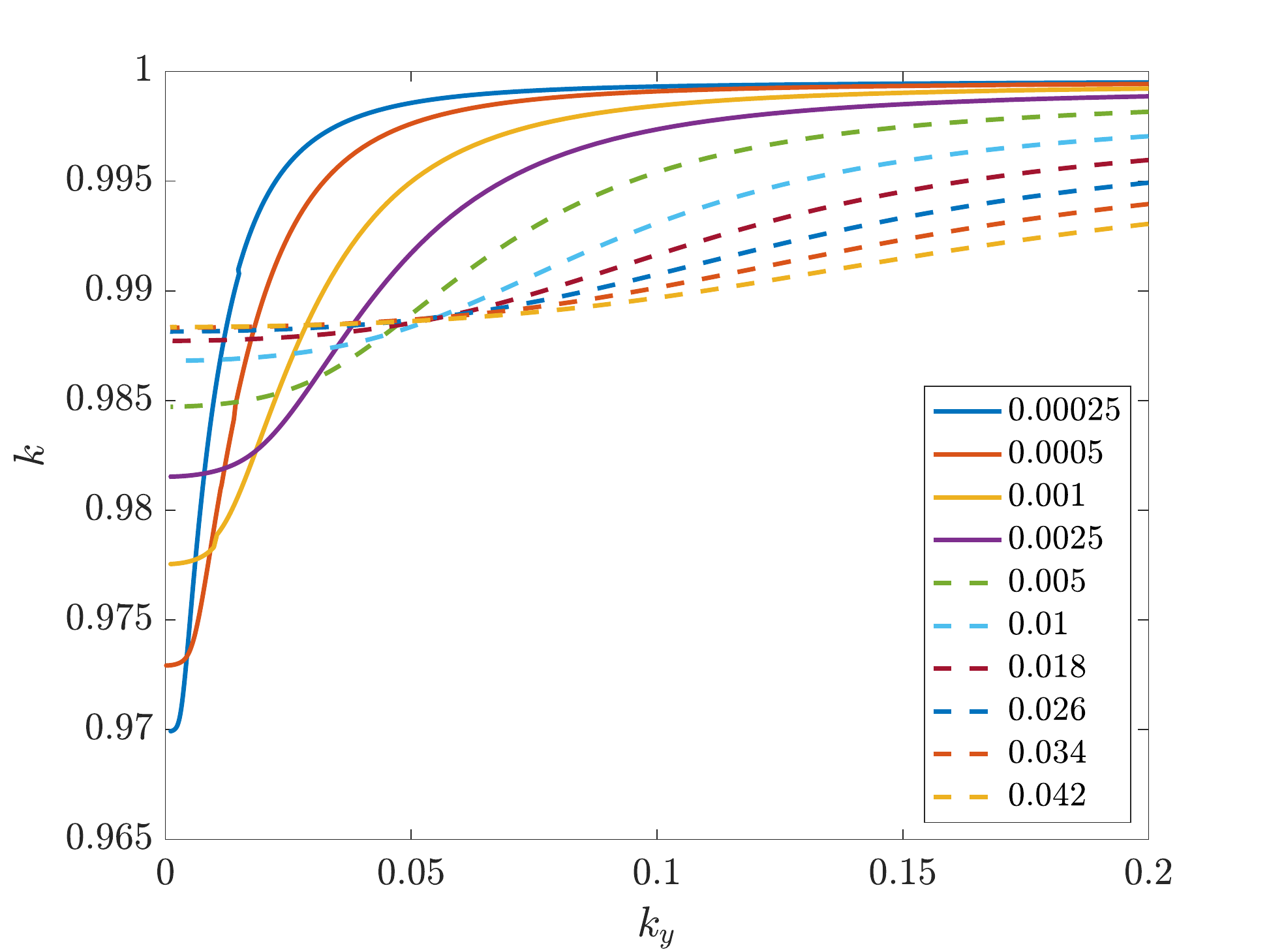}\hspace{-0.2in}
\includegraphics[trim = 0.cm 0.0cm 0.0cm 0.0cm,clip,width=0.33\textwidth]{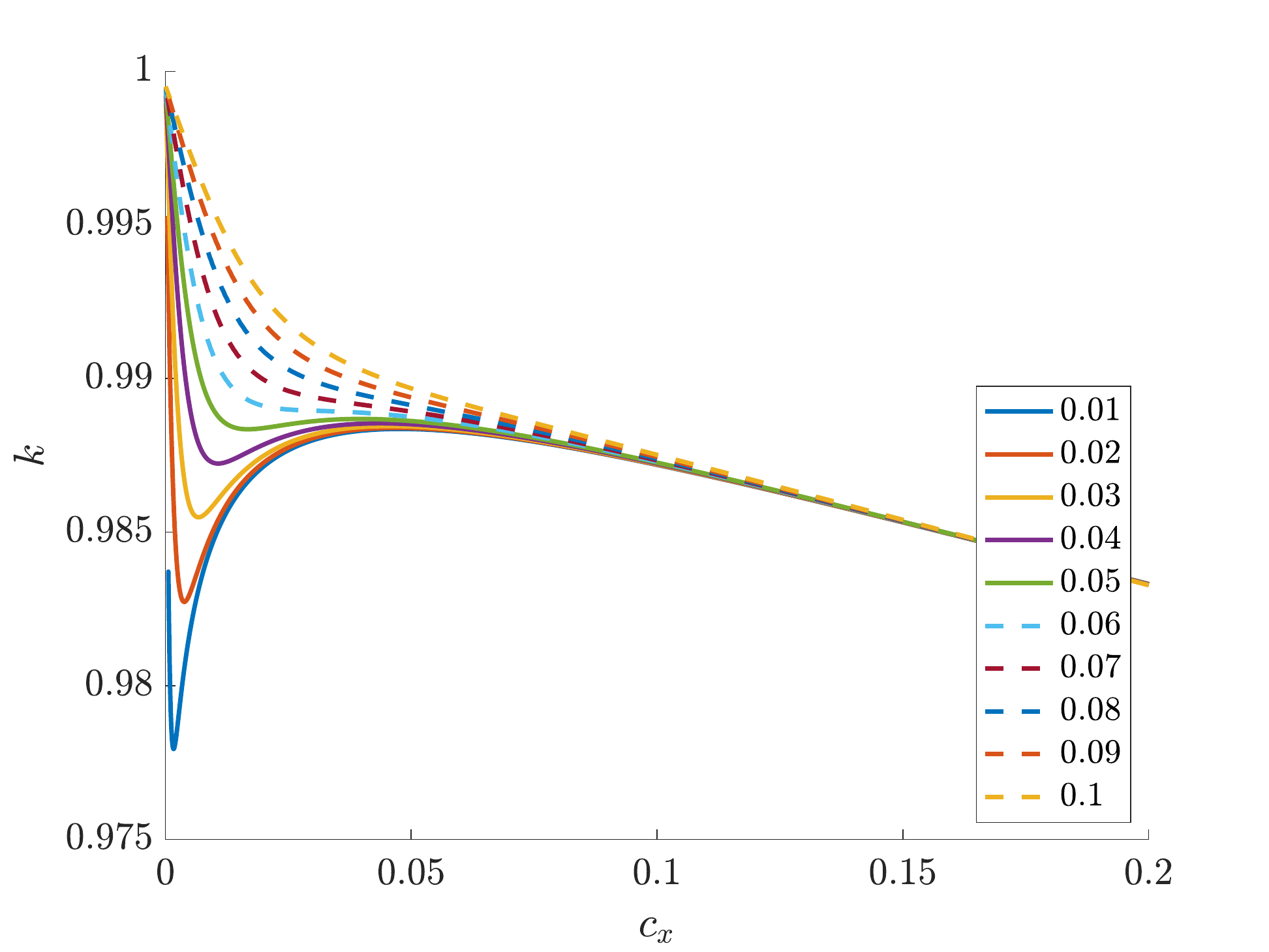}
\caption{ Left: Top-down view of moduli space associated with \eqref{e:sh-an}, once again for $\mu = 3/4, \beta = 1$, color denotes the bulk wavenumber $k = \sqrt{k_x^2+k_y^2}$ value; Center: slices of bulk wavenumber $k(c_x,k_y)$ for a range of fixed $c_x$ values. Right:  slices of $k(c_x,k_y)$ for a range of fixed $k_y$ values small}\label{f:an-mod-k}
\end{figure}

\subsubsection{Phase diffusion approximation}\label{sss:pdax}
Continuing to focus on the slowly growing weakly oblique stripe regime, $c_x,k_y\sim0$ we note that bulk wavenumbers $k$ in the range $(k_\mathrm{sd,min},k_\mathrm{sd,max})$ of the strain-displacement curve are stabilized by the suppression of the zigzag instability. One can then understand wavenumber selection dynamics using a phase-diffusion approximation similar to the one-dimensional case described in Section \ref{ss:slow}.  In particular, following a similar multiple-scales analysis for phase dynamics $u = u_\mathrm{p}(\varphi)$ with $|\nabla \varphi|\sim 1$ slowly varying, one can describe patterned fronts using a linear phase diffusion equation with nonlinear boundary condition given by the one-dimensional strain-displacement relation,
\begin{align}
\varphi_t &= \Delta_{x,y} \varphi + c_x \varphi_x, \quad x<0, y\in \R\notag\\
\varphi_x &= g_\mathrm{SH}(\varphi),\quad x = 0, y\in \R,\label{e:pd-2d}
\end{align}
see \cite{chen2021strain} for more detail.

\begin{figure}[!ht]
\centering
\includegraphics[trim = 0.5cm 0.0cm 0.5cm 0.5cm,clip,width=0.33\textwidth]{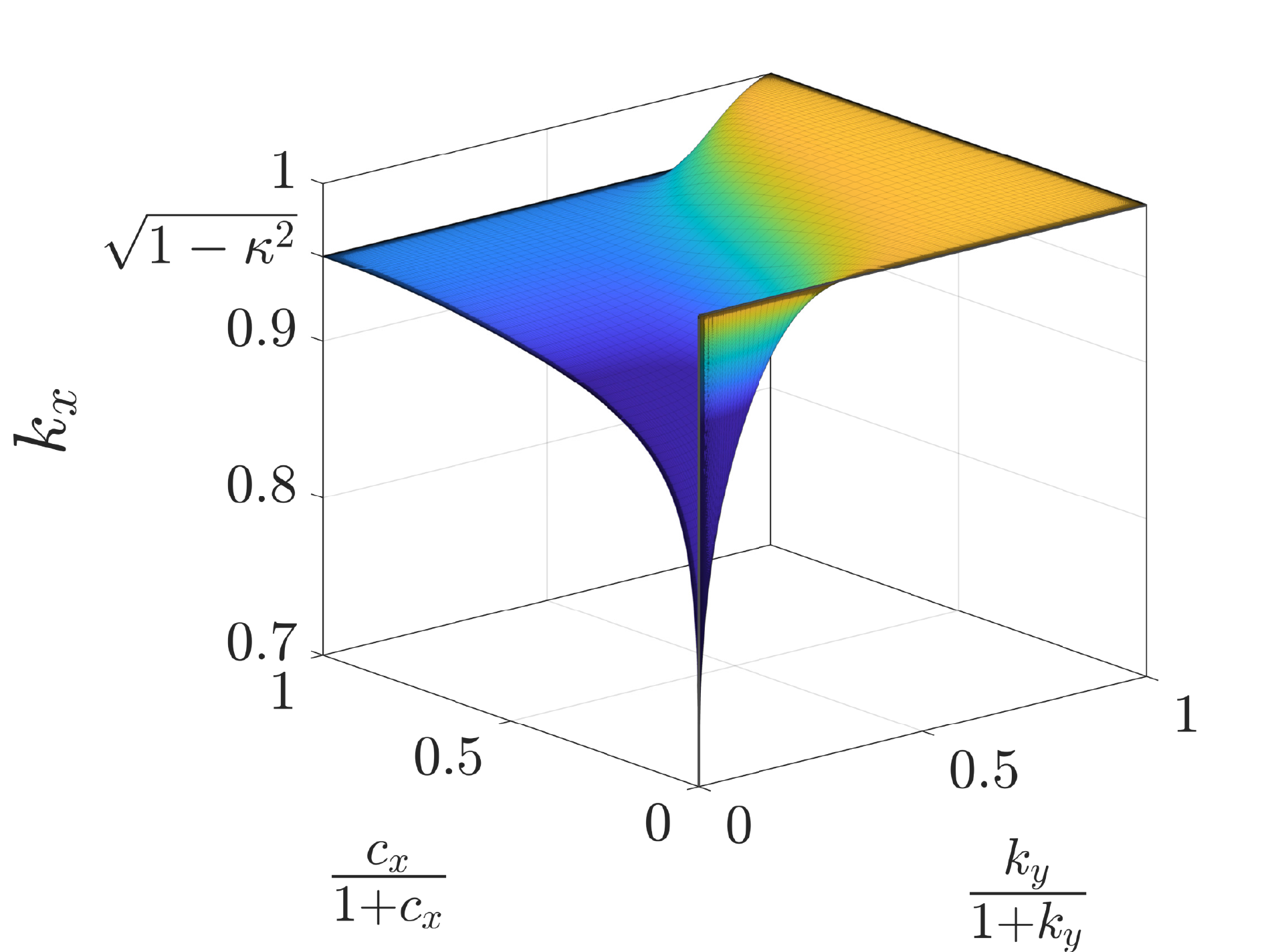}
\includegraphics[trim = 0.7cm 0.7cm 0.5cm 0.5cm,clip,width=0.33\textwidth]{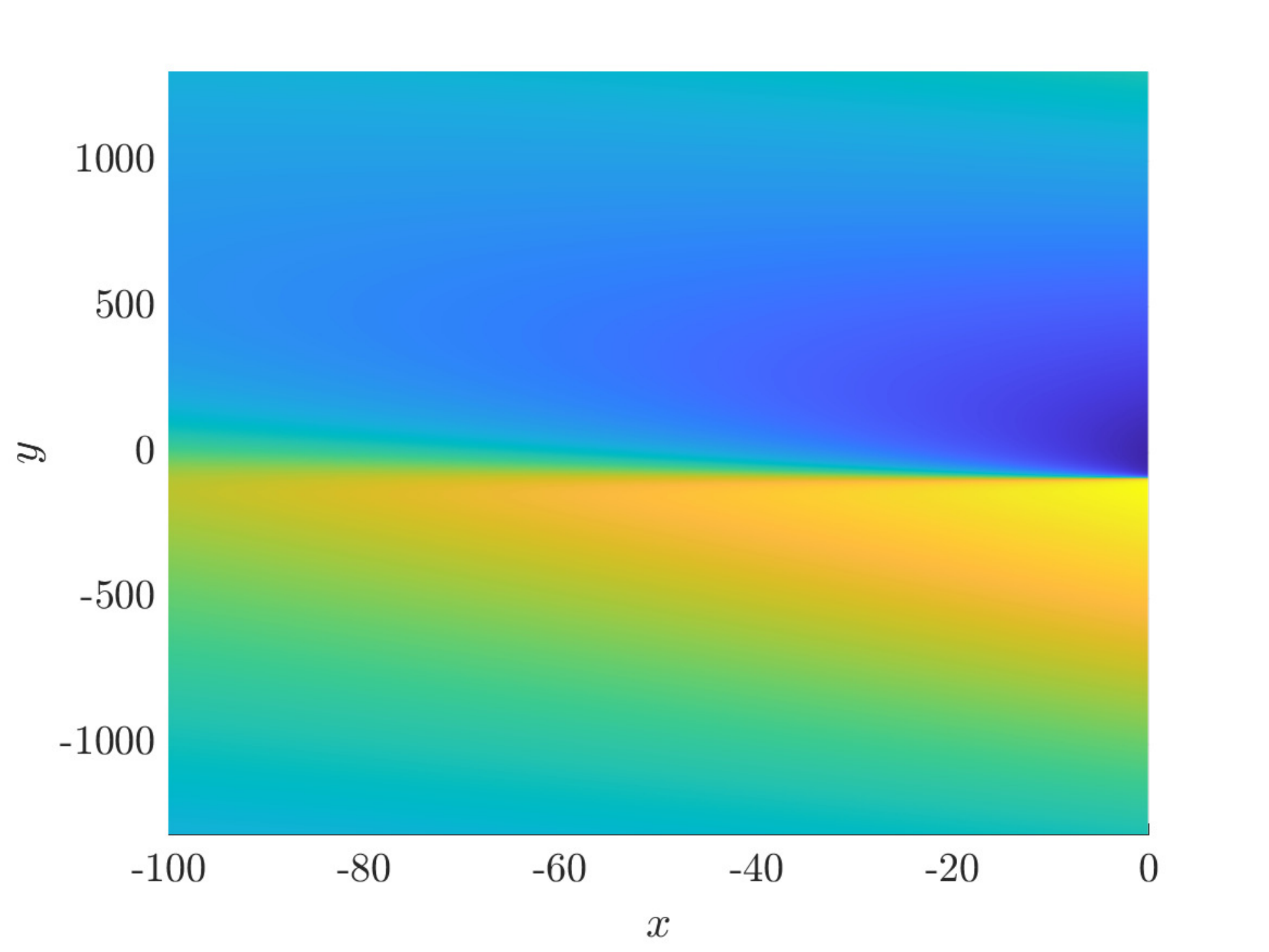}\hspace{-0.5cm}
\includegraphics[trim = 0.7cm 0.7cm 0.5cm 0.5cm,clip,width=0.33\textwidth]{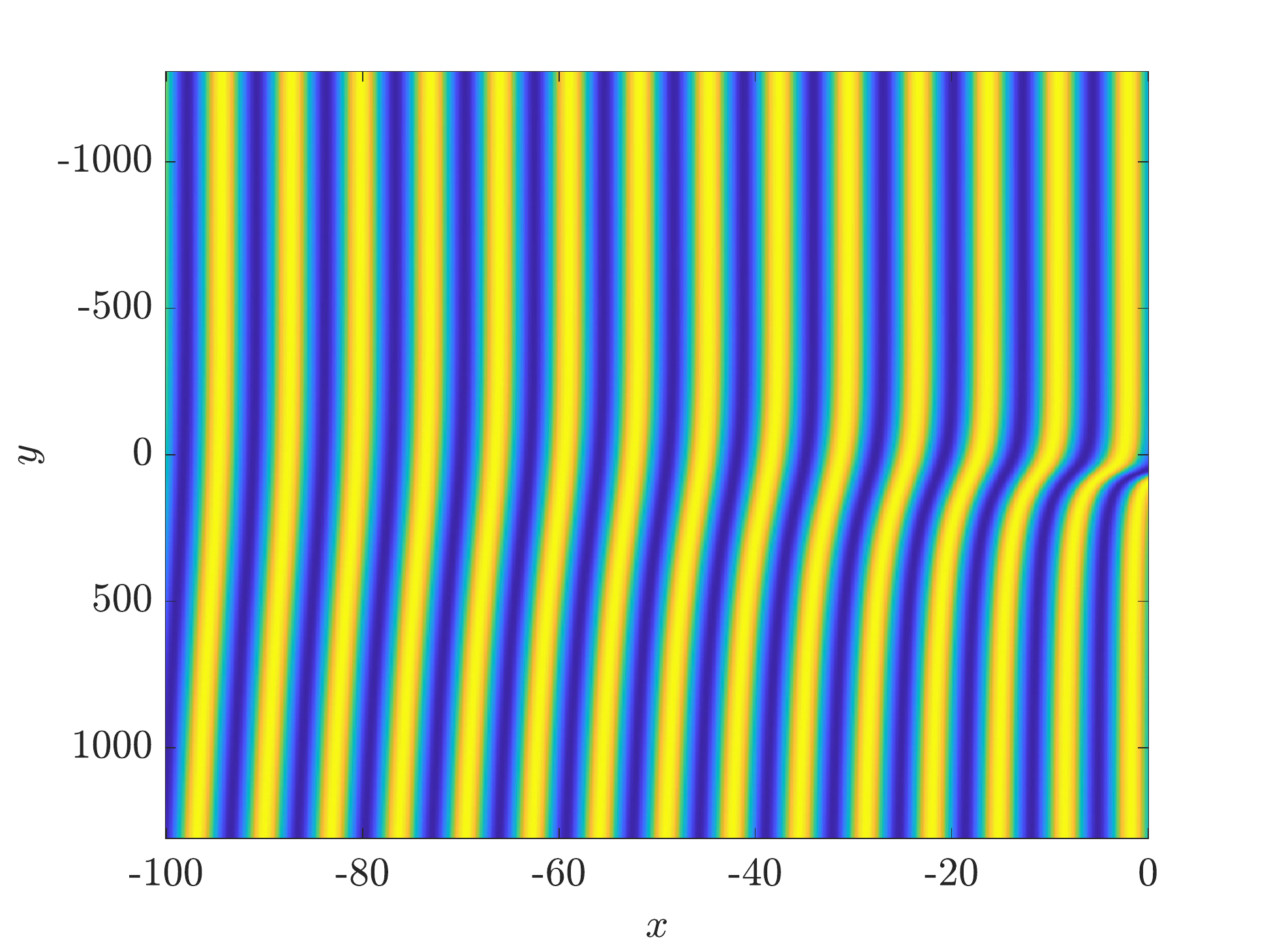}\hspace{-0.5cm}
\caption{Left: Compactified plot of moduli space for \eqref{e:pd-2d}; Reproduced with permission from \cite[Fig. 2]{chen2021strain}. Copyrighted by SIAM. Center: Profile of $\psi := \phi - (k_x x + k_y \tl y)$ in co-rotating frame for $0<c_y,k_y\ll1$; Right: Plot of $\cos(\phi(x,\tl y))$. }\label{f:slow} 
\end{figure}

Stripe-forming front solutions are then represented by solutions with $\varphi \sim k_xx + k_xc_x t + k_y y= k_xx +  k_y( y - c_y t)$, with $c_y = -k_xc_x/k_y$, in the far-field $x\rightarrow-\infty$, and which are periodic in the $y$ variable up to the gauge symmetry induced by the periodicity of the strain-displacement relation $g_\mathrm{SH}$. In particular, one restricts to solutions which are traveling waves in the $y$ direction, $\varphi = \varphi(x,k_y(y - c_y t))$, with $\zeta =k_y(y - c_y t)$ and $\varphi(\cdot,\zeta+2\pi) = \varphi(\cdot,\zeta)+2\pi$. Defining a new variable which subtracts off the desired asymptotic state $\psi:= \varphi - (k_x x +\zeta)$, one obtains the following system of equations
\begin{align}
0&= \psi_{xx} + k_y^2 \phi_{\zeta\zeta} + c_x\psi_x - k_x c_x\psi_\zeta,\quad x<0,\zeta\in \R,&\label{e:psi1}\\
0&= \psi(x,\zeta+2\pi) - \psi(x,\zeta),\qquad x\leq0,\zeta\in\R&\label{e:psi2}\\
0 &= \psi_x - g_\mathrm{SH}(\psi+\zeta) + k_x,\quad x = 0,\zeta\in \R&\label{e:psi3}\\
0 &= \lim_{x\rightarrow-\infty} \psi(x,\zeta),\qquad \zeta\in\R.&\label{e:psi4}
\end{align}
Since $\psi$ is periodic in $\zeta$ and linear in the bulk domain $x<0$, one can decompose $\psi(x,\zeta) = \sum_{\ell\in\Z}\psi^\ell(x) \re^{\ri\ell \zeta}$ and map the equation onto the boundary by solving each decoupled linear second-order equation for $\psi_\ell$ and obtaining a boundary integral equation. The work \cite{chen2021strain} establishes existence of solutions to this system, using a priori bounds, maximum principle arguments, and Fredholm properties of the linearization. It also gives results of numerical continuation which explore the moduli surface $\mc{M}_{PD}:= \{ (c_x,k_y,k_x): \text{ \eqref{e:psi1}-\eqref{e:psi4} has a solution }\}$ for this system (see Figure \ref{f:slow} left), and derives formal leading-order expansions near various limits in $c_x$ and $k_y$. After suitable scaling, in the $c_x,k_y\sim 0$ regime, good agreement was found between the moduli space of the phase-diffusion system and that of the anisotropic Swift-Hohenberg equation \eqref{e:sh-an} above. In this work, it was also found that wavenumber selection for slowly grown, nearly parallel stripes is governed by the glide-motion of a dislocation defect along the boundary $\{x = 0\}$ of the domain; see Figure \ref{f:slow} center and right for a depiction. For extremely slow speeds, this defect relieves local strain on the striped phase at the quenching interface causing a decrease in the wavenumber. Then for yet larger but still small speeds, strain dynamics behave like in the one-dimensional case, with wavenumber increasing in $c_x$. 


\subsection{Directionally quenched complex Ginzburg-Landau equation}\label{ss:cgl}
Beyond the spatial striped patterns explored thus far, one can also investigate the effect of quenches, or other spatial inhomogeneities, on temporal oscillations. A universal model for temporal oscillations in spatially extended systems near onset is the cubic complex Ginzburg-Landau equation, which we consider with  a directional quenching parameter as in the Swift-Hohenberg equation above. 
\begin{align}\label{e:cgl}
A_t = (1+\ri\alpha) \Delta A + \rho(x - c_xt) A - (1+\ri \gamma)A|A|^2,\qquad \rho(x)= -\mathrm{sign}(x).
\end{align}
Here, when $\rho\equiv -1$ the trivial state $A\equiv0$ is stable, while for $\rho\equiv1$ the trivial state is unstable and there exists an explicit family of periodic wave trains $ r\re^{\ri\omega t} \re^{\ri (k_x x+k_y y)}$ where $\omega, k_x,k_y$ satisfy a nonlinear dispersion relation
\beq\label{e:cgl-disp}
r^2 =1-k^2,\qquad  \omega = k^2(\gamma - \alpha) - \gamma,\qquad k^2 = k_x^2 + k_y^2.
\eeq
These periodic solutions are relative equilibria with respect to the gauge action $A\mapsto \re^{\ri \theta} A,\, \theta\in [0,2\pi)$. Within this setting, existence and stability of quenched fronts were studied in \cite{gs1,goh2020spectral}.

Focusing first on $y$-independent, parallel stripes $A = A(x,t)$, one looks for pattern forming fronts by decomposing 
\beq\label{e:cgl-1d}
A(x,t) = \re^{\ri\omega t} A_{\mathrm{tf}}(x - c_xt),\qquad\qquad 
\eeq
so that $A_\mathrm{tf}$ solves
\begin{align}
0 &= (1+\ri\alpha)A_{\tl x\tl x} +c_x A_{\tl x} + (\rho - \ri\omega)A -(1+\ri\gamma)A|A|^2, \qquad \tl x = x - c_x t, \label{e:cgl-tw}\\
0 &= \lim_{\tl x\rightarrow-\infty}  A(\tl x) - \sqrt{1-k_x^2}\re^{\ri k_x x} ,\qquad\quad
0=\lim_{\tl x\rightarrow+\infty} A(\tl x),
\end{align}
for parametert pairs $(\omega,c_x)$. Recall that $\omega$ determines $k_x$ through the shifted nonlinear dispersion relation $\omega = k_x^2(\gamma - \alpha) +\ri c_x k_x - \gamma.$

For fast quench speeds $c_x$ near the stripe ``detachment" speed, the front selection mechanism is the same as described in Section \ref{ss:csel} in the Swift-Hohenberg equation. This is due to the fact that the free invasion front for the homogeneous system with $\rho\equiv1$ is once again pulled.
 Thus, the linear spreading speed $c_\rlin$ determines the speed at which patterns ``detach" from  the quenching interface and and the absolute spectrum determines leading order wavenumber selection properties in the quenched system for $c_x\lesssim c_\rlin$.  

Performing a pinched double root analysis similar to \eqref{e:msc}--\eqref{e:klinsh} as in Sec. \ref{ss:fg} on the linear dispersion relation $0 = d(\lambda,\nu,c_x) = (1+\ri\alpha)\nu^2 +c\nu+ (1-\ri\omega) $ one finds the linear spreading speeds, frequencies, and wavenumbers as
\beq
c_\rlin = 2\sqrt{1+\alpha^2},\qquad \omega_\rlin = \alpha,\qquad 
k_\rlin = \begin{cases}
\frac{\sqrt{1+\alpha^2}-\sqrt{1+\gamma^2}}{\gamma - \alpha},&\,\, \text{for } \gamma\neq\alpha\\
 -\frac{\alpha}{\sqrt{1+\alpha^2}}&\,\, \text{ for } \gamma = \alpha
\end{cases}.
\eeq
Additionally, the frequency given by the intersection of the absolute spectrum with the imaginary axis for $c_x\lesssim c_\rlin$ can be explicitly calculated as $\omega_\mathrm{abs} = -\alpha + \frac{\alpha c_x^2}{2(1+\alpha^2)}$.

In this speed regime, fronts for quenching speeds $c_x\lesssim c_\rlin$ were constructed rigorously using heteroclinic bifurcation and desingularization techniques \cite{gs1}. In particular  the selected wavenumber $k_x(c_x)$ and the location of the interface $\tl x_f = \inf\left\{ \tl x\,:\, \sup_{\tl x'>x}|A(\tl x)| <\delta \right\}$, for some fixed $0<\delta\ll1$, of the pattern forming front, have the expansions
\begin{align}
k_x(c_x) &= k_\rlin + g_1(\alpha,\gamma) (c_\rlin - c_x) - g_2(\alpha,\gamma)|\Delta Z_i| (c_\rlin - c_x)^{3/2} + \mathcal{O}((c_\rlin - c_x)^2),\label{e:kcgl_sel1}\\
\tl x_f &= \pi(1+\alpha^2)^{1/4} (c_\rlin - c_x)^{-1/2} + (1+\alpha^2)^{1/2} \Delta Z_r + \mathcal{O}((c_\rlin - c_x)^{1/2}),\label{e:kcgl_sel2}
\end{align}
where $g_1$ and $g_2$ are continuous functions of $\gamma$ and $\alpha$; see \cite[Thm. 1]{gs1} for more detail.  

\begin{figure}[!ht]
\centering
\includegraphics[trim = 0.25cm 0.25cm 0.25cm 0.25cm,clip,width=0.4\textwidth]{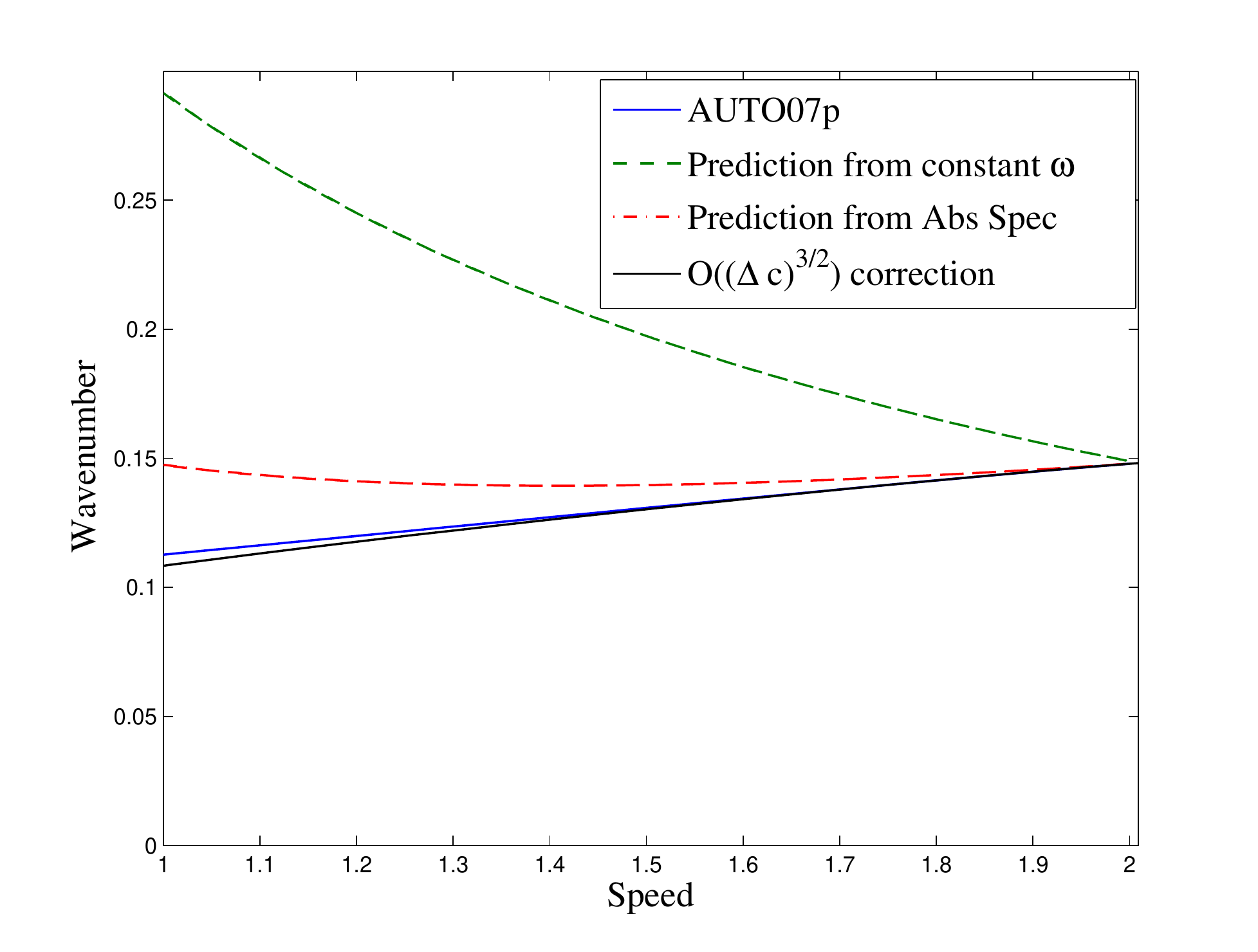}\hspace{-0.0in}
\includegraphics[trim = 0.0cm 0.0cm 0.05cm 0.25cm,clip,width=0.3\textwidth]{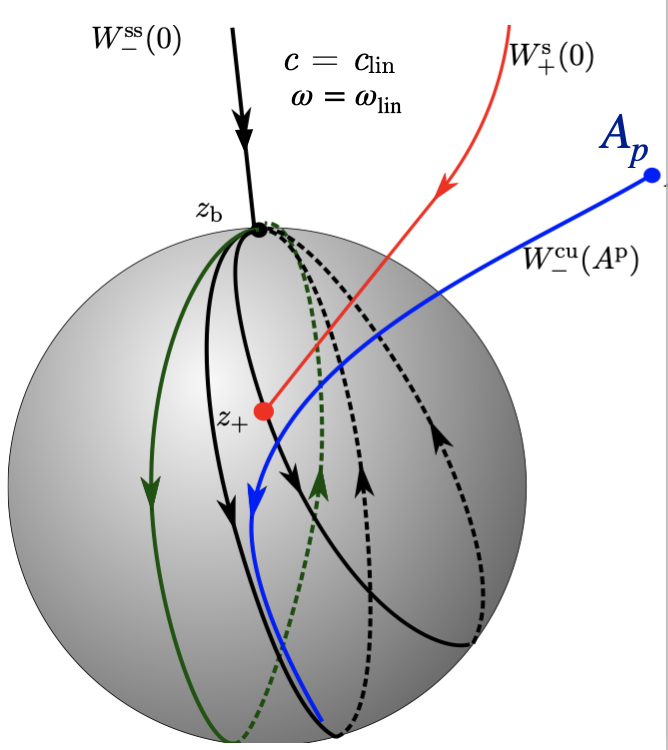}\hspace{-0.2in}
\caption{ Left: Comparison of selected wavenumber $k_x(c_x)$ from numerical continuation using AUTO07p (blue) with various truncations of the expansion \eqref{e:kcgl_sel1} for $\alpha = -0.1, \gamma = -0.2$, including prediction using a fixed $\omega = \omega_\rlin$ (dashed green), the prediction using the leading order prediction using $\omega = \omega_\mathrm{abs}$ (dashed red), and the higher order prediction using the projectivized distance $\Delta Z$;    Right: depiction of the heteroclinic matching problem in the reduced phase portrait variables $(z,R)$. Reproduced with permission from \cite[Fig. 3.1, 4.1]{gs1}. Copyrighted by Springer. }\label{f:cgl-0}
\end{figure}

Technically, these results rely on first factoring out orbits of the gauge symmetry in the phase space  $\C^2$ of the traveling wave equation \eqref{e:cgl-tw}, by using directional blow-up, with coordinate charts $z = A_{\tl x}/A, R = |A|^2$ and $\tl z = A/A_{\tl x}, S = |B|^2$.  This coordinate change, reduces the phase space to $\C^2/S^1 \approx \R^+ \times \mb{S}^2$, where $\mb{S}^n$ is the unit $n$-sphere. Additionally, these coordinates desingularize a Jordan block which arises in the linearized system for $(c_x,\omega) = (c_\rlin,\omega_\rlin)$ by  ``blowing it up" into the sphere $\{0\}\times \mb{S}^2$. The dynamics on this sphere give the evolution of one-dimensional complex linear subspaces under the linearized flow near the origin. Furthermore, the periodic orbits formed by $A_p$ collapse to equilibria, while its unstable manifold (blue curve in Fig. \ref{f:cgl-0} right), as well as the stable manifold of the origin for $\rho\equiv -1$ (red curve in Fig. \ref{f:cgl-0} right), are reduced to one-dimensional manifolds. Pattern forming fronts can be obtained as heteroclinic orbits bifurcating from the free invasion front as the parameters $(c_x,\omega)$ are unfolded near $(c_\rlin,\omega_\rlin).$  The parameter $\Delta Z$ gives the leading order projective distance between the tangent spaces of the relevant unstable and stable manifold.

 We also remark here that the work \cite{goh2020spectral}, under further assumptions on $\alpha$ and $\gamma$ to guarantee diffusive stability of the asymptotic pattern and existence of the freely invading front, proved that these fronts are spectrally stable in a suitably defined exponentially weighted space. The main technical barrier in this result is caused in the region $\tilde x \in (\tl x_f,0)$, where the front solution lies near the absolutely unstable trivial state. In the limit $c_x\rightarrow c_\rlin^-$ this causes point spectrum to accumulate on the weakly unstable absolute spectrum of the trivial state. Projective blow-up techniques allow for detailed tracking of eigenvalues and the somewhat surprising fact that the front is spectrally stable.

\paragraph{Oblique and perpendicular stripes.}

\begin{figure}[ht!]
\centering
\includegraphics[trim = 0.25cm 0.25cm 0.5cm 0.5cm,clip,width=0.49\textwidth]{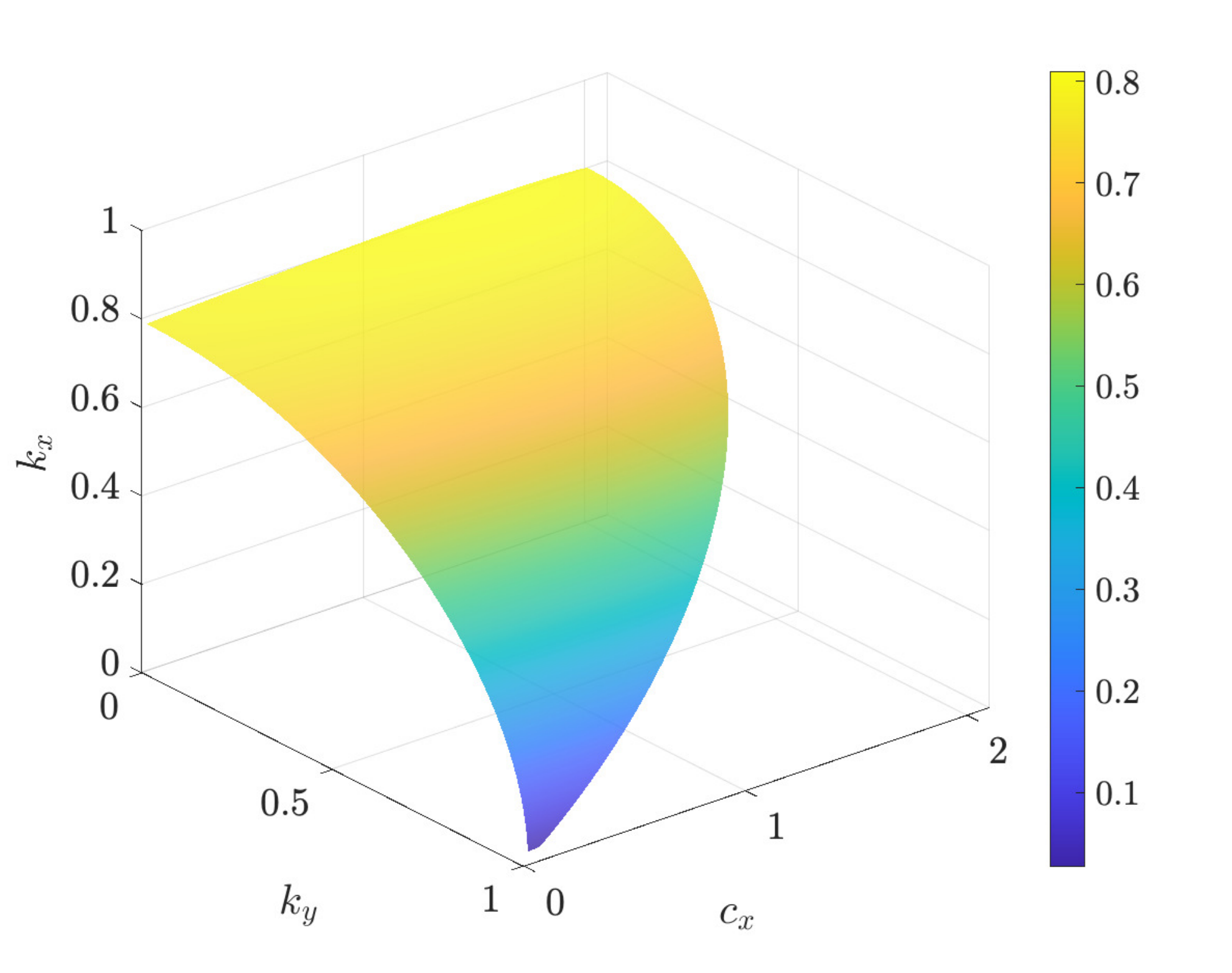}\hspace{-0.0in}
\includegraphics[trim = 0.5cm 0.25cm 0.5cm 0.5cm,clip,width=0.49\textwidth]{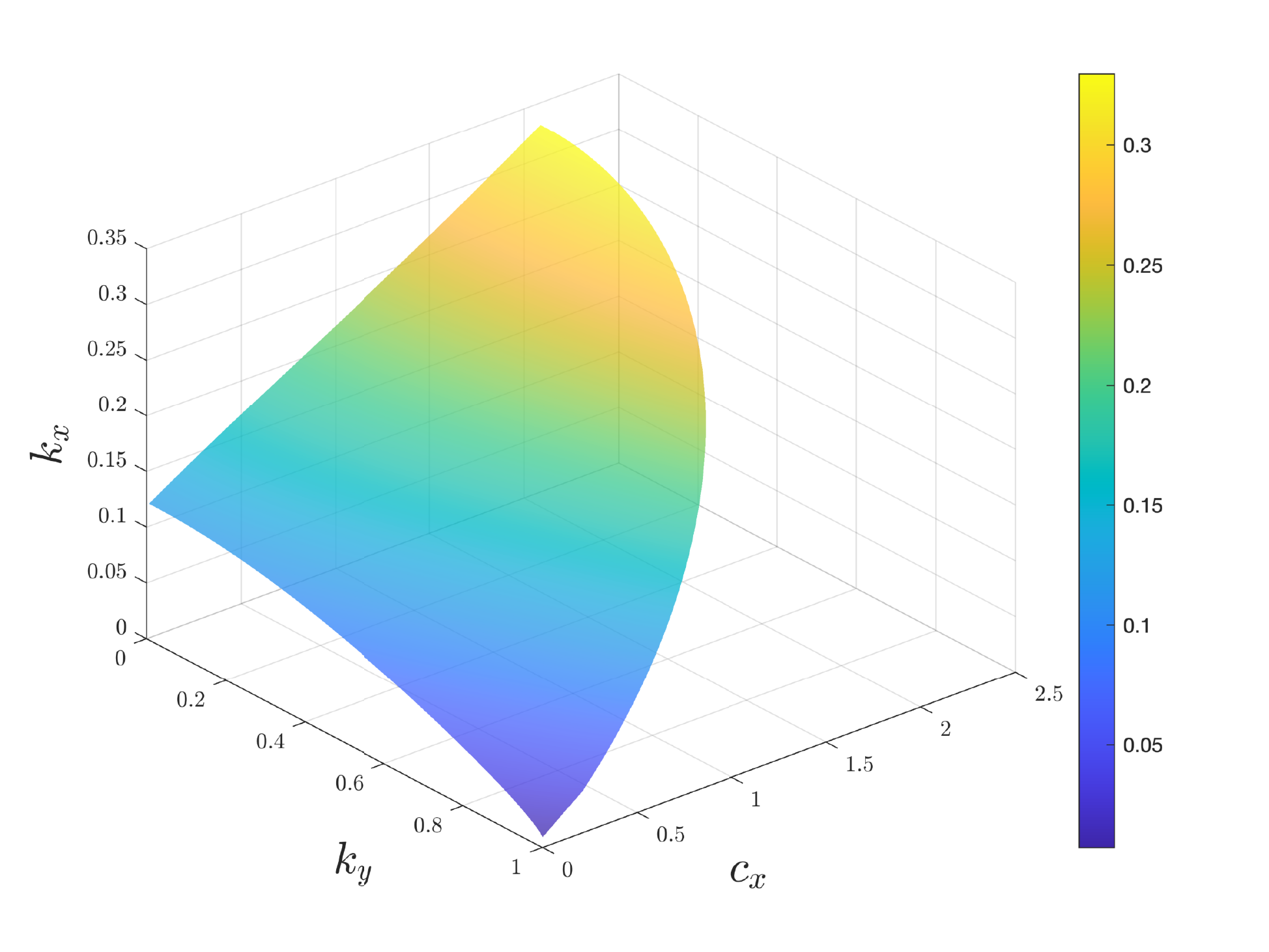}\hspace{-0.2in}
\caption{ Moduli surfaces for \eqref{e:cgl-tw2} for $(\alpha,\gamma) = (0.1,0.4)$ (left) and $(\alpha,\gamma) = (0.3,0.4)$ (right) }\label{f:cgl-mod}
\end{figure}

The simplest $y$-dependent pattern forming fronts can be obtained by including an oscillatory factor in $y$ to the solution decomposition \eqref{e:cgl-1d} in \eqref{e:cgl}, setting $A = \re^{\ri(\omega t + k_y y)} A(x - c_x t)$ to obtain
\begin{align}
0 &= (1+\ri\alpha)A_{\tl x\tl x} +c_x A_{\tl x} + (\rho-k_y^2 - \ri(\omega+\alpha k_y^2))A -(1+\ri\gamma)A|A|^2 \label{e:cgl-tw2}\\
0 &= \lim_{\tl x\rightarrow-\infty}  A(\tl x) - \sqrt{1-k_x^2-k_y^2}\re^{\ri k_x x} ,\qquad\quad
0=\lim_{\tl x\rightarrow+\infty} A(\tl x).
\end{align}
Here the transverse wavenumber $k_y$ shifts the parameter step in the complex plane.   Continuing numerically gives a surface over the $c_x,k_y$-plane. Figure \ref{f:cgl-mod} depicts this portion of the moduli surface $\mc{M}_{CGL}$ for two pairs of $(\gamma,\alpha)$. We note that the range of $k_x$ values vary widely between the two parameter cases. Furthermore, we note that in the left plot curves are nearly constant in $c_x$ with $k_y$ fixed, while for the right plot the moduli surfaces varies more in $c_x$ than in $k_y$. Also note the difference in wavenumber range between the two cases. The approach  described above applies here with only minor modifications due to a rescaling of $\rho$ and $\omega$ yielding $k_x = k_x(c_x,k_y)$ as selected by the quenching speed $c_x$ and vertical wavenumber $k_y$.  Indeed by rescaling $A\mapsto m A$ and $x\mapsto x/\ell$, setting $\ell^2 =m^2= (1-k_y^2)$ for $|k_y|<1$ and redefining the speed $\tl c_x := c_x(1-k_y^2)^{-1/2}$ and frequency $\tl \omega = \frac{\omega+\alpha k_y^2}{1-k_y^2}$, one obtains \eqref{e:cgl-tw} in the pattern-forming region $x<0$. Note, in the stable region $x>0$, this leaves a perturbed coefficient  $-(1+k_y^2)(1-k_y^2)^{-1}-\ri \tl \omega $ on the linear $A$ term. As this term contributes to the direction of the tangent space of the stable manifold for $x>0$, non-scaling related $k_y$-contributions enter into to the wavenumber expansion only at third-order, via the $\Delta Z_i$ term in \eqref{e:kcgl_sel1}. Thus, we expect oblique stripe selection to be governed by a rescaling of the parallel stripe selection curves at leading- and second-order.  

The upper boundary in $c_x$, that is the fast speed boundary, of $\mc{M}_{CGL}$, is governed by the linear spreading dynamics of each $k_y$-mode, see Section \ref{ss:fg}. Following the calculation in \eqref{e:clinsh}-\eqref{e:klinsh} gives  spreading speed, frequency, and horizontal wavenumber,  for the $y$-dependent invasion front which is $2\pi/k_y$-periodic in $y$,
\begin{equation}\label{e:clincgl}
c_\rlin(k_y) = 2\sqrt{1-k_y^2}\sqrt{1+\alpha^2},\quad
\omega_\rlin(k_y) = (1-2k_y^2)\alpha,\qquad
k_{x,\rlin}(k_y) =\sqrt{1-k_y^2} \begin{cases}
\frac{\sqrt{1+\alpha^2}-\sqrt{1+\gamma^2}}{\gamma - \alpha},&\,\, \text{for } \gamma\neq\alpha\\
 -\frac{\alpha  }{\sqrt{1+\alpha^2}}&\,\, \text{ for } \gamma = \alpha
\end{cases}.\quad
\end{equation}
Leading order predictions from the $k_y$-dependent absolute spectrum (not shown) agree well along the stripe-detachment boundary of $\mc{M}_{CGL}.$ 
We remark that the system does not appear to support quenched fronts with perpendicular stripes, $k_x=0$. In fact, this case is not robust since patterns in the wake lack the reflection symmetry present in the Swift-Hohenberg case and one would expect this situation for isolated values of $k_y$, only. 
 Fronts of this form are solutions of the traveling wave equation \eqref{e:cgl-tw2} with $\lim_{\tilde x\rightarrow+\infty} A(\tilde x) = 0$ and $\lim_{\tilde x\rightarrow-\infty} A(\tilde x) = A_-$, where $A_-$ is an asymptotic equilibrium for $\tilde x = -\infty$ with amplitude $|A_-| = \sqrt{1-k_y^2}, \, |k_y|<1$. Note, the equilibrium condition fixes the temporal frequency $\omega =  k_y^2(\gamma - \alpha) - \gamma$ for a given $k_y$.  In the simplest case,  $\alpha = \gamma = 0$, of real coefficient Ginzburg-Landau, one can restrict to $A\in \R$ and perform a straight-forward two-dimensional phase portrait analysis to obtain front existence for all $c_x\leq c_\rlin(k_y),\, k_y\in[0,1);$ see also the Section \ref{ss:acch} for the moduli space of Allen-Cahn. Despite this, since $\omega$ is fixed by $k_y$, one does not expect to such fronts to generically perturb for $\gamma\neq \alpha$. Indeed, moving into projective coordinates $z = A_{\tl x}/A, R = |A|^2$ in \eqref{e:cgl-tw2}, one obtains a real three-dimensional system where the stable manifold of $0$ for $\tl x>0$ and unstable manifold for $A_-$ for $\tl x<0$ are both one-dimensional. Hence, flowing the former backwards under the $\tl x<0$ flow, one searches for the intersection of one-dimensional and two-dimensional manifolds. Thus we only expect the existence of intersections for isolated points in parameter space.

\subsection{Allen-Cahn and Cahn-Hilliard}\label{ss:acch}
Quenching is of particular interest in phase separation problems. The simplest models here are the Allen-Cahn or Nagumo equation, 
\begin{equation}\label{e:ac}
\frac{d u}{dt} = \Delta u + f(u), \qquad f(u)=\mu u - u^3 + a, 
\end{equation}
or the Cahn-Hilliard equation 
\begin{equation}\label{e:ch}
\frac{d u}{dt} = -\Delta (\Delta u + \mu u-u^3). 
\end{equation}
Both equations are gradient flows to the energy $E[u]=\int \frac{1}{2}|\nabla u|^2 + W(u)$, $W'(u)=-(\mu u - u^3)$ and support ''phase-separating" interfaces between $u=\pm 1$ when $\mu=1$, $a=0$. Interfaces propagate in \eqref{e:ac} when $a\neq 0$. For $\mu<0$, the energy is convex with a unique stable, global minimizer. A simple quench would change $\mu=\rho(x-c_x t)$, mimicking for instance a progressive change in temperature that induces phase separation, with $\rho(\tl x)\to \pm 1$ for $\tl x\to \mp \infty$. 

\begin{figure}
    \centering
    \includegraphics[width=0.45\textwidth]{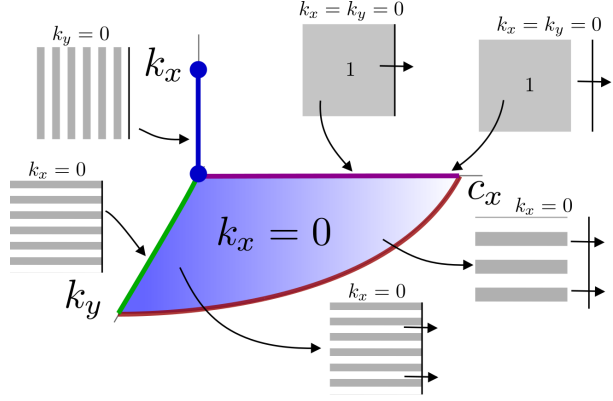}
    \caption{Schematics of moduli space in Allen-Cahn with step-quench; all striped formed for $c_x>0$ are perpendicular and stripes at $c_x=0$ are either perpendicular or parallel. The boundary in the $c_x-k_y$ plane is given by the linear spreading speed $c_x=2\sqrt{10k_y^2}$; see \cite{monteiro2017phase}}
    \label{f:modac}
\end{figure}

In the case where $\rho(\tl x)=-\,\mathrm{sign}\,(\tl x)$, the moduli space was completely described in \cite{monteiro2017phase}; see Figure \ref{f:modac}. One can show that stripes are either parallel or perpendicular to the interface and propagating stripes are constructed by exploiting comparison principles of the system. 

Results at $c_x=0$ also yield results in the Cahn-Hilliard equation with step quench, but little is known about fronts in this equation beyond small speeds. We note that the limit $k_y\to 0$ in Figure \ref{f:modac} (as in the Swift-Hohenberg case) yields solutions with $k_y=0$, but, depending on the spatial translate, may also converge to a solution heteroclinic in $y$, exhibiting a nodal line $u=0$ along $y=0,\, \tl x<0$. These solutions were also constructed in \cite{monteiro2017phase} and elucidate the role of single defects in this limit in other systems. A wealth of interesting phenomena arises when this nodal line, corresponding to the interface between the two possible selected states $u=\pm1$ in the wake of the quench, develops intrinsic dynamics for $a=0$, which then interact with the quench. The analysis in \cite{monteiro_2018} shows how small values of $a$ lead to effective boundary conditions for this interface at the quench, or in other words, the selection of an angle relative to the quenching line. We also mention \cite{gh1} which studies oblique stripes and checkerboard type patterns for a quench where $\mu$ is unstable only on a bounded interval in $\tl x$.

We conclude noticing that other quenches may be of interest, in particular the case of mass deposition in the Cahn-Hilliard equation, where
\begin{equation}\label{e:ch2}
\frac{d u}{dt} = -\Delta (\Delta u + \mu u-u^3)+c_x m \delta(x-c_x t),
\end{equation}
where $\delta$ is the Dirac delta distribution and $m$ is the amount of mass deposited at the location $x=c_x t$. Boundary conditions $u=-1$ at $x\to\infty$ and deposition of mass $\sim 1$ ramps the system into the spinodal unstable state leading to phase separation in the wake.

\subsection{CDIMA reaction diffusion system}

\begin{figure}[ht!]
\centering
\includegraphics[trim = 0.0cm 0.0cm 0.0cm 0.0cm,clip,width=0.28\textwidth]{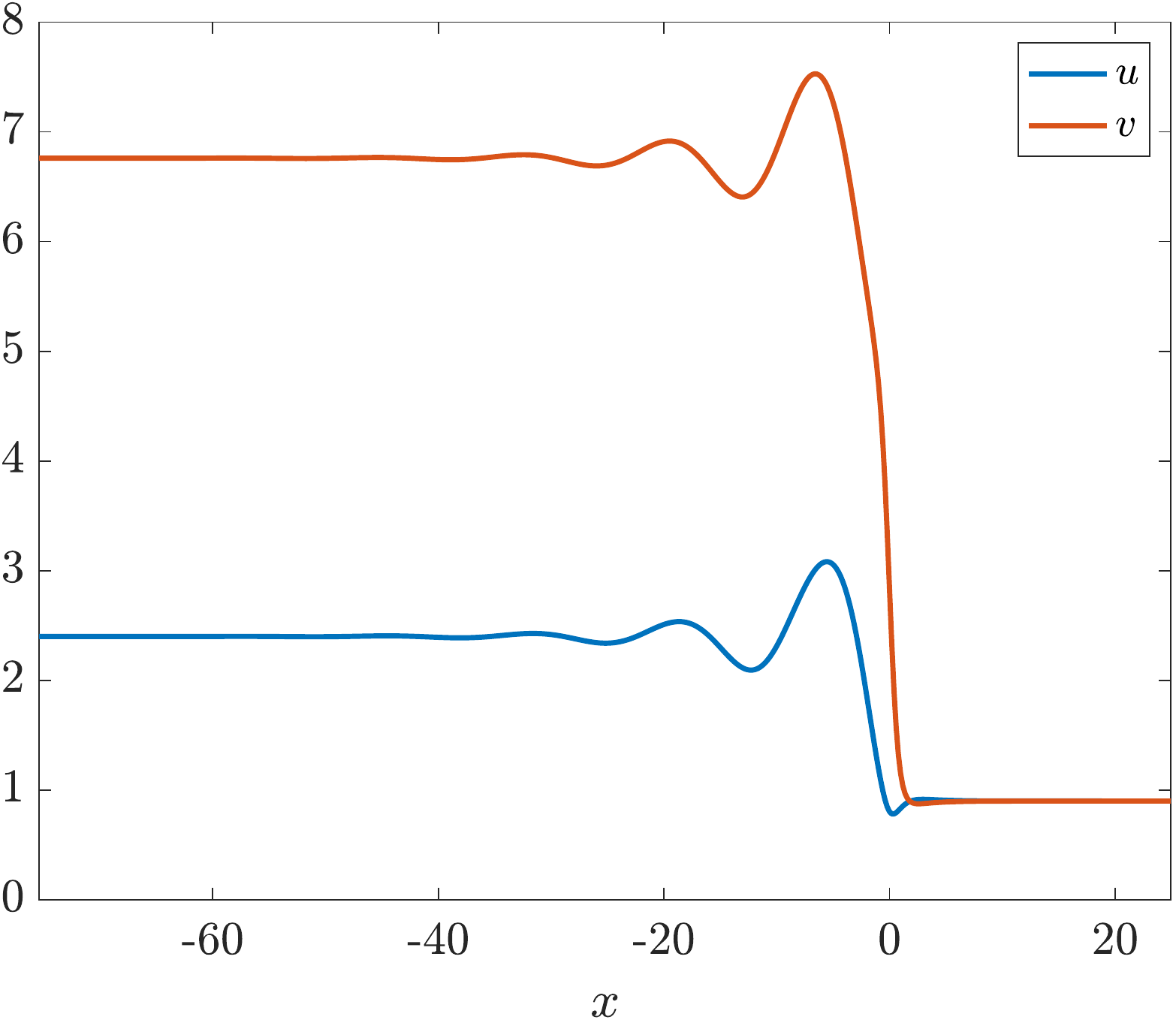}\hspace{-0.05in}
\includegraphics[trim = 0.6cm 0.25cm 0.5cm 0.5cm,clip,width=0.42\textwidth]{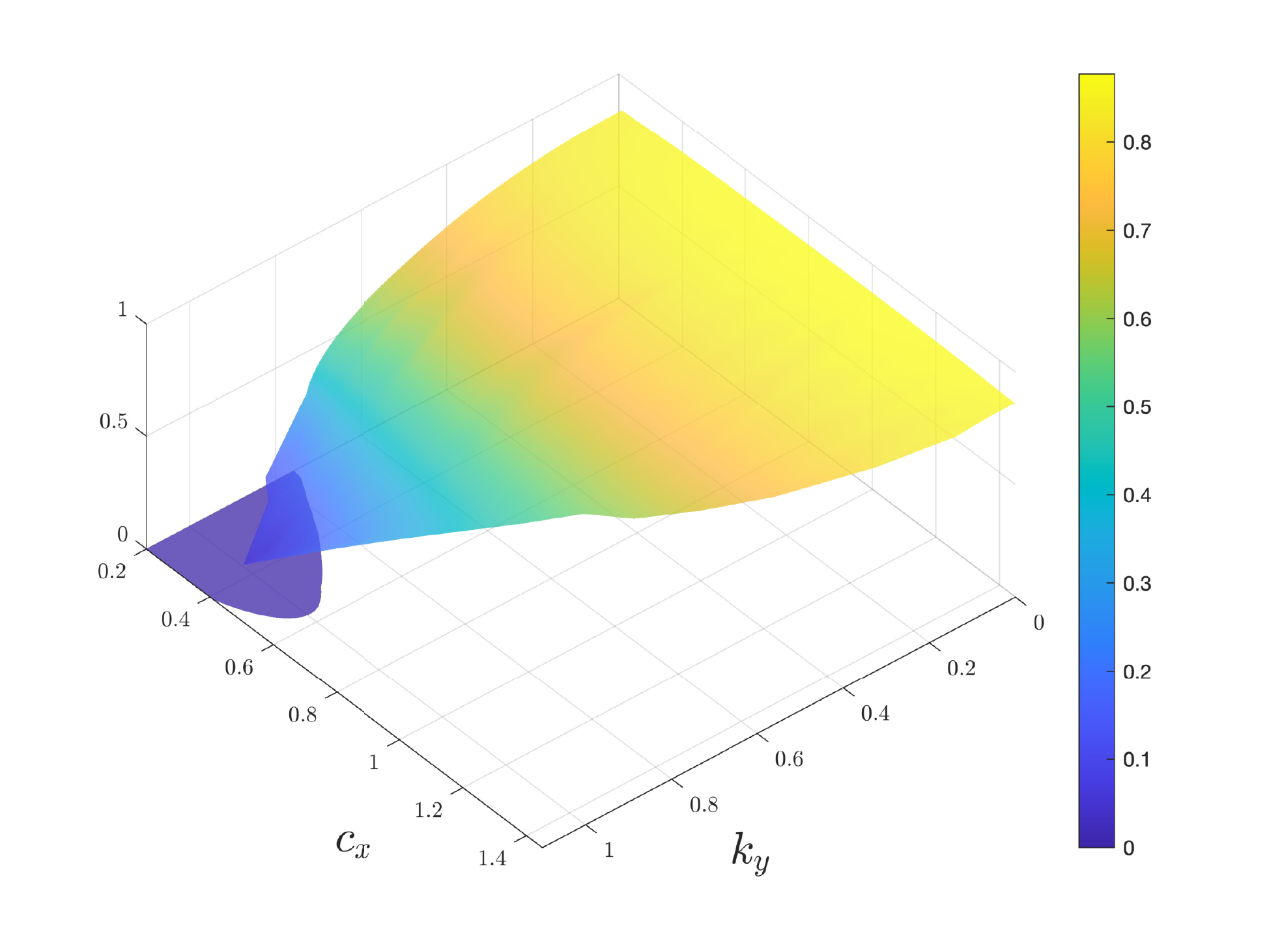}\hspace{-0.2in}
\includegraphics[trim = 0.0cm 0.0cm 0.5cm 0.5cm,clip,width=0.32\textwidth]{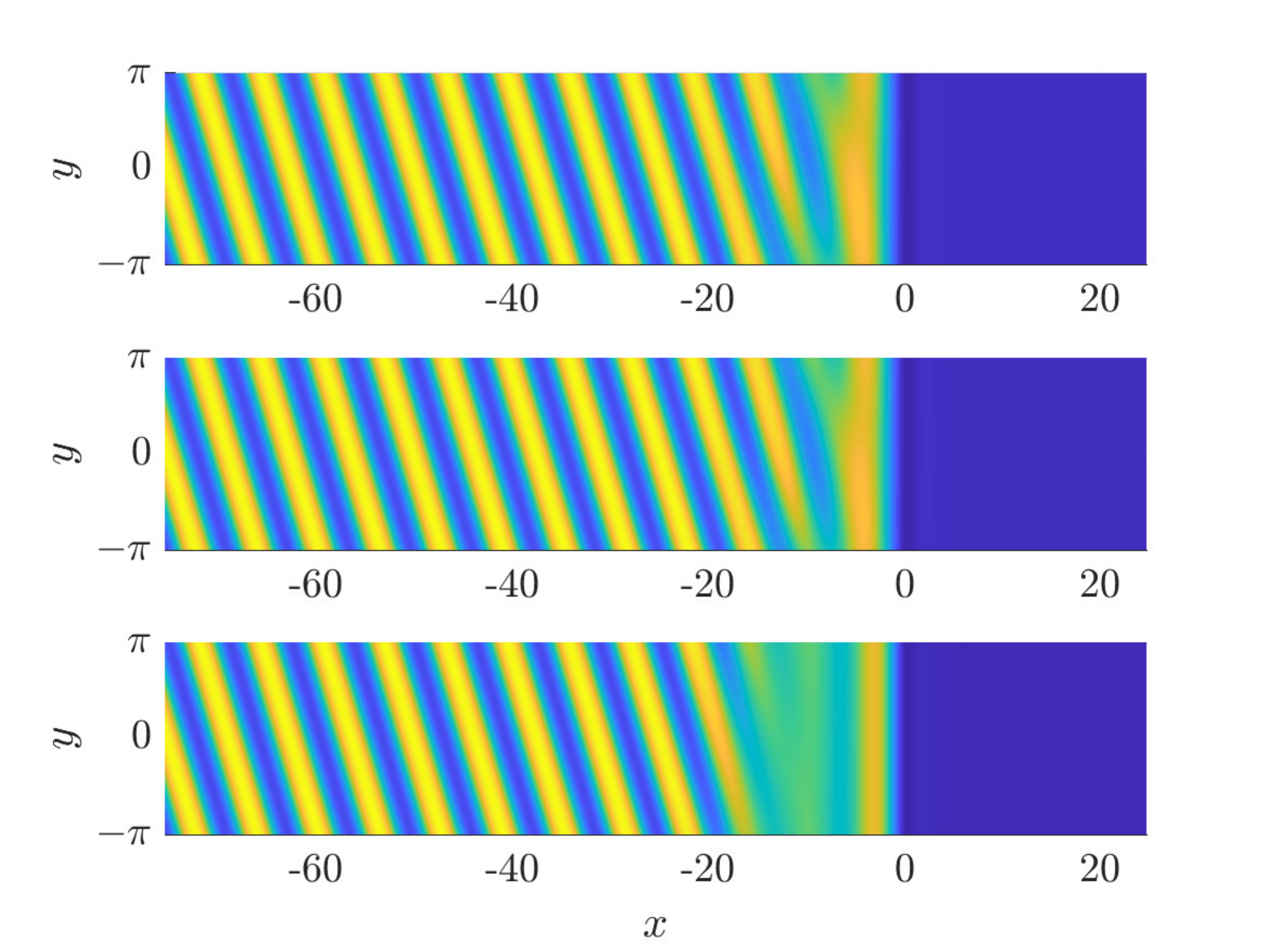}\hspace{-0.15in}
\caption{Left: Plot of front solution connecting stable and unstable equilibria $(u_+,v_+), (u_-,v_-)$ respectively as $x$ decreases across the quenching interface at $x = 0$;  Parameters $a = 12$, $b = 0.32$, $\sigma = 50$, $d = 1$, $\phi = 1.5$;  Center: Numerical exploration of the $k_x$-moduli space for system \eqref{e:cdima_mod}; Right: $u$ solution profiles for $c_x = 1$ fixed, and $(k_y,k_x) \approx (0.003,0.8778),(0.305,0.821),(0.670,0.574)$ from top to bottom.  }\label{f:rd-mod}
\end{figure}

To conclude this section, we show some computational results for the moduli surface in the two-component Lengyel-Epstein reaction-diffusion equation
\begin{align}
\frac{d u}{dt} =& \Delta u + a - u - \frac{4uv}{1+u^2} - W,\\
\sigma^{-1}\frac{dv}{dt} =& d \Delta v + b(u - \frac{uv}{1+u^2} + W).\label{e:le-rd}
\end{align}
which models the CDIMA light-sensing chemical reaction diffusion system \cite{lengyel1992chemical},  a key experimental system in the study of Turing patterns. Experimentally, one observes the formation of spatial patterns in a gel substrate fed by a continuously stirred reaction vessel. The system allows for control of patterns as high-intensity light suppresses the formation of patterns. The formulation \eqref{e:le-rd} is dimensionless, with $u$ and $v$ representing the activator and inhibitor ion concentrations, $a,b,d,$ and $\sigma$ are dimensionless parameters, and the constant $W$ represents the effect of light on the reaction: $W >0$ represents  illuminated dynamics and  $W = 0  $ absence of illumination. Recent works \cite{miguez2006effect,konow2019turing,dolnik2022effect} have studied the effect of moving masks experimentally, blocking illumination and hence dynamically exciting patterns as they move across the domain. Various quenching geometries were explored, including one-dimensional, two-dimensional directional, as well as radial quenches. 
%
A directional quench for this equation in two dimensions is represented by the function
$$
W(x, t) = \begin{cases}
\phi,\qquad &x>c_x t\\
0,\qquad &x\leq c_xt. 
\end{cases}
$$
The system with $W\equiv \phi$ possesses a stable equilibrium $(u_+,v_+)$, while $W\equiv0$ has a Turing-unstable homogeneous equilibrium $(u_-,v_-)$ and in fact a family of stable periodic stripe solutions, $(u_p,v_p)$ for a range of wavenumbers.  A standard traveling wave analysis shows that there is a family of traveling front solutions $(u_*,v_*)(x)$ connecting $(u_-,v_-)$ to $(u_+,v_+)$ as $x$ increases, for a range of positive speeds. See Figure \ref{f:rd-mod} left for a numerical depiction of this front.

\begin{figure}[ht!]
\centering
\includegraphics[trim = 0.0cm 0.0cm 0.0cm 0.0cm,clip,width=0.25\textwidth]{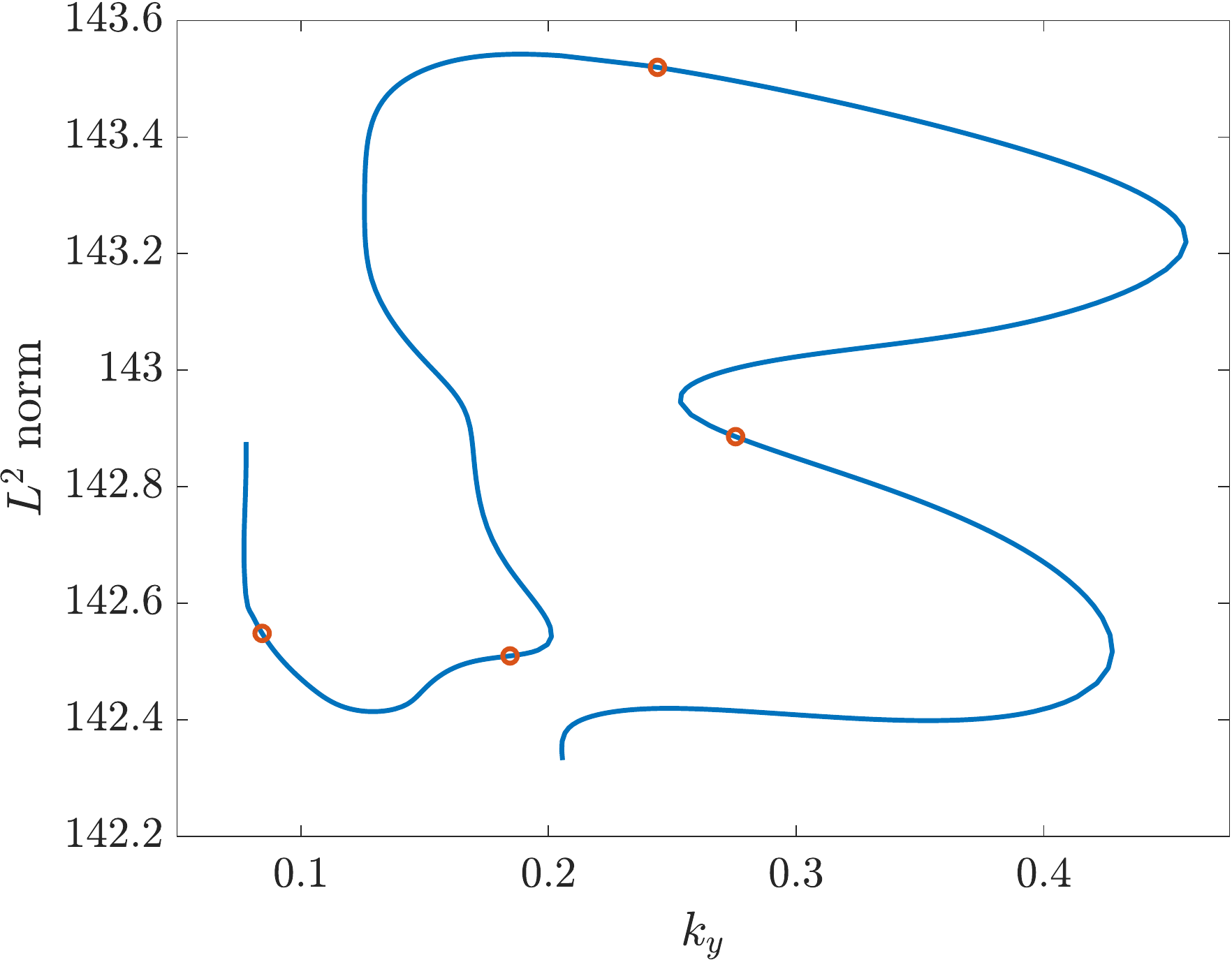}\hspace{-0.05in}
\includegraphics[trim = 0.0cm 0.0cm 0.5cm 0.0cm,clip,width=0.25\textwidth]{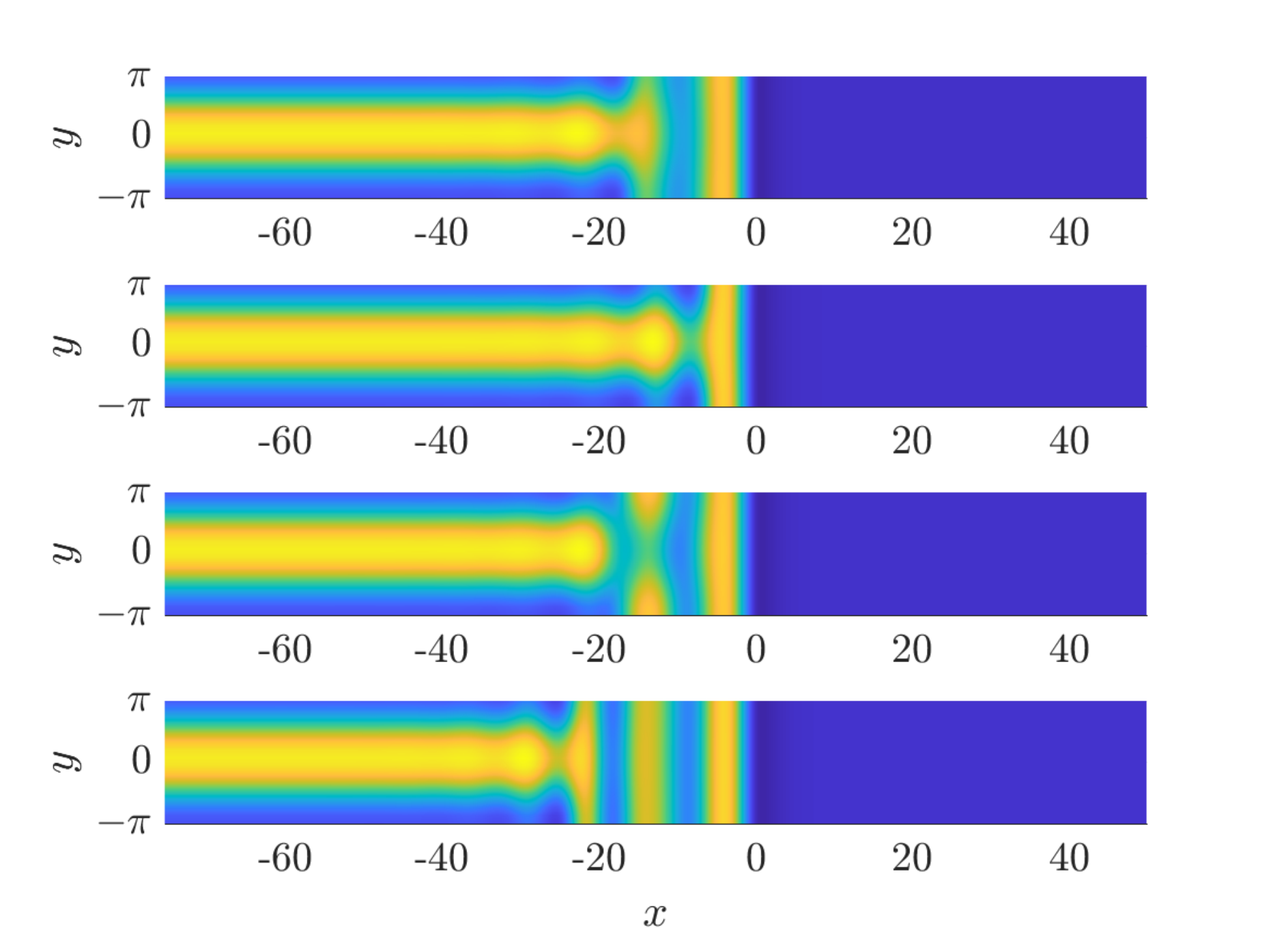}
\hspace{-0.15in}
\includegraphics[trim = 0.0cm 0.0cm 0.0cm 0.0cm,clip,width=0.25\textwidth]{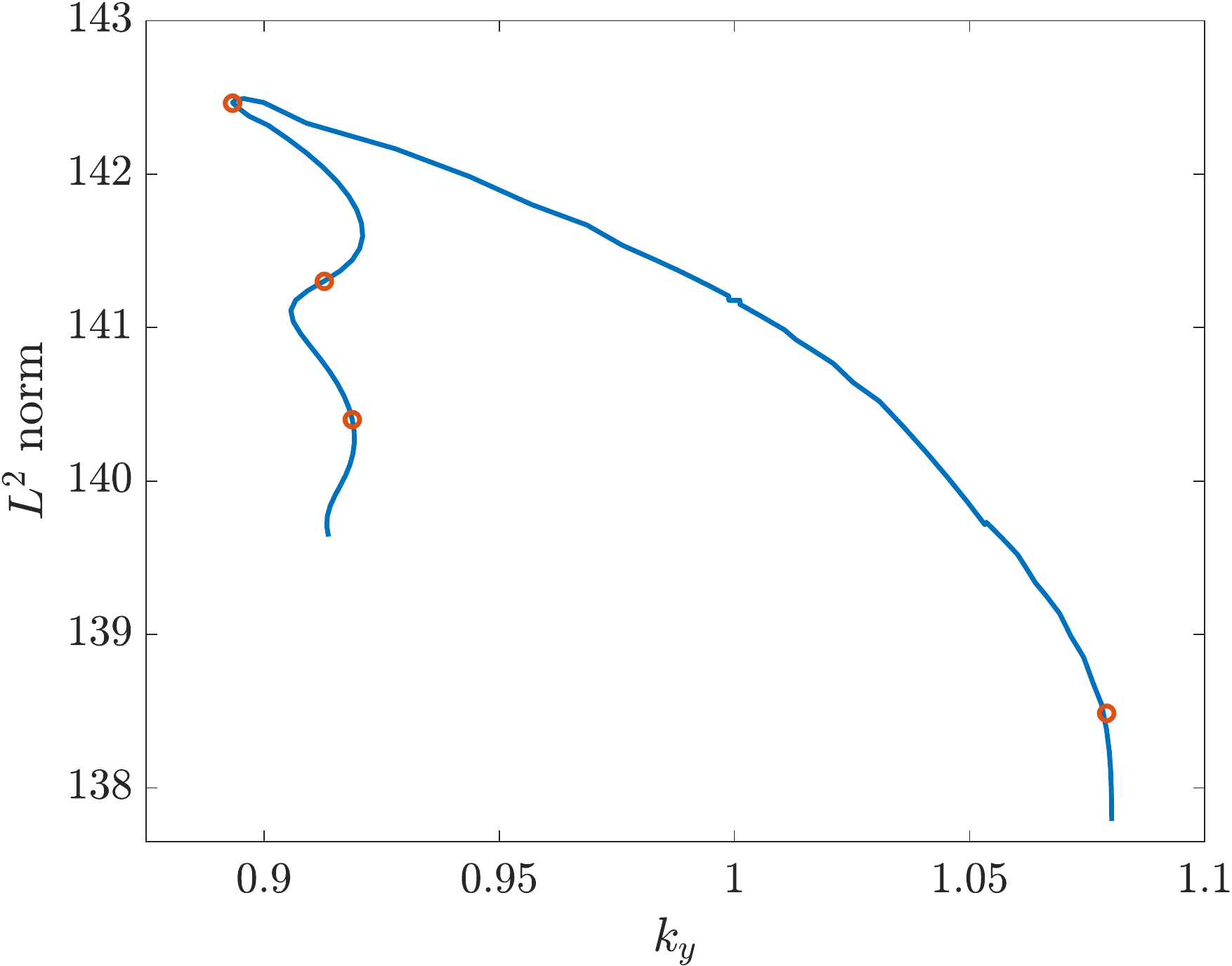}\hspace{-0.05in}
\includegraphics[trim = 0.0cm 0.0cm 0.5cm 0.0cm,clip,width=0.25\textwidth]{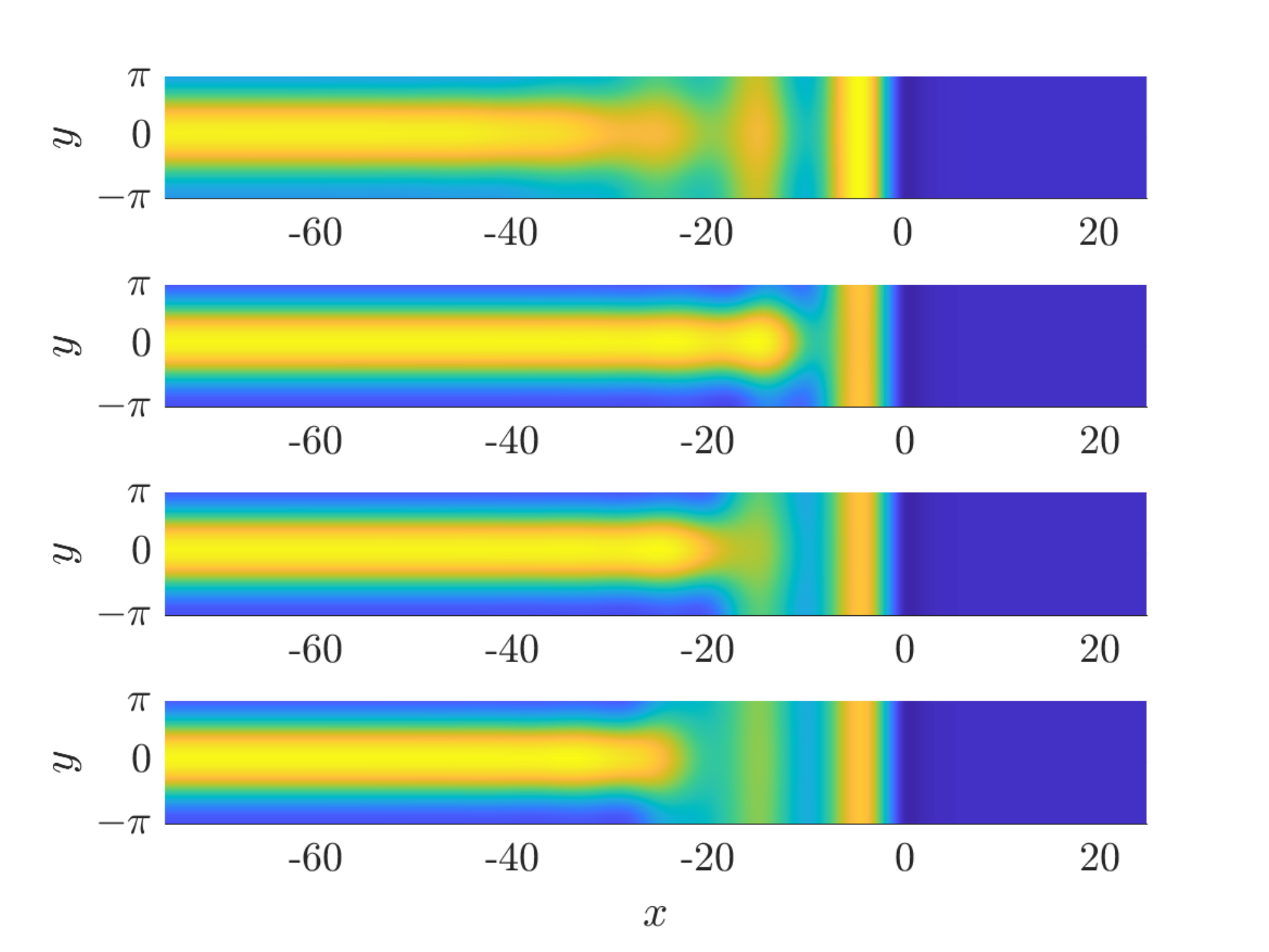}\hspace{-0.2in}

\caption{ Parameters $a = 12 , b = 0.32, \sigma = 50,d = 1 ,\phi = 1.5$; 
Left two plots: continuation of perpendicular stripes of the moduli space for \eqref{e:cdima_mod} with $k_x = 0$ and $k_y = 0.881$ fixed, plotted against the $L^2$-norm of the numerical solution $u(x,y)$; $u$-profiles for select points starting from the bottom left of the $L^2$ norm. 
Right two plots: similar but continuing in $k_y$ with $c_x = 0.457$ fixed. }\label{f:pd-prof}
\end{figure}

Looking for modulated traveling wave solutions $(u,v)(x,y,t) = (u,v)(x - c_x t, k_y y - \omega t)$ as above, we seek to numerically continue solutions of the following traveling-wave system 
\begin{align}
0&= u_{\tl x \tl x}+k_y^2u_{\tl y\tl y} + c_x\left(u_{\tl x} + k_xu_{\tl y}\right) +  a - u - \frac{4uv}{1+u^2} - W(\tl x),\notag\\
0&=  d\left(v_{\tl x \tl x}+k_y^2v_{\tl y\tl y} \right) + c_x\left( v_{\tl x} + k_x\sigma^{-1}v_{\tl y}\right) +  b\left(u - \frac{uv}{1+u^2} + W(\tl x)\right), \notag\\
0&=(u,v)(\tl x,\tl y+2\pi) = (u,v)(\tl x,\tl y),\notag\\
0&= \lim_{\tl x\rightarrow-\infty} |(u,v)(\tl x,\cdot) - (u_p,v_p)(\tl x,\cdot)|,\qquad 0 = \lim_{\tl x\rightarrow+\infty} \left((u,v)(\tl x,\cdot) - (u_+,v_+)\right).\label{e:cdima_mod}
\end{align}
Mimicking the numerical approach for the Swift-Hohenberg equation, we continued front solutions connecting $(u_p,v_p)$ to $(u_+,v_+)$ to determine the dependence of $k_x$ on parameters $k_y$ and $c_x$.  For more detail see Appendix \ref{s:num}.
 
Figure \ref{f:rd-mod} depicts a rough exploration of the moduli space $\mc{M}$ for a fixed set of parameters as well as $u$-solution profiles for a set of points on $\mc{M}$ all with $c_x = 1$. We remark that the existence of one parallel stripe at the leading-edge of the patterned front solution comes from the unstable asymptotically constant front between $(u_+,v_+)$ and $(u_-,v_-)$ on which the stripes are built. We find in the continuation of both parallel and oblique stripes that, as parameters increase towards the boundary of $\mc{M}$, stripes recede from the quench interface.
 
Figure \ref{f:pd-prof} gives slices through a few parts of the perpendicular stripe region of $\mc{M}$ with corresponding solution profiles. For perpendicular stripes we see the pattern detatches from the quench for large $k_y$. Continuing perpendicular stripes downward in $k_y$, we find that the perpendicular stripe front undergoes a series of fold bifurcations for which the stripe interface loses its spot-like end-cap and passes through consecutively smaller oscillations in the base front $(u_*,v_*)$. 

We expect that many of the above heuristics and selection mechanisms discussed for the Swift-Hohenberg equation above could be used to characterize various boundaries and regions of this equation. In particular, preliminary experiments indicate pattern-forming fronts in the homogeneous system with $W\equiv0$ are pulled so that the linear spreading speed of patterns invading the homogeneous state $(u_-,v_-)$ should give the upper boundary of $\mc{M}$ in $(k_y,c_x)$.



\section{Discussion and outlook}\label{s:disc}

Pattern formation in the wake of directional quenching yields a wealth of novel physical behaviors, is relevant in numerous application areas, and is a fruitful mathematical playground to develop deeper understanding (at both the rigorous and heuristic level) of how patterns interact with spatial heterogeneity, growth, and geometry.  Even in a standard pattern-forming scalar partial differential equation, such as the supercritical Swift-Hohenberg equation highlighted above, quenches can create a wealth of spatio-temporal behaviors.  As highlighted throughout this review there are many areas which are not yet fully understood, even at the heuristic level. These include the interaction of perpendicular and oblique striped fronts, the limit of slowly-grown, and weakly oblique stripes,  stability and instability of quenched patterns in various growth regimes, as well as the dynamics of defects and modulations of quenched fronts. Beyond heuristics, rigorous existence and selection results exist only for a few growth regimes. 

\paragraph{Moduli spaces of quenched patterns.}
To organize and encode the pattern-forming dynamics in a given quenched system, we introduced the \emph{moduli space} as an effective means to organize phenomena, easily predict quantitative and qualitative changes, build towards more complex phenomena and growth mechanisms, but also to catalog universal features in growth processes. 


\paragraph{Quenched spots and more frequencies.}
The cross-roll and wrinkling instabilities already introduced dynamics not described in the moduli space. Those dynamics are not ''simplest", but possess an additional frequency compared to the simple quenched fronts that we focused on here. As solutions to the PDE, they depend on 3 independent variables rather than 2 for fronts pertaining to points in the moduli space. This complexity is intrinsic when "growing" crystalline phases of higher complexity. As a next step, one may wish to consider lattices of spots, for instance the common hexagonal arrangement. In the simplest case of growing such hexagonal patterns in the direction of one of the symmetry axes, one already expects dependence of solutions in a periodic fashion on both time and a periodic variable across the quenching interface. Some preliminary analysis in the case of stationary quenches was initiated in  \cite{weinburd2019patterns} where the Swift-Hohenberg equation with nonlinearity $f(u;\tilde x) = \rho(\tilde x) u + \nu u^2 - u^3,\quad \nu\in \R$ is studied in a regime favoring hexagonal patterns. Rigorous existence is obtained in some parameter regimes. In a slowly growing regime, one observes periodic stretch-slip dynamics similar to those discussed in Section \ref{sss:slow-para} as rows of spots are added behind the quenching line. Similar to the wrinkling transition, where kink-type defects are shed as the lateral aspect ratio is changed, we expect similar transitions for hexagons, as lattice orientations and parameters change. Even understanding simple singularities in the associated moduli space, which now features two wave vectors and the speed, should give valuable insight into pinning and depinning effects. We suspect that many of the tools described here, such as numerical continuation based on farfield-core decompositions, amplitude and phase modulation approximation, and heteroclinic gluing methods will be valuable tools in such an endeavor. 


  \begin{Acknowledgment}
  The authors were partially supported by the National Science Foundation through grants NSF-DMS-2006887 (RG), NSF DMS-1907391 and DMS-2205663 (AS). 
  \end{Acknowledgment}

\appendix

\section{Fredholm indices, group velocities, and generic points on $\mc{M}$}\label{ss:fred} 

We outline how Fredholm properties of $\mb{L}$, the linearization about a solution $u_*(x,y;k_x,k_y,c)$ of \eqref{e:mtw}, can be understood. Similar arguments were used in \cite{gs3,avery2019growing,lloydscheel,morrissey,ssmorse} and we recall the main ideas here for convenience and to set the stage for our description of numerical algorithms. Recall that a linear operator between Banach spaces is \emph{Fredholm} if it has closed range, and finite-dimensional kernel and co-kernel. The \emph{Fredholm index} of $\mc{L}$ is then defined as $\mathrm{ind} \,\mc{L} = \dim \ker \mc{L} - \dim \ker \mc{L}$.  First we note that $\mathbb{L}$ is closed and densely defined, but it is not Fredholm in $L^2$. To see this last fact, note that $\mathbb{L} \partial_y u^*=0$ due to the $y$-translation invariance of \eqref{e:mtw} but the bounded function $\partial_y u^*$ does not converge to zero as $x\rightarrow-\infty$ and hence is not in $L^2$. A Weyl sequence construction \cite[\S3]{kapitula13} readily shows then that the range is not closed.

Fredholm properties can be regained by posing the operator on exponentially weighted function spaces,
\begin{align}
 L^2_{\eta,<}&(\R\times \mb{T}):=\{w\in L^2(\R\times \mb{T})\,:\, \|w\|_{L^2_{\eta,<}}<\infty \},\quad
 \|w\|_{L^2_{\eta,<}}^2:= \int_{\R\times\mb{T}} |\re^{-\eta\xi} w(x,y)|^2dx dy.\label{e:l2eta}
\end{align}
or
\begin{equation}
L^2_\eta(\R\times \mb{T}):=\{w\in L^2(\R\times \mb{T})\,:\, \|w\|_{L^2_\eta}<\infty \},\quad
 \|w\|_{L^2_\eta}^2:= \int_{\R\times\mb{T}} |\re^{\eta|\xi|} w(x,y)|^2dx dy. \label{e:l2etaf}
\end{equation}
Next, a compactness argument can be used \cite{ssmorse} to show that $\mb{L}$ is Fredholm if and only if the  linearizations at the asymptotic states at $x\rightarrow\pm\infty$,
$$
\mb{L}_-:= -(1+\p_x^2 + k_y^2 \p_y^2)^2  + \mu  - 3 u_*^2  + c_x(\p_x - k_x\p_y) ,\quad
\mb{L}_+:= -(1+\p_x^2 + k_y^2 \p_y^2)^2  - \mu   + c_x(\p_x - k_x\p_y) ,
$$  
are invertible, possibly after conjugating with the exponential weight function. 
Since $\mb{L}_+$ has constant coefficients, its spectrum is purely continuous spectrum, 
$$\mathrm{spec}_{L^2}(\mb{L}_+):= \{\lambda\in \C\,:\, \lambda:= -(1-\ell_x^2 - k_y^2 \ell_y^2)^2 - \mu + \ri c_x(\ell_x + k_x \ell_y),\, \ell_x\in \R, \ell_y\in \Z\}.$$
Since $\mu >0$ we find that the spectrum is contained in the region $\{\Re \lambda \leq -\mu\}$ which implies that $\mb{L}_+$ is invertible in $L^2_\eta$ for any $\eta\sim0$.

To study the Fredholm properties of $\mb{L}_-$, it helpful to first consider the stability of a pure stripe solution $u_p(k x;k)$, in the original homogeneous equation \eqref{e:sh0}, posed in the unbounded plane $(x,y)\in \R^2$. That is consider the linear operator
$$
\mb{L}_p:= -(1+\p_x^2 + \partial_y^2)^2  + \mu  - 3 u_p(k \,\cdot)^2,
$$ 
As this operator has periodic coefficients, one can use a Bloch wave decomposition \cite{mielke} of the perturbation $w(x,y) = \re^{\nu_x x}\re^{\nu_y y} W(x)$ with $(\nu_x,\nu_y)\in \ri[0,2\pi/k)\times \ri\R$, and $W$ a $2\pi/k$-periodic function, to study the spectrum.
 Inserting this decomposition into the eigenvalue equation $\mb{L}_p w = \lambda w$ one obtains a family of eigenvalue problems posed on a compact domain.
$$
\lambda W = \widehat{\mb{L}}_p(\nu_x,\nu_y)W:= -(1+(\p_x+\nu_x)^2 + \nu_y^2)^2W  + \mu W  - 3 u_p(k_x \, \cdot\,)^2W,\quad x\in (0,2\pi/k_x),
$$
so that the spectrum of $\mc{L}_p$ in $L^2(\R^2)$ is decomposed as
$$
\mathrm{spec}_{L^2} \mb{L}_p = \bigcup_{\nu_x\in \ri[0,2\pi/k),\nu_y\in \ri\R} \mathrm{spec}_{L^2}\widehat{\mb{L}}_p(\nu_x,\nu_y).
$$
For wavenumbers $k\in (k_\mathrm{zz},k_\mathrm{eck})$ and $\nu_x,\nu_y\sim0$, $\mathrm{spec}_{L^2} \widehat{\mc{L}}_p(\nu_x,\nu_y)$ is contained in the closed left-half plane and bounded away from the imaginary axis except for a simple eigenvalue curve $\lambda(\nu_x,\nu_y)$ which touches the imaginary axis in a quadratic tangency
$$
\lambda(\nu_x,\nu_y) = d_{||} \nu_x^2 + d_\perp \nu_y^2 + \mc{O}(|\nu_x|^4+|\nu_y|^4), \qquad d_{||},d_\perp >0;
$$
see for instance \cite{mielke}. 
Conjugating with the weight, one finds the spectrum in $L_{\eta,>}^2$ with $0<\eta\ll1$ that the small eigenvalue takes the form
$$
\lambda(\nu_x,\nu_y) = d_{||} (\eta+\nu_x)^2 + d_\perp (\eta+\nu_y)^2 + \mc{O}(|\nu_x|^4+|\nu_y|^4), \qquad d_{||},d_\perp >0,
$$
shifting the curve to the right for $\eta>0$. 

Now let us return to our asymptotic operator, $\mb{L}_-$, posed on $L^2_\eta(\R\times \mb{T})$. Because of the periodic domain in the $y$-direction, we use a Floquet-Bloch decomposition $w(x,y) = \re^{\ri \ell_y y}\re^{\nu_x x} W(k_x x + y)$ with $\ell_y\in\Z, \nu_x\in \ri[0,2\pi/k_x)$ and $W(\cdot) = W(\cdot+2\pi)$. One then obtains the family of operators
$$
\widehat{\mb{L}}_-(\nu_x,\ell_y):= -(1+(\nu_x+k_x\partial_\theta)^2 + k_y^2(\p_\theta + \ri\ell_y)^2)^2 + \mu-3u_p(\cdot)^2+c_x \nu_x - k_x c_x \ri \ell_y,\qquad \theta\in [0,2\pi). 
$$
Hence, the eigenvalues $\tl\lambda$ of $\widehat{\mb{L}}_p(\nu_x,\ri\ell_y)$ give eigenvalues $\lambda = \tl\lambda - c_x(\nu_x+k_x\ri\ell_y)$ for $\widehat{\mb{L}}_- (\nu_x,\ell_y)$. Note for $\eta = 0$, this comoving and co-rotating frame only shifts the imaginary part of $\tl \lambda(\nu_x,\ell_y)$. Furthermore, for $\eta>0$ the neutral eigenvalue curve is shifted away from the origin.

These curves of spectra denote the values of $\lambda$ for which the operator $\mb{L}_- - \lambda$ is not invertible and bound the regions where the operator is Fredholm with a constant index.  In order to compute the Fredholm index of $\mb{L}$, one uses homotopy invariance of the index when adding the spectral parameter $\lambda$. For the values to the right of the right-most curve $\lambda(\nu_x,\nu_y)$ one readily calculates the Fredholm index of $\mb{L}$ to be 0, since  the linearization does not have spectrum near $\lambda=+\infty$.  Next, one can use the group velocity of the striped pattern to calculate the change in index as $\lambda$ moves from the right to left across $\lambda(\nu_x,\ell_y)$. In general, the relative sign of the group velocity for a wave with dispersion relation $\omega(k)$ counts the change in Fredholm index as one moves from right to left across the Fredholm boundary \cite{ssmorse,fiedlerscheel}. In our case, since striped patterns $u_p$ are stationary in the stationary frame, they have trivial dispersion relation $\omega(k_x) = -c_x k_x$ in the frame moving with the quench at speed $c_x$ so that the phase velocity, $c_p = \omega/k_x$, and group velocity, $c_g = \omega'(k_x)$, are both equal to $-c_x$, pointing away from the quenching interface. In other words the heterogeneity acts like a source defect shedding waves \cite{sandstede2004defects}. Thus, for generic points along $\mc{M}$, we find that $\mb{L} - \lambda$ has Fredholm index -1 to the left of the curve $\lambda(\nu_x,\ell_y)$ near the origin. Finally, we may conclude that $\mb{L}$ is index 0 for $\eta<0$ and index -1 for $\eta>0$. 


The negative Fredholm index suggests that, generically  we expect $\mc{M}$ to be determined by a locally two-dimensional graph with $k_x$ in terms $(c_x,k_y)$. Note that working in a space of exponentially localized functions precludes the presence of a kernel induced by the $y$-derivative, which yields bounded, non-localized functions near $-\infty$.

\section{Farfield-core numerics}\label{s:num}

The numerical continuation results given above use a far-field core decomposition approach to represent and approximate heteroclinic-type front solutions of the unbounded domain problem in a bounded computational domain. The approach was developed in  \cite{morrissey} for striped patterns in a general one-dimensional pattern-forming system on both semi-bounded and bounded domains, and in \cite{lloydscheel} to study grain-boundaries. In the present context, this approach was adapted in \cite{avery2019growing,chen2021strain} to  the quenched Swift-Hohenberg equation \eqref{e:mtw} to obtain the results described in Sections \ref{s:edge} and \ref{ss:an-sh} above. We briefly outline the computational approach here in the context of the directionally quenched Swift-Hohenberg equation \eqref{e:mtw}.

We sets $x = k_x \tilde x$ in \eqref{e:mtw} and, following the functional analytic approach outlined in Section \ref{ss:obl} above, decompose front solutions into
$$
u(x,y) = w(x,y) + \chi(x) u_p(x + y;k), \qquad k = \sqrt{k_x^2+k_y^2},
$$ 
where $w(x,y)$ is the \emph{core} perturbation which matches the stable stationary state $u = 0$ ahead of the quench with the periodic state, $u_p$, which solves the stripe equation \eqref{e:shroll}, and $\chi(x)$ a smooth monotonic step function with $\chi(x) = 1$ for $x<-d$, $\mathrm{supp} \chi \subset (-\infty,-d-1)$, and $\chi'$ exponentially localized. Here $w$ corrects and connects the asymptotic state $u_p$ at $x = -\infty$ with the trivial equilibrium state at $x = +\infty$.

We let $$\mc{L}(k_x,k_y,c_x):= -(1+k_x^2\p_x^2+k_y^2\p_y^2)^2  + c_xk_x(\p_x -\p_y),\qquad \mc{N}(u):=\rho(x) u  -u^3,$$ and use the fact that $\chi\left( \mc{L}(k_x,k_y,c_x) u_p(\cdot;k) + \mc{N}(u_p)\right) \equiv 0$ for suitable wavenumbers $k$ to obtain a nonlinear equation for the core variable with $k_x,k_y$ controlling the far-field
\begin{align}
0 &= \mc{F}(w,k_x;k_y,c_x) := \mc{L}(k_x,k_y,c_x) w + \lp[ \mc{L},\chi\rp] u_p + N(w+\chi u_p(k)) - \chi N(u_p(k)),\\ 
&\quad\quad \lp[ \mc{L},\chi\rp] v:= \mc{L}(\chi v) - \chi\mc{L} v.\notag
\end{align} 
Note that the linear   $\lp[ \mc{L},\chi\rp] u_p$  and the nonlinear commutator $ N(w+\chi u_p(k)) - \chi N(u_p(k))$ are exponentially localized. The linearization in this space is Fredholm of index -1 as discussed in Section \ref{ss:fred}, so that the additional free variable $k_x$ gives a Fredholm 0 linearization on the unbounded domain. 

We then truncate this problem onto a bounded computational domain $(x,y)\in [-3L_x/2, L_x/2]\times [0,2\pi]$, with $w$ periodic in both variables $x$ and $y$. Alternative approaches with $w$ satisfying Dirichlet boundary conditions in $x$ are also sufficient, but periodic boundary conditions are used to take advantage of spectral discretizations and the efficiency of the Fast Fourier Transform. 
To enforce exponential localization of the solution, we add a phase condition that prohibits neutral growth according to the derivative of the wave train at $-\infty$, $0 = \int_{I\times [0,2\pi]} w\cdot u'_p dxdy$ where $I\subset [-3L_x/2,-d]$ is an interval of length $2\pi$.

In sum, we obtain the truncated problem
\begin{alignat}{2}
0&= \mc{F}(w,k_x;k_y,c_x),\quad &&(x,y)\in (-3L_x/2, L_x/2)\times (0,2\pi)\label{e:f_tr1}\\
0&= \partial_x^j(w(-3L_x/2,y) - w(L_x/2,y)),\quad &&y\in [0,2\pi],\ j=0,1,2,3\label{e:f_tr2}\\
0&= \partial_x^j(w(x,0) - w(x,2\pi)),\quad &&x\in [-3L_x/2, L_x/2],\ j=0,1,2,3\label{e:f_tr3}\\
0&=  \int_{I\times [0,2\pi]} w\cdot u'_p dxdy\quad &&\label{e:f_tr4}\\
0&=-(1+k^2\partial_\theta^2)^2 u_p + \mu u_p - u_p^3,\quad &&\theta\in[0,2\pi], \quad 0 = u_p(0) - u_p(2\pi)\label{e:f_tr5}
\end{alignat}
Of course, the linearization in any bounded domain would be Fredholm of index zero without farfield-core decomposition. However, the condition number of the linearization grows rapidly as $L_x$ increases. The farfield-core decomposition described here avoids this issue and we find that linear solvers work well roughly independent of the domain size \cite{lloydscheel}. 

Numerically, we discretize both the core and far-field equations \eqref{e:f_tr1} and \eqref{e:f_tr4} using a Fourier-Galerkin approximation
$$
w(x,y)\sim \sum_{\ell_x = -N_x/2,\ell_y = -N_y/2}^{N_x/2, N_y/2} \widehat{w}_{\ell_x,\ell_y}
\re^{\ri(\ell_x x + \ell_y y)}, $$
and apply of the linear operator $\mc{L}$ as well as the linearization $\p_w \mathcal{F}$  as pointwise multiplication in spectral space using the fast Fourier transform:
$$
\widehat{\mathcal{L}}: \left\{\widehat{w}_{\ell_x,\ell_y}\right\}\mapsto \left\{\lp[-(1-k_x^2\ell_x^2 - k_y^2\ell_y^2)^2 + c_xk_x\ri(\ell_x - \ell_y) \rp] \widehat{w}_{\ell_x,\ell_y}\right\}.
$$ 
 We use pseudo-arclength continuation with a Newton-GMRES nonlinear solver to continue solutions in either $k_y$ or $c_x$.

We find as expected that bounded domain solutions converge exponentially  to front solutions on the unbounded domain as $L_x\rightarrow\infty$ and work with moderate $L_x\sim 200\ldots 800$. To fully exploit Fredholm well-posedness in the unbounded domain, we conjugate the nonlinear equation \eqref{e:f_tr1} with an exponential weight $\tilde{\mathcal{F}}(w,\cdot) := h(x)\cdot \mathcal{F}(h(x)^{-1} w,\cdot)$ where $h(x) = \re^{\eta (x-L_x/2)}$ before using the iterative linear solver. We also improve convergence of the GMRES solver for each Newton step by preconditioning the linear solves with the constant coefficient operator $\mc{P}:=(\mc{L}(k_x,k_y,c_x) - I)^{-1}$, which acts as a simple Fourier multiplier.


Discretization and domain size of the scaled system were adaptively controlled, expanding the domain when the tails of $w(x,y)$ grew beyond a threshold value, and refining the mesh size in either $x$ or $y$ if the tails of the Fourier space variable $\hat{w}(\ell_x,\ell_y)$ grew above a threshold value. Typical mesh sizes were $N_x = 4096, N_y = 128$ Fourier modes in the $x$ and $y$ directions respectively. Computations were performed on an NVIDIA GV100 GPU using the MATLAB 2021a software package to take advantage of the massive parallelization of the ``fft", and ``gmres" functions as well as pointwise and matrix arithmetic operations.



\bibliography{moduli-review}
\bibliographystyle{abbrv}

\end{document}